\documentclass[twocolumn, preprint, 12pt]{aastex631}

\usepackage{color}

\usepackage{upgreek}
\usepackage{graphicx}
\usepackage{tabularx}
\usepackage{natbib}
\usepackage{rotating}
\graphicspath{ {./images/} }
\usepackage{appendix}
\newcommand{\Hii}{H{\sc ii}}
\newcommand{\Hi}{H{\sc i}}

\newcommand{\h} {^{\mathrm{h}}}
\newcommand{\m} {^{\mathrm{m}}}
\newcommand{\s} {^{\mathrm{s}}}

\newcommand{\kms}	{\mbox{km s}^{-1}}

\newcommand{\K}	{{\rm K}}

\begin{document}
\received{13 July 2024}
\revised{11 October 2024}

\submitjournal{ApJ}


\title{The JWST-NIRCam View of Sagittarius C. I. Massive Star Formation and Protostellar Outflows}

\correspondingauthor{Samuel Crowe}
\email{fec5fg@virginia.edu}
\author[0009-0005-0394-3754]{Samuel Crowe}
\affiliation{Dept. of Astronomy, University of Virginia, Charlottesville, Virginia 22904, USA}

\author[0000-0003-4040-4934]{Rub\'en Fedriani}
\affiliation{Instituto de Astrof\'isica de Andaluc\'ia, CSIC, Glorieta de la Astronom\'ia s/n, E-18008 Granada, Spain}

\author[0000-0002-3389-9142]{Jonathan C. Tan}
\affiliation{Department of Space, Earth \& Environment, Chalmers University of Technology, 412 93 Gothenburg, Sweden}
\affiliation{Dept. of Astronomy, University of Virginia, Charlottesville, Virginia 22904, USA}

\author[0009-0008-6570-9287]{Alva Kinman}
\affiliation{Department of Space, Earth \& Environment, Chalmers University of Technology, 412 93 Gothenburg, Sweden}

\author[0000-0001-7511-0034]{Yichen Zhang}
\affiliation{Department of Astronomy, Shanghai Jiao Tong University, 800 Dongchuan Rd., Minhang, Shanghai 200240, China}

\author[0000-0002-5306-4089]{Morten Andersen}
\affiliation{European Southern Observatory, Karl-Schwarzschild-Strasse 2, D-85748 Garching bei München, Germany}

\author[0009-0004-4390-7937]{Luc\'ia Bravo Ferres}
\affiliation{Instituto de Astrof\'isica de Andaluc\'ia, CSIC, Glorieta de la Astronom\'ia s/n, E-18008 Granada, Spain}

\author[0000-0002-6379-7593]{Francisco Nogueras-Lara}
\affiliation{European Southern Observatory, Karl-Schwarzschild-Strasse 2, D-85748 Garching bei München, Germany}

\author[0000-0001-5404-797X]{Rainer Sch\"odel}
\affiliation{Instituto de Astrof\'isica de Andaluc\'ia, CSIC, Glorieta de la Astronom\'ia s/n, E-18008 Granada, Spain}

\author[0000-0001-8135-6612]{John Bally}
\affiliation{Center for Astrophysics and Space Astronomy, Astrophysical and Planetary Sciences Department, University of Colorado, UCB 389 Boulder, CO 80309, USA}

\author[0000-0001-6431-9633]{Adam Ginsburg}
\affiliation{Department of Astronomy, University of Florida, P.O. Box 112055, Gainesville, FL, USA}

\author[0000-0002-8691-4588]{Yu Cheng}
\affiliation{National Astronomical Observatory of Japan, 2-21-1 Osawa, Mitaka, Tokyo 181-8588, Japan}

\author[0000-0001-8227-2816]{Yao-Lun Yang}
\affiliation{Star and Planet Formation Laboratory, RIKEN Cluster for Pioneering Research, Wako, Saitama 351-0198, Japan}

\author[0000-0002-7612-0469]{Sarah Kendrew}
\affiliation{European Space Agency, Space Telescope Science Institute, Baltimore, MD, USA}

\author[0000-0003-1964-970X]{Chi-Yan Law}
\affiliation{Osservatorio Astrofisico di Arcetri, Largo Enrico Fermi, 5, 50125 Firenze FI, Italy}

\author[0000-0002-4855-1325]{Joseph Armstrong}
\affiliation{Department of Space, Earth \& Environment, Chalmers University of Technology, 412 93 Gothenburg, Sweden}

\author[0000-0002-7402-6487]{Zhi-Yun Li}
\affiliation{Dept. of Astronomy, University of Virginia, Charlottesville, Virginia 22904, USA}



\begin{abstract}
We present James Webb Space Telescope (JWST)-NIRCam observations of the massive star-forming molecular cloud Sagittarius C (Sgr C) in the Central Molecular Zone (CMZ). In conjunction with ancillary mid-IR and far-IR data, we characterize the two most massive protostars in Sgr C via spectral energy distribution (SED) fitting, estimating that they each have current masses of $m_* \sim 20\:M_\odot$ and surrounding envelope masses of $\sim 100\:M_\odot$. We report a census of lower-mass protostars in Sgr C via a search for infrared counterparts to mm continuum dust cores found with ALMA. We identify 88 molecular hydrogen outflow knot candidates originating from outflows from protostars in Sgr C, the first such unambiguous detections in the infrared in the CMZ. About a quarter of these are associated with flows from the two massive protostars in Sgr C; these extend for over 1 pc and are associated with outflows detected in ALMA SiO line data. An additional $\sim 40$ features likely trace shocks in outflows powered by lower-mass protostars throughout the cloud. We report the discovery of a new star-forming region hosting two prominent bow shocks and several other line-emitting features driven by at least two protostars. We infer that one of these is forming a high-mass star given an SED-derived mass of $m_* \sim 9\:M_\odot$ and associated massive ($\sim 90\:M_\odot$) mm core and water maser. Finally, we identify a population of miscellaneous Molecular Hydrogen Objects (MHOs) that do not appear to be associated with protostellar outflows. 

\end{abstract}


\section{Introduction}\label{sec:intro}

Massive stars ($m_*>8\:M_{\odot}$) play a central role in star and planet formation, dispersing the natal clouds of their birth clusters and ionizing the interstellar medium through their radiation and winds. During the course of their stellar evolution, they drive much of the physical and chemical evolution of galaxies. The precise formation mechanism for massive stars, however, remains poorly understood, specifically whether they form in a similar, but scaled-up, manner as low-mass stars, i.e., Core Accretion models \citep[e.g.,][]{mckee03}, or whether they form more chaotically at the center of dense clusters, i.e., Competitive Accretion models \citep[e.g.,][]{bonnell01,wang10,grudic22} \citep[see, e.g.,][for reviews]{tan14, rosen20}. 

One of the most extreme star-forming environments in our Milky Way is the Central Molecular Zone (CMZ), which we define as the region within a Galactocentric radius of $\sim 300\:$pc \citep[see, e.g.,][]{henshaw23}. This is a region of high metallicity \citep[i.e., about $2\times Z_{\odot}$;][]{giveon02,schodel20,schultheis21,nogueras-lara22b}, thermal pressures \citep[$P_{\rm th}\sim10^{-10}\,{\rm erg\:cm}^{-3}$;][]{morris96}, molecular gas temperatures \citep[e.g., $\gtrsim60\:\mathrm{K}$;][]{ginsburg16}, cosmic ray ionization rates \citep[$\zeta \gtrsim 10^{-15}\:{\rm s}^{-1}$;][]{carlson16}, and magnetic field strengths \citep[$\geq 50 {\rm \mu G}$ over large scales and up to $\sim5\,\mathrm{mG}$ in dense molecular clouds;][]{crocker10,pillai15}. In addition, the CMZ region holds a large amount of molecular gas \citep[$>10^7\:M_{\odot}$;][]{ferriere07}, but has a relatively low (by about an order of magnitude) star formation rate compared to empirical scaling relations derived from main disk regions; observational methods yield a current star formation rate of $\sim 0.08\:M_{\odot}\:{\rm yr}^{-1}$ \citep{henshaw23}.
Furthermore, there is an asymmetry of dense gas distribution (and therefore star formation) in the CMZ, with around $2/3$ of the dense gas being located at positive (eastern) Galactic longitudes \citep{yusef-zadeh09,sormani18}.

Sagittarius C (hereafter Sgr C) is an active star-forming region in the CMZ, and therefore one of the key laboratories for testing theories of star formation in this extreme environment \citep{yusef-zadeh84,law04,kendrew13,lu16,lu19a,lu19b,lu22}. It is the most massive and luminous star-forming region in the western (negative longitude) side of the CMZ \citep{kendrew13}, and it has furthermore been suggested to be a connection point to a stream of gas and dust linking the CMZ and Nuclear Stellar Disk to the Galactic bar \citep{molinari11,henshaw23}, potentially explaining the strong 24 $\mathrm{\mu m}$ emission in this region \citep{carey09} as high amounts of warm dust heated by ionizing stars. This also explains the presence of large reserves of gas in this region that are necessary to produce the observed star formation activity.

Despite confusion surrounding the distance to the cloud due to its line of sight velocity, $\sim60 \mathrm{km\:s^{-1}}$, which may imply a foreground distance of $\sim5$ kpc, previous studies have affirmed the cloud's location at the CMZ distance based on its consistency with CMZ gas kinematics \citep{kendrew13,kruijssen15,lu19a}. Therefore, we adopt a distance to Sgr C as the general galactocentric distance measured by \citet{reid19} to be $8.15\pm0.15$ kpc based on water maser parallaxes.
    
Much attention has been paid to the most luminous source in the cloud, G359.44$-$0.102 \citep{kendrew13}, and its immediate surroundings. The source itself is a prominent ``Extended Green Object'' \citep[][]{cyganowski08,chen13}, a class of objects visualized in Spitzer-IRAC images (identified in the $4.5 \mu m$ IRAC2 filter) and associated with massive star formation, methanol maser emission and protostellar outflows. Accordingly, the source is associated with a Class II 6.7 GHz CH$_3$OH maser; additionally, a neighboring source is associated with both a CH$_3$OH maser \citep{caswell10} and a 1665 MHz
OH maser \citep{cotton16}, indicating that the region is forming at least two massive stars \citep{breen13,lu19b}. \citet{lu22} analyzed Atacama Large Millimeter Array (ALMA) Band 6 ($\sim 1.3$ mm) observations of G359.44$-$0.102 to reveal that it is a massive protostar (measuring a stellar mass of $m_*\sim30\:M_{\odot}$) with a Keplerian disk with spiral features, potentially induced by dynamical interactions. 
The ALMA Band 6 molecular line data also indicate the presence of several protostellar outflows throughout the cloud, including a collimated jet proposed to have originated from G359.44$-$0.102 \citep[see Fig. 21 of][]{lu21}. \citet{kendrew13} also discuss the possibility of an outflow from G359.44-0.102 on the basis of detection of faint atomic hydrogen (Br-$\gamma$) and molecular hydrogen emission features in the region using infrared spectroscopy.
    
There has been some discussion if the rest of the Sgr C cloud is quiescent \citep[e.g.,][]{kendrew13} or star-forming \citep[e.g.,][]{lu16}. From analysis of ALMA Band 6 continuum data, \citet{lu20} and \citet{kinman24} have found $>100$ low- to high-mass mm cores distributed throughout Sgr C, indicating that star formation is likely more widespread.

    
In this paper, we present JWST-NIRCam observations of Sgr C, which provide an infrared view of its massive protostars, its lower-mass star-forming activity, and the wider environment surrounding the molecular cloud. Observations and reduction of the data used in this study are presented in \S\ref{sec:observations}. Our results are presented in \S\ref{sec:results}. Discussion of the broader implications of these results is made in \S\ref{sec:discussion}. A summary and conclusion are given in \S\ref{sec:conclusions}.
    
\section{Observations and Data Reduction}\label{sec:observations}
    
\subsection{James Webb Space Telescope}\label{sec:JWST_obs}

Observations with the James Webb Space Telescope (JWST) were taken on the 22$^\mathrm{nd}$ of September 2023 (Program ID 4147, PI: S. Crowe)\footnote{The JWST data presented in this article were obtained from the Mikulski Archive for Space Telescopes (MAST) at the Space Telescope Science Institute. The specific observations analyzed can be accessed via \dataset[doi: 10.17909/p8dv-8593]{https://doi.org/10.17909/p8dv-8593}.}. The  Near Infrared Camera \citep[NIRCam;][]{rieke05} was used with Primary Dither Type \textit{FULLBOX} and Primary Dithers \textit{6TIGHT}; i.e., a total of 6 dither positions. The final Field of View (FOV) using this observing strategy is $\sim2\arcmin\times6\arcmin$.
The filters used were paired in Short Wavelength (SW) and Long Wavelength (LW) filters, as listed in Table \ref{tab:observations}. Each SW filter has a pixel scale of 0.031$\arcsec$/pixel, while each LW filter has a pixel scale of 0.063$\arcsec$/pixel.
The readout pattern for all filter pairs was \textit{SHALLOW2} with Groups/Int 6 and Integrations/Exp 1 resulting in a Total Exposure Time of 1739.357\,s (except for the filter pair F115W \& F360M where Groups/Int 3 and Integrations/Exp 1 were used, resulting in a Total Exposure Time of 773.047\,s). The position angle with respect to the V3 axis (PA\_V3) was $91.42^\circ{}$. See Table\,\ref{tab:observations} for a summary of the observations.
        
        \begin{table*}[ht!]
            \caption{Summary of the JWST/NIRCam observations.}             
            \label{tab:observations}      
            \centering          
            \begin{tabular}{c c c c c c c c}     
            \hline\hline       
            \noalign{\smallskip}
            Filter Pair & Wavelength & Tracing & Readout & Groups & Integrations & Total Exp. Time\\ 
            (SW \& LW) & ($\mu$m) & Emission & Pattern & /Int & /Exp & (s)\\
            \noalign{\smallskip}
            \hline              
            \noalign{\smallskip}
            F115W \& F360M & 1.15 \& 3.60 & cont. & SHALLOW2 & 3 & 1 & 773.047\\   
            F162M \& F405N & 1.62 \& 4.05 & H$_2$O \& Br$\alpha$ & SHALLOW2 & 6 & 1 & 1739.357\\
            F182M \& F480M & 1.82 \& 4.80 & H$_2$O \& cont. & SHALLOW2 & 6 & 1 & 1739.357\\
            F212N \& F470N & 2.12 \& 4.69 & H$_2$ & SHALLOW2 & 6 & 1 & 1739.357\\
            \noalign{\smallskip}
            \hline                  
            \end{tabular}
        \end{table*}

The data were reduced using the Python package JWST pipeline \citep[version = \texttt{1.12.5};][]{jwst_calib_1_12_5} and the Calibration Reference Data System (CRDS) Context = \verb+jwst_1217.pmap+. The raw data, i.e., the UNCAL frames, were downloaded from the MAST archive using the Python package astroquery \citep{astroquery,mast}. 
We then ran the standard JWST reduction in the three stages, but modifying a number of default parameters, which we summarize below:

\begin{itemize}
\item Stage 1: This involves applying the basic detector-level corrections to our imaging uncalibrated mutilaccum ramp products. We set the \verb+suppress_one_group=False+ in the input \verb+ramp_fit+ to recover saturated pixels between the first and second read. The rest was set as default.
            
\item Stage 2: This involves applying instrumental corrections and calibrations to the output products from Stage 1. All input parameters were set to the default values.
            
\item $1/f$ noise removal: $1/f$ noise is a pixel-to-pixel correlated noise, causing horizontal banding in JWST NIRCam images. The distinctive pattern of $1/f$ noise varies with each readout of the detector. Despite the ramp-fitting step in the pipeline (Stage 1) reducing this noise, traces of it remain visible in the resulting rate images. This effect is not removed in the standard JWST reduction. We used the script \verb+image1overf.py+
\footnote{\url{https://github.com/chriswillott/jwst/blob/master/image1overf.py}}. The tool is designed to run on the calibrated files that are output from the Stage 2 pipeline. In summary, the script subtracts a background and masks sources to determine $1/f$ stripes. The input parameters used were \verb+sigma_bgmask = 3.0+, \verb+sigma_1fmask = 2.0+, \verb+splitamps = True+, and \verb+usesegmask = True+. The former input assumes that these stripes can be split whereas the latter input allows us to use a segmentation image as the mask before fitting the $1/f$ stripes. The output from this script can be used through the Stage 3 pipeline as usual. We only applied this correction to the SW filters, i.e., F115W, F162M, F182M, and F212N, which improved the image quality significantly. On the other hand, the LW filters showed no improvement, or even a slight decrease in the image quality, when the $1/f$ noise removal was applied. Therefore we used the LW images without applying this correction.

\item Stage 3:  This involves taking the calibrated slope images and combining them into the final mosaics. The \verb+pmap+ version used provided us with accurately aligned mosaics between modules A and B. Here we also forced the final shape of the image (i.e., the number of pixels in the $x$ and $y$ axes) to be the same between filters. We set a final shape for the LW of $5660\times2280$ and the SW filters of $11450\times4740$. To do this, we set, in the \verb+resample+ step, the argument \verb+output_shape: (2280,5660)+ for the LW filters and \verb+output_shape: (4740,11450)+ for the SW filters. This facilitated the registration between filters. We also set celestial North up and East to the left, aligning with the pixel axes $y$ and $x$, respectively, by setting in the \verb+resample+ step, the argument \verb+rotation: 0.0+.

            
\item Registration: The output images, after the three stages of reduction, were aligned in the World Coordinate System (WCS) to the Gaia DR3 reference frame \citep{gaia2023}. However, in the pixel frame, there were sub-pixel shifts between the images. Therefore, we registered the images to the F470N and F212N filters for the LW and SW filters, respectively. We did this using the python package \verb+astroalign+. \citep{astroalign}. To do this, we applied an affine transformation using common stars in the images.
\end{itemize}

    \subsubsection{Continuum-Subtracted Images}\label{sec:continuum-sub}

        To characterize the line-emitting structures in the narrow-band images, particularly shocked emission from protostellar outflows, continuum-subtracted images were made following the methods detailed in \citet{long20} and used in \citet{crowe24}. Briefly, the method involves sampling the flux from $\sim15-20$ isolated, non-saturated point sources throughout a given region in both continuum- and narrow-band images in order to constrain the ratio between their fluxes, which is used to scale the continuum data to subtract from the narrow-band images pixel-by-pixel. Although the NIRCam images are already normalized for the differences in bandwidth of each filter (having units of MJy/sr), and could be subtracted directly, the former method is effective at accounting for differences between the continuum and line image fluxes due to the high levels of extinction in this region, which \citet{nogueras-lara24} has shown to be $\sim2.6\:A_{K_s}$ on the basis of red giant color analysis \citep[see][for a description of the methods]{nogueras-lara22a}.
            
        Due to the large size of the images ($2\arcmin\times6\arcmin$), we split each image into four main parts for the creation of continuum-subtracted images using the above method: one part containing the Sgr C protocluster and extended molecular cloud, a second part enclosing the star-forming region G359.42-0.104 (see \S\ref{sec:G359.42-0.104}) and the western-most side of the image, a third part enclosing the rest of the southern half of the image (i.e., most of the bright \Hii\: region), and a fourth part that covers the rest of the northern half of the image (see Fig.~\ref{fig:RGB}). We note that the flux ratio between the line and continuum images varied by $<10\%$ over all parts of the image.
            
        The above method was carried out directly for the F470N image, from which we subtracted the F480M image. We acknowledge the caveat that, given the overlap between the filters, there will inevitably be some H$_2$ $4.7\:{\rm \mu m}$ line emission in F480M that is inadvertently subtracted from the F470N image; however, given that the bandpass of F480M is relatively large compared to that of F470N ($0.303\:{\rm \mu m}$ for F480M compared to $0.051 \mu m$ for F470N; i.e., 6$\times$ higher bandpass\footnote{https://jwst-docs.stsci.edu/jwst-near-infrared-camera/nircam-instrumentation/nircam-filters}), this subtle ``over-subtraction" poses
        negligible impact to the fluxes in the F470N continuum-subtracted images, meaning that at most $1/6$ of the H$_2$ flux is over subtracted in the F470N image. The same general method was used to subtract the F212N image using the F182M image.
        
        The F405N image was subtracted by linearly interpolating the F360M and F480M image fluxes at each pixel to estimate the continuum flux at $4.05\: {\rm \mu m}$. This interpolation is considered more robust than taking flux from a single filter, since F360M and F480M provide an estimate of the continuum emission on both the blue and red sides of F405N without themselves being contaminated with any Br-$\alpha$ flux. Similar methods have been adopted for continuum-subtraction of the F335M filter to isolate Polycyclic Aromatic Hydrocarbon (PAH) emission in nearby star-forming galaxies \citep[see, e.g., \S3.1 of ][]{gregg24}.
        
        We note that the continuum-subtraction performed imperfectly on some of the stars in the image, producing residuals as a combination of dark and bright pixels. These are mainly due to the fact that the stellar Point Spread Functions (PSFs) between NIRCam filters are diffraction-limited and therefore wavelength-dependent. We also note imperfections in some regions of nebulosity in the image resulting from color differences in these regions with respect to the stars. These features are typical in continuum-subtracted images \citep[e.g.,][]{reiter22,bally23} and do not pose any significant effect on our characterization of the authentic features in the images.

\subsection{Archival Data}\label{sec:archival_data}

The mosaicked 25 and 37 $\mu$m images provided by the Stratospheric Observatory for Infrared Astronomy (SOFIA)-FORCAST Galactic Center Legacy Survey \citep[SOFIA Level 4 data products;][]{hankins20} were used to provide mid-infrared (MIR) \& far-infrared (FIR) fluxes for Spectral Energy Distribution (SED) fitting. Note that the survey covers a total area of 403 arcmin$^2$ (2180 pc$^2$), including regions such as the Sgr A, B, C complexes and the 50 $\kms$ molecular cloud. However, in this paper we limit our analysis to the region observed with JWST, Sgr C.
In addition, archival images from the Spitzer Space Telescope GALCEN survey \citep{stolovy06} at 3.6, 4.5, 5.6, and 8.0 $\mathrm{\mu m}$ and from the Herschel infrared Galactic Plane Survey \citep[Hi-GAL;][]{molinari16} at 70, 160, 250, 350, and 500 $\mathrm{\mu m}$, were used for SED fitting.

\subsection{Atacama Large Millimeter/Submillimeter Array}\label{sec:alma_obs}

We used ALMA Band 6 (1.3 mm) continuum and spectral line data (Program ID 2016.1.00243.S) in Sgr C \citep{lu20} to compare to our JWST data.
The synthesized beam size of these observations is $0.25\arcsec\times0.17\arcsec$ and the rms noise is in the range $40-60\,\mu$Jy\,beam$^{-1}$. 

We also utilized ALMA Band 3 (3 mm) continuum data from the ALMA Central Molecular Zone Exploration Survey (ACES) program (Program ID 2021.1.00172.L; PI: Longmore). The beam size of these observations is $1.97\arcsec\times1.52\arcsec$ and the rms noise is $\sim$0.2 $\mathrm{mJy~beam^{-1}}$. We use the Band 3 data solely to explore the region G359.42-0.104 (see \S\,\ref{sec:G359.42-0.104}). A full description of the reduction of this data will be given in Ginsburg et al., Walker et al., and Longmore et al. (in prep.), however some additional details about the survey and data reduction can be found in \citet{nonhebel24,ginsburg24}.

\section{Results}\label{sec:results}

Here we present the near-infrared (NIR) data obtained from JWST-NIRCam on the Sgr C molecular cloud, featuring broad and deep coverage of the entire cloud, its associated \Hii\:region, and surrounding environment. We use this imaging to directly characterize three luminous massive protostars via SED fitting, identify as YSO candidates 5 highly reddeened IR sources that have matching ALMA Band 6 dust cores, and survey 88 bright line emission features, most likely knots of shocked outflow emission produced by over a dozen protostellar outflows. Some of the star formation activity is located in a newly discovered region, $\sim1\arcmin$ to the south of Sgr C.


\begin{figure*}
\centering        
\includegraphics[width=0.933\textwidth,height=0.933\textheight,keepaspectratio]{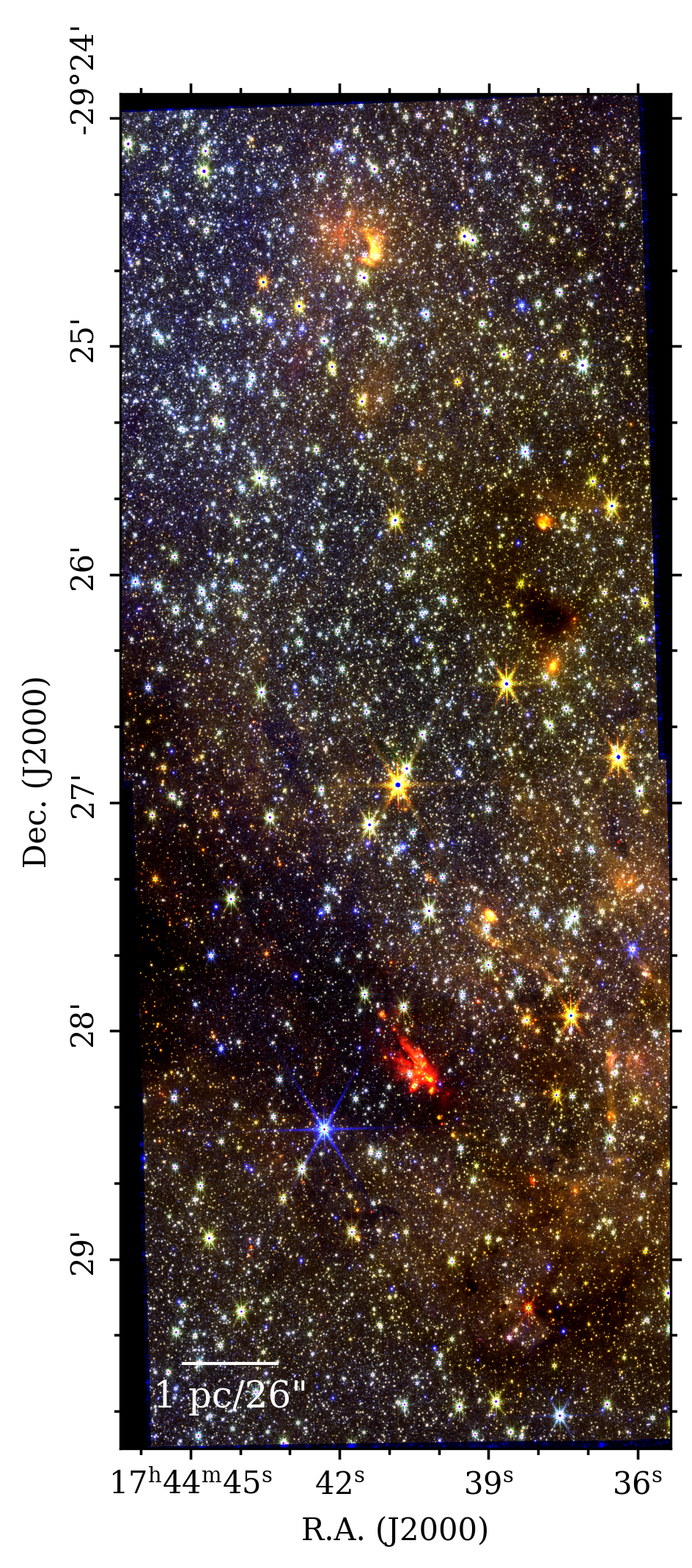}
\includegraphics[width=0.89\textwidth,height=0.89\textheight,keepaspectratio]{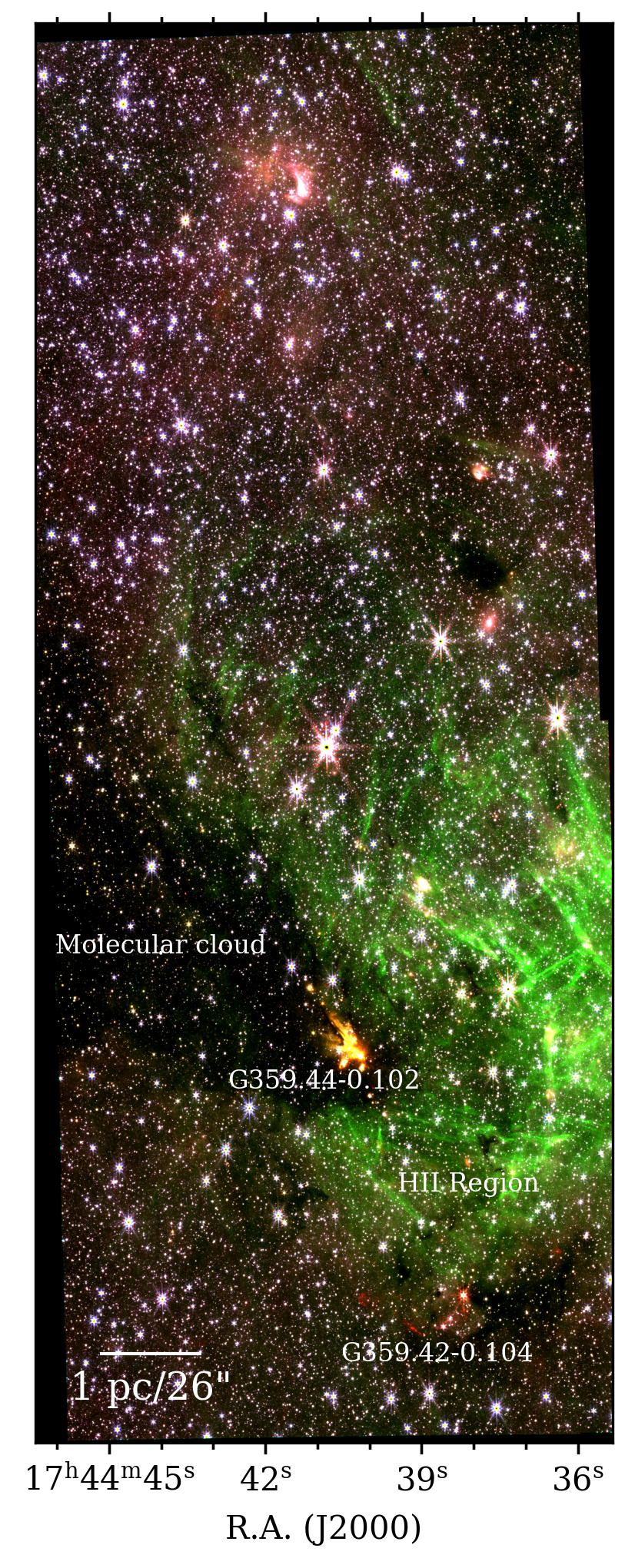}
\caption{\label{fig:RGB} 
RGB image of Sgr C showing two different color schemes. N is up and E is left in both panels. \textit{Left:} ``Continuum" image with F480M shown in red, F360M in green, and F182M in blue. \textit{Right:} ``Emission line" image with F470N shown in red, F405N in green, and F360M in blue. Notable features discussed in the text are labeled.}
\end{figure*}

\subsection{Overview of NIRCam Imaging of the Sgr C Region}\label{sec:NIR_presentation}

The left panel of Fig. \ref{fig:RGB} shows an RGB image of Sgr C and its surroundings covering a total FOV of $2\arcmin\times6\arcmin$. Red is used for F480M, green for F360M, and blue for F182M emission. We expect these filters mainly trace continuum emission, especially from stars. Glowing red in the lower middle of the image is the primary target of the observation, the massive protostellar source G359.44-0.102 in Sgr C, and its associated protocluster. This source is surrounded by a relatively dark region, which is inferred to be a dusty molecular cloud, which blocks much of the light from background stars. This cloud extends mainly to the east and north of the protocluster. Other dark clouds are also apparent in the image, including prominent examples at RA$\sim$17h44m37s, Dec$\sim-29^{\circ}26\arcmin10\arcsec$, to the NW of G359.44-0.102, and at RA$\sim$17h44m38s Dec$\sim-29^{\circ}29\arcmin30\arcsec$, which we will see is associated with the star-forming region G359.42-0.104 (see \S\ref{sec:G359.42-0.104}).

The right panel of Fig. \ref{fig:RGB} shows the same region, but now red is used for the F470N filter (covering the H$_2$ 0-0 S(9) line at $4.6947\,\mathrm{\mu m}$), green for F405N (covering the \Hi\:line Br-$\alpha$ at $4.05\,\mathrm{\mu m}$), and blue for F360M, which mostly covers continuum emission at $3.6\,\mathrm{\mu m}$. Most prominent in this image is the glow of hydrogen recombination line emission Br-$\alpha$, which traces ionized gas surrounding G359.44-0.102 in northern, western and southern directions. We note that the lack of Br-$\alpha$ emission to the east could potentially be explained by extinction of the main Sgr C dark cloud, i.e., an infrared dark cloud. The Br-$\alpha$ nebula displays remarkable linear, ``needle''-like features, which have a variety of orientations. We discuss this Br-$\alpha$ emission further in \S\ref{sec:discussion}, while its main analysis will be presented in forthcoming paper.


Also present in the right panel of Fig. \ref{fig:RGB} are ``knots'' of emission that appear red and trace H$_2$ 0-0 S(9) line emission, i.e., excited molecular hydrogen, that may be powered by shocks from protostellar outflows. These are key tracers of star formation activity, which we discuss extensively in \S\ref{sec:knot_ID}.


\subsection{The Massive Protostars in Sgr C}\label{sec:SED_fitting}


\begin{figure*}
        \centering
        \includegraphics[width=0.4\textwidth,keepaspectratio]{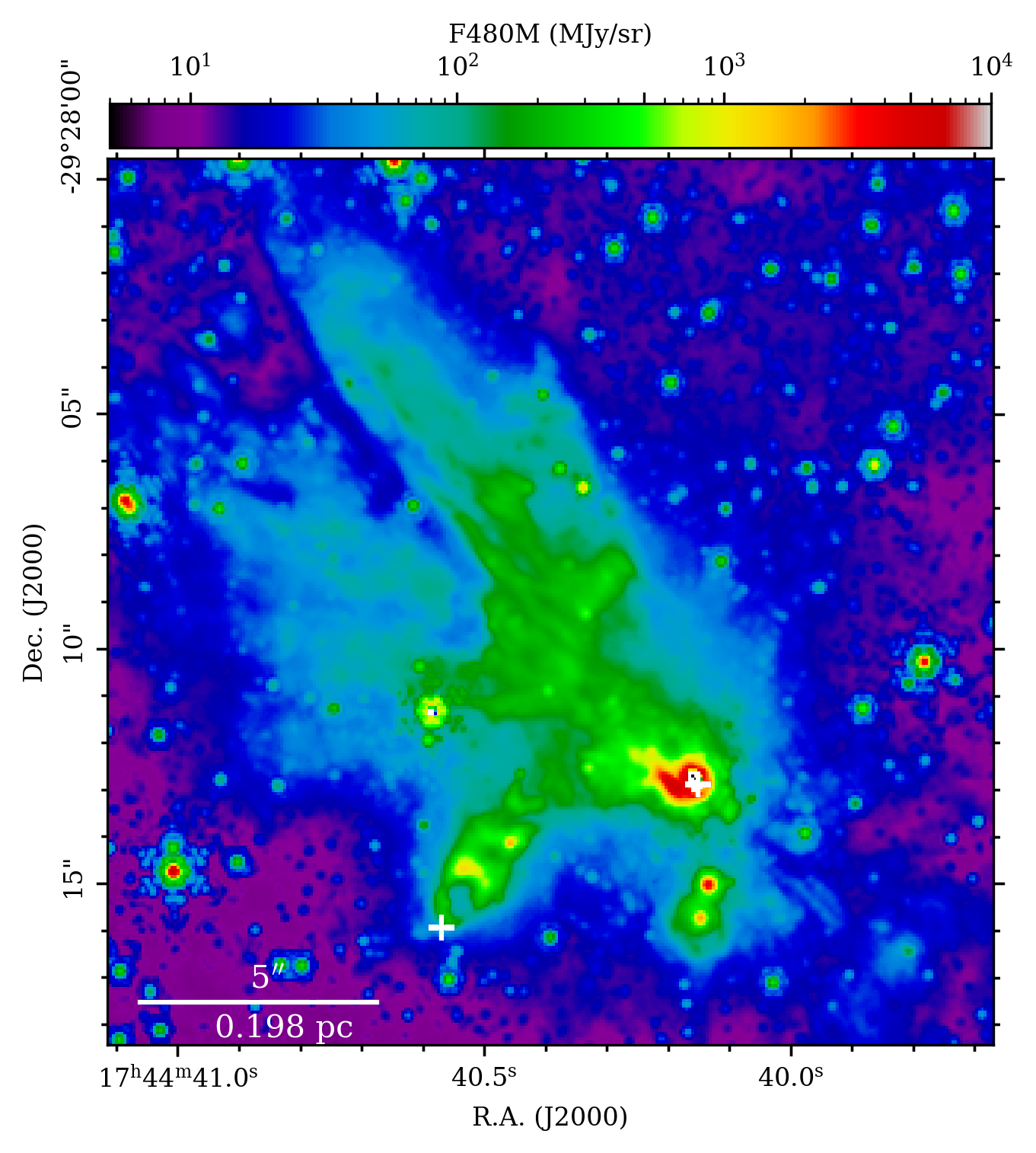}
        \includegraphics[width=0.4\textwidth,keepaspectratio]{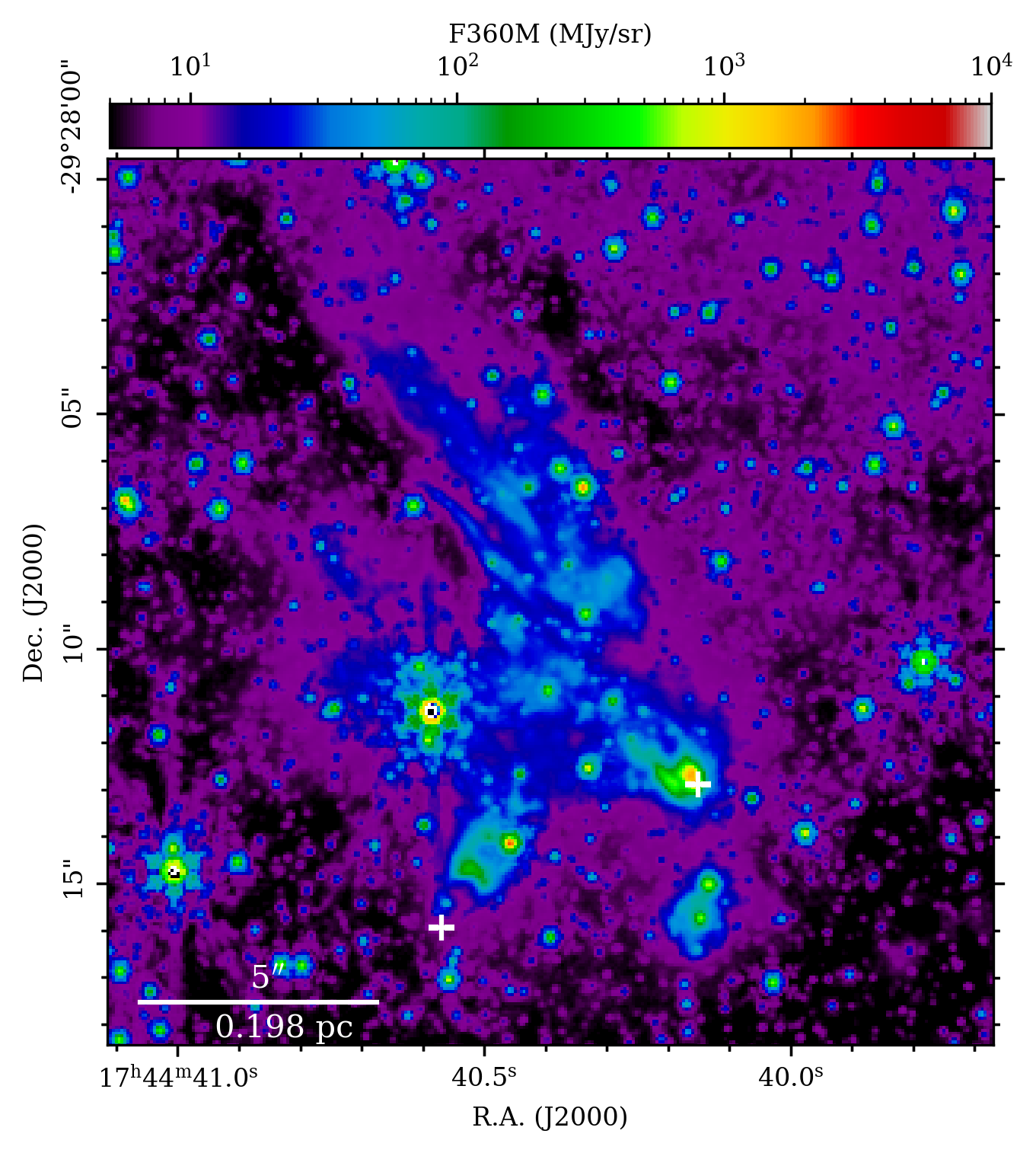}
        \includegraphics[width=0.4\textwidth,keepaspectratio]{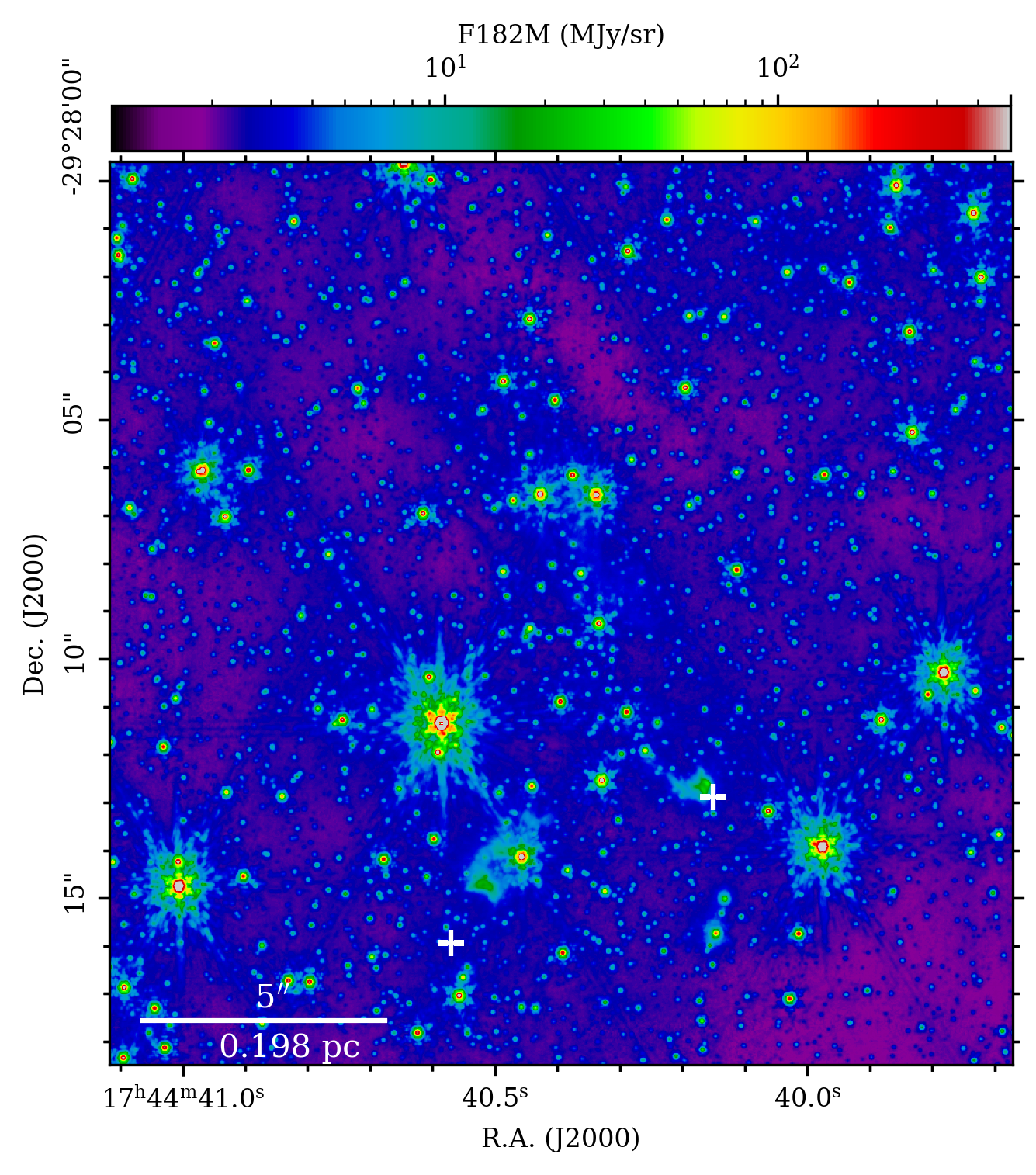}
        \includegraphics[width=0.4\textwidth,keepaspectratio]{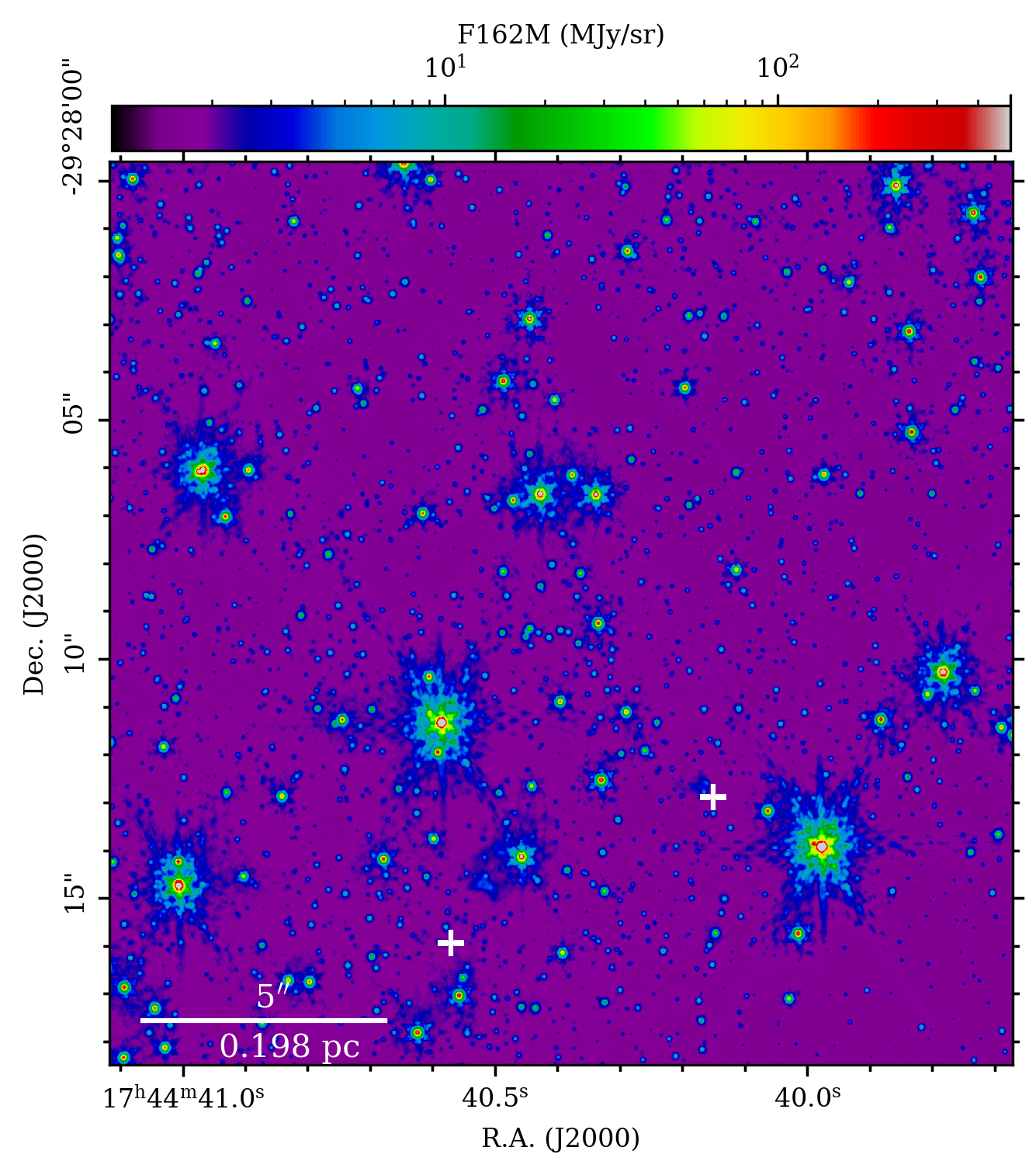}

        \begin{minipage}[b]{0.4\textwidth}%
        \caption{\label{fig:sgrc_main_zoom} 5-panel view of the heart of the Sgr C protocluster in several NIRCam filters, shown and labeled in descending order of wavelength from left to right, top to bottom. N is up and E is left in all panels. Indicated as white plus markers are the massive protostars G359.44a (western marker) and G359.44b (eastern marker; see text). Coordinates for each protostar are taken from \citep{lu20}. A logarithmic scale is used to highlight both the bright emission features (e.g., the outflow cone surrounding G359.44a) as well as the dimmer nebulosity and sources surrounding the massive protostars.}
        
        \end{minipage}        
        \includegraphics[width=0.4\textwidth,keepaspectratio]{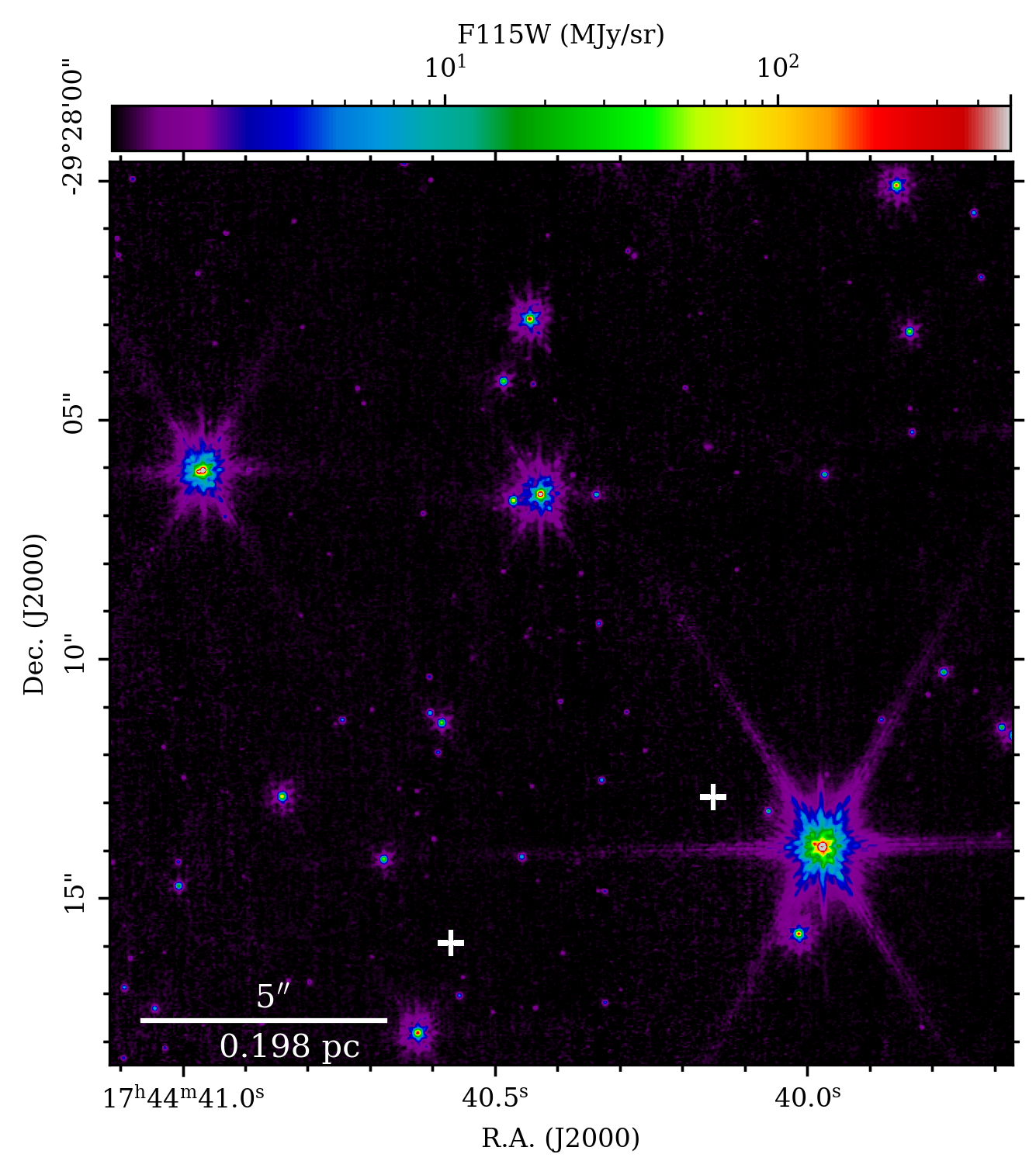}

    \end{figure*}
    
There are two main massive protostars in Sgr C, which we refer to as G359.44a (G359.44-0.102; see \S\ref{sec:intro}) and G359.44b. Fig. \ref{fig:sgrc_main_zoom} shows the inner region surrounding both protostars in several NIRCam filters, all of which trace predominantly continuum emission from the protostars, including scattered light in their outflow cones. The protostar G359.44a itself can be seen in F480M (upper left panel of Fig. \ref{fig:sgrc_main_zoom}), along with its outflow cone, which extends in ``green'' emission towards the north-east part of the image. A bright limb of the inner cone near G359.44a (marked with a white ``$+$'') can be seen in bright ``red'' emission. Nebulosity associated with the outflow axis of G359.44b can also be seen in ``green'' to the east, although the protostar itself is not a prominent infrared source. Going to shorter wavelengths, the emission from both protostars drops off dramatically, likely due to the effects of high extinction towards the heart of the protocluster and towards Sgr C in general \citep[$\sim2.6\:A_{K_s}$;][]{nogueras-lara24}. In F360M, G359.44a still appears to be visible as a point source, and extended continuum emission is seen tracing its outflow cone; additionally, some emission can be seen in F182M towards the NE of G359.44a and towards the NW of G359.44b, likely reflected light off of the outflow cavity walls very close to each protostar. Both protostars and their associated outflows are, however, essentially invisible in both F162M and F115W.

Spectral Energy Distribution (SED) fitting was performed on these two massive protostars to place constraints on their physical properties.
We note that these sources are co-located with mm continuum cores sgrc65 and sgrc22, respectively, that were identified by \citet{lu20}. The central coordinates of G359.44a and G359.44b are (RA=17h44m40.23s, Dec=-29$^\circ28\arcmin14.903\arcsec$) and (RA=17h44m40.63s, Dec=-29$^\circ28\arcmin15.458\arcsec$), respectively, with these positions defined from the local emission peaks of the SOFIA-FORCAST 37~$\mathrm{\mu m}$ image.

The SEDs were constructed using JWST-NIRCam 1.62, 1.82, 3.60, and 4.80 $\mathrm{\mu m}$, Spitzer IRAC 3.6, 4.5, 5.6, and 8.0 $\mathrm{\mu m}$, SOFIA-FORCAST 25 and 37 $\mathrm{\mu m}$, and Herschel PACS \& SPIRE 70, 160, 250, 350, and 500 $\mathrm{\mu m}$ data (see \S\ref{sec:observations}). Aperture photometry and SED construction were performed using the open source python package \textit{sedcreator} \citep{fedriani23a}, generally following methods developed in the SOFIA Massive (SOMA) Star Formation Survey \citep{debuizer17,liu19,liu20,fedriani23a}. However, the Sgr C massive protostars are relatively closely separated, i.e., by $\sim5\arcsec$. This is smaller than the minimum separation normally treated in the SOMA Survey of $10\arcsec$, which allows each source to have a minimum aperture radius of $5\arcsec$ (the resolution of Herschel at 70 $\mathrm{\mu m}$). Thus, rather than the standard SOMA method of using the Herschel 70~$\mathrm{\mu m}$ image to define source apertures, we define these from the longest wavelength image at which they are clearly resolved, i.e., SOFIA-FORCAST 37 $\mathrm{\mu m}$. The aperture radii are derived using the automated algorithm of \citep{fedriani23a} applied to this image, in which the aperture size is increased from a minimum size of $3\arcsec$ until background-subtracted flux attains an approximate plateau, with the background evaluated in an annulus from one to two aperture radii, excluding overlap with nearby source apertures (Telkamp et al., in prep.). The fitted aperture radii for the two sources are 4.25$\arcsec$. Pixels overlapping both apertures are distributed to the closer source. These aperture geometries are held fixed for evaluating the fluxes at wavelengths of 37~$\mathrm{\mu m}$ and shorter. For longer wavelengths, where the sources are unresolved, we use the same algorithm to define a new joint aperture for the two sources, based on the 70~$\mathrm{\mu m}$ image, of 9$\arcsec$, centered in the 70 $\mathrm{\mu m}$ emission peak via visual inspection at (RA=17h44m40.19s, Dec=-29$^\circ28\arcmin15.344\arcsec$). We then distribute the fluxes and errors to G359.44a and G359.44b in a 3:1 ratio for all wavelengths $\geq 70 \mu m$, based on the flux ratio of the sources in the 37~$\mathrm{\mu m}$ image (the longest wavelength image in which the sources are still resolved). This approximation appears to be valid in the 70~$\mathrm{\mu m}$ image, where G359.44a appears to remain the brighter source.



Photometry was conducted in all filters with the above apertures, and errors are assigned for each flux; for fluxes below 100 $\mathrm{\mu m}$ this error is derived from fluctuations in the background flux taken from the annulus \citep[see][]{fedriani23a}, and for fluxes above 100 $\mathrm{\mu m}$ this error is taken as the integrated background flux over the annulus (from one to two aperture radii). All errors are taken in quadrature with an assumed systematic error of 10\%. All fluxes and their associated error for each wavelength for each source are given in Appendix \ref{sec:source_photometry_table}.


The observed SEDs were then fit to the model grid of radiative transfer models of massive star formation via monolithic core collapse from \citet{zhang18}, which has primary parameters of mass surface density of the clump environment $\Sigma_\mathrm{{cl}}$, initial core mass $M_c$, current stellar mass $m_*$ (which measures the extent of evolution of a given protostellar model), inclination angle $\theta_{\rm view}$, and level of foreground extinction $A_V$. The goodness of fit of models was then assessed via a reduced $\chi^2$ parameter, and, since SED fitting results are subject to degeneracies \citep[e.g., between $\mathrm{A_V}$ and $\theta_{\mathrm{view}}$, and between $\Sigma_{\mathrm{cl}}$ and $\theta_{\mathrm{view}}$, see][for more information]{zhang18}, we present averages of the results of ``good'' model fits, following the methods of \citet{fedriani23a}, which we describe here. If the minimum $\chi^2$ of all models in the fitting routine, $\chi^2_{\mathrm{min}}$, is less than 1, we average together all models with $\chi^2<2$; if $\chi^2_{\mathrm{min}}>1$, we average all models with $\chi^2<2\times\chi^2_{\mathrm{min}}$. We note that, following methods from \citet{debuizer17,liu19,liu20,fedriani23a}, all fluxes below 10 $\mathrm{\mu m}$ were designated as upper limits due to the addition of strong contaminants to the emission at these wavelengths (e.g., from Polycyclic Aromatic Hydrocarbon bands) that are not modeled by the \citet{zhang18} model grid. We also note that we exclude models that predict a core size that is greater than twice the aperture used to define the SED, i.e., $R_{c}>2\times R_{\rm ap}$.



\begin{figure*}
\centering
\includegraphics[width=0.98\textwidth,height=0.98\textheight,keepaspectratio]{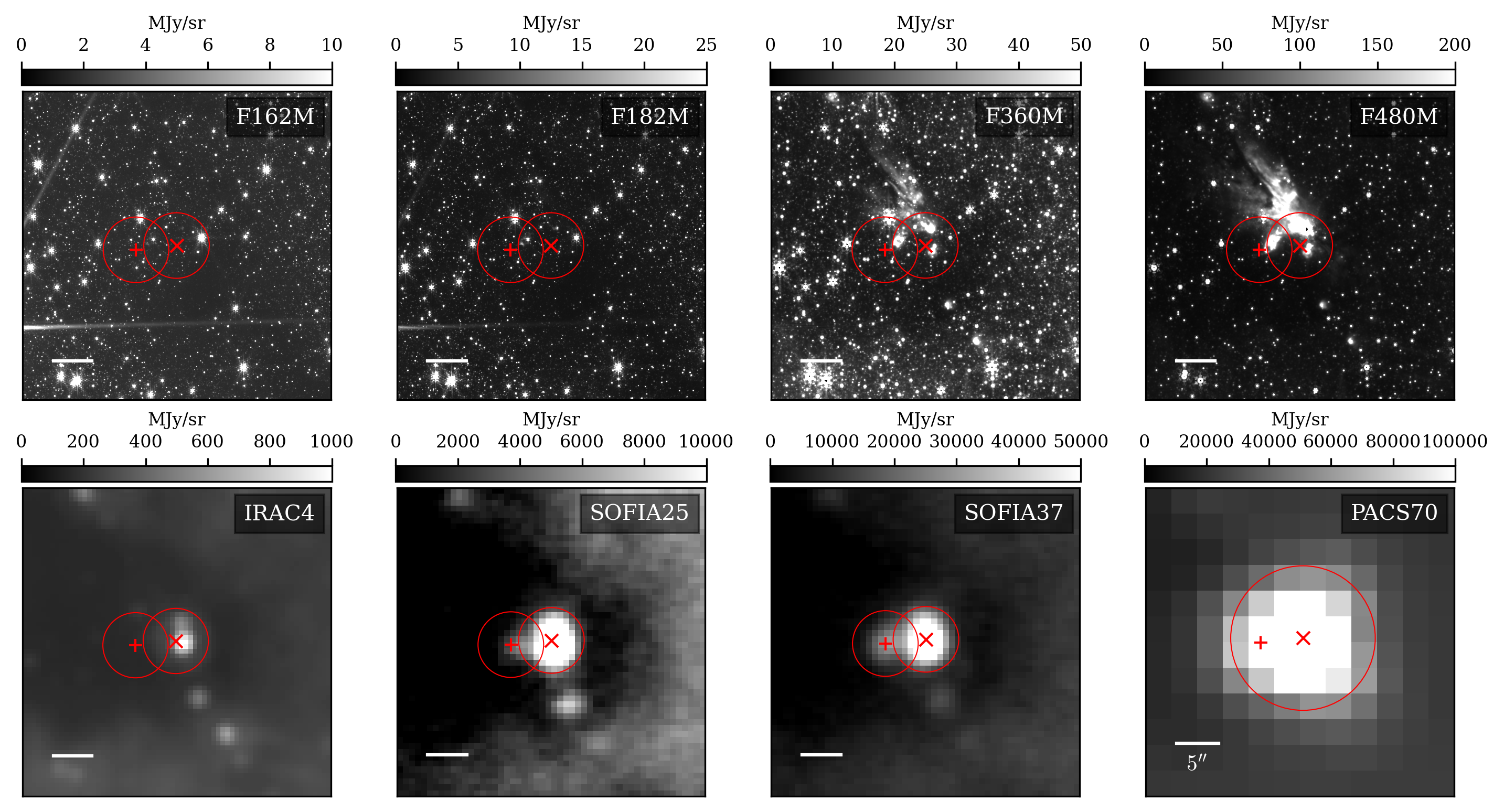}

\includegraphics[width=0.49\textwidth,height=0.49\textheight,keepaspectratio]{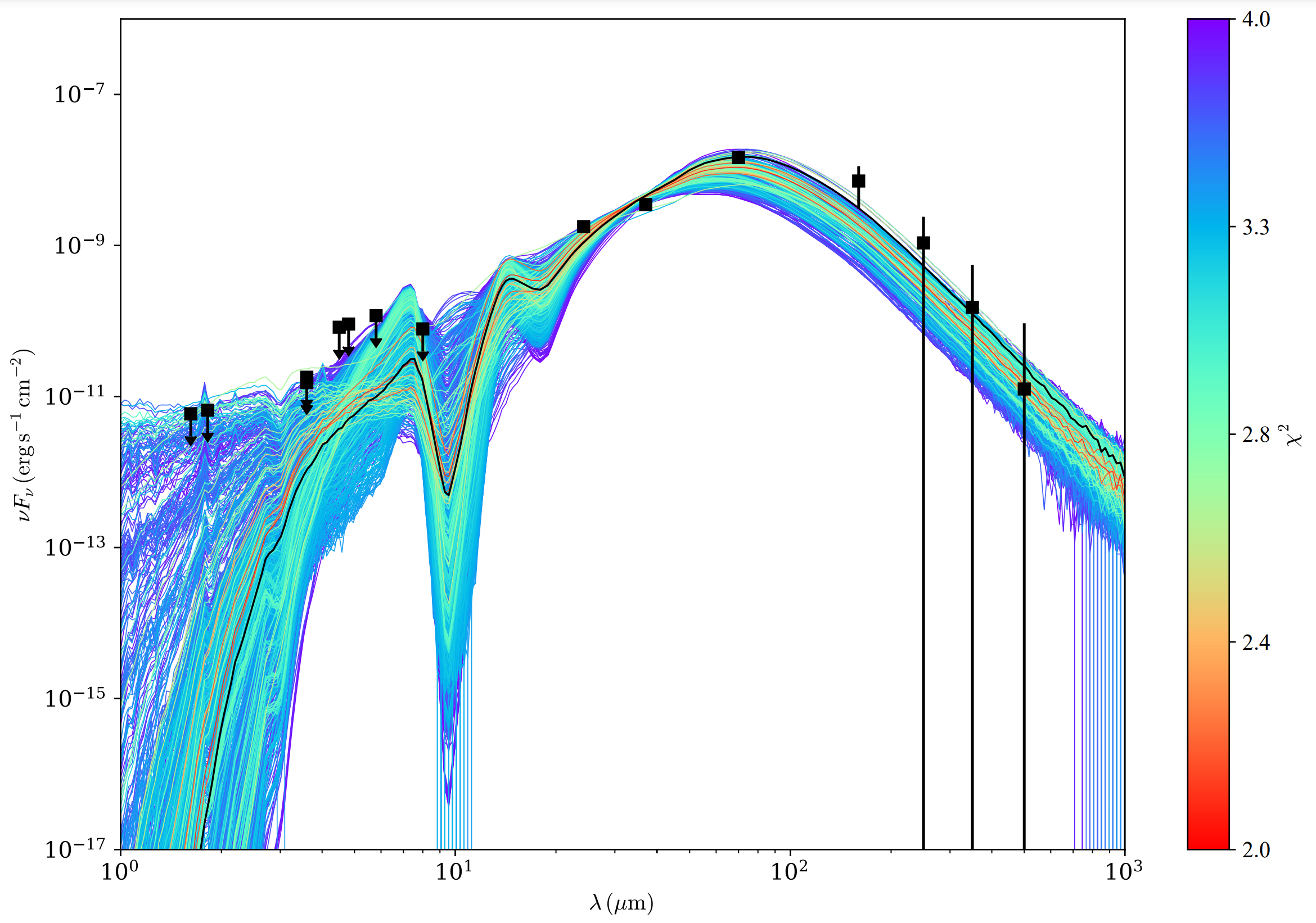}
\includegraphics[width=0.49\textwidth,height=0.49\textheight,keepaspectratio]{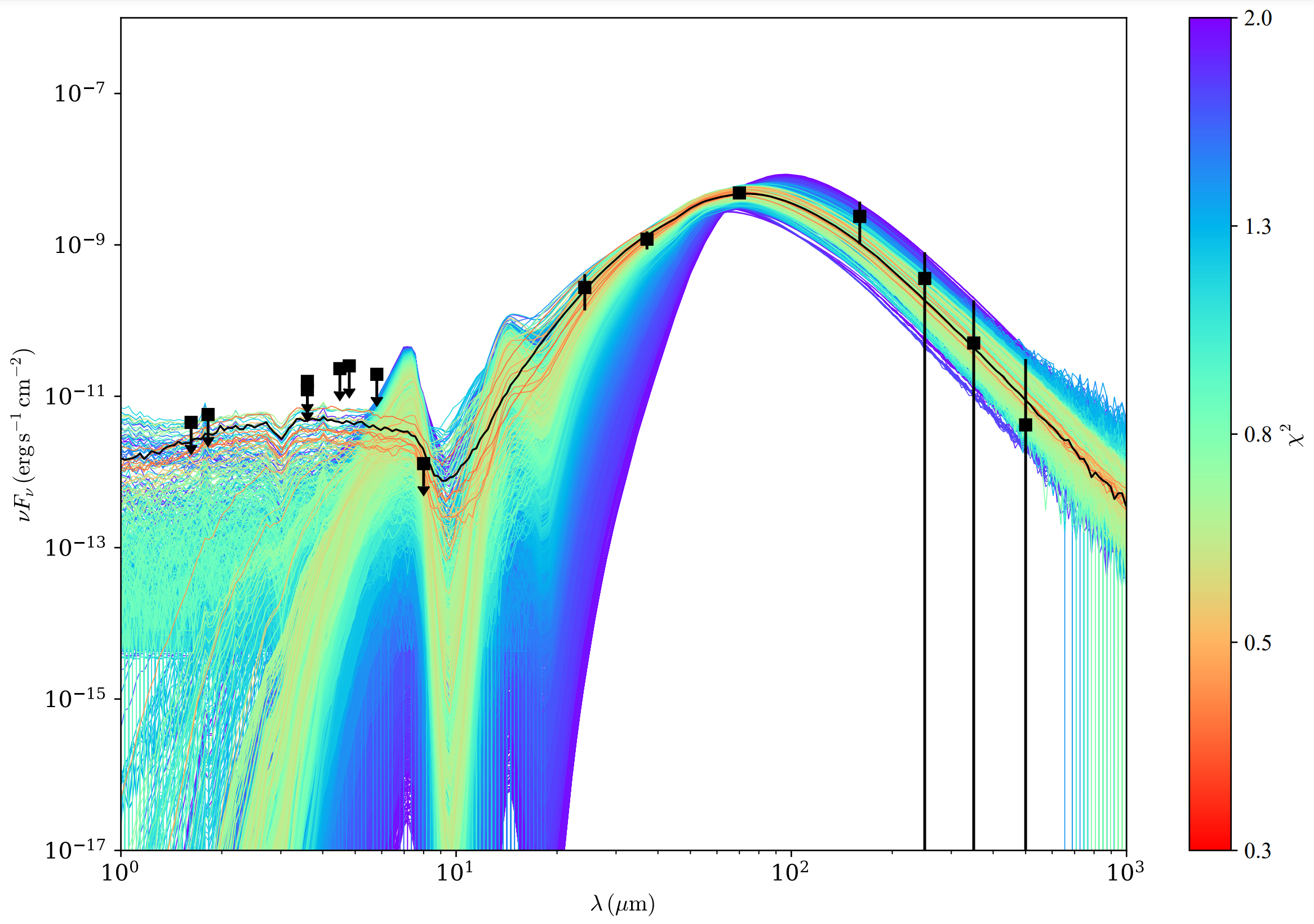}
\includegraphics[width=0.49\textwidth,height=0.49\textheight,keepaspectratio]{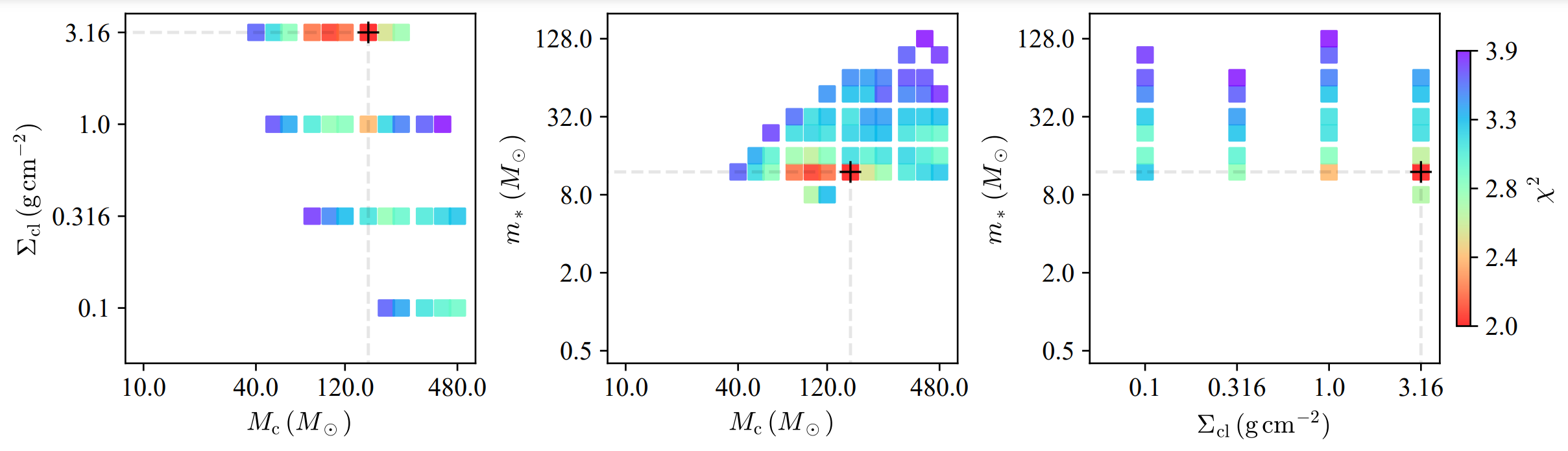}
\includegraphics[width=0.49\textwidth,height=0.49\textheight,keepaspectratio]{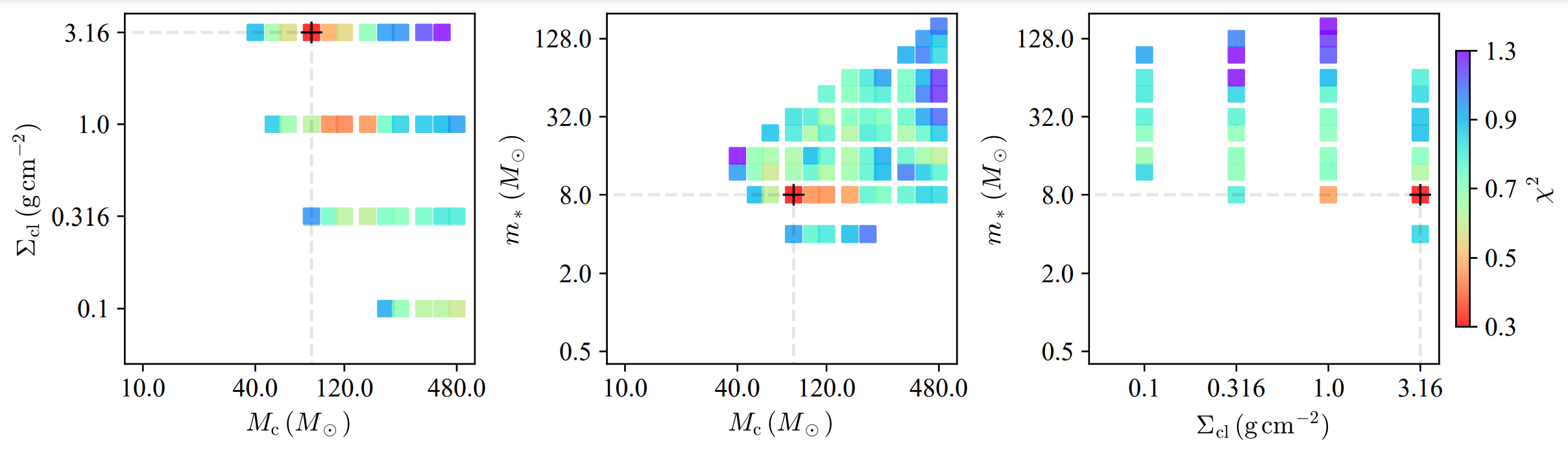}
\caption{\label{fig:SED_main_protostars}
\textit{Top:} 8-panel image of the protostars G359.44a (red cross) and G359.44b (red plus) in NIR and MIR filters. Note that the plotted coordinates of each protostar have been chosen from their peak emission in the SOFIA 37 $\mu$m image, and are therefore different from the coordinates shown in Fig. \ref{fig:sgrc_main_zoom}. A physical bar of 5\arcsec~is shown in the bottom left corner of each panel. The aperture(s) used for photometry at each wavelength, as discussed in the text, are shown. \textit{Middle:} Protostellar SED for G359.44a (left) and G359.44b (right), with model fittings overlaid and color-coded by $\chi^2$, with the best-fit SED as a black line. \textit{Bottom:} 2-d parameter space plots from the SED fitting results for G359.44a (left) and G359.44b (right). Various parameter pairings are given on the x- and y-axis of each plot, with each pairing of parameter values color-coded by $\chi^2$. Dashed gray lines indicate the lowest $\chi^2$ (i.e., best-fitting) pair of parameter values.
}
\end{figure*}

G359.44a, G359.44b, their SEDs and the 2d parameter space plots of their fitting are shown in Fig. \ref{fig:SED_main_protostars}.  A summary of key physical parameters from the fitting results is given in Table \ref{tab:fitting_results}. Note that this table contains fitting results for an additional massive protostar candidate, G359.42a, which is outside the main Sgr C cloud and will be discussed in \S\ref{sec:G359.42-0.104}.

We note a good agreement in the SED for each protostar between the Spitzer and NIRCam data points (see the data points below $10 \mu m$ in their SEDs), and note that the additional NIRCam points do not appear to have significantly altered the results of the fitting. A peak in each SED at $70 \mu m$ is also seen, as is typical of the SEDs of massive protostars \citep[see, e.g.,][]{fedriani23a}. 


\begin{sidewaystable*}       
\centering
\caption{\label{tab:fitting_results}SED fitting results, averaged over all ``good'' model fits (see text), for massive protostars in Sgr C and G359.42-0.104.}
\begin{tabular}{c c c c c c c c c c c} 
\hline\hline  
Name & $M_c (M_\odot)$  & $\Sigma_\mathrm{{cl}}$ (g/cm$^2$) & $m_* (M_\odot)$ & $\theta_{\mathrm{view}}$ ($^\circ$) & $M_\mathrm{{env}} \:(M_\odot)$ & $L_\mathrm{{bol}}\:(L_\odot)$ & $M_\mathrm{{disk}} \:(M_\odot)$ & $\dot{M} \:(M_\odot/yr)$ & Age (yr) & $\mathrm{A_V}$\\
\hline
G359.44a & $188.5^{+165.5}_{-88.1}$ & $0.46^{+1.14}_{-0.33}$ & $20.7^{+14.1}_{-8.4}$ & $63\pm17$ & $119.6^{+168.7}_{-70.0}$ & $9.7^{+15.6}_{-6.0} \times 10^4$ & $6.9^{+4.7}_{-2.8}$ & $3.0^{+3.7}_{-1.6} \times 10^{-4}$ & $1.2^{+1.6}_{-0.7} \times 10^5$ & $91\pm69$\\
G359.44b & $202.9^{+179.0}_{-95.1}$ & $0.72^{+1.60}_{-0.50}$ & $20.4^{+24.1}_{-11.0}$ & $60\pm20$ & $132.3^{+159.1}_{-72.2}$& $1.0^{+3.1}_{-0.8} \times 10^5$& $6.8^{+8.0}_{-3.7}$ & $4.2^{+5.0}_{-2.2} \times 10^{-4}$ & $8.6^{+13.9}_{-5.3} \times 10^4$ & $423\pm296$\\
G359.42a & $32.9^{+40.7}_{-18.2}$ & $0.26^{+0.50}_{-0.17}$ & $8.5^{+11.9}_{-5.0}$ & $49\pm25$ & $5.4^{+9.3}_{-3.4}$& $8.2^{+41.7}_{-6.9} \times 10^3$& $2.8^{+4.0}_{-1.7}$ & $4.3^{+3.9}_{-2.1} \times 10^{-5}$ & $2.1^{+3.0}_{-1.2} \times 10^5$ & $35\pm25$\\
\hline   
\end{tabular}
\end{sidewaystable*}
\subsection{A Search for Low-Mass Protostars}\label{sec:jwst_alma_crossmatch}

To search for lower-mass protostars in the Sgr C region, we start with the 1.3~mm emission cores detected by ALMA.
We ran dendrogram source finding algorithm \citep{rosolowsky08} 
 using the following parameters: minimum flux density of 4$\sigma$, minimum significance of a structure of 1$\sigma$, and minimum area of a structure of one beam. These values were adopted to match those used by \citet{lu20}, and we verify we recover the same sources as presented in their study.
Of the 274 cores in \citet{lu20}, 267 are within the field of view of the NIRCam observations. 

We next used the \textit{Starfinder} tool to extract PSFs and perform photo-astrometry on the NIRCam filters F162M, F360M, and F480M \citep{diolaiti00}. In brief, we constructed the PSFs for each mosaic by selecting bright, unsaturated, isolated stars, running source detection and iterating on the procedure. The cores of all but the most strongly saturated stars were repaired via PSF fitting. Finally, we performed photo-astrometry with a five-sigma threshold (using the error maps provided by NIRCam as noise). This algorithm was run on a $150\arcsec\times90\arcsec$ box centered at the coordinates (RA=17h44m42.85s, Dec=-29$^\circ27\arcmin20.055\arcsec$), enclosing the entire part of the Sgr C cloud covered by the NIRCam data, resulting in the identification of 91,642 sources in F162M, 40,287 sources in F360M, and 40,382 sources in F480M. We refer further details to a follow-up paper focused on the global stellar population in the NIRCam images.


Cross-matching between IR sources and ALMA cores was carried out for those sources identified in the longest NIRCam wavelength of $4.8\:{\rm \mu m}$ (F480M), where we expect embedded protostars to be relatively bright and dust extinction to be minimized.
A source was considered to match an ALMA core if its position was within 0.10$\arcsec$, i.e., half an ALMA beam, from the peak of the dust emission. This corresponds to a physical projected separation of about 800~au.
Using this criterion, 23 of the ALMA cores were found to have a matching F480M source. 


Due to the high stellar density of the CMZ, there is a possibility that some of these 23 mm/NIR matches are aligned by chance. We estimate the expected number of spurious matches in the following way. First, we calculate the stellar density of the field by dividing the number of F480M sources in the \textit{Starfinder} catalog with the area of the field. The source density is 3.0 sources/(\arcsec)$^2$, or $\lambda=0.10$ sources per FWHM ALMA beam. If we assume that the number of sources in a given ALMA beam follows a Poisson distribution, we can calculate the probability that an ALMA beam (i.e., the area within matching distance of a core peak) contains at least one source by chance. The probability is $1 - P(\text{0 sources in beam}) = 1 - \text{e}^{-\lambda} = 0.095$. This means that we expect $\sim25$ cores to have NIR matches by chance. The detected number of matches, 23, is almost identical. Therefore, we cannot rely on spatial coincidence alone to find YSOs, but need to use an additional metric, reddening, to identify YSO candidates.

        
Therefore, the 23 identified F480M sources were matched to sources in the filters F360M and F162M in order to investigate their reddening. Reddening could give an indication of whether the source is likely to be a foreground star or if it is at the galactocentric distance \citep{nogueras-lara21}, and additionally if it is an ordinary star in the CMZ or if it may be an embedded (highly reddened) YSO. 

        First, the small offsets between the coordinates of F162M, F360M, and F480M, were compensated for by subtracting the median $x$ and $y$ separation between F162M/F360M source positions and source positions in F480M (-0.5483 LW pixels in x, -0.6718 LW pixels in y for F162M; +0.0340 LW pixels in x, -0.7740 LW pixels in y for F360M). The median separation was obtained by a preliminary cross-matching to 1000 randomly selected sources in the F480M source catalog. After offset correction, sources in the F360M and F162M filters were cross-matched to the F480M sources with matching ALMA cores. If the distance between positions was less than one long-wavelength pixel, i.e., 0.063\arcsec, the sources were considered to be matching. Out of the 23 F480M sources with ALMA counterparts, this resulted in 13 matches with F360M and 14 matches with F162M. 
        
        The magnitude of the sources was calculated in all three filters, as well as the color index F162M-F480M. In the case where a source was undetected at the shorter wavelength, the completeness limit of the source catalog ($m_{\rm F162M}=23.2$, dominated by crowding) was used to obtain a lower limit on the color index. A list of all ALMA cores with matching F480M sources is shown in Table \ref{tab:crossmatching}. The sources are listed in order of increasing distance between the ALMA peak and the F480M source. 
        
\begin{figure}
            \centering
            \includegraphics[width=\linewidth]{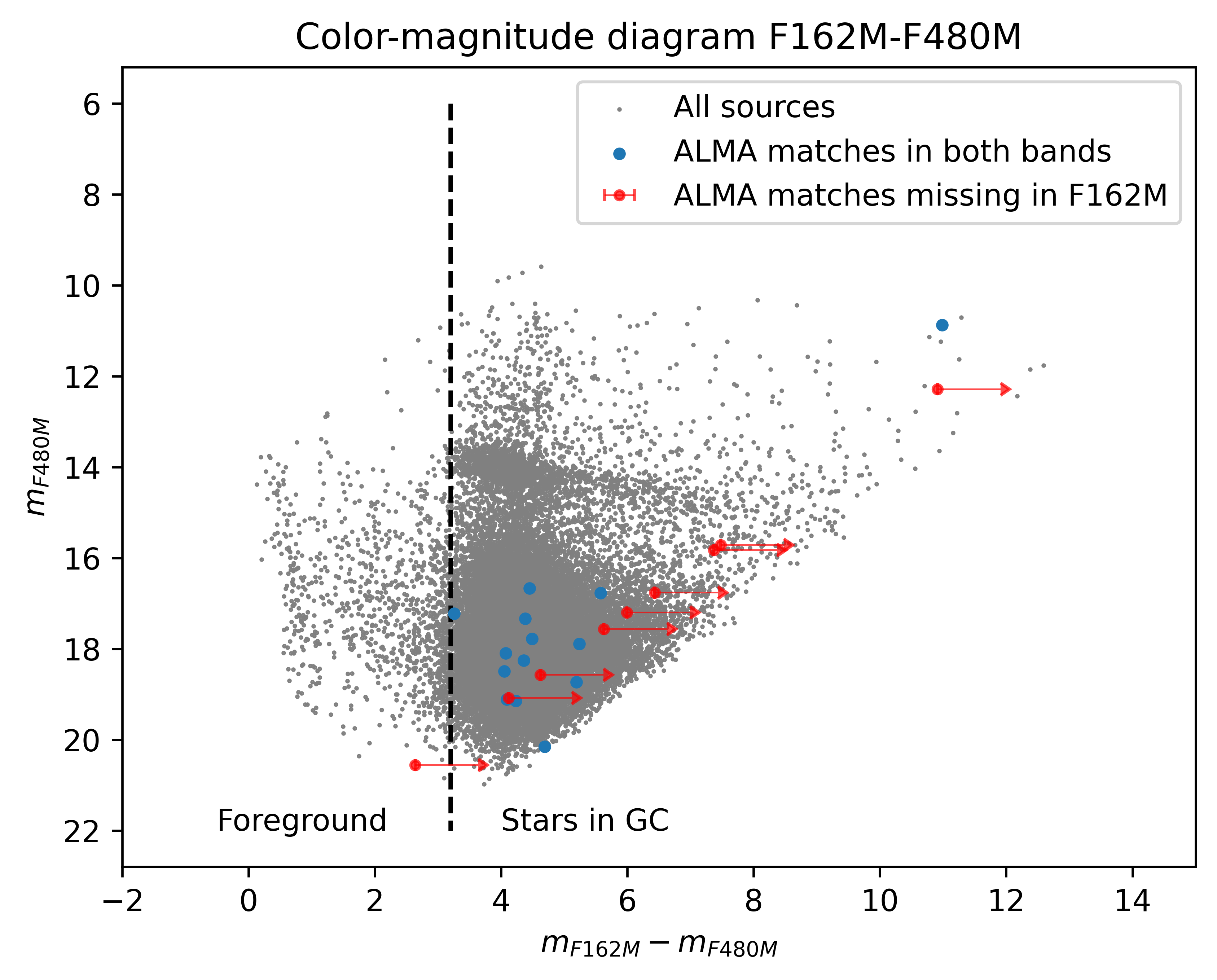}
            \caption{\label{fig:CMD} Color-magnitude diagram showing the F162M-F480M color index versus F480M magnitude. Grey dots represent the sources in the F480M catalog that have a F162M match. Blue dots represent the subset of these that match an ALMA peak. The red dots with arrows represent the F480M sources that have an ALMA match but are undetected in F162M. Their F162M magnitude has been replaced with the completeness limit $m=23.2$ to obtain a lower limit on the color index. The dashed line at $m_{\rm F162M}-m_{\rm F480M}=3.2$ represents the color cut separating foreground stars from CMZ stars \citep{nogueras-lara24}.}
        \end{figure}

        
        To determine which sources were significantly reddened, their color index were compared to the mean and standard deviation of the color index in the full F480M source catalog. A color-magnitude diagram of all F480M sources with a F162M counterpart can be seen in Fig. \ref{fig:CMD}. By similar analysis as \citet{nogueras-lara24}, we apply a color cut of $\sim 3.2$ to separate foreground stars from stars in the CMZ. Excluding the foreground stars, the mean color index of these sources was 4.54, and the standard deviation was 0.81. Five of the ALMA-matching sources have a reddening more than two standard deviations above the mean. 
        These are taken to be our sample of strong YSO candidates. Out of these five sources, only one (sgrc41) is detected in all bands (F162M, F360M, F480M). We note that all five of these sources have associated mm core masses between $\sim1-2~\mathrm{M_{\odot}}$ (see Table \ref{tab:crossmatching}), implying that they may be low-mass YSOs, although it must be noted that these masses do not represent the masses of the YSOs themselves. Additionally, none of these five sources is visible in the MIR/FIR emission (e.g., from SOFIA, Herschel), which implies that they are much dimmer than the massive protostars in the cloud, G359.44a and G359.44b, to be below the sensitivity limit of those observations. This would also imply that these are low-mass rather than high-mass protostar candidates.

\begin{figure*}
            \gridline{
            \includegraphics[height=0.6\linewidth]{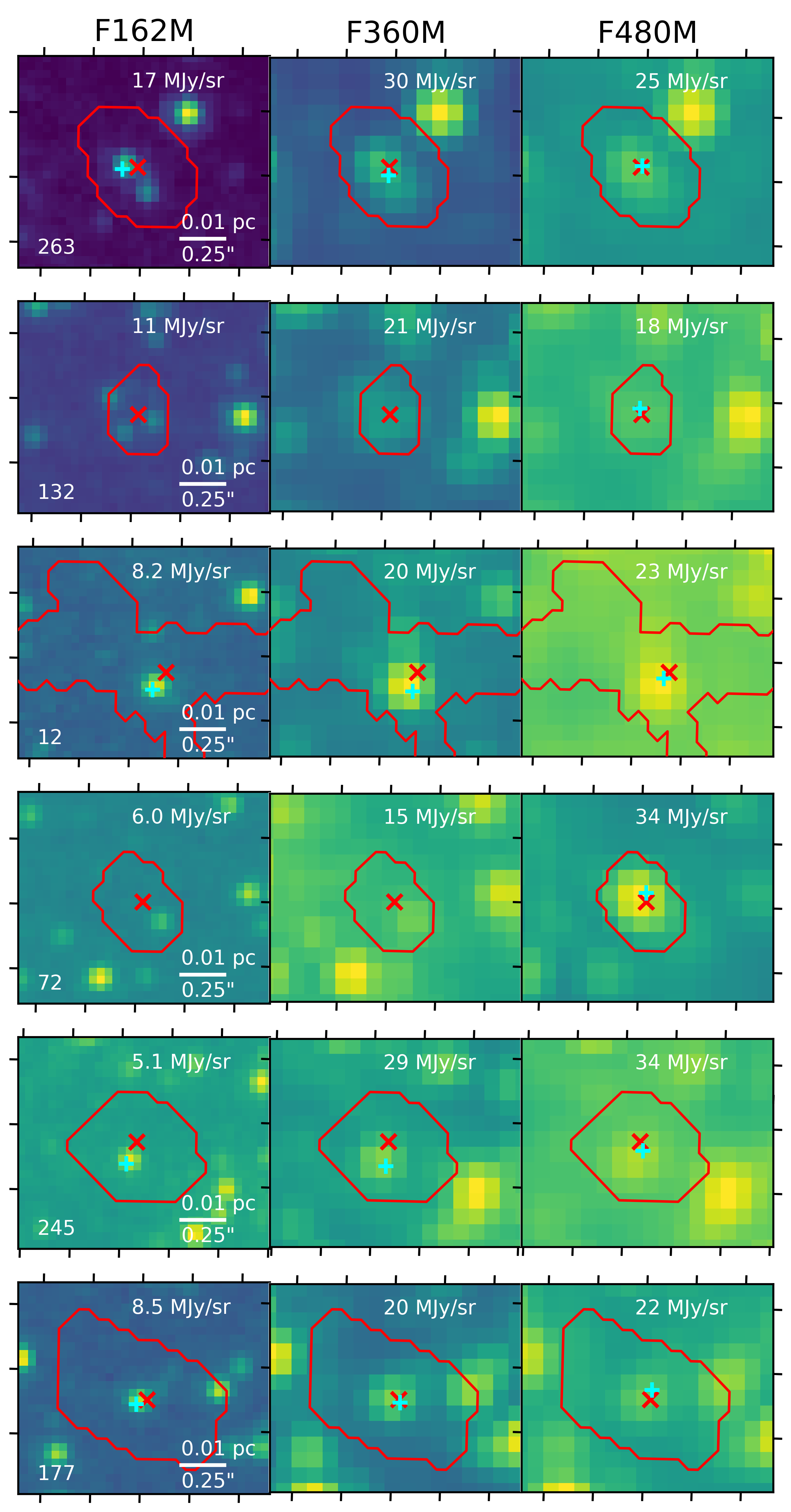}
             \includegraphics[height=0.6\linewidth]{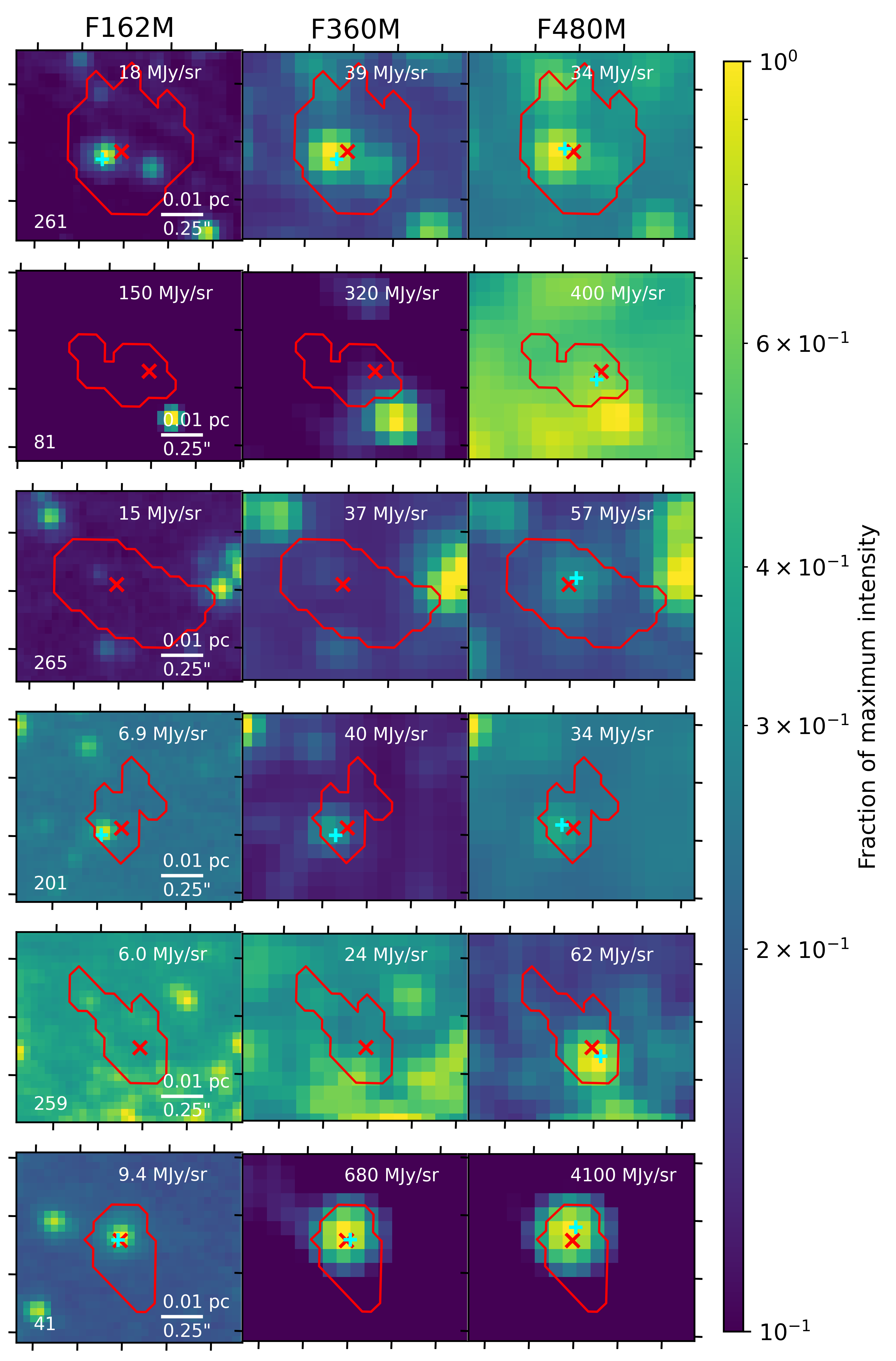}
               }
        \gridline{
             \includegraphics[height=0.6\linewidth]{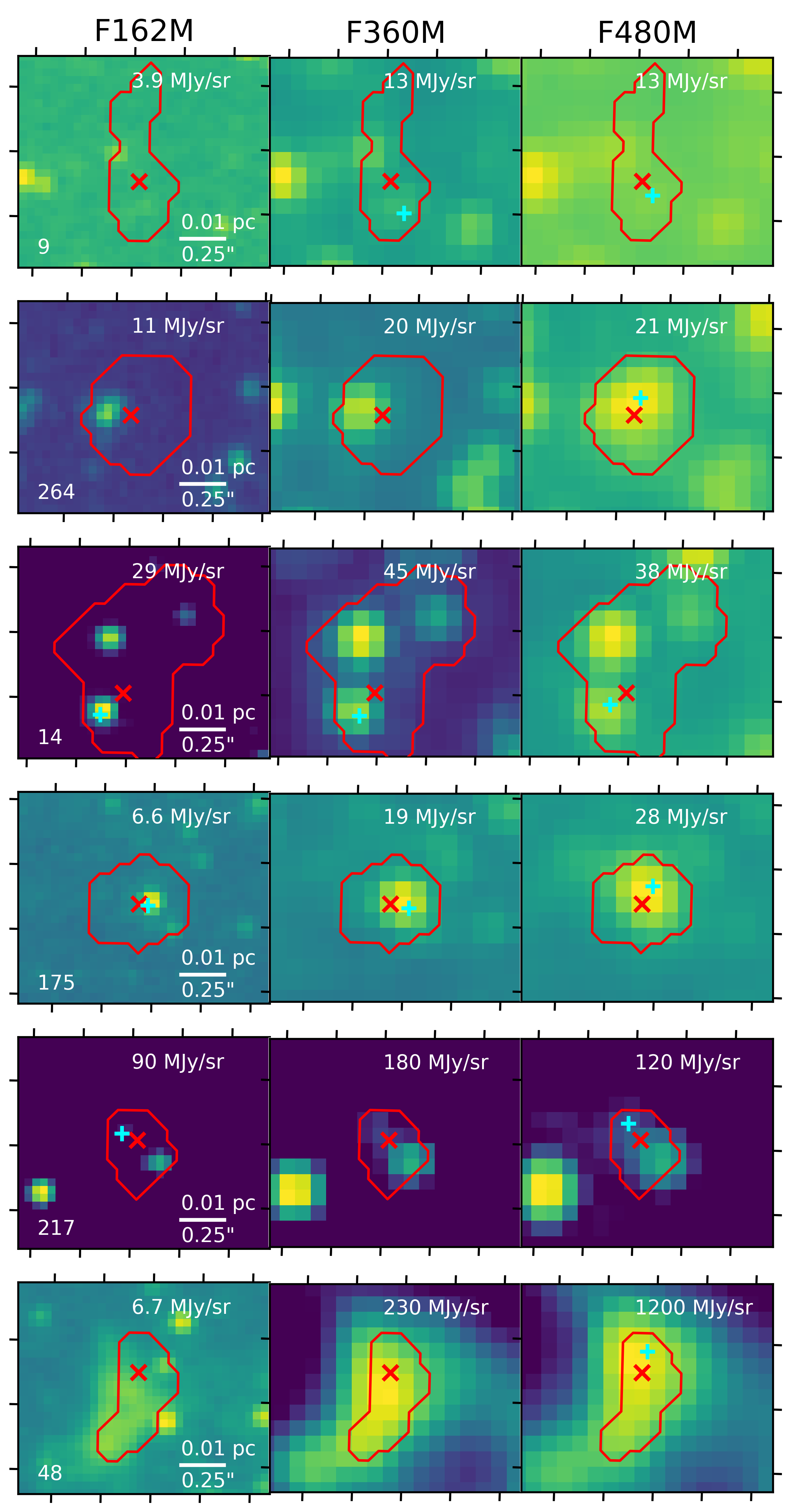}
             \includegraphics[height=0.6\linewidth]{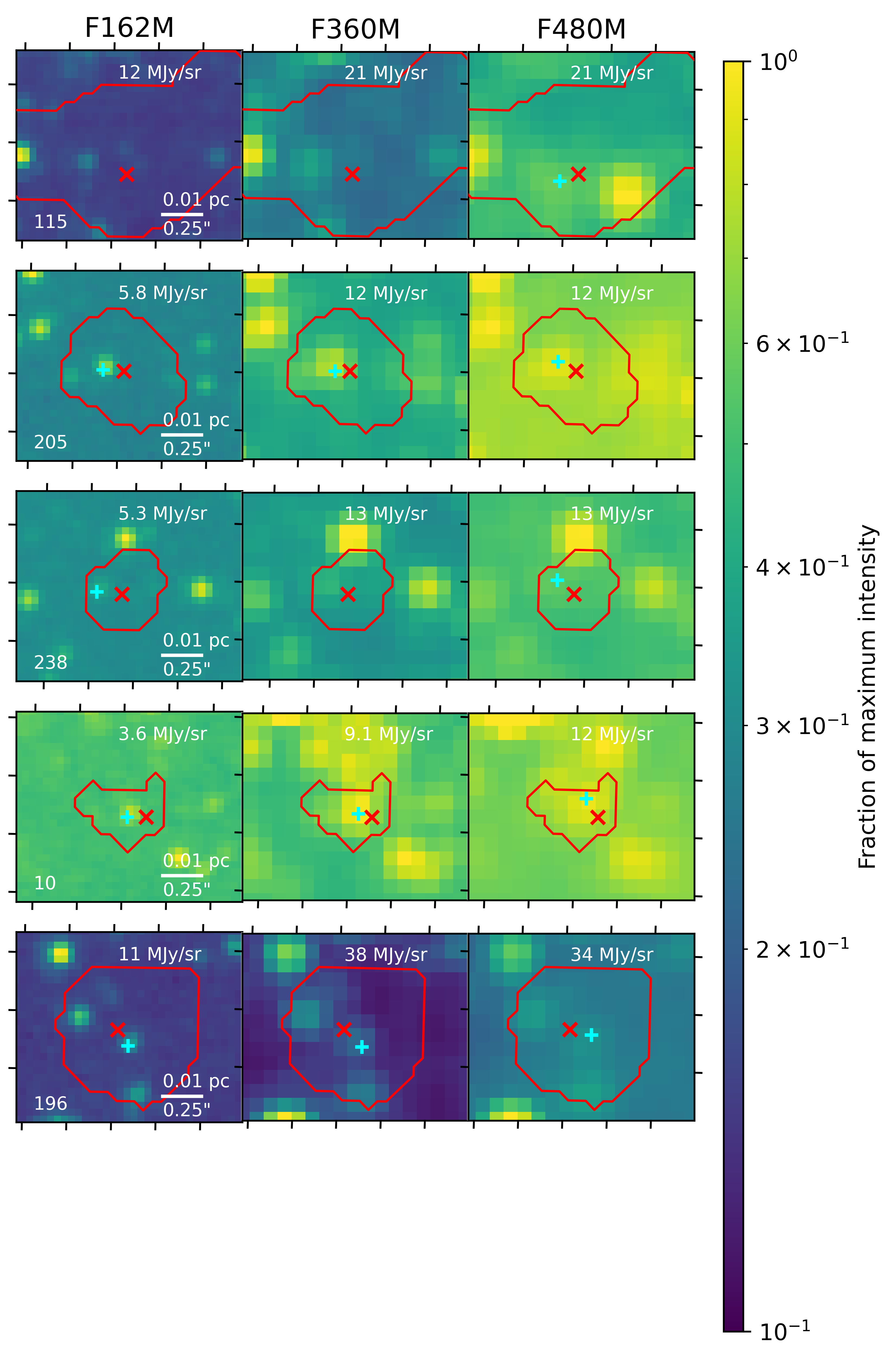}
               }
        \caption{\label{fig:crossmatching} All ALMA cores with matching NIR sources. The background images are NIRCam images in the F162M, F360M and F480M bands. Red contours and crosses represent the dendrogram contours and 1.3 mm peaks respectively. The cyan plus signs mark matching NIR sources in the respective images. The indices in the lower left corners are the core IDs listed in Table \ref{tab:crossmatching}. The colorscale has been normalized to the maximum intensity in each image, which is shown in the top right of each panel.}
        \end{figure*}

        Fig. \ref{fig:crossmatching} shows all 23 ALMA cores that have a matching F480M source.
        Red dendrogram contours and mm peaks are overlaid on the F162M, F360M and F480M NIRCam images. The matching NIR source detected in the respective images is marked with a cyan plus sign. Fig. \ref{fig:big_diagram} (b) also shows the spatial locations of these sources in the Sgr C cloud.

        \begin{deluxetable*}{lllllllllll}
            \tabletypesize{\footnotesize}
            \tablewidth{\linewidth}
            \tablecaption{Cores from \citet{lu20} with a matching NIR source. \label{tab:crossmatching}}
            \tablehead{\colhead{ALMA ID} & \colhead{$M_{mm}$} & \colhead{R.A.} & \colhead{Dec.} & \colhead{$\Delta\theta$} & \colhead{$m_{\rm F480M}$} & \colhead{$m_{\rm F360M}$} & \colhead{$m_{\rm F162M}$} & \colhead{$m_{\rm F162M}$-}  \\ [-2ex]
             & \colhead{($M_\odot$)} & \colhead{(J2000)} & \colhead{(J2000)} & \colhead{(\arcsec)} &  &  & &  \colhead{$m_{\rm F480M}$}}
            \startdata
                sgrc263     & 3.44             & 17:44:42.684 & -29:27:35.21 & 0.007        & 18.098 ± 0.010 & 18.421 ± 0.008 & 22.172 ± 0.014 & 4.074                 \\
sgrc132     & 1.53             & 17:44:41.797 & -29:28:02.18 & 0.025        & 19.081 ± 0.017 & $\cdots$            & $\cdots$            & \textgreater{}4.119   \\
sgrc12      & 3.63             & 17:44:40.274 & -29:28:17.28 & 0.032        & 17.781 ± 0.008 & 18.113 ± 0.005 & 22.272 ± 0.015 & 4.491                 \\
sgrc72$^{\dagger}$      & 1.33             & 17:44:43.584 & -29:28:12.18 & 0.036        & 16.768 ± 0.004 & $\cdots$            & $\cdots$            & \textgreater{}6.432   \\
sgrc245     & 2.47             & 17:44:40.852 & -29:27:44.32 & 0.038        & 17.890 ± 0.009 & 18.606 ± 0.009 & 23.127 ± 0.020 & 5.237                 \\
sgrc177     & 13.59            & 17:44:42.074 & -29:27:56.37 & 0.040        & 18.495 ± 0.011 & 18.772 ± 0.008 & 22.549 ± 0.015 & 4.054                 \\
sgrc261     & 7.57             & 17:44:42.678 & -29:27:36.86 & 0.041        & 16.673 ± 0.004 & 17.139 ± 0.004 & 21.127 ± 0.007 & 4.454                 \\
sgrc81$^{\dagger}$      & 0.69             & 17:44:40.304 & -29:28:11.23 & 0.042        & 15.721 ± 0.009 & $\cdots$            & $\cdots$            & \textgreater{}7.479   \\
sgrc265     & 9.38             & 17:44:43.525 & -29:27:32.58 & 0.044        & 17.570 ± 0.007 & $\cdots$            & $\cdots$            & \textgreater{}5.630   \\
sgrc201     & 0.41             & 17:44:42.084 & -29:27:54.55 & 0.054        & 18.259 ± 0.009 & 18.565 ± 0.007 & 22.619 ± 0.017 & 4.360                 \\
sgrc259$^{\dagger}$     & 1.06             & 17:44:41.927 & -29:27:37.77 & 0.056        & 15.829 ± 0.003 & $\cdots$            & $\cdots$            & \textgreater{}7.371   \\
sgrc41$^{\dagger}$      & 2.15             & 17:44:40.137 & -29:28:15.00 & 0.062        & 10.876 ± 0.001 & 13.531 ± 0.001 & 21.861 ± 0.012 & 10.985                \\
sgrc9       & 0.90             & 17:44:40.667 & -29:28:19.13 & 0.072        & 20.562 ± 0.068 & 20.616 ± 0.035 & $\cdots$            & \textgreater{}2.638   \\
sgrc264     & 5.25             & 17:44:43.498 & -29:27:33.11 & 0.074        & 17.204 ± 0.005 & $\cdots$            & $\cdots$            & \textgreater{}5.996   \\
sgrc14      & 7.34             & 17:44:40.438 & -29:28:17.28 & 0.081        & 17.224 ± 0.005 & 17.461 ± 0.004 & 20.482 ± 0.004 & 3.258                 \\
sgrc175     & 1.26             & 17:44:41.354 & -29:27:56.61 & 0.084        & 16.770 ± 0.004 & 18.169 ± 0.005 & 22.346 ± 0.015 & 5.576                 \\
sgrc217     & 0.98             & 17:44:42.327 & -29:27:53.23 & 0.084        & 17.337 ± 0.006 & $\cdots$            & 21.720 ± 0.008 & 4.383                 \\
sgrc48$^{\dagger}$     & 1.09             & 17:44:40.537 & -29:28:14.63 & 0.088        & 12.289 ± 0.001 & $\cdots$            & $\cdots$            & \textgreater{}10.911  \\
sgrc115     & 57.71            & 17:44:41.707 & -29:28:03.42 & 0.089        & 18.578 ± 0.012 & $\cdots$            & $\cdots$            & \textgreater{}4.622   \\
sgrc205     & 11.91            & 17:44:42.184 & -29:27:54.26 & 0.091        & 19.117 ± 0.019 & 19.213 ± 0.010 & 23.206 ± 0.024 & 4.089                 \\
sgrc238     & 0.90             & 17:44:42.541 & -29:27:49.13 & 0.099        & 20.153 ± 0.041 & $\cdots$            & 24.844 ± 0.076 & 4.691                 \\
sgrc10      & 0.54             & 17:44:40.733 & -29:28:18.95 & 0.099        & 18.730 ± 0.014 & 19.576 ± 0.015 & 23.921 ± 0.045 & 5.191                 \\
sgrc196     & 22.57            & 17:44:42.152 & -29:27:54.68 & 0.100        & 19.150 ± 0.020 & 19.706 ± 0.017 & 23.386 ± 0.028 & 4.236                 \\     
                \enddata
            \tablecomments{ALMA ID refers to the ID assigned in \citet{lu20}. The sources marked with daggers are those with a F162M-F480M color index that deviates from the mean by more than 2$\sigma$. 
            $M_{mm}$ represents the estimated core mass from \citet{lu20}.
            Coordinates represent the position of the F480M source. $\Delta\theta$ is the angular separation between the ALMA peak and the F480M source. The magnitudes of sources that are undetected in a given filter are indicated by $\cdots$. Lower limits on color were calculated using the completeness limit for the F162M catalog: $m_{\rm F162M}=23.2$.}
        \end{deluxetable*}

    \subsection{Outflow Knot and Outflow Source Identification}\label{sec:knot_ID}

    \begin{figure*}
        \centering
        \includegraphics[width=0.98\textwidth,keepaspectratio]{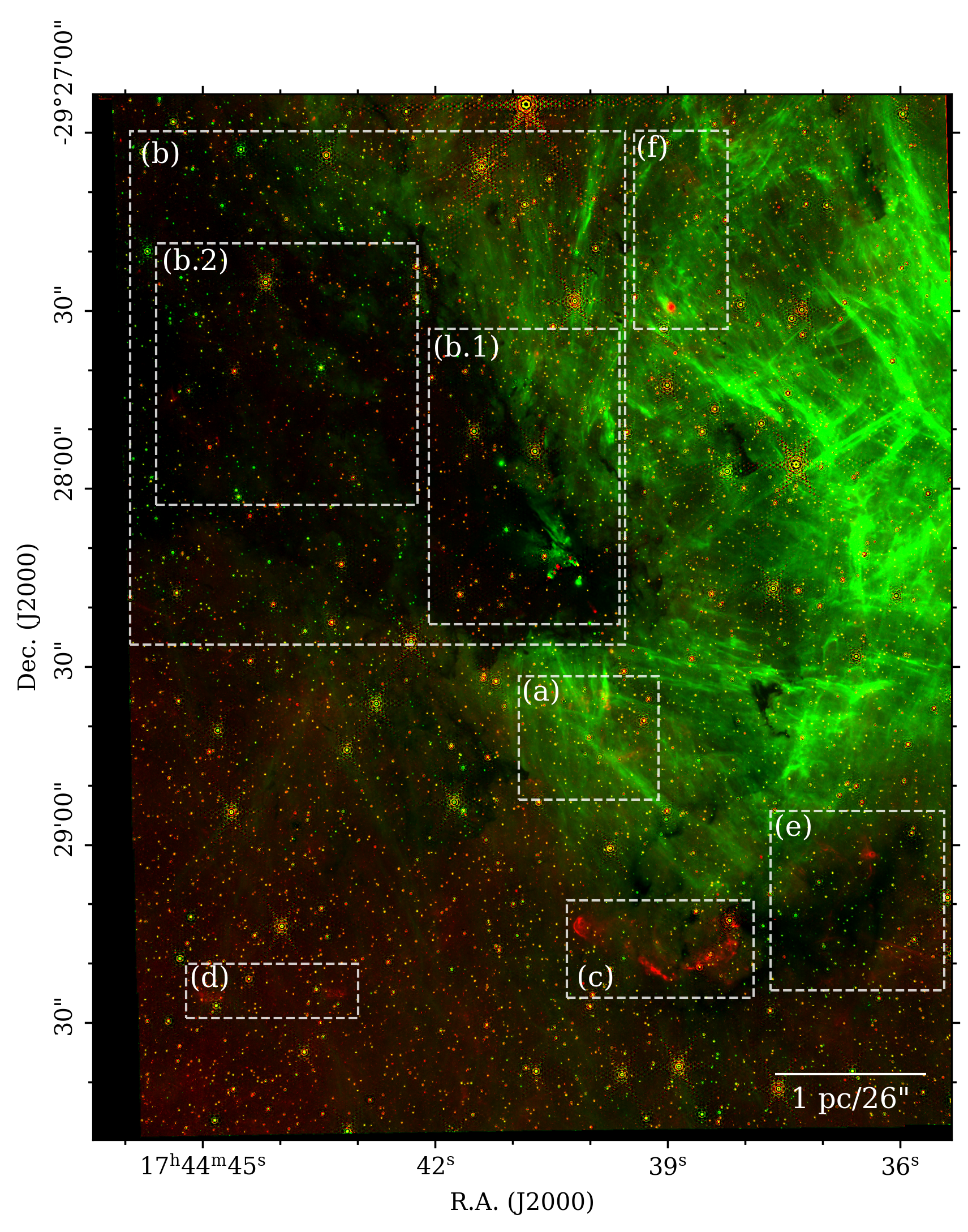}
        \caption{\label{fig:big_diagram} Diagram of the NIRCam data showing the Sgr C cloud and its surroundings. Red represents F470N and green F405N, both continuum-subtracted. The boxes with letters represent cutaways presented in the following figures. N is up and E is left in all panels.}
    \end{figure*}
    \renewcommand{\thefigure}{\arabic{figure}}
    \addtocounter{figure}{-1}
    \begin{figure*}
        \centering
        \includegraphics[width=0.95\textwidth,keepaspectratio]{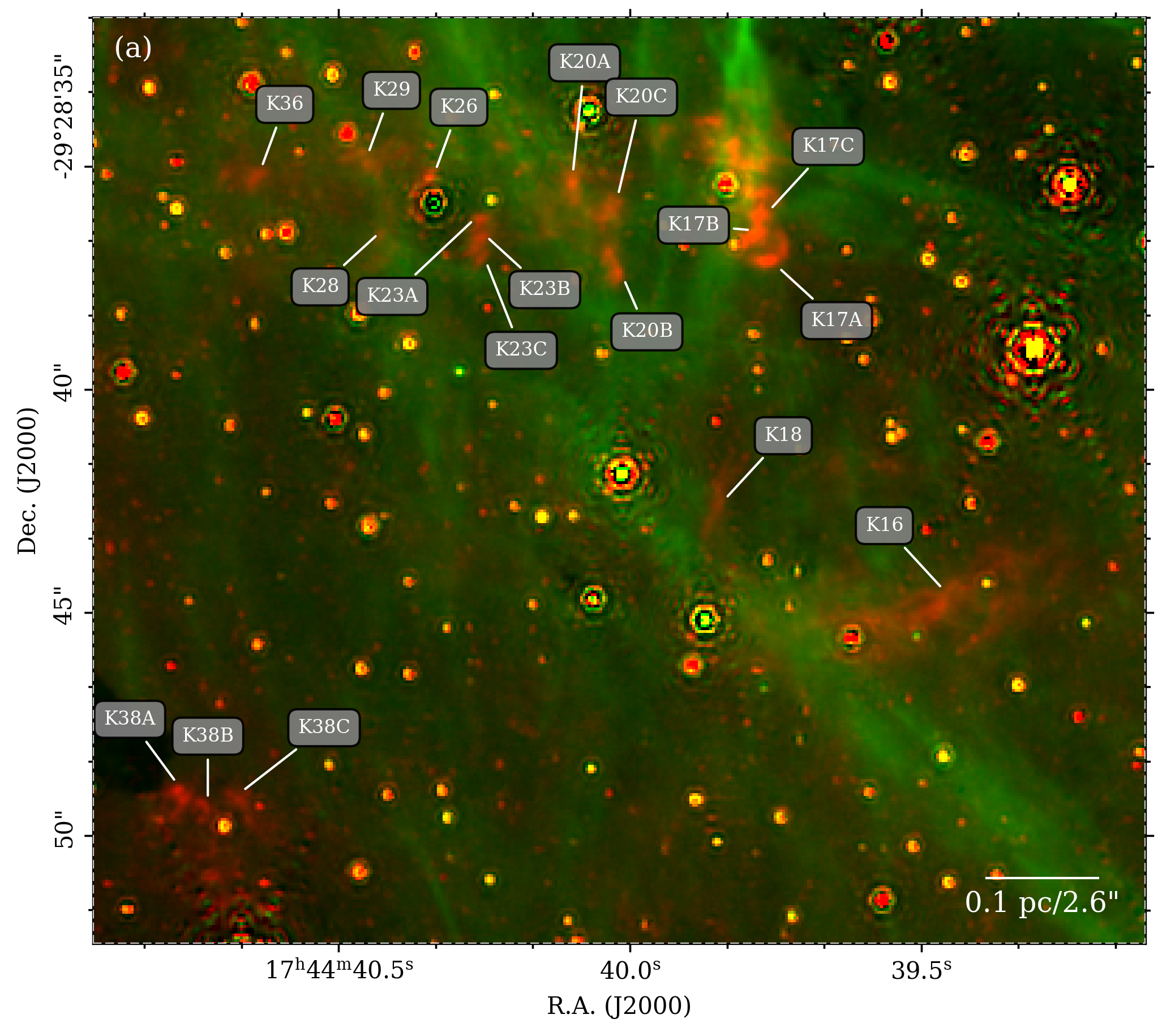}        
        \caption{Continued. (a) Magnification of the region to the south of the main protocluster.}
    \end{figure*}
    \begin{figure*}
        \centering
        \includegraphics[width=0.98\textwidth,keepaspectratio]{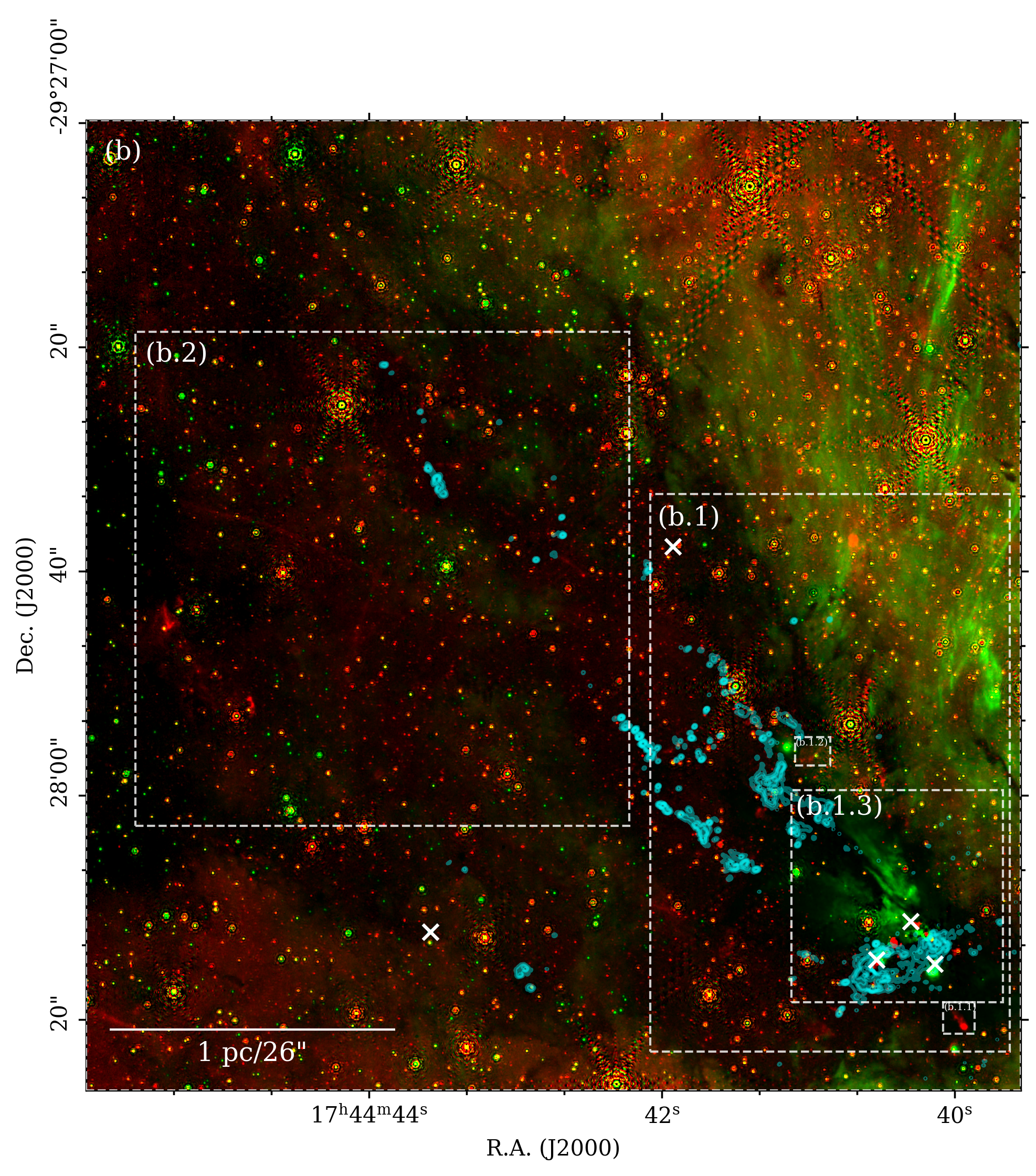}
        \caption{\label{fig:sgrc_cloud_diagram} Diagram of the Sgr C molecular cloud. (b) Magnification of the entire cloud, with ALMA Band 6 continuum \citep{lu20} contours (masked for primary beam response $<0.4$) with 10, 20, 30, 50, 100, 200$\times$rms noise of the image. White crosses mark the positions of the five YSO candidates identified in \S\ref{sec:jwst_alma_crossmatch}.}
    \end{figure*}
    \renewcommand{\thefigure}{\arabic{figure}}
    \addtocounter{figure}{-1}
    \begin{figure*}
        \centering
        \includegraphics[width=0.98\textwidth,keepaspectratio]{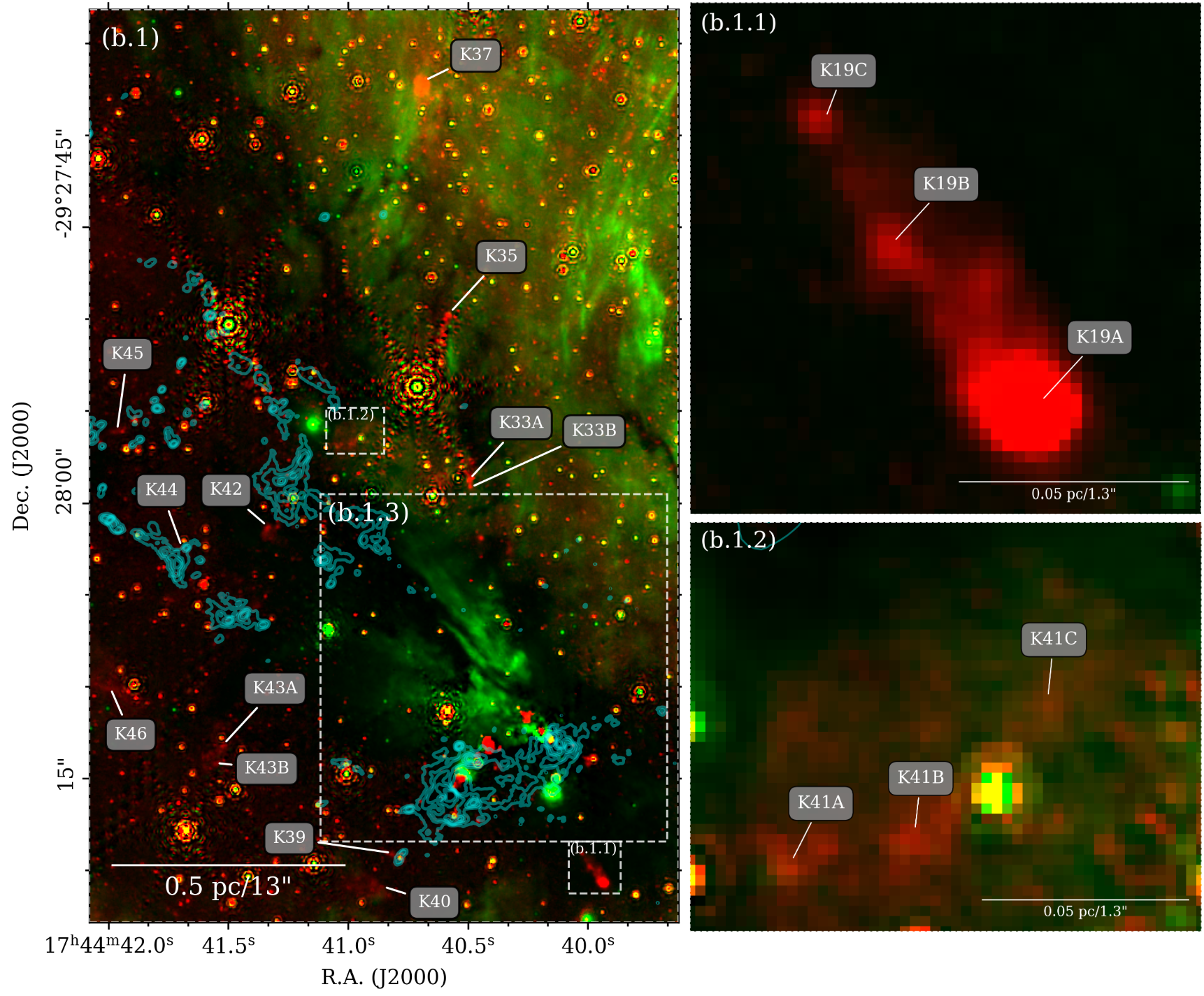}  
        \caption{Continued. (b.1) Magnification of the Sgr C protocluster, with outflow knots labeled. The cyan contours are the same as in the previous panels. (b.1.1) Magnification of the knot 19 complex. (b.1.2) Magnification of the knot 41 complex.}
    \end{figure*}
    \renewcommand{\thefigure}{\arabic{figure}}
    \addtocounter{figure}{-1}
    \begin{figure*}
        \centering
        \includegraphics[width=0.98\textwidth,keepaspectratio]{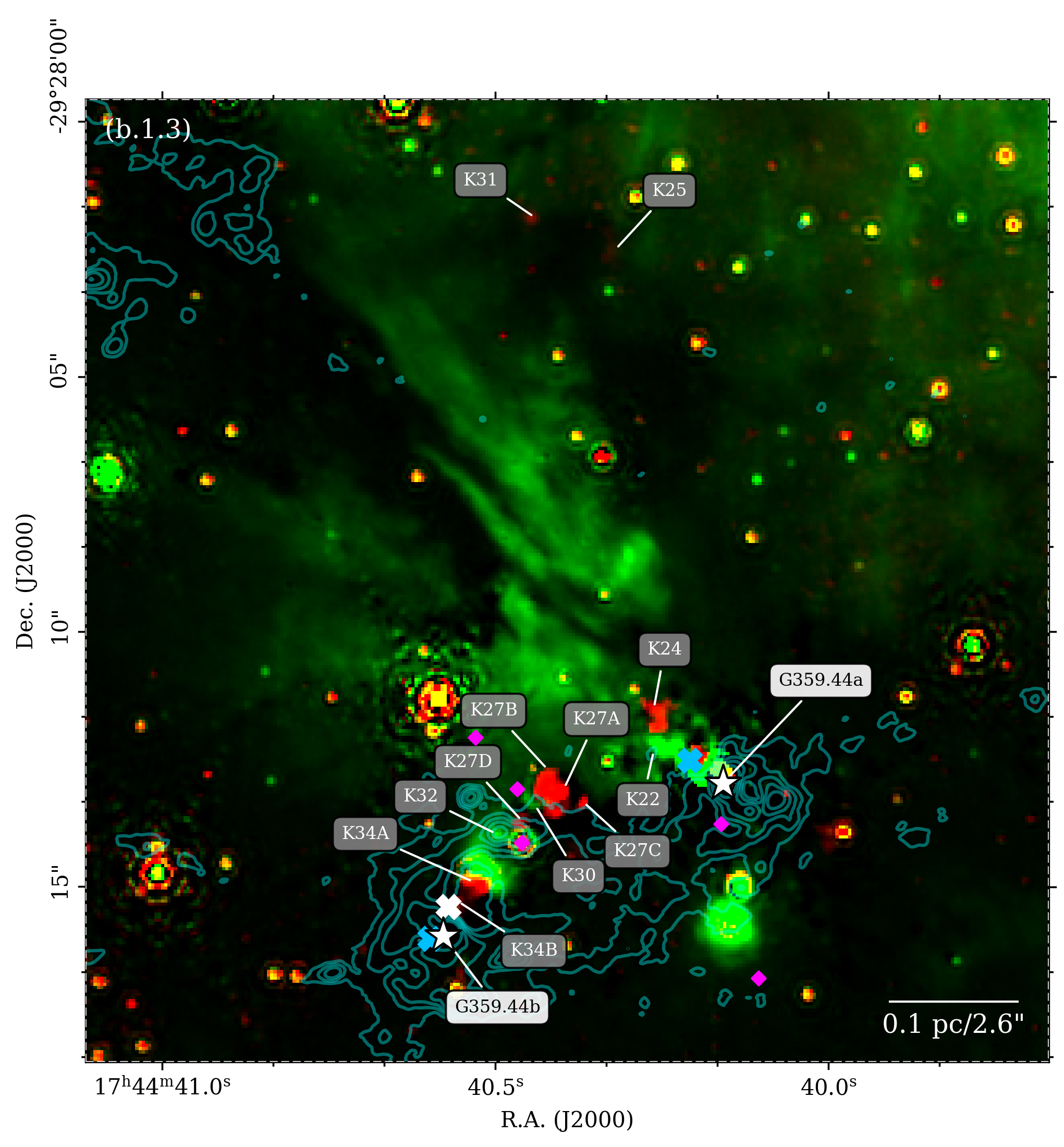}     
        \caption{Continued. (b.1.3) Magnification of the inner part of the Sgr C protocluster, with knots and sources labeled. The FOV is the same as Fig. \ref{fig:sgrc_main_zoom}. Note the ``yellow'' pixels (e.g., around G359.44a) that are the result of saturation in the F480M image used for subtraction. White star symbols mark the positions of the massive protostars discussed in the text. White and blue crosses mark the positions of OH masers from \citet{cotton16} and CH$_3$OH masers from \citet{caswell10}, respectively. Magenta diamonds represent $\mathrm{H_2}$ and $\mathrm{Br{\gamma}}$ line features observed by \citet{kendrew13}.}
    \end{figure*}
    \renewcommand{\thefigure}{\arabic{figure}}
    \addtocounter{figure}{-1}
    \begin{figure*}
        \centering
        \includegraphics[width=0.98\textwidth,keepaspectratio]{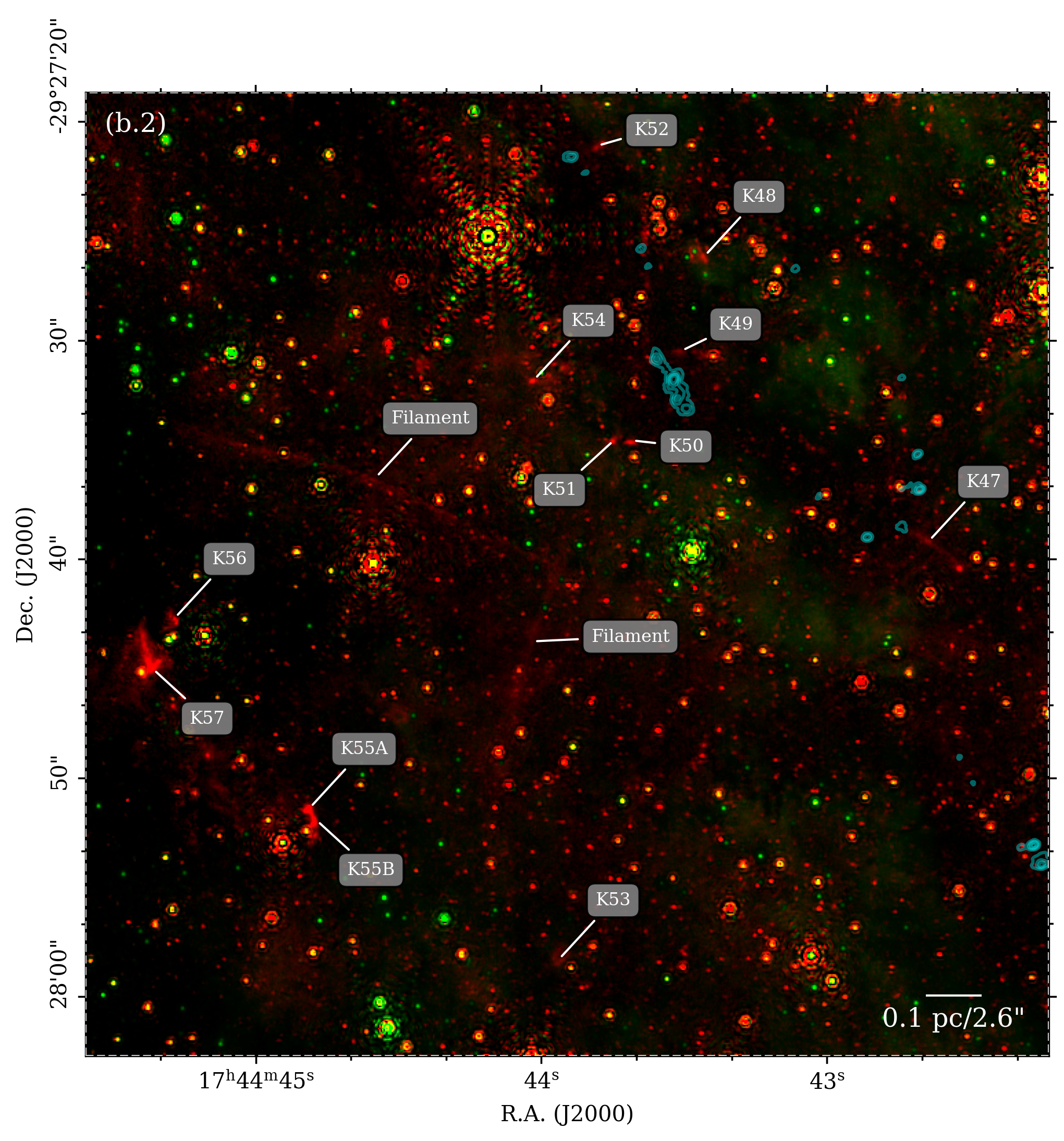}     
        \caption{Continued. (b.2) Upper part of the Sgr C dark cloud, with outflow knots and filamentary structures labeled. The contours are the same as in previous panels.}
    \end{figure*}

    \begin{sidewaystable*}
        \centering
        \vspace{10cm}
        \caption{\label{tab:knot_table} Knot features. Coordinates and peak fluxes are given from the highest value pixel of the knot. Candidate driving sources are given in Tables \ref{tab:crossmatching} and \ref{tab:outflow_source_table}. Knots that are not at a high enough $\sigma$ peak above their local background (see \S\ref{sec:knot_ID}) to constitute a detection (or are unresolvable at a given wavelength) are indicated by $\cdots$. Sources are grouped by driving source and are ordered in ascending order of R.A. for each source. The name of the box showing each knot in Figs. \ref{fig:big_diagram}, \ref{fig:sgrc_cloud_diagram} and \ref{fig:G359.42-0.104} is given.}
        \begin{tabular}{c c c c c c c c c c c c c}
        \hline \hline
        Feature & R.A. & Dec. & $F_{470}$ & $F_{212}$ & $F_{405}$ & Source & P.A. & $\Delta\theta$ & $\sigma_{470}$ & $\sigma_{212}$ & $\sigma_{405}$ & Box\\
        & (J2000) & (J2000) & (MJy/sr) & (MJy/sr) & (MJy/sr)&  & $(^{\circ})$ & (\arcsec) & (MJy/sr) & (MJy/sr) & (MJy/sr) &\\
        \hline
        K19A & 17:44:39.93 & -29:28:20.75 & 49.56 & $\cdots$ & $\cdots$ & G359.44a & 201 & 8.32 & 0.32 & $\cdots$ & $\cdots$& (b.1.1)\\
        K19B & 17:44:39.99 & -29:28:19.83 & 7.53 & $\cdots$ & $\cdots$ & " & 198 & 7.19 & 0.32 & $\cdots$ & $\cdots$& (b.1.1) \\
        K19C & 17:44:40.02 & -29:28:19.08 & 6.68 & $\cdots$ & $\cdots$ & " & 196 & 6.36 & 0.32 & $\cdots$ & $\cdots$& (b.1.1) \\
        K22 & 17:44:40.25 & -29:28:12.18 & $\cdots$ & 13.19 & 155.89 & " & 58 & 1.46 & $\cdots$ & 0.37 & 0.41 & (b.1.3)\\
        K24 & 17:44:40.25 & -29:28:11.54 & 27.09 & $\cdots$ & $\cdots$ & " & 41 & 1.90 & 0.32 & $\cdots$ & $\cdots$ & (b.1.3)\\
        K31 & 17:44:40.45 & -29:28:01.91 & 10.99 & $\cdots$ & $\cdots$ & " & 19 & 11.68 & 0.32 & $\cdots$ & $\cdots$ & (b.1.3)\\
        K33A & 17:44:40.50 & -29:27:58.78 & 13.89 & $\cdots$ & $\cdots$ & " & 17 & 14.86 & 0.32 & $\cdots$ & $\cdots$ & (b.1)\\
        K33B & 17:44:40.49 & -29:27:59.15 & 11.28 & $\cdots$ & $\cdots$ & " & 17 & 14.48 & 0.32 & $\cdots$ & $\cdots$ & (b.1)\\
        K35 & 17:44:40.58 & -29:27:49.80 & 18.65 & $\cdots$ & $\cdots$ & " & 13 & 23.81 & 0.32 & $\cdots$ & $\cdots$ & (b.1)\\
        K37 & 17:44:40.68 & -29:27:37.12 & 28.59 & $\cdots$ & $\cdots$ & " & 11 & 36.49 & 0.32 & $\cdots$ & $\cdots$ & (b.1)\\
        & & & & & & & & & & & & \\
        K27A & 17:44:40.41 & -29:28:13.17 & 88.43 & $\cdots$ & $\cdots$ & G359.44b & 321 & 3.56 & 0.32 & $\cdots$ & $\cdots$ & (b.1.3)\\
        K27B & 17:44:40.42 & -29:28:12.92 & 81.49 & $\cdots$ & $\cdots$ & " & 326 & 3.65 & 0.32 & $\cdots$ & $\cdots$ & (b.1.3)\\
        K27C & 17:44:40.37 & -29:28:13.34 & 27.63 & $\cdots$ & $\cdots$ & " & 313 & 3.82 & 0.32 & $\cdots$ & $\cdots$ & (b.1.3)\\
        K27D & 17:44:40.47 & -29:28:13.75 & 23.24 & $\cdots$ & $\cdots$ & " & 326 & 2.65 & 0.32 & $\cdots$ & $\cdots$ & (b.1.3)\\
        K30 & 17:44:40.45 & -29:28:13.29 & $\cdots$ & 15.17 & 44.88 & " & 327 & 3.17 & $\cdots$ & 0.37 & 0.41 & (b.1.3)\\
        K32 & 17:44:40.48 & -29:28:14.07 & $\cdots$ & 40.01 & 67.29 & " & 325 & 2.29 & $\cdots$ & 0.37 & 0.41 & (b.1.3)\\
        K34A & 17:44:40.54 & -29:28:14.91 & 76.79 & 61.68 & 210.23 & " & 334 & 1.16 & 0.32 & 0.37 & 0.41 & (b.1.3)\\
        K34B & 17:44:40.56 & -29:28:15.42 & 20.95 & $\cdots$ & $\cdots$ & " & 334 & 0.60 & 0.32 & $\cdots$ & $\cdots$ & (b.1.3)\\
        K39 & 17:44:40.82 & -29:28:19.09 & 5.83 & $\cdots$ & $\cdots$ & " & 135 & 4.42 & 0.27 & $\cdots$ & $\cdots$ & (b.1)\\
        K40 & 17:44:40.90 & -29:28:21.01 & 5.72 & $\cdots$ & $\cdots$ & " & 141 & 6.53 & 0.27 & $\cdots$ & $\cdots$ & (b.1)\\
        & & & & & & & & & & & & \\
        K25 & 17:44:40.32 & -29:28:02.56 & 5.32 & $\cdots$ & $\cdots$ & sgrc81* & 2 & 8.69 & 0.32 & $\cdots$ & $\cdots$ & (b.1.3)\\
        K41A & 17:44:41.04 & -29:27:56.95 & 6.81 & $\cdots$ & $\cdots$ & " & 34 & 17.28 & 0.32 & $\cdots$ & $\cdots$ & (b.1.2)\\
        K41B & 17:44:40.99 & -29:27:56.74 & 6.46 & $\cdots$ & $\cdots$ & " & 32 & 17.04 & 0.32 & $\cdots$ & $\cdots$ & (b.1.2)\\
        K41C & 17:44:40.92 & -29:27:55.96 & 5.24 & $\cdots$ & $\cdots$ & " & 28 & 17.29 & 0.32 & $\cdots$ & $\cdots$ & (b.1.2)\\
        K42 & 17:44:41.34 & -29:28:01.32 & 7.36 & $\cdots$ & $\cdots$ & " & 54 & 16.79 & 0.27 & $\cdots$ & $\cdots$ & (b.1)\\
        & & & & & & & & & & & & \\
        K43A & 17:44:41.52 & -29:28:13.22 & 6.88 & $\cdots$ & $\cdots$ & sgrc48* & 84 & 12.99 & 0.27 & $\cdots$ & $\cdots$ & (b.1)\\
        K43B & 17:44:41.54 & -29:28:14.17 & 5.92 & $\cdots$ & $\cdots$ & " & 88 & 13.14 & 0.27 & $\cdots$ & $\cdots$ & (b.1)\\
        K44 & 17:44:41.69 & -29:28:02.38 & 5.11 & $\cdots$ & $\cdots$ & " & 51 & 19.48 & 0.27 & $\cdots$ & $\cdots$ & (b.1)\\
        K46 & 17:44:41.95 & -29:28:10.40 & 5.15 & $\cdots$ & $\cdots$ & " & 77 & 19.00 & 0.27 & $\cdots$ & $\cdots$ & (b.1)\\
        \hline
        \end{tabular}
    \end{sidewaystable*}
    \renewcommand{\thetable}{\arabic{table}}
    \addtocounter{table}{-1}

    Protostellar outflow features were identified in F212N, F405N, and F470N continuum-subtracted images (\S\ref{sec:continuum-sub}). F212N and F470N trace two bright shock-excited H$_2$ lines ($\nu = 1-0\:\mathrm{S(1)}$; $\lambda=2.1218\mathrm{\mu m}$ for the former, and $\nu = 0-0\:\mathrm{S(9)}$; $\lambda=4.6947\mathrm{\mu m}$ for the latter), generally associated with shocked material from both collimated jets and wide-angle outflows from protostars \citep[see, e.g.,][]{ray23}. F405N traces atomic hydrogen recombination emission, particularly the Br$-\alpha$ ($n = 5-4$; $\lambda=4.05\mathrm{\mu m}$) line. Br-$\alpha$, as well as Br-$\gamma$, another line in the Brackett series at $\lambda=2.16\mathrm{\mu m}$, are strongly associated with both accretion and ejection processes in YSOs \citep{muzerolle98c,caratti16, fedriani18}. Recently, \citet{federman23} have shown extended Br-$\alpha$ knots directly associated with outflows from multiple YSOs using JWST-NIRSpec IFU observations.
        
    Individual outflow knots are identified by eye as extended features that stand out among the local noise in the continuum-subtracted images, estimated by sampling the standard deviation in a square aperture capturing blank sky in the vicinity of each knot candidate. We impose a minimum resolvable size for each knot candidate based on the angular resolution, which is $\sim 3$ pixels for each image (F212N, F405N, F470N).
    
    
    The minimum intensity above the local background to constitute a detection across all filters was set at $10\sigma$ due to the high background emission levels throughout the image. The value of $\sigma$ varies between $\sim0.3-0.5$ MJy/sr (see Table \ref{tab:knot_table}) depending on the region and filter. Knots that appear to have distinct substructure are split into parts (e.g., A, B, C, etc.), named in descending order of brightness.
    
    Due to the inherently crowded nature of the CMZ star field, knot candidates were thoroughly cross-matched via visual inspection across all continuum filters to confirm that they are authentic line-emitting structures rather than stars. Table \ref{tab:knot_table} presents a list of all identified knots and the coordinate and flux at their peak pixel.

    Table \ref{tab:knot_table} also provides a prediction of the driving source for most knots, including its position angle 
    and angular separation with respect to this source. Associations between individual outflow knots and driving sources were made first based on inspection of the continuum-subtracted images to identify linear patterns that would indicate strings of knots emanating from a single source. These individual associations are used to reconstruct the outflow axis for each source. This analysis revealed more YSOs via associations with outflow knots that are not in Table \ref{tab:crossmatching} (i.e., they do not possess a reddened IR counterpart or are not in the ALMA Band 6 data); these are presented in Table \ref{tab:outflow_source_table}. Knots within the main Sgr C cloud that cannot be assigned a driving source based on linear associations with other knots are simply assigned the closest source from Table \ref{tab:crossmatching} that satisfy the reddening requirements described in \S\ref{sec:jwst_alma_crossmatch}. These are marked with an asterisk (*) in Table \ref{tab:knot_table}. Knots outside the main Sgr C cloud that still appear to be associated with protostars within the cloud are designated ``SgrC" as their driving source. A similar approach is taken for outflow knots in the star-forming region G359.42-0.104 (see \S\ref{sec:G359.42-0.104}) that cannot be traced back to either of the two identified protostars in that region, G359.42a and G359.42b: these are designated ``G359.42-0.104" as their driving source.
    \begin{sidewaystable*}[t]
        \vspace{10cm}
        \centering
        \caption{\label{tab:knot_table_cont}Continued.}
        \begin{tabular}{c c c c c c c c c c c c c}
        \hline \hline
        Feature & R.A.(J2000) & Dec.(J2000) & $\rm F_{470}$ & $\rm F_{212}$ & $\rm F_{405}$ & Source & P.A. & Sep. & $\sigma_{470}$ & $\sigma_{212}$ & $\sigma_{405}$ & Box\\
        & (J2000) & (J2000) & (MJy/sr) & (MJy/sr) & (MJy/sr)&  & $(^{\circ})$ & (\arcsec) & (MJy/sr) & (MJy/sr) & (MJy/sr) & \\
        \hline
        K45 & 17:44:41.94 & -29:27:56.11 & 7.29 & $\cdots$ & $\cdots$ & sgrc259* & 179 & 18.32 & 0.27 & $\cdots$ & $\cdots$ & (b.1)\\
        K47 & 17:44:42.64 & -29:27:39.15 & 5.06 & $\cdots$ & $\cdots$ & " & 98 & 9.50 & 0.25 & $\cdots$ & $\cdots$ & (b.2)\\
        K48 & 17:44:43.44 & -29:27:26.13 & 6.98 & $\cdots$ & $\cdots$ & " & 59 & 22.92 & 0.25 & $\cdots$ & $\cdots$ & (b.2)\\
        K49 & 17:44:43.52 & -29:27:30.50 & 4.50 & $\cdots$ & $\cdots$ & " & 71 & 22.06 & 0.25 & $\cdots$ & $\cdots$ & (b.2)\\
        K50 & 17:44:43.69 & -29:27:34.65 & 9.04 & $\cdots$ & $\cdots$ & " & 82 & 23.26 & 0.25 & $\cdots$ & $\cdots$ & (b.2)\\
        K51 & 17:44:43.75 & -29:27:34.60 & 8.40 & $\cdots$ & $\cdots$ & " & 82 & 24.01 & 0.25 & $\cdots$ & $\cdots$ & (b.2)\\
        K52 & 17:44:43.79 & -29:27:20.96 & 6.31 & $\cdots$ & $\cdots$ & " & 55 & 29.59 & 0.25 & $\cdots$ & $\cdots$ & (b.2)\\
        K54 & 17:44:44.03 & -29:27:31.86 & 10.20 & $\cdots$ & $\cdots$ & " & 78 & 28.15 & 0.25 & $\cdots$ & $\cdots$ & (b.2)\\
        & & & & & & & & & & & & \\
        K53 & 17:44:43.93 & -29:27:58.33 & 4.45 & $\cdots$ & $\cdots$ & sgrc72* & 18 & 14.60 & 0.25 & $\cdots$ & $\cdots$ & (b.2)\\
        K55A & 17:44:44.81 & -29:27:51.45 & 19.19 & $\cdots$ & $\cdots$ & " & 38 & 26.26 & 0.24 & $\cdots$ & $\cdots$ & (b.2)\\
        K55B & 17:44:44.79 & -29:27:51.94 & 12.07 & $\cdots$ & $\cdots$ & " & 38 & 25.71 & 0.24 & $\cdots$ & $\cdots$ & (b.2)\\
        K56 & 17:44:45.28 & -29:27:42.72 & 7.96 & $\cdots$ & $\cdots$ & "& 37 & 36.92 & 0.24 & $\cdots$ & $\cdots$ & (b.2)\\
        K57 & 17:44:45.37 & -29:27:45.02 & 14.56 & $\cdots$ & $\cdots$ &" & 41 & 35.80 & 0.24 & $\cdots$ & $\cdots$ & (b.2)\\
        & & & & & & & & & & & & \\
        K16 & 17:44:39.47 & -29:28:44.84 & 15.36 & $\cdots$ & $\cdots$ & SgrC & ? & ? & 0.49 & $\cdots$ & $\cdots$ & (a)\\
        K17A & 17:44:39.77 & -29:28:37.13 & 29.71 & $\cdots$ & $\cdots$ & " & ? & ? & 0.49 & $\cdots$ & $\cdots$ & (a)\\
        K17B & 17:44:39.79 & -29:28:36.57 & 26.40 & $\cdots$ & $\cdots$ & " & ? & ? & 0.49 & $\cdots$ & $\cdots$& (a)\\
        K17C & 17:44:39.78 & -29:28:36.13 & 23.36 & $\cdots$ & $\cdots$ & " & ? & ? & 0.49 & $\cdots$ & $\cdots$& (a)\\
        K18 & 17:44:39.85 & -29:28:42.25 & 11.08 & $\cdots$ & $\cdots$ & " & ? & ? & 0.49 & $\cdots$ & $\cdots$& (a)\\
        K20A & 17:44:40.10 & -29:28:35.41 & 18.85 & $\cdots$ & $\cdots$ & " & ? & ? & 0.49 & $\cdots$ & $\cdots$& (a)\\
        K20B & 17:44:40.02 & -29:28:37.40 & 17.77 & $\cdots$ & $\cdots$ & " & ? & ? & 0.49 & $\cdots$ & $\cdots$& (a)\\
        K20C & 17:44:40.03 & -29:28:35.96 & 15.19 & $\cdots$ & $\cdots$ & " & ? & ? & 0.49 & $\cdots$ & $\cdots$& (a)\\
        K23A & 17:44:40.25 & -29:28:36.16 & 15.91 & $\cdots$ & $\cdots$ & " & ? & ? & 0.49 & $\cdots$ & $\cdots$ & (a)\\
        K23B & 17:44:40.25 & -29:28:36.60 & 15.31 & $\cdots$ & $\cdots$ & " & ? & ? & 0.49 & $\cdots$ & $\cdots$ & (a)\\
        K23C & 17:44:40.26 & -29:28:37.10 & 13.04 & $\cdots$ & $\cdots$ & " & ? & ? & 0.49 & $\cdots$ & $\cdots$ & (a)\\
        K26 & 17:44:40.34 & -29:28:35.24 & 20.10 & $\cdots$ & $\cdots$ & " & ? & ? & 0.49 & $\cdots$ & $\cdots$ & (a)\\
        K28 & 17:44:40.43 & -29:28:36.53 & 11.02 & $\cdots$ & $\cdots$ & " & ? & ? & 0.49 & $\cdots$ & $\cdots$ & (a)\\
        K29 & 17:44:40.45 & -29:28:34.96 & 13.68 & $\cdots$ & $\cdots$ & " & ? & ? & 0.49 & $\cdots$ & $\cdots$ & (a)\\
        K36 & 17:44:40.65 & -29:28:35.34 & 13.55 & $\cdots$ & $\cdots$ & " & ? & ? & 0.49 & $\cdots$ & $\cdots$ & (a)\\
        K38A & 17:44:40.77 & -29:28:48.98 & 17.14 & $\cdots$ & $\cdots$ & " & ? & ? & 0.49 & $\cdots$ & $\cdots$ & (a)\\
        K38B & 17:44:40.74 & -29:28:49.28 & 13.21 & $\cdots$ & $\cdots$ & " & ? & ? & 0.49 & $\cdots$ & $\cdots$ & (a)\\
        K38C & 17:44:40.67 & -29:28:49.20 & 12.21 & $\cdots$ & $\cdots$ & " & ? & ? & 0.49 & $\cdots$ & $\cdots$ & (a)\\
        \hline
        \end{tabular}
    \end{sidewaystable*}

    \renewcommand{\thetable}{\arabic{table}}
    \addtocounter{table}{-1}
    \begin{sidewaystable*}[t]
        \vspace{-10cm}
        \centering
        \caption{Continued.}
        \begin{tabular}{c c c c c c c c c c c c c}
        \hline \hline
        Feature & R.A.(J2000) & Dec.(J2000) & $\rm F_{470}$ & $\rm F_{212}$ & $\rm F_{405}$ & Source & P.A. & Sep. & $\sigma_{470}$ & $\sigma_{212}$ & $\sigma_{405}$ & Box\\
        & (J2000) & (J2000) & (MJy/sr) & (MJy/sr) & (MJy/sr)&  & $(^{\circ})$ & (\arcsec) & (MJy/sr) & (MJy/sr) & (MJy/sr) & \\
        \hline
        K2 & 17:44:38.13 & -29:29:13.67 & 54.38 & 17.00 & $\cdots$ & G359.42a & 208 & 1.01 & 0.43 & 0.28 & $\cdots$ & (c)\\
        K15 & 17:44:39.46 & -29:29:13.04 & 23.21 & 10.25 & $\cdots$ & " & 91 & 16.87 & 0.43 & 0.28 & $\cdots$ & (c)\\
        K21A & 17:44:40.13 & -29:29:12.50 & 33.76 & 11.91 & $\cdots$ & " & 89 & 25.64 & 0.43 & 0.28 & $\cdots$& (c.3) \\
        K21B & 17:44:40.18 & -29:29:14.59 & 31.96 & 15.80 & $\cdots$ & " & 94 & 26.28 & 0.43 & 0.28 & $\cdots$& (c.3) \\
        & & & & & & & & & & & & \\
        K9 & 17:44:38.91 & -29:29:19.84 & 11.14 & $\cdots$ & $\cdots$ & G359.42b & 290 & 3.56 & 0.43 & $\cdots$ & $\cdots$ & (c)\\
        K10A & 17:44:39.04 & -29:29:18.88 & 10.32 & $\cdots$ & $\cdots$ & " & 322 & 2.76 & 0.43 & $\cdots$ & $\cdots$ & (c.2)\\
        K10B & 17:44:39.14 & -29:29:18.97 & 9.56 & $\cdots$ & $\cdots$ & " & 349 & 2.13 & 0.43 & $\cdots$ & $\cdots$ & (c.2)\\
        K11 & 17:44:39.04 & -29:29:22.15 & 33.03 & 5.72 & $\cdots$ & " & 236 & 1.97 & $\cdots$ & 0.28 & $\cdots$ & (c.2)\\
        K12A & 17:44:39.12 & -29:29:21.48 & 52.94 & 20.39 & $\cdots$ & " & 238 & 0.80 & 0.43 & 0.28 & $\cdots$ & (c.2)\\
        K12B & 17:44:39.11 & -29:29:21.66 & $\cdots$ & 19.41 & $\cdots$ & " & 233 & 0.98 & $\cdots$ & 0.28 & $\cdots$ & (c.2)\\
        K12C & 17:44:39.14 & -29:29:21.20 & $\cdots$ & 12.80 & $\cdots$ & " & 249 & 0.39 & $\cdots$ & 0.28 & $\cdots$ & (c.2)\\
        K13 & 17:44:39.17 & -29:29:21.06 & 168.48 & 144.64 & $\cdots$ & " & 102 & 0.09 & 0.43 & 0.28 & $\cdots$ & (c.2)\\
        K14 & 17:44:39.32 & -29:29:19.92 & 22.57 & 5.58 & $\cdots$ & " & 60 & 2.31 & 0.43 & 0.28 & $\cdots$ & (c.2)\\
        & & & & & & & & & & & & \\
        K1 & 17:44:37.99 & -29:29:20.92 & 12.76 & $\cdots$ & $\cdots$ & G359.42-0.104 & ? & ? & 0.43 & $\cdots$ & $\cdots$ & (c.1)\\
        K3A & 17:44:38.14 & -29:29:18.20 & 24.65 & 7.14 & $\cdots$ & " & ? & ? & 0.43 & 0.28 & $\cdots$ & (c.1)\\
        K3B & 17:44:38.14 & -29:29:18.14 & $\cdots$ & 6.17 & $\cdots$ & " & ? & ? & $\cdots$ & 0.28 & $\cdots$ & (c.1)\\
        K4 & 17:44:38.19 & -29:29:23.32 & 11.10 & $\cdots$ & $\cdots$ & " & ? & ? & 0.43 & $\cdots$ & $\cdots$ & (c.1)\\
        K5A & 17:44:38.20 & -29:29:16.27 & 28.81 & 15.60 & $\cdots$ & " & ? & ? & 0.43 & 0.28 & $\cdots$ & (c.1)\\
        K5B & 17:44:38.22 & -29:29:16.59 & 24.02 & 5.84 & $\cdots$ & " & ? & ? & 0.43 & 0.28 & $\cdots$ & (c.1)\\
        K5C & 17:44:38.15 & -29:29:16.32 & 19.11 & $\cdots$ & $\cdots$ & " & ? & ? & 0.43 & $\cdots$ & $\cdots$ & (c.1)\\
        K5D & 17:44:38.19 & -29:29:16.74 & $\cdots$ & 6.30 & $\cdots$ & " & ? & ? & $\cdots$ & 0.28 & $\cdots$ & (c.1)\\
        K5E & 17:44:38.19 & -29:29:17.06 & $\cdots$ & 6.54 & $\cdots$ & " & ? & ? & $\cdots$ & 0.28 & $\cdots$ & (c.1)\\
        K6A & 17:44:38.30 & -29:29:18.07 & 19.68 & 8.57 & $\cdots$ & " & ? & ? & 0.43 & 0.28 & $\cdots$ & (c.1)\\
        K6B & 17:44:38.36 & -29:29:18.91 & 18.97 & $\cdots$ & $\cdots$ & " & ? & ? & 0.43 & $\cdots$ & $\cdots$ & (c.1)\\
        K6C & 17:44:38.31 & -29:29:19.20 & 14.14 & $\cdots$ & $\cdots$ & " & ? & ? & 0.43 & $\cdots$ & $\cdots$ & (c.1)\\
        K7A & 17:44:38.49 & -29:29:19.20 & 26.99 & 7.47 & $\cdots$ & " & ? & ? & 0.43 & 0.28 & $\cdots$ & (c.1)\\
        K7B & 17:44:38.57 & -29:29:18.53 & 14.71 & $\cdots$ & $\cdots$ & " & ? & ? & 0.43 & $\cdots$ & $\cdots$ & (c.1)\\
        K8 & 17:44:38.70 & -29:29:20.84 & 25.08 & $\cdots$ & $\cdots$ & " & ? & ? & 0.43 & $\cdots$ & $\cdots$ & (c.1)\\
        
        \hline
        \end{tabular}
    \end{sidewaystable*}

    \begin{table*}[t]
        \centering
        \caption{\label{tab:outflow_source_table} Outflow sources that were not identified in the JWST + ALMA source cross-matching (\S\ref{sec:jwst_alma_crossmatch}), but rather by association to nearby outflow knots in atomic/molecular hydrogen emission (see Table \ref{tab:knot_table}). Coordinates for G359.44a, 
        G359.44b, and G359.42a are given from 2D-Gaussian fitting in the Band 3 continuum image; coordinates for G359.42b, which is not visible in the Band 3 image, are given from the estimated geometric center of its outflow cone visible in the NIRCam image. The associated coordinates, flux, and mass of G359.44a and G359.44b from 1.3 mm Band 6 data analyzed by \citet{lu20} are also given; the masses are given first from Band 3 and second from Band 6, and are separated by a slash ( / ).}           
        \begin{tabular}{c c c c c c c}     
        \hline\hline       
        Source Name &R.A.&Dec.& $S_3$ & $S_6$ & $M_{\rm mm}$ & Area \\
        & (J2000) & (J2000) & (mJy) & (mJy) & ($M_{\odot}$) & (arcsec$^2$) \\
        \hline  
        $^\dagger$G359.44a & 17:44:40.11 & -29:28:13.05 & 19.4 & 102.7 & 629.7/154.8 & 9.8 \\
        $^\ddagger$G359.44b & 17:44:40.55 & -29:28:16.08 & 40.8 & 201.53 & 1321.4/303.7 & 17.7\\
        G359.42a & 17:44:38.17 & -29:29:12.75 & 2.8 & $\cdots$ & 89.7 & 6.6\\
        G359.42b & 17:44:39.15 & -29:29:21.14 & $\cdots$ & $\cdots$ & $\cdots$ & $\cdots$\\
            
        \hline
        \multicolumn{7}{l}{$^\dagger$ RA and Dec from band 6 data: 
        17:44:40.16 -29:28:12.95}\\
        \multicolumn{7}{l}{$^\ddagger$ RA and Dec from band 6 data: 
        17:44:40.60 -29:28:15.91}\\
        \end{tabular}
    \end{table*}         
    Three sources presented in Table \ref{tab:outflow_source_table}, G359.44a, G359.44b, and G359.42a, are also characterized in 3 mm Band 3 ALMA continuum data (§\ref{sec:alma_obs}), and the source coordinates, fluxes, and areas are derived via 2D Gaussian fitting. The fluxes are derived by integrating within $1.5\times$FWHM of the fitted Gaussian ellipse. We also estimate masses for each source using standard assumptions of optically thin thermal emission from dust, i.e.,
    \begin{equation}\label{eq:core_mass}
        M_{\rm mm} = \frac{R_{g/d}S_{\nu}d^2} {\kappa_{\nu}B_{\nu}(T_{\rm dust})} \,,
    \end{equation}
    where $S_{\nu}$ is the dust continuum flux at the frequency $\nu$ (100 GHz for Band 3), $d$ is the distance to Sgr C, which we take to be 8.15 kpc \citep{reid19}, $\kappa_{\nu}$ is the dust opacity per unit mass, which we take to be 0.18 $\mathrm{cm^{2}\:g^{-1}}$ for Band 3 based on the ``OH5'' model \citep{ossenkopf94} that has been extended to longer wavelengths in \citet{young05}, and $B_{\nu}$ is the spectral radiance (Planck function) at the frequency $\nu$ and dust temperature $T_{\rm dust}$, which is taken to be 20 K, similar to other studies of cores in Sgr C \citep[][]{lu20,kinman24}. A caveat to note is that the mass estimate is relatively sensitive to the value of temperature, which is uncertain at these small scales \citep[see][for further discussion]{lu20}; if a value of 50 K is used, the mass estimates would decrease by a factor of $\sim3$.
    A gas-to-dust ratio of 100 was adopted following \citet{ossenkopf94}, i.e. a value of $R_{g/d}=101$ (to account for the additional 1 part in 100 of dust). It is important to note that these values represent primarily the masses of the envelopes of the protostar candidates, rather than the masses of the protostars themselves. Additionally, the mass estimates for Band 3 are likely overestimated due to the strong presence of free-free emission at 3 mm, which we have ignored by assuming all emission around the protostars is thermal emission from dust. Therefore, the Band 3 mass estimates presented in Table \ref{tab:outflow_source_table} should be interpreted as upper limits.

    Figs. \ref{fig:big_diagram}, \ref{fig:sgrc_cloud_diagram}, and \ref{fig:G359.42-0.104} show diagrams of the locations of each identified knot. Appendix \ref{sec:Knot_significance} presents a highly-magnified ``significance map" of each knot, where contours are shown representing increments of the standard deviation of the local background, indicating the statistical significance of each knot. There are also a number of line-emitting objects that appear unrelated to star formation activity. Appendix \ref{sec:misc_mhos} presents a compilation of such miscellaneous Molecular Hydrogen Objects (MHOs) that are not spatially coincident with any star-forming regions. 
    
    \subsection{A Newly-Discovered Star-Forming Region: G359.42-0.104}\label{sec:G359.42-0.104}

\begin{figure*}
        \centering
        \includegraphics[width=0.98\textwidth,keepaspectratio]{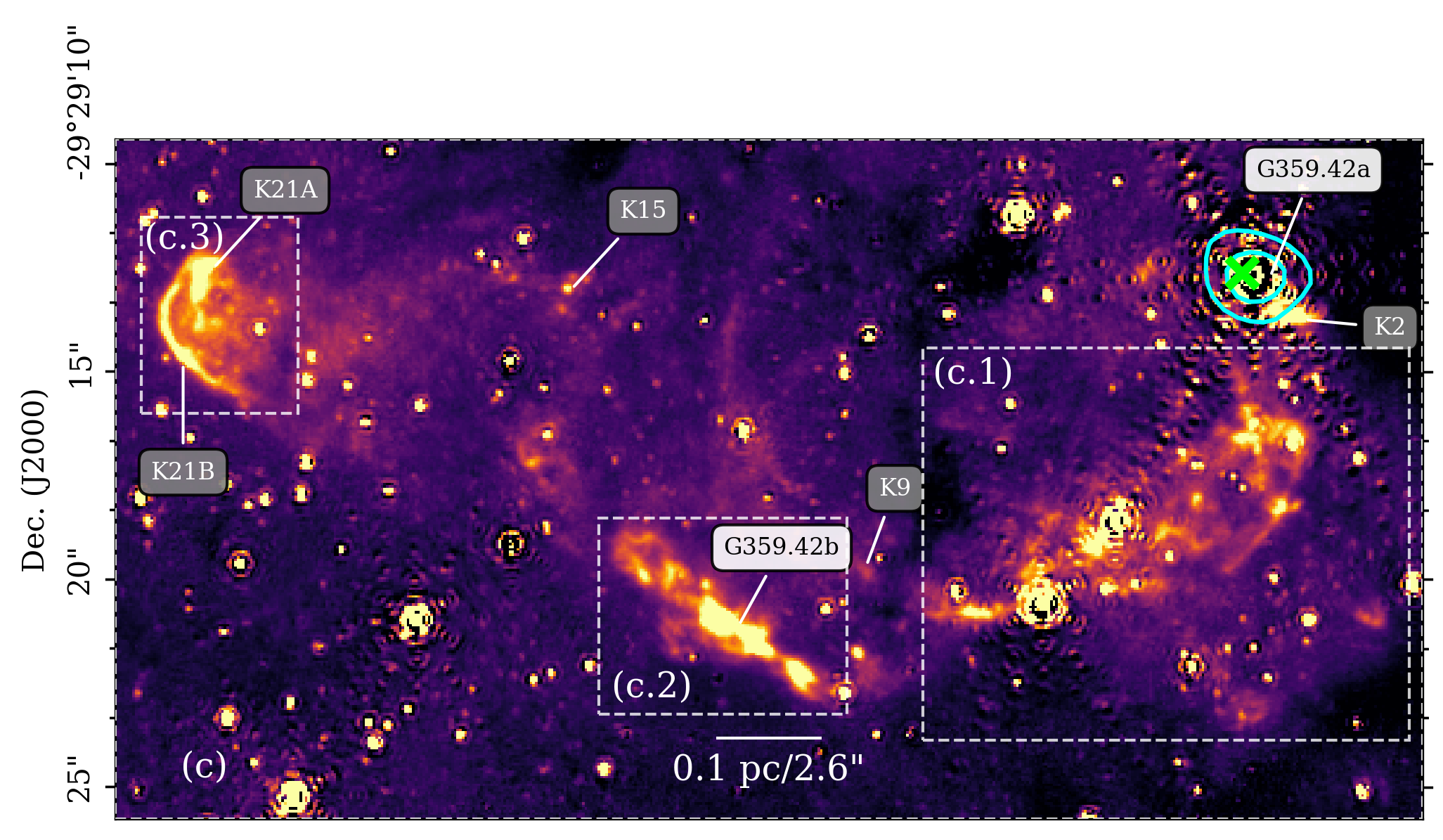}
        \includegraphics[width=0.98\textwidth]{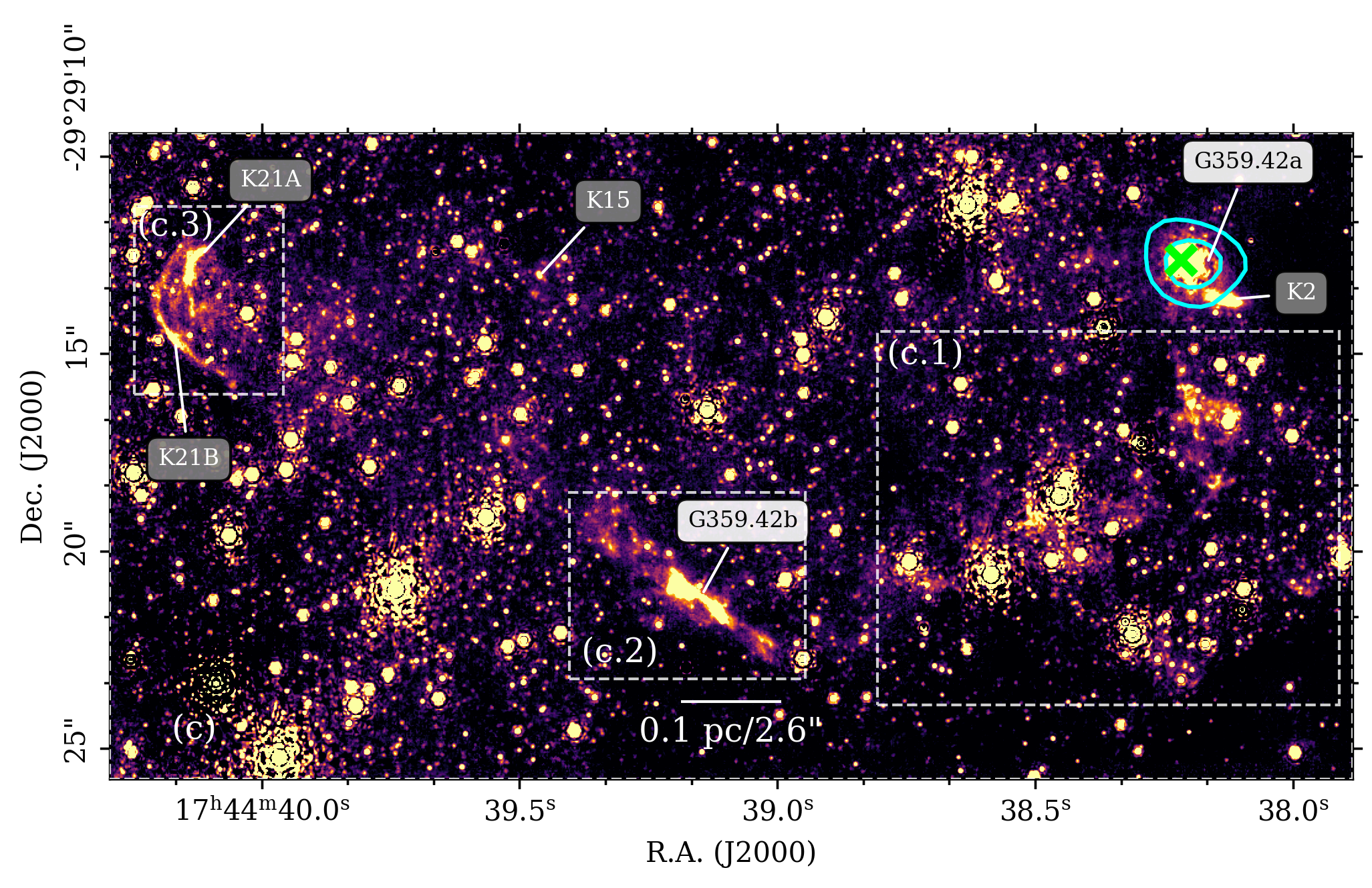}
        \caption{\label{fig:G359.42-0.104}  Diagram of the G359.42-0.104 star-forming complex with knots and sources identified and labeled. The FOV is the same as indicated in the first panel of Fig. \ref{fig:big_diagram}. \textit{Top:} continuum-subtracted F470N data. The cyan contours show 3 mm ALMA continuum data (§\ref{sec:alma_obs}) and represent 5 and 10 $\times$ rms noise of the data. The green cross represents the site of a water maser detected in VLA Band C data \citep{lu19a}. \textit{Bottom:} F212N continuum-subtracted image. The contours and green cross are the same as in the left panel.}
    \end{figure*}

        The JWST data reveal H$_2$ line emission features indicating star formation activity $\sim1\arcmin$ to the South of the main Sgr C protocluster (see Fig. \ref{fig:RGB} and region (c) in Fig. \ref{fig:big_diagram}). We name this star-forming region G359.42-0.104.
        Fig. \ref{fig:G359.42-0.104} shows a magnification of this region, revealing several bright H$_2$ features, including a large bow shock, knot 21, to the East. 
        In general, these shocks tend to ``curve back'' in the direction of their driving source; in this case, knot 21B (shown magnified in box (c.3) of Fig. \ref{fig:G359.42-0.104}) points directly back (along the same line as knot 15) to 
        a driving source, which we name G359.42-0.104-a (hereafter G359.42a).

        G359.42a is the site of a bright ALMA Band 3 continuum peak ($M_{\rm mm}\simeq 90 M_{\odot}$; Table \ref{tab:outflow_source_table}) as well as a water maser emission detected in \citet{lu19a}, labeled by those authors as W16.
        This water maser has a peak velocity close to $0\:\mathrm{kms^{-1}}$, which contrasts with the velocity of the masers in the Sgr C cloud at approximately $-60\:\mathrm{kms^{-1}}$. This difference may indicate that this source is not part of the Sgr C cloud, although the velocity is still consistent with being within the CMZ.
            
        Knot 21A may be an overdensity in flux in the main bow shock, knot 21B, or it may represent another bow shock entirely that points toward the bright emission in the South, magnified in panel (c.2) of Fig. \ref{fig:G359.42-0.104}. We speculate that a driving source, G359.42-0.104-b (hereafter G359.42b), is located in this complex, although it is not associated with an ALMA Band 3 dust continuum peak. Given that the mass limit of the ACES observations is $\sim 10 M_{\odot}$ (Steven Longmore, private communication),
        it is altogether possible that the mm counterpart is simply too low-mass to be detected in the Band 3 data. The coordinates of G359.42b given in Table \ref{tab:outflow_source_table} and indicated in Fig. \ref{fig:G359.42-0.104} are taken from the estimated geometric center of the nebulosity/outflow cone in the F212N image. 
        
        Additionally, many outflow knots are seen in both F470N and F212N to the south-east of G359.42a (see Fig. \ref{fig:G359.42-0.104_cont} (c.1)). These features likely represent shocked outflow knots from further unresolved YSOs in the region, as they do not appear to lie along the outflow axis of either G359.42a or G359.42b. Inspection of the rest of the short-wavelength filters (i.e., F182M, F162M, and F115W; see Fig. \ref{fig:g359.42-0.102_protocluster_zoomin}) reveals a morphology reminiscent of a young, nebulous protocluster, i.e., bright extended emission enshrouding a cluster of (proto)stars, with what may be a cloud to the north-west that is being internally illuminated by the cluster members. Therefore, due to the clustering of line-emitting features in this region, as well as its morphology across the short-wavelength filters, we predict that this is the most likely location of additional YSOs in G359.42-0.104, although it is difficult to predict their exact positions. 
        


        It is uncertain whether this region is at the galactocentric distance or if it is foreground. Although there are several stars with distance estimates from Gaia \citep{gaia2023} in the vicinity of the cluster shown in Fig. \ref{fig:g359.42-0.102_protocluster_zoomin}, it is impossible to disentangle foreground stars from authentic cluster members, especially since there is no noticeable enhancement of Gaia stars in the cluster region. Reddening/extinction measurements are similarly uncertain due to an inability to reliably differentiate cluster members in the crowded field. Proper motions, which will be available in the coming months from the GALACTICNUCLEUS survey \citep{nogueras-lara18}, will be useful in discerning the movement of the cluster members to compare with typical CMZ kinematics.
        
        Therefore, we tentatively adopt a distance to the this region as the galactocentric distance of 8.15 kpc \citep{reid19}, placing G359.42-0.104 in the CMZ until further evidence is available. Using this distance, SED fitting was conducted on the protostar G359.42a via similar methods as described in \S\ref{sec:SED_fitting}, but with a single $3.5\arcsec$ aperture that was algorithmically fit to the source in the SOFIA 37 $\mathrm{\mu m}$ image \citep[see][for further details]{fedriani23a}. This aperture was used for photometry across all wavelengths. An 8-panel figure showing G359.42a from the NIR to FIR, along with its SED and 2d parameter space plot, are shown in Fig. \ref{fig:SED_second_p}.

    \renewcommand{\thefigure}{\arabic{figure}}
    \addtocounter{figure}{-1}
    \begin{figure*}
        \centering
        \includegraphics[width=0.49\textwidth,keepaspectratio]{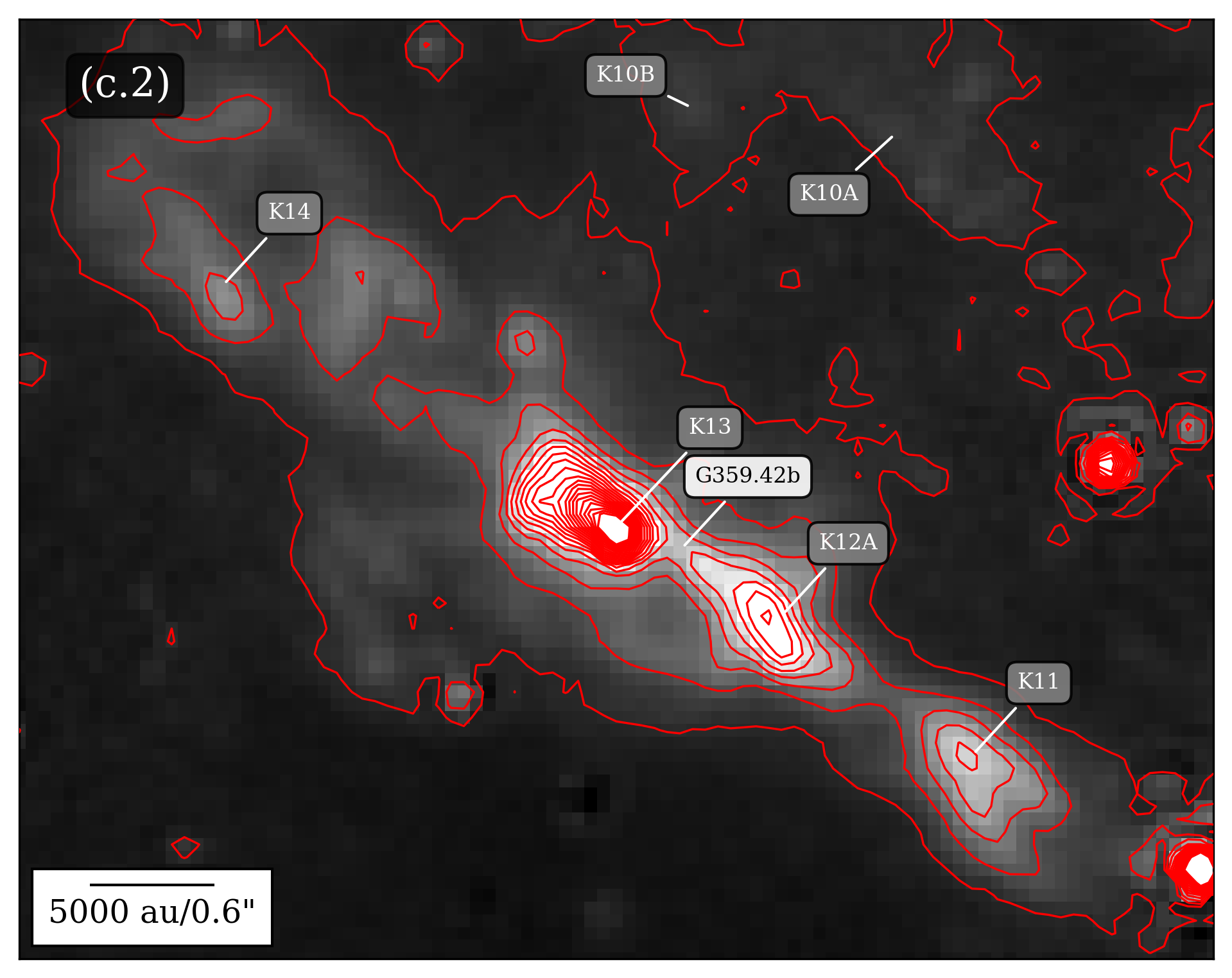}
        \includegraphics[width=0.48\textwidth,keepaspectratio]{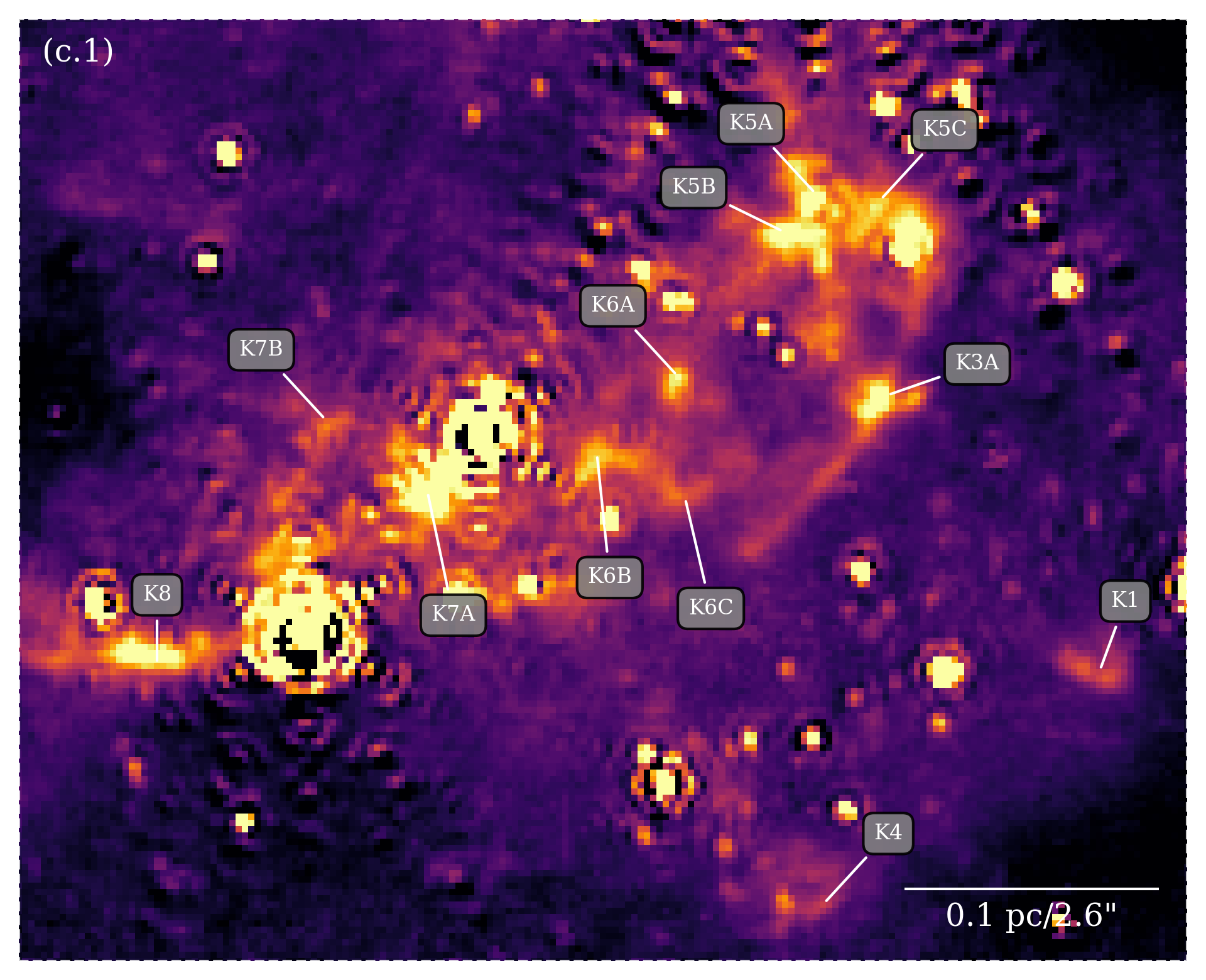}

        \includegraphics[width=0.49\textwidth,keepaspectratio]{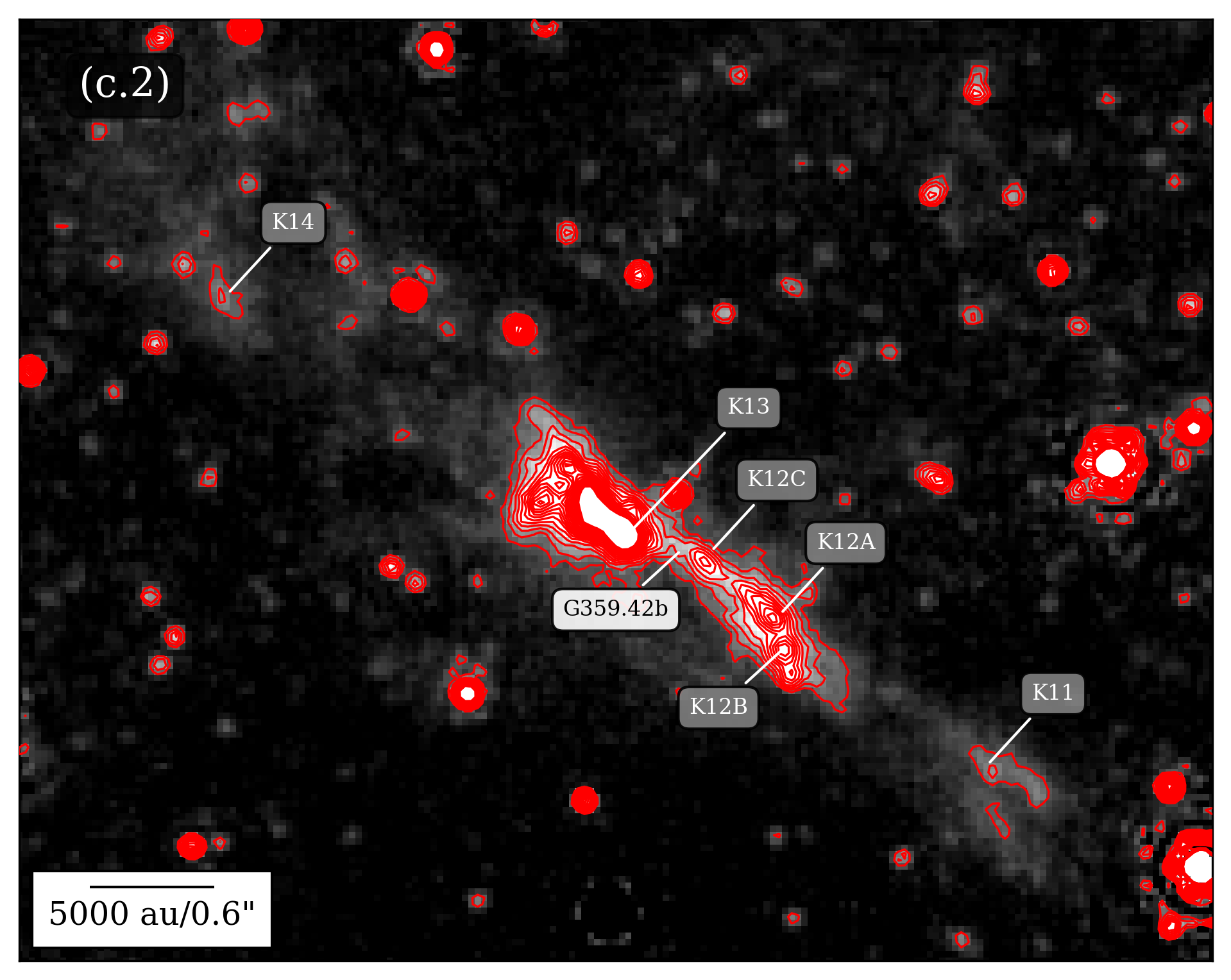}
        \includegraphics[width=0.48\textwidth,keepaspectratio]{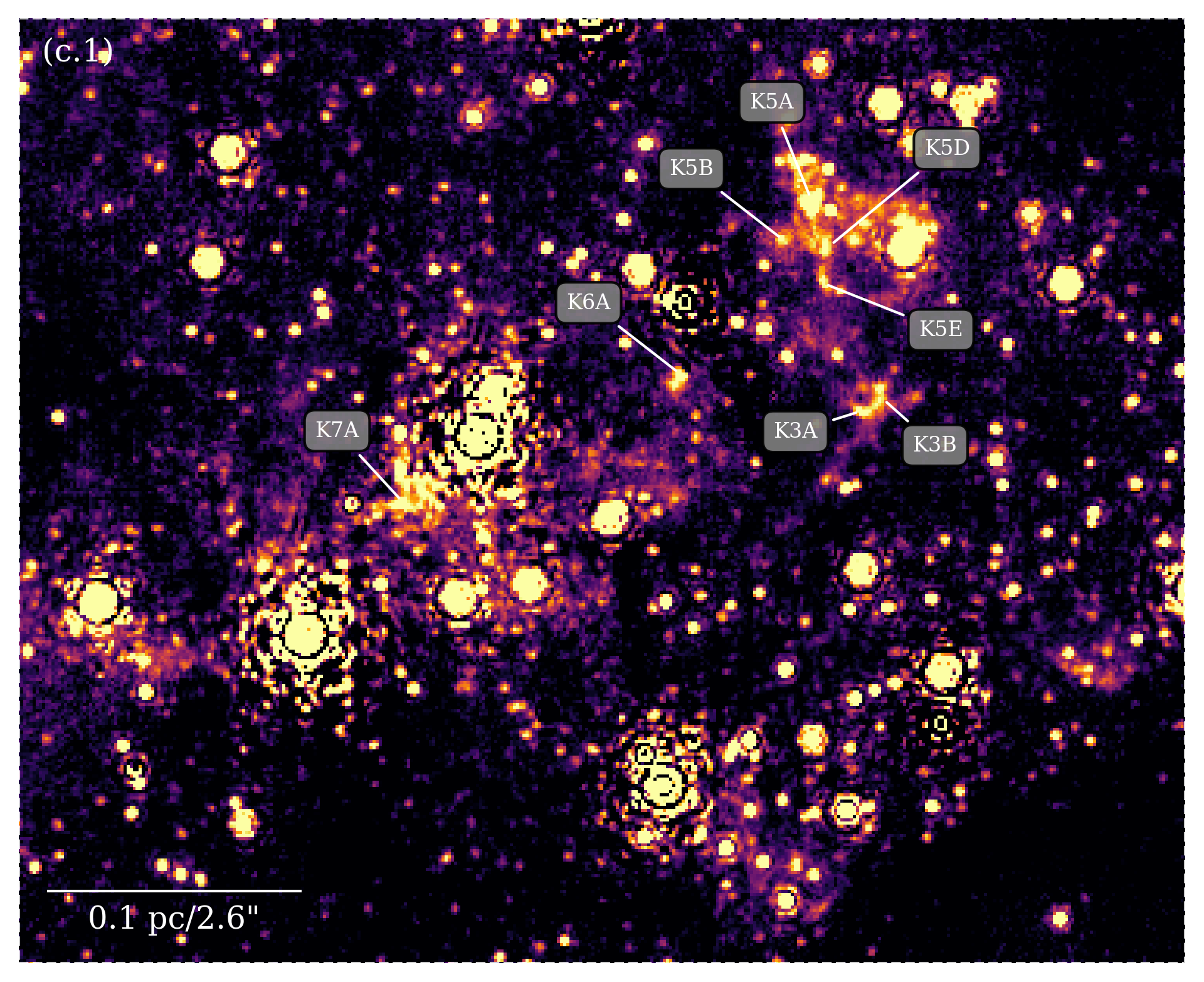}
        
        \caption{\label{fig:G359.42-0.104_cont} Continued. \textit{Left:} Magnification of the G359.42b complex as a ``significance map'' in the style of Appendix \ref{sec:Knot_significance}; contours in F470N (top) represent 15 to 100 $\sigma$ in steps of 5 $\sigma$ above the local background; contours in F212N (bottom) represent 10 to 100 $\sigma$ in steps of 5 $\sigma$ above the local background. \textit{Right:} Magnification of the south-west region of G359.42-0.104, with several bright line-emitting features labeled in both F470N (top) and F212N (bottom).}
    \end{figure*}

    \begin{figure*}
        \centering
        \includegraphics[width=0.49\textwidth,keepaspectratio]{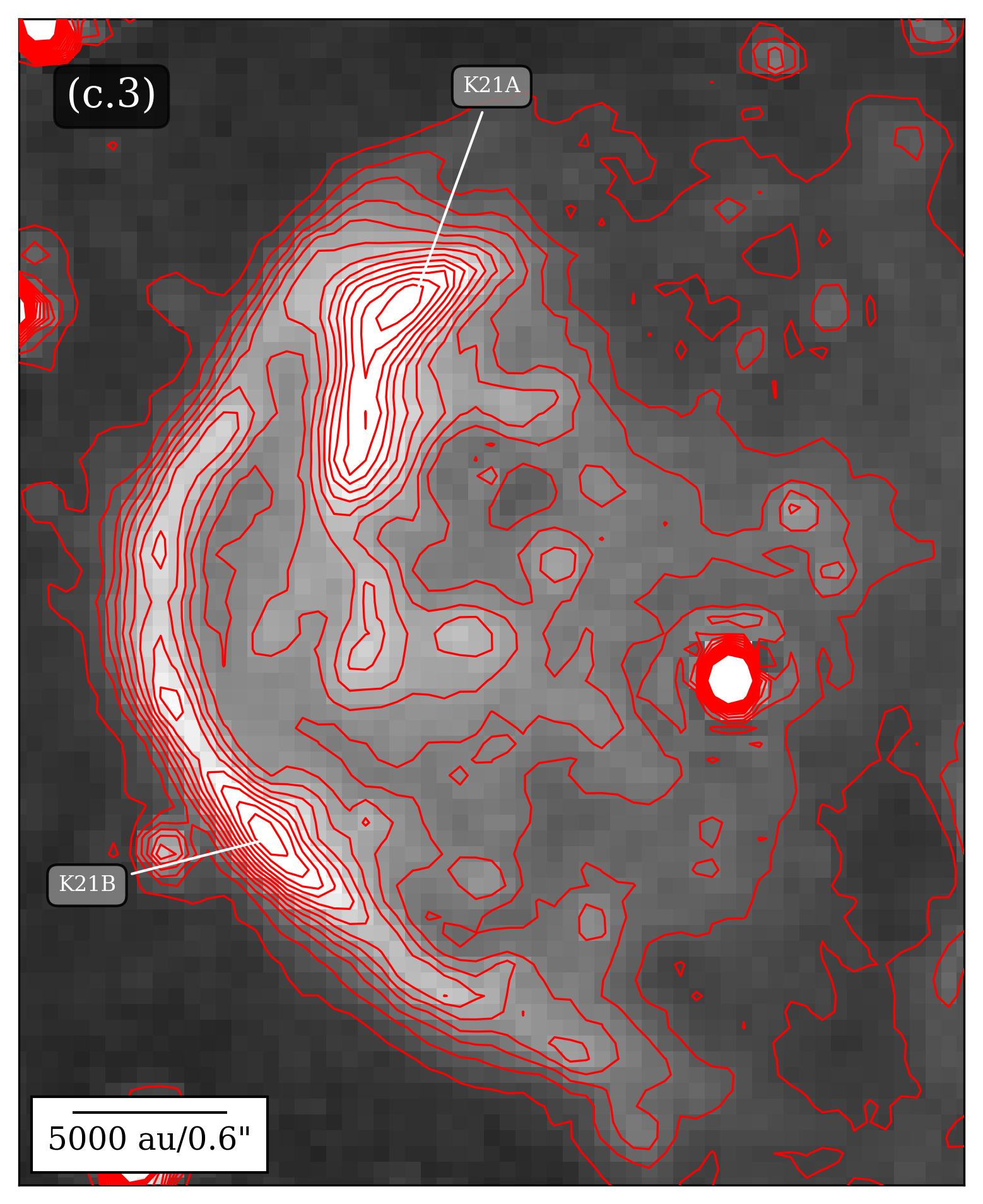}
        \includegraphics[width=0.48\textwidth,keepaspectratio]{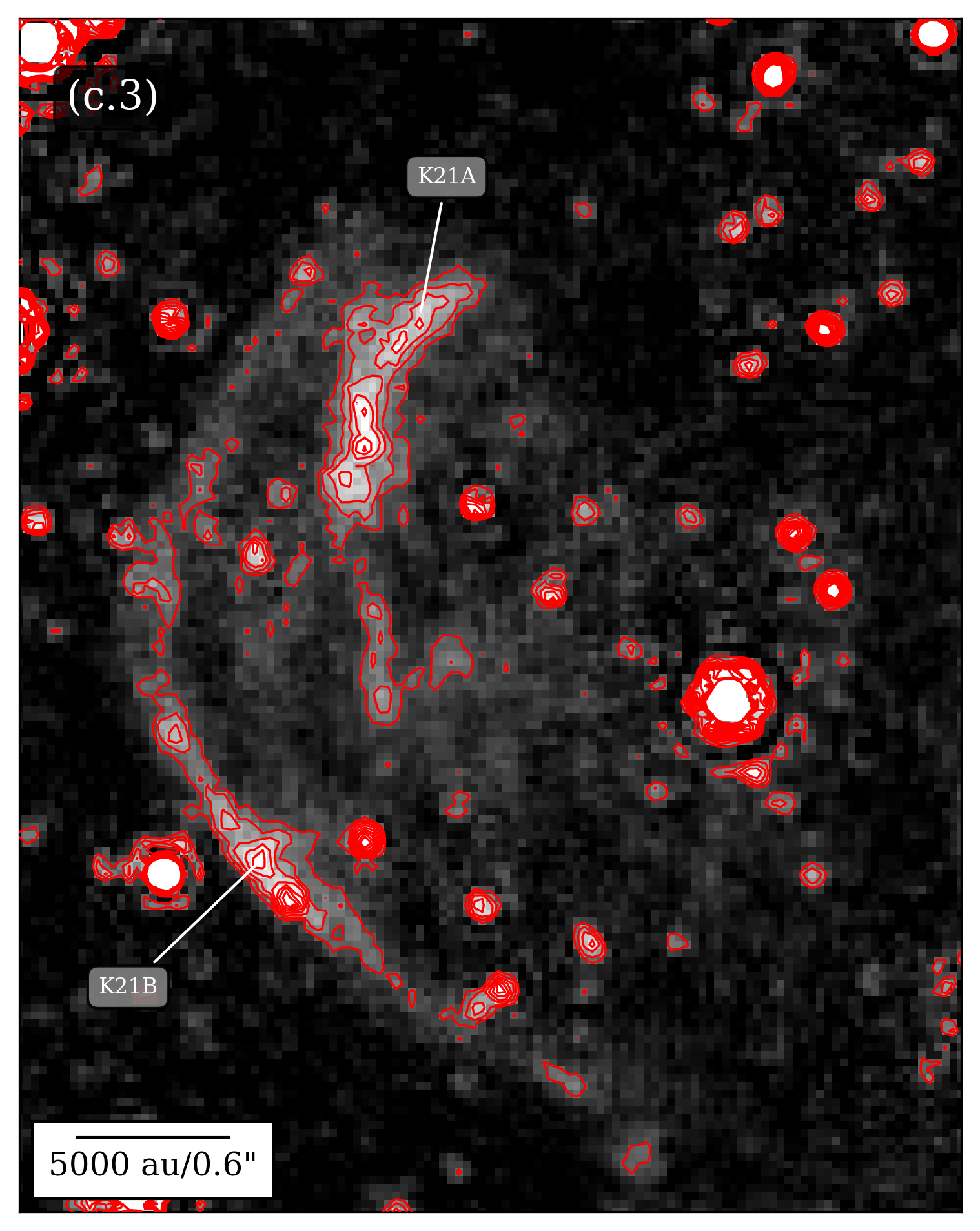}
        \renewcommand{\thefigure}{\arabic{figure}}
        \addtocounter{figure}{-1}
        \caption{\label{fig:G359.42-0.104_cont2} Continued. Magnification of the two prominent bow shocks in G359.42-0.104, knots 21A and 21B, with the same contours in F470N (left) and F212N (right) as in panel (c.2).}
    \end{figure*}
    
    \begin{figure*}
        \centering
        \includegraphics[width=0.495\textwidth,keepaspectratio]{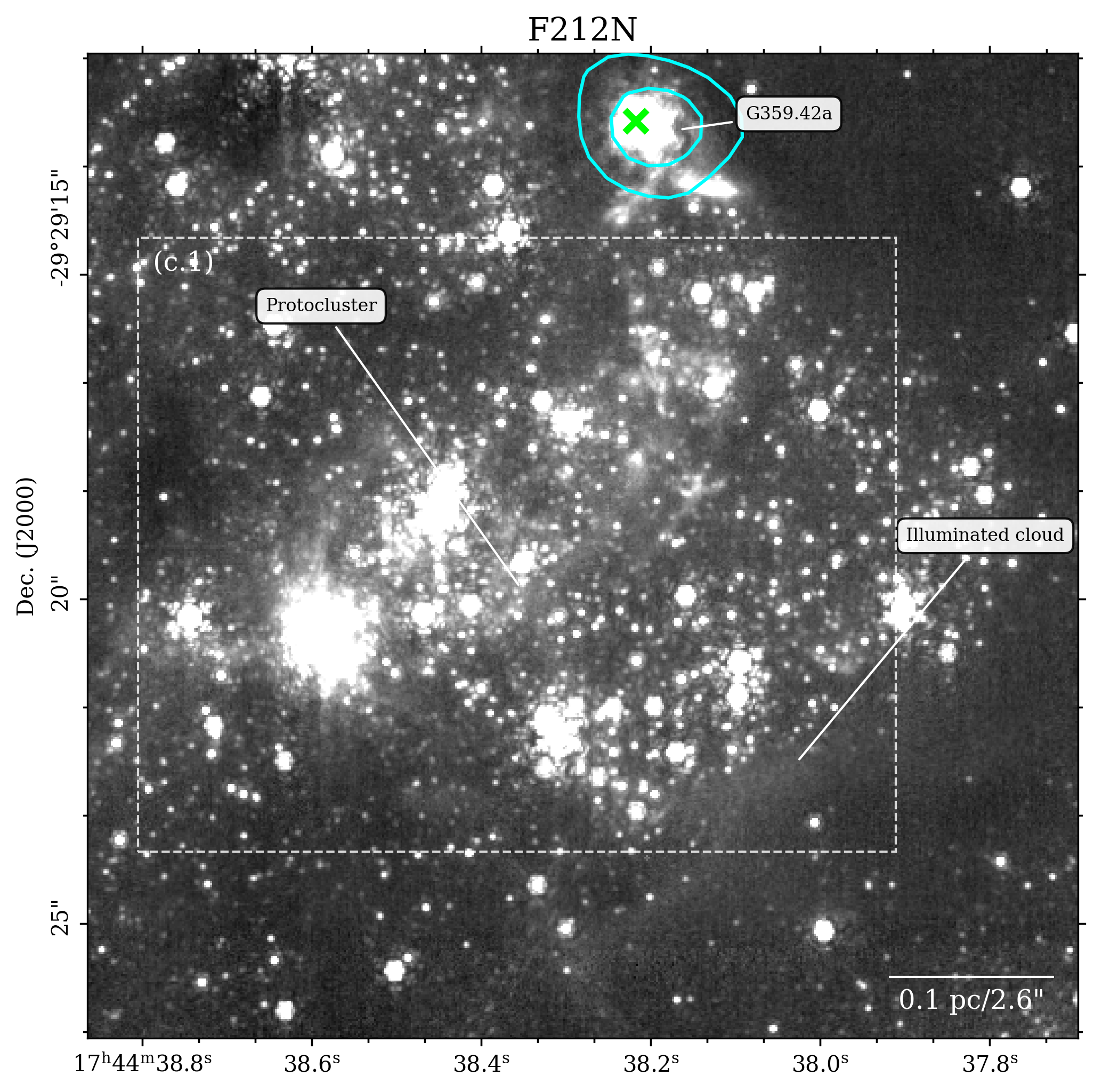}
        \includegraphics[width=0.48\textwidth,keepaspectratio]{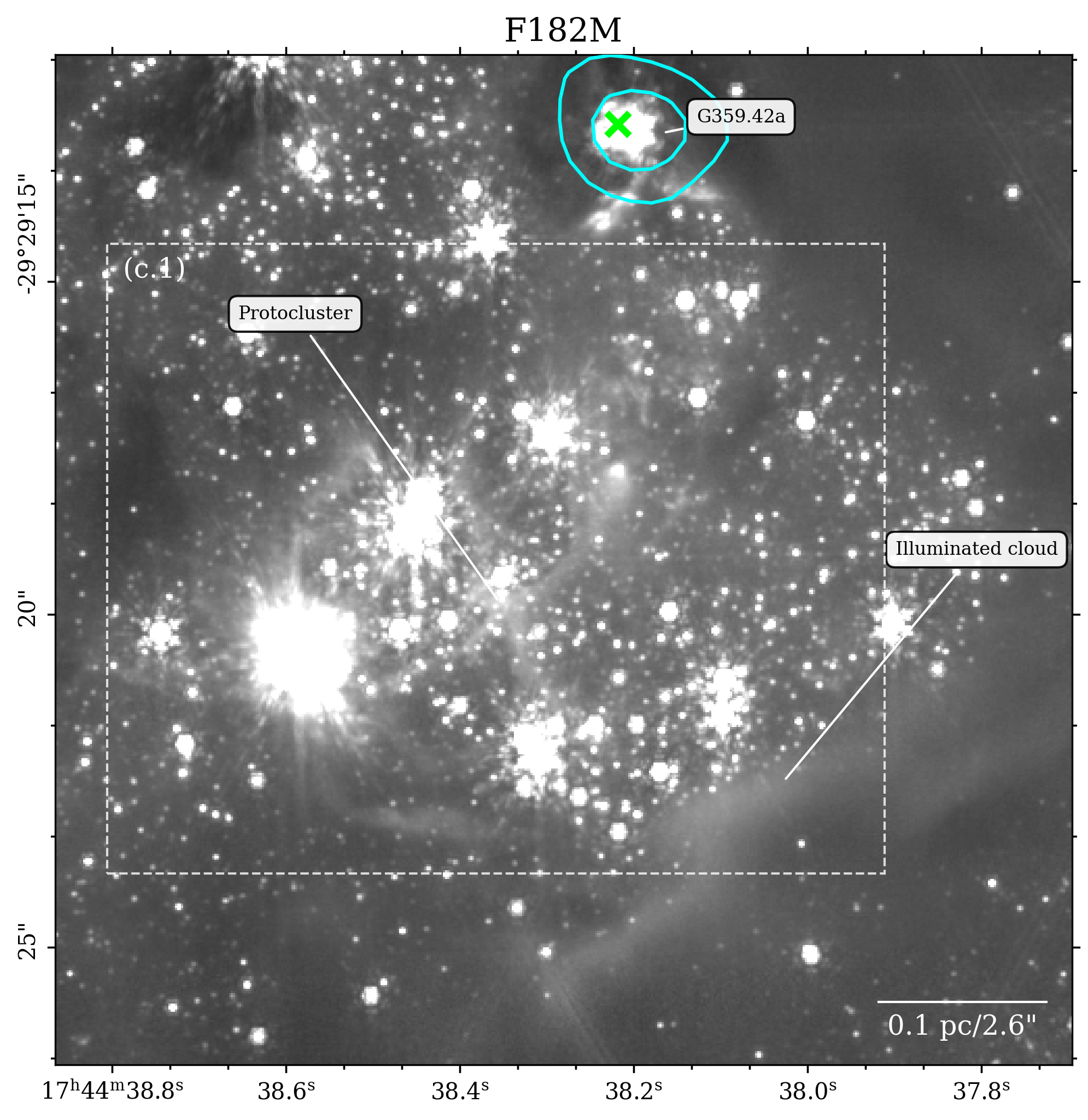}
        \includegraphics[width=0.495\textwidth,keepaspectratio]{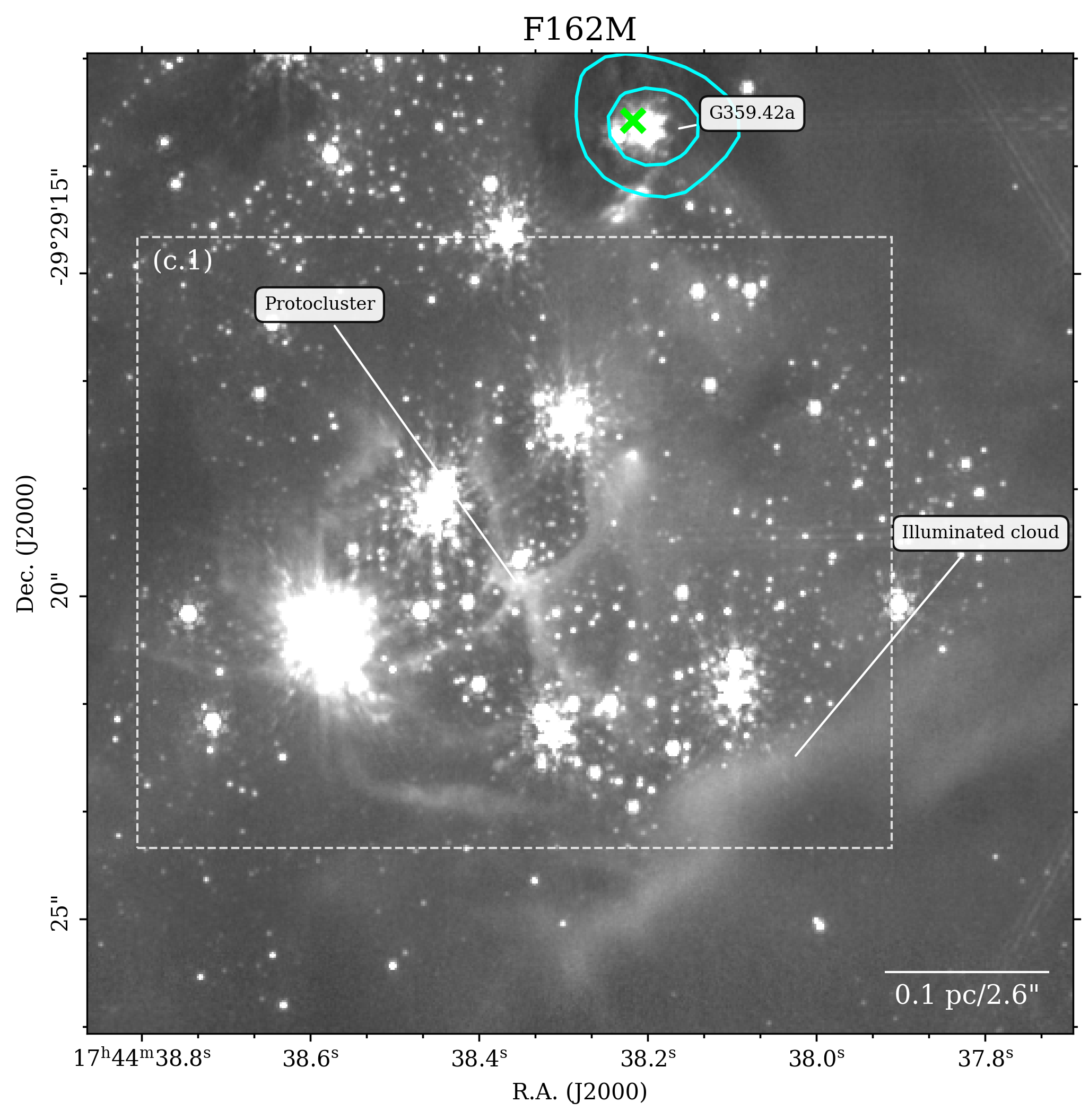}
        \includegraphics[width=0.48\textwidth,keepaspectratio]{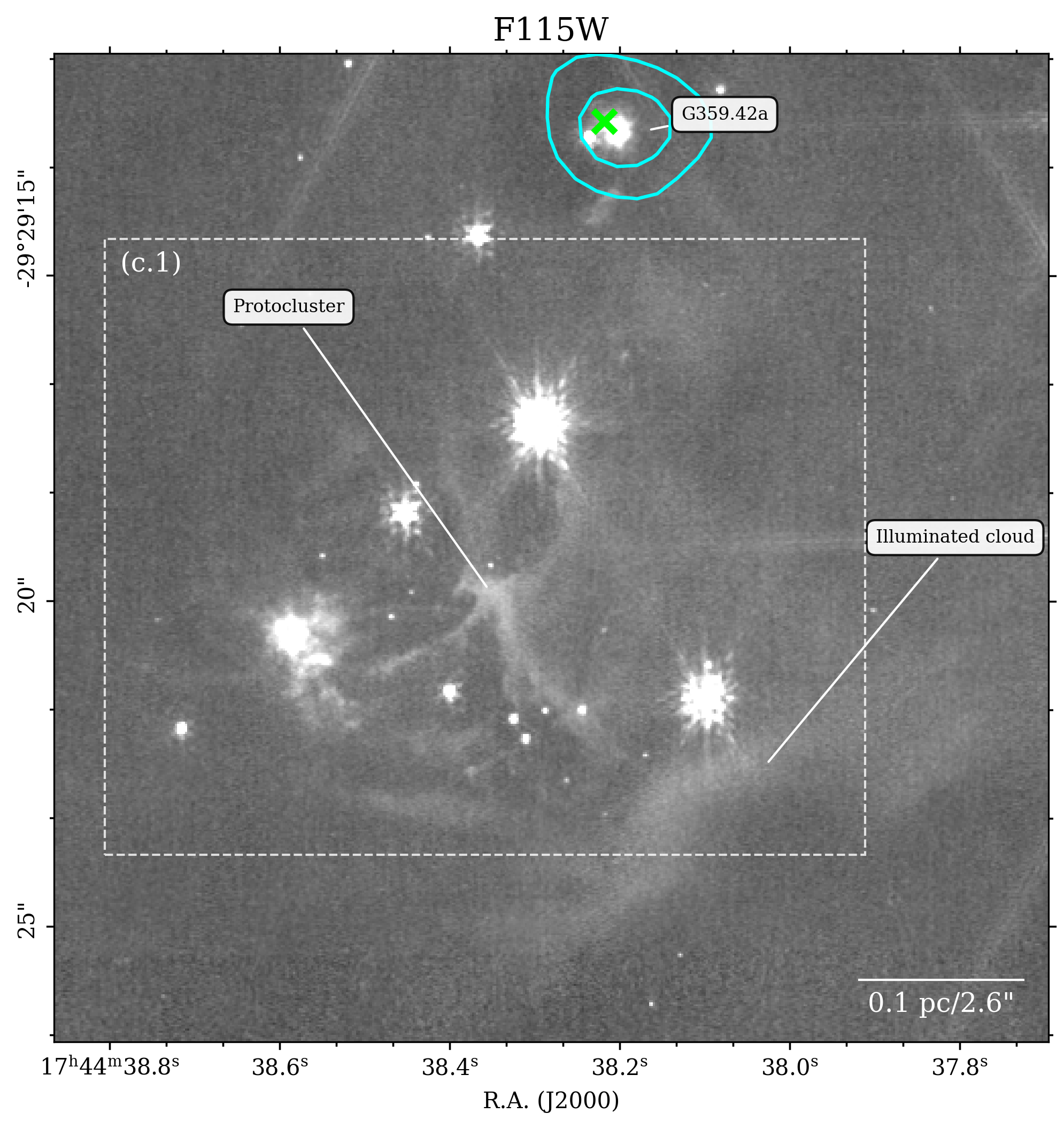}
        \caption{\label{fig:g359.42-0.102_protocluster_zoomin} Magnification of the protocluster in G359.42-0.104. The cyan contours show 3 mm ALMA continuum data (§\ref{sec:alma_obs}) and represent 5 and 10 $\times$ rms noise of the data. The green cross represents the site of a water maser detected in VLA Band C data \citep{lu19a}. The box (c.1) from Fig. \ref{fig:G359.42-0.104} is shown for reference. The protocluster candidate and illuminated cloud discussed in the text are indicated.}
    \end{figure*}

    \begin{figure*}
        \centering
        \includegraphics[width=0.98\textwidth,height=0.59\textheight,keepaspectratio]{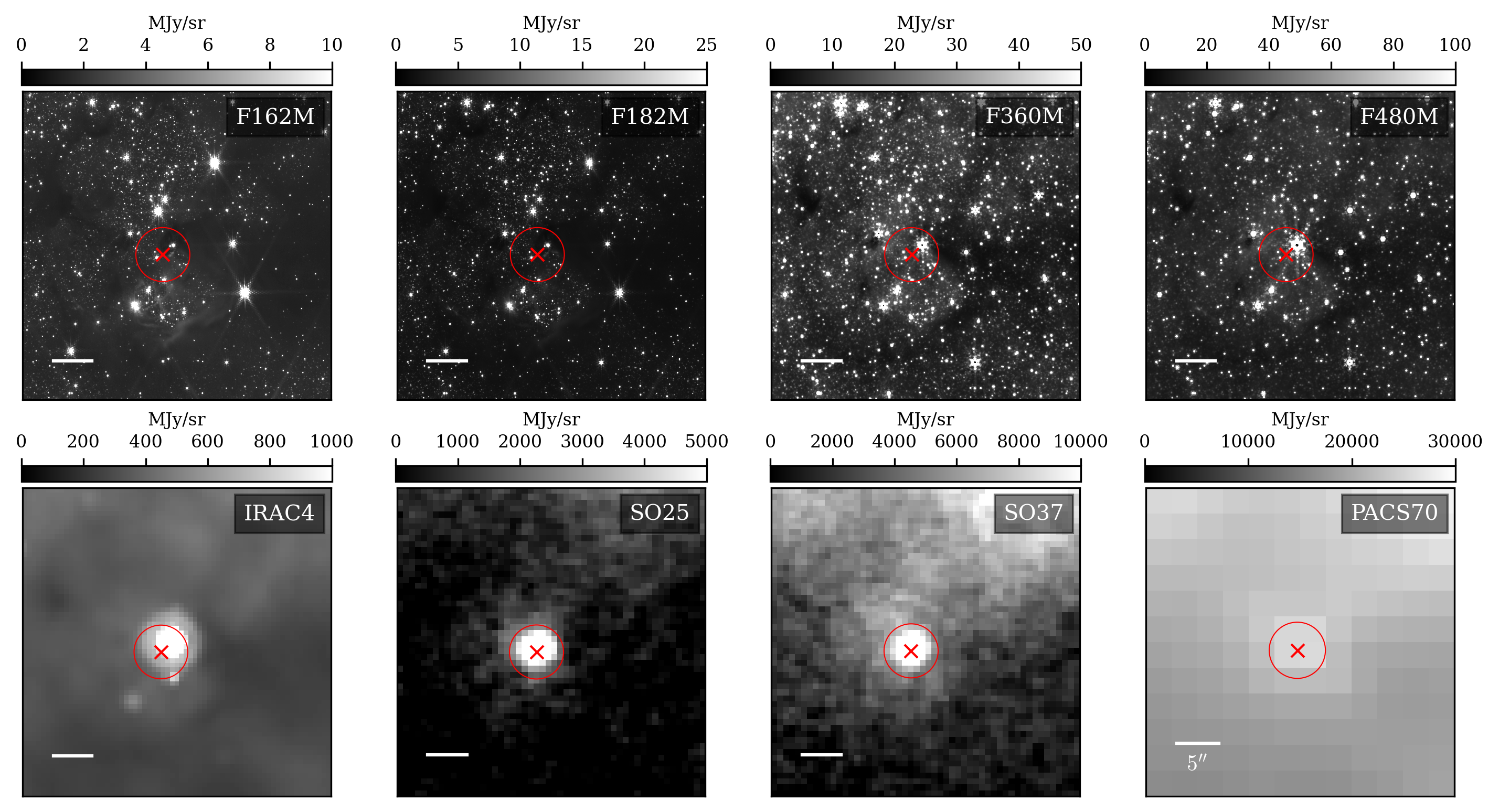}
        \includegraphics[width=0.35\textwidth,height=0.39\textheight,keepaspectratio]{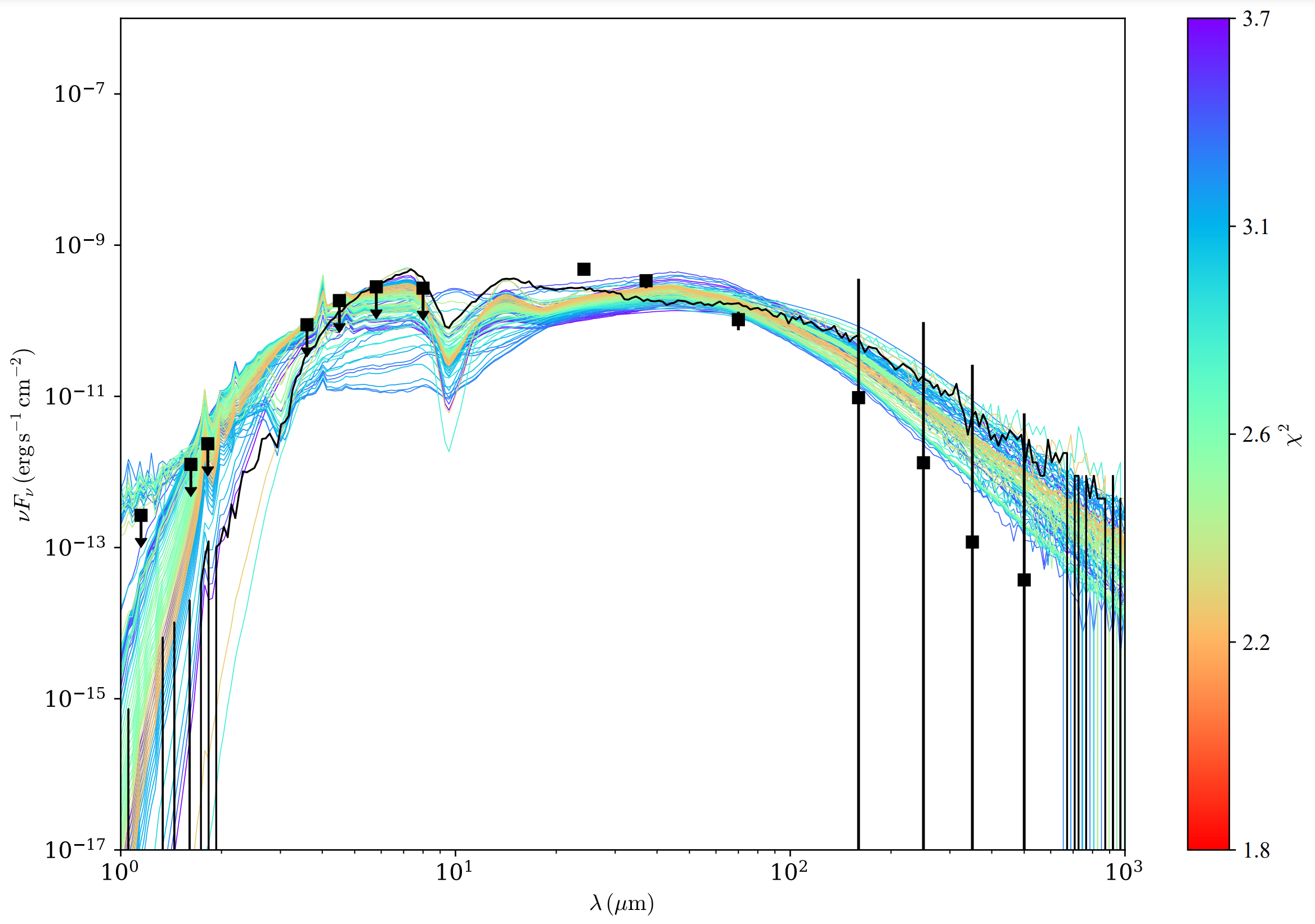}
        \includegraphics[width=0.60\textwidth,height=0.7\textheight,keepaspectratio]{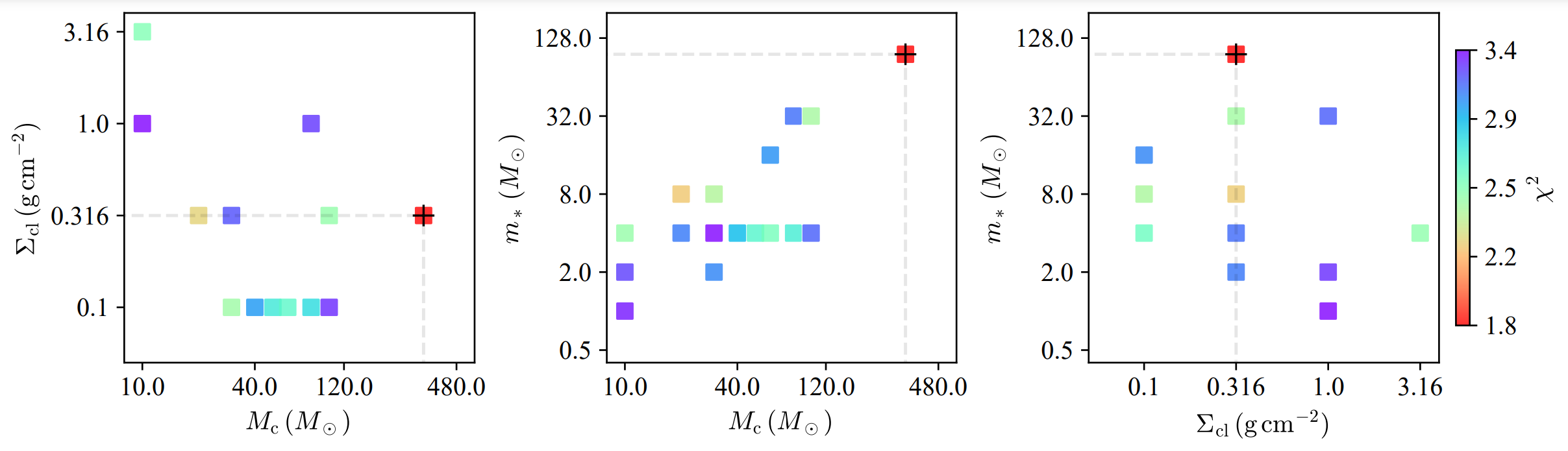}
        \caption{\label{fig:SED_second_p}\textit{Top:} 8-panel image of the protostar G359.42a in NIR and MIR filters. In all panels N is up and E to the left. \textit{Bottom left:} Protostellar SED for G359.42a, with SED fittings overlaid and color-coded by $\chi^2$, with the best-fit SED as a black line. \textit{Bottom right:} 2d parameter space plots for the fitting results on G359.42a. Various parameter pairings are given on the x- and y-axis of each plot, with each pairing of parameter values color-coded by $\chi^2$. Dashed gray lines indicate the lowest $\chi^2$ (i.e., best-fitting) pair of parameter values.}
    \end{figure*}
    
\section{Discussion}\label{sec:discussion}
    \subsection{Massive Star Formation in Sgr C}\label{sec:massive_sf_sgr_c}

    Star formation in the Sgr C cloud is dominated by two massive protostars, G359.44a and G359.44b, separated by $\sim5\arcsec$ (i.e., $\sim0.2$\,pc). The SED-fitted mass for G359.44a, $20.7^{+14.1}_{-8.4} M_{\odot}$ (see \S\ref{sec:SED_fitting} and Table \ref{tab:fitting_results}), is in good agreement, within uncertainties, with the measurement of $m_*=31.7\pm4.7 M_{\odot}$ made by \citet{lu22} based on the dynamical mass inferred by rotation velocities in its protostellar disk. The bolometric luminosity, $L_{bol}=9.7^{+15.6}_{-6.0} \times 10^4 L_{\odot}$, and disk mass, $M_{disk}=6.9^{+4.7}_{-2.8} M_{\odot}$, for G359.44a predicted by SED fitting are also in good agreement with the measurements made by \citet{lu22} with ALMA of $L_{bol}\gtrsim10^5 L_{\odot}$ and $M_{disk}=4.7^{+4.7}_{-2.35} M_{\odot}$, corroborating that G359.44a is indeed a massive protostar and placing independent constraints on its physical parameters.

    G359.44b, on the other hand, appears to be a much more deeply embedded, though perhaps eventually more massive, protostar. Although the NIR to FIR emission in Sgr C is dominated by G359.44a, itself a bright infrared source (see, e.g., Fig. \ref{fig:sgrc_main_zoom}), G359.44b is much brighter in mm wavelengths, being associated with a millimeter core that has almost double the ALMA Band 6 1.3 mm flux (and therefore inferred mass; $\mathrm{M_{mm}}=303.7\:M_\odot$) as G359.44a ($\mathrm{M_{mm}}=154.8\:M_\odot$). The same trend is seen in ALMA Band 3 (3 mm) data (Table \ref{tab:outflow_source_table}). G359.44b is also associated with a methanol maser almost two orders of magnitude brighter than that associated with G359.44a \citep{caswell10}, as well as an OH maser \citep{cotton16}. SED fitting on both protostars (\S\ref{sec:SED_fitting}) indicates that although they have nearly identical masses, G359.44b is more luminous, younger, embedded in a more massive envelope and surrounded by a denser environment than G359.44a (see Table \ref{tab:fitting_results}). These differences are within the error bounds on each measurement, however they are consistent with the other available evidence in indicating that G359.44b is either currently more massive than G359.44a or will be after its formation.

    There is also evidence for other, slightly more evolved, YSOs in the vicinity of G359.44a and G359.44b. The F405N (Br-$\alpha$) filter shows three bright ultracompact (UC) \Hii~regions in the vicinity of the protostars, each surrounding a point source in the NIRCam data (see Fig. \ref{fig:master_fig_outflows}). Two of these match with Very Large Array Band C (6 cm) UC \Hii~regions in \citet{lu19b}, C102 and C103. The presence of UC \Hii~regions surrounding young, ionizing massive stars in the vicinity of these massive protostars implies that multiple generations of star formation have occurred and are coexisting in Sgr C; this is further corroborated by the extended, mature Sgr C \Hii~region \citep[with an age of $\sim4 \mathrm{Myr}$;][]{simpson18a} and the detection of $>10^5 M_{\odot}$ of young stars in the Sgr C region by \citet{nogueras-lara24}, which have an estimated age of $\sim 20$ Myr.

    The stellar density in the region can also be estimated. This has been put forward as a potential metric for assessing the formation conditions of massive protostars, in particular in determining the degree of YSO crowding in the vicinity of these protostars to compare with massive star formation models \citep[see, e.g.,][]{crowe24}. Within a 9$\arcsec$ (0.35 pc) radius of G359.44a, the source which appears to be central in the nebulosity defining the protocluster region, the surface density of sources identified in F480M that are invisible in F115W (and therefore more likely to be protocluster members due to the high levels of extinction towards Sgr C) is $\sim2600\:\mathrm{pc^{-2}}$. The surface density of mm dust cores identified by \citep{lu20}, which represent additional YSO candidates not visible in the IR, is $\sim400\:\mathrm{pc^{-2}}$. Therefore, the total YSO candidate surface density surrounding G359.44a is $\sim3000\:\mathrm{pc^{-2}}$. However, it must be acknowledged that many of the IR sources identified are likely spurious detections (i.e., foreground/background sources), rather than YSOs associated with the Sgr C protocluster, due to the highly crowded nature of CMZ star fields. Furthermore, the bright and extended nebulosity surrounding G359.44a limits the detection of sources in this region to only the brightest objects, further limiting the completeness of the source counts (and meaning that the mass sensitivity also varies across the region). More sophisticated methods for distinguishing protocluster members (YSOs) from contaminants, as well as for conducting completeness corrections to account for sources missed in the bright nebulosity, will be needed to place accurate constraints on the YSO surface density in the vicinity of G359.44a, and to properly compare it with massive star formation models.

        \subsubsection{Outflows in the main Sgr C molecular cloud}\label{sec:outflows_sgr_c_cloud}
        \begin{figure*}
            \centering
            \includegraphics[width=0.95\textwidth,height=0.95\textheight,keepaspectratio]{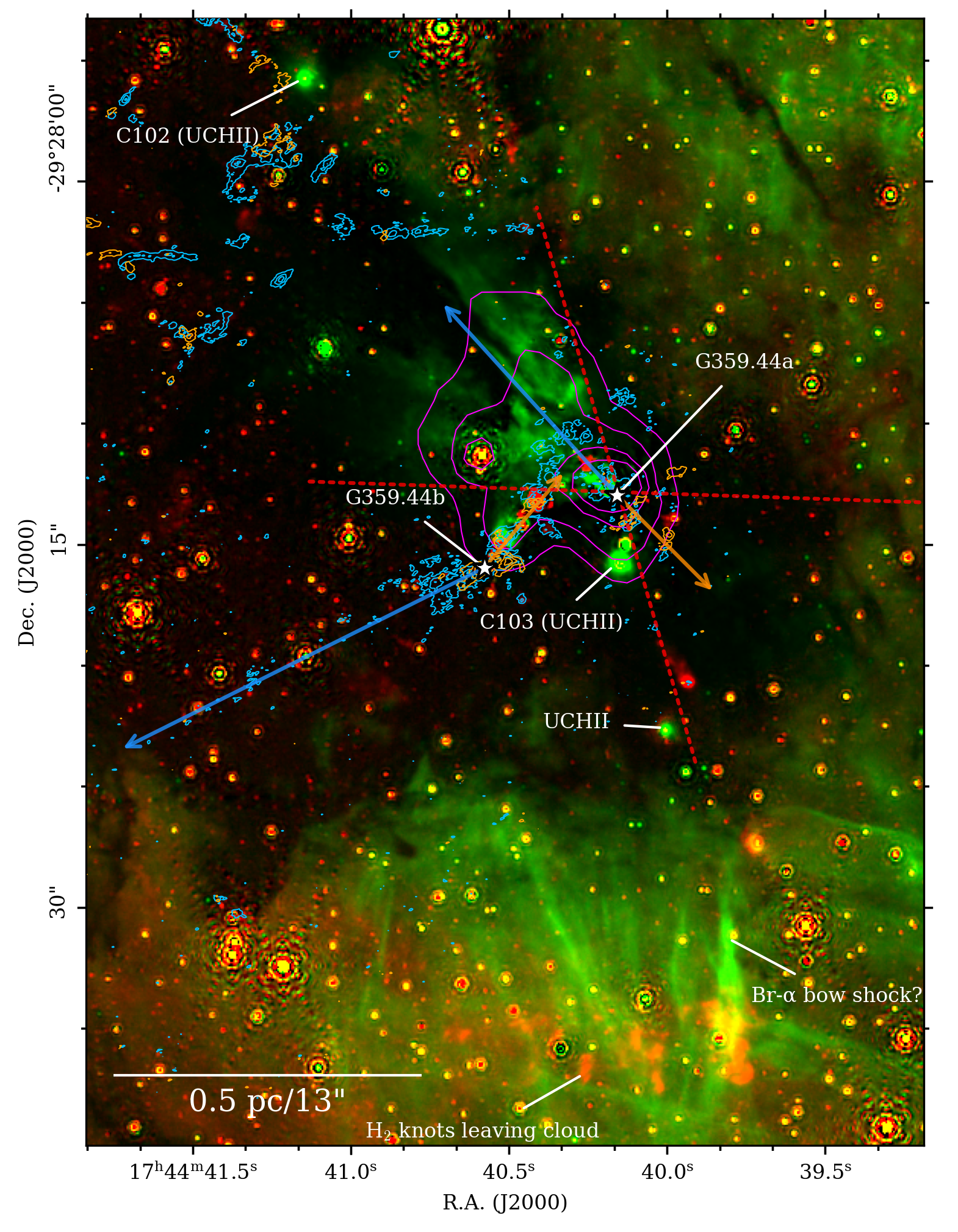}
            \caption{\label{fig:master_fig_outflows} The region around G359.44a and G359.44b. F470N is shown in red and F405N in green, both continuum-subtracted. The protostars are labeled and indicated by white star symbols. Blue and orange contours show ALMA Band 6 SiO 5-4 data from \citet{lu21} in blueshifted emission (integrated from -80 to -51 $\rm kms^{-1}$) and redshifted emission (integrated from -48 to -25 $\rm kms^{-1}$), respectively \citep[similar to Figure 21 of][]{lu21}. Both sets of contours are shown with levels of 0.06, 0.12, and 0.18 Jy/beam, and pixels with primary beam response $<0.6$ have been masked. The blue- and red-shifted outflow axis of each source is indicated by the blue and orange arrows, respectively. Purple contours show Spitzer IRAC2 4.5 $\mathrm{\mu m}$ emission around G359.44a with levels of 50, 100, 200, and 300 MJy/sr. The outflow cone of G359.44a measured from the NIRCam data, with an opening angle of $\sim70^{\circ}$, is indicated by the dotted red lines. Other features discussed in the text are labeled.}
            
        \end{figure*}
    
            The outflow axes for G359.44a and G359.44b implied by the identified shocked emission features in this work correspond well with those identified using ALMA molecular line data in \citet{lu21}. Fig. \ref{fig:master_fig_outflows} shows a comparison between the ALMA Band 6 SiO 5-4 line data and the NIRCam data of outflows from each protostar. 

            The redshifted SiO emission from G359.44b, in particular, aligns very well with the line of outflow knots comprising knots 34, 32, and 27 (see Fig. \ref{fig:sgrc_cloud_diagram}\:(b.1.3)), which extends for $\sim5\arcsec$ ($\sim$0.2 pc). However, the blueshifted end of this source's outflow is entirely invisible in the NIRCam data, despite its prominence in SiO. This could be explained as the redshifted outflow from G359.44b entering into a region of much lower extinction than its blueshifted lobe, potentially in this case the outflow cavity cleared out by other sources in the region (e.g., G359.44a). This phenomenon, the redshifted ouflow lobe being brighter than the blueshifted lobe, is well-documented in other systems, such as the HH 80-81 system \citep[][and references therein]{bally23} and HH135-136 system \citep[see, e.g., Fig. 1 and Fig. 5 of][]{fedriani20}, both of which host outflows with a significantly brighter redshifted lobe.
            
            
            A similar process may be occurring with the bright chains of shocked H$_2$ emission just south of the main dark cloud i.e., knots 17, 20, 23, 28, 29, 36, 16, 18, and 38 (see Fig. \ref{fig:big_diagram} (a) and Fig. \ref{fig:master_fig_outflows}). The brightness of these knots compared to material in the cloud could be explained by this material having just exited the high-extinction medium of the cloud, causing it to appear much brighter along our line-of-sight compared to the knots deep in the cloud. Some of these knots may be attributable to G359.44a and/or G359.44b; however, such an association is unclear, and they may have instead originated from other protostars in the cloud. Spatially coincident with the bright H$_2$ knots is what may be a bow shock in Br-$\alpha$ (F405N) emission, labeled in Fig. \ref{fig:master_fig_outflows}, which may have been driven by protostars in the cloud. Despite the brightness of this object, it may instead be a fragment of the filamentary \Hii\:region, which dominates the F405N emission outside of the cloud (see the first panel of Fig. \ref{fig:big_diagram}).  
            
            The SiO emission measured by \citet{lu21} for G359.44a corresponds well with that seen in the infrared, particularly in encompassing knots 22 and 24. The blueshifted wide-angle outflow cone of G359.44a can also clearly be seen with NIRCam (see e.g. Fig. \ref{fig:sgrc_main_zoom} and Fig.  \ref{fig:master_fig_outflows}), with knot 22 appearing to comprise part of the outflow cavity wall close to the source and knot 24 comprising part of the source's jet, which flows through the middle of the wide-angle outflow cone (see Fig. \ref{fig:sgrc_cloud_diagram} (b.1.3)). This is consistent with the understanding of jets and outflow cones in local, low-mass star-forming regions \citep[see][]{bally16}. The outflow cone detected in NIRCam is also in good agreement with Spitzer IRAC2 $4.5 \mu m$ emission (shown as purple contours in Fig. \ref{fig:master_fig_outflows}). This filter is known to trace both molecular hydrogen and CO emission from outflows from ``Extended Green Objects'' \citep[EGOs;][]{cyganowski08,ray23}, a class of objects of which G359.44a is a prominent example \citep{chen13}. From the NIRCam data, we measure an opening angle of the outflow cone of G359.44a of $\sim 70^{\circ}$. 
                

            There are other outflow knots further away that also appear to be associated with G359.44a: knots 19 (see Fig. \ref{fig:sgrc_cloud_diagram}, panel (b.1.1)), 25 (b.1.3), 31 (b.1.3), 33 (b.1), 35 (b.1), and 37 (b.1). These knots trace out a curve that bends over its full extent, potentially indicating some precession of the protostar's outflow axis over time. The chain of knots (from knot 37 to knot 19) is about $45\arcsec$ (or 1.75 pc at  distance of 8.15 kpc) long. We can also make a rough estimation of the dynamical age for knot 37, which is likely the furthest knot associated with G359.44a. Assuming a knot velocity of $50-100\,\mathrm{km\,s^{-1}}$ \citep[see e.g.,][for H$_2$ knot velocity estimations]{fedriani18,fedriani20} and taking its angular separation ($\sim36\arcsec$, see Table\,\ref{tab:knot_table}) at a distance of 8.15 kpc, we calculate a dynamical age of $\sim14000-28000$\,yr. This indicates that the G359.44a system is at least $\gtrsim10^4$\,yr old to have generated this object.
                
            
            \citet{kendrew13} identified NIR emission in the vicinity of G359.44a and G359.44b in H$_2$ at 2.12 $\mu$m and in Br-$\gamma$ (2.16 $\mu$m), which they proposed to represent shocked outflow material. Our data match well with the five line features identified in \citet{kendrew13} tracing their origin to G359.44a and G359.44b (see Fig. \ref{fig:sgrc_cloud_diagram} (b.1.3)). In particular, the NIRCam data would indicate that their features 1 and 2 correspond with emission from G359.44a, whereas features 3, 4, and 5 can be traced back to G359.44b. 
            

            Previous authors \citep[e.g.,][]{kendrew13} found no evidence of star formation in the Sgr C cloud outside of the protostars G359.44a and G359.44b. The advances in sensitivity and spatial resolution brought by ALMA and JWST have revealed a population of dense sub-mm cores \citep{lu20,kinman24} and widespread signs of outflow activity
            \citep[][]{lu21},
            indicating the cloud is harboring significant star formation after all. We have identified $\sim 40$ line-emitting features that we associate with star formation activity in Sgr C besides G359.44a and G359.44b (see Table \ref{tab:knot_table}). Although there is a possibility that some of these objects could be shocked emission unrelated to star formation activity, or foreground to the CMZ and unrelated to Sgr C entirely, a survey of the line-emitting objects outside of star-forming regions in the data (presented in Appendix \ref{sec:misc_mhos}) indicates that there are relatively few, and those that are present are extended, dim, and clearly distinct from the concentrated knots of shocked H$_2$ emission associated with the main Sgr C cloud.
            
            There is also the possibility that some of the identified knots could have been produced by large-scale shocks passing through Sgr C and interacting with its dense gas. However, it is noteworthy that many of these line-emitting features coincide with bright regions of ALMA Band 6 continuum data, which traces warm protostellar cores in the region \citep{lu20}; see, e.g., panels (b.1), (b.2) of Fig. \ref{fig:sgrc_cloud_diagram}. There is also a close correspondence between the features identified in F470N and the SiO 5-4 molecular line data presented in \citet{lu21}; this will be presented and discussed further in a forthcoming paper.

        \subsubsection{Comparison with JWST studies of massive star formation in the Galactic Disk} \label{sec:iras23385_comparison}

            \citet{beuther23} present JWST MIRI-MRS spectro-imaging data on the massive star-forming complex IRAS 23385+6053 (d$\sim5$ kpc), which acts as a useful Galactic Disk counterpart to Sgr C. The MIRI data reveals a complicated system with multiple outflow structures, centered on the primary massive protostar ($\mathrm{m_{*}\sim9M_{\odot}}$), in several MIR atomic and molecular lines ($\mathrm{H_2}$, [FeII], [NeII]). These structures are corroborated by emission seen in 3.5 mm SiO 2-1 data taken with the IRAM NOrthern Extended Millimeter Array (NOEMA). Unlike IRAS 23385+6053, which displays at least three separate outflow axes centered at its main massive protostar in MIR $5.511\:\mathrm{\mu m}\:\mathrm{H_2(0-0)S(7)\:emission}$ \citep[see Fig. 5 of][]{beuther23}, Sgr C does not show evidence of multiple outflow axes for either of its main massive protostars, G359.44a and G359.44b, particularly in the MIR $\mathrm{H_2}$ line filter F470N. Rather, single outflow axes for both G359.44a and G359.44b are well-defined in both IR and mm lines (see Fig. \ref{fig:master_fig_outflows}), implying that they are forming relatively unperturbed by nearby YSOs, although G359.44a does demonstrate some indication of precession in its outflow that may have been caused by dynamical interactions with nearby sources (\S\ref{sec:outflows_sgr_c_cloud}). This finding is somewhat counter-intuitive, given the extremity of CMZ conditions compared to the Galactic Disk, however it may indicate that massive star formation can occur in the CMZ in a largely similar manner as in the Galactic Disk.
            
            However, as the results of \citet{beuther23} indicate, multiple infrared tracers of jet/outflow emission, such as [FeII] and [NeII], which are not included in the present study, are needed to obtain a full picture of the different components of outflows in these regions. Additionally, measurements of accretion/ejection rates, the former of which \citet{beuther23} make with the Humphreys $\alpha$ HI(7-6) line at 12.37 $\mu$m, and probes of the warm/hot gas content ($\sim200-2000$ K) in the wider environment, can place further constraint on massive star formation \citep[][]{gieser23}. Therefore, high-resolution MIR spectroscopy measurements with JWST, on G359.44a/b and other CMZ massive protostars in other clouds \citep[e.g. Sgr B, 50 km$\mathrm{s^{-1}}$, 20km$\mathrm{s^{-1}}$ and the Brick; see][for more information about the Brick in particular]{ginsburg23}, along with further studies on Galactic Disk massive protostars \citep[from, e.g., the IPA survey,][]{megeath21,federman23}, will be needed to construct a proper comparison between massive star formation in the CMZ as opposed to the Galactic Disk.

        \subsubsection{Outflows in G359.42-0.104}\label{sec:outflows_G359.42-0.104}

            Another promising result of analysis of the F470N continuum-subtracted emission near Sgr C is the discovery and tentative placement at the galactocentric distance of a new star-forming region, G359.42-0.104 (\S\ref{sec:G359.42-0.104}). The source G359.42a, the brightest infrared source in the region, may be a massive ($m_*>8 M_{\odot}$) protostar, with a current stellar mass predicted by SED fitting of $m_*\simeq9\:M_{\odot}$ (Table \ref{tab:fitting_results}). This claim is supported by its massive associated mm core ($M_{\rm mm}\sim90 M_{\odot}$) and the relatively large luminosity of its associated water maser, $\sim 1\times10^{-5} L_{\odot}$ \citep{lu19a}, which implies massive star formation activity as suggested by the relationship between H$_2$O maser luminosity and host source luminosity observed by \citet{urquhart11}. It is worth noting, however, that the trend observed by \citet{urquhart11} experiences quite a substantial spread, and that water masers are known to vary by orders of magnitude over relatively short timescales \citep{lu19a}. The lower bound on the current stellar mass of G359.42a, $m_*\simeq3.5\:M_{\odot}$, is also outside of the normal mass range for massive protostars ($m_*>8 M_{\odot}$), and would instead be indicative of an intermediate-mass protostar. Additionally, the SED-fitting derived parameters and measured mm core mass depend heavily on the adopted distance to the region, which we take to be at the galactocentric distance of 8.15 kpc, but which may be lower if the region is foreground to the CMZ. For example, if the distance to G359.42-0.104 is instead 4 kpc, the SED-fitting derived mass and mm core mass estimate would be $m_*\simeq5\:M_{\odot}$ and $M_{\rm mm}\sim22\:M_{\odot}$, respectively.

            A dynamical age estimate can be derived for G359.42a, as for G359.44a. Assuming a knot velocity of $50-100\,\mathrm{km\,s^{-1}}$, and with an angular separation between the bright bow shock K21B and G359.42a of $\sim26\arcsec$, and with our adopted distance to the source of 8.15 kpc, we estimate a dynamical age of 10,000-20,000 yr, placing a lower limit on the age of G359.42a of $\sim10^4$\,yr old, depending on its distance.

            Although G359.42a is the only protostar in G359.42-0.104 associated with both a mm core and water maser, there is evidence for other protostars in the region. We call the most promising candidate G359.42b, which was identified based on the distinctive morphology of its surrounding emission (see Fig. \ref{fig:G359.42-0.104_cont} (c.2)). One of its associated emission line objects, knot 13, has a morphology reminiscent of an outflow cone emanating from G359.42b; in this case, we estimate this outflow cone to have an opening angle of $\sim40^{\circ}$. We also note that some of the outflow knots originating from G359.42b come in pairs on opposing sides of the source, e.g., knots 14 and 11 ($2.31\arcsec\:\mathrm{and}\:1.97\arcsec$ from G359.42b, respectively). This may provide an indication of episodic accretion resulting in the ejection of outflow knots from the source, a phenomenon that has been previously observed in massive star-forming regions \citep[e.g.,][]{caratti17,cesaroni18,fedriani23b}. From end to end, the ouflow from G359.44b appears to be $\sim18\arcsec$ ($\sim0.7$ pc at a distance of 8.15 kpc), including knot 21A.
            
            Ultimately, higher sensitivity and resolution observations of G359.42-0.104, especially in the mm and sub-mm, where individual protostellar cores can be identified and traced back to potential outflow knots, will be needed to disentangle the full picture of star formation in this region. Potential relationships with the Sgr C cloud, and especially with the Sgr C \Hii\:region, which is directly adjacent in projection to G359.42-0.104 (see Fig. \ref{fig:big_diagram}), may also be revealed with further observations and study of this star-forming region and its definitive placement at the CMZ distance (or not) with further data.

        \subsubsection{Outflows in the CMZ}\label{sec:outflows_GC}
            The first unambiguous detection of protostellar outflows in the CMZ in the Sgr C cloud, and their corroboration directly with ALMA, presented in this study, has significant implications for future infrared studies of the CMZ, especially in the infrared. Infrared observations of the CMZ, particularly those with the aim of resolving individual protostars and their associated outflow features, have been historically difficult for a number of reasons, most notably a lack of resolving power and the effects of extinction and crowding \citep{nogueras-lara18,nogueras-lara19}. Studies that attempt to provide comprehensive identification and characterization of CMZ massive protostars \citep[see, e.g.,][]{yusef-zadeh09},
            have significant limitations in confirming protostellar detections and in turn, constructing a robust catalog. For example, follow-up studies on the global sample of CMZ massive YSOs from \citet{yusef-zadeh09} prove, using spectroscopic follow-up, that many of the proposed detections are spurious and confused with cool late-type stars \citep[see][]{an11,koepferl15,nandakumar18}. The detection of outflows from candidate protostars in the infrared, finally possible with JWST, presents another robust way of differentiating between authentic protostars and contaminants using imaging alone. 
            
            Furthermore, individual infrared-luminous YSO counting is a key diagnostic for placing constraints on significant star formation parameters in the CMZ, such as the star formation rate and efficiency, that allows the CMZ to be appropriately compared and contrasted with regions in the main Galactic disk \citep{henshaw23}. In this respect, the ability of JWST-NIRCam to resolve individual knots of shocked emission comprising outflows from individual protostars is promising for future studies of CMZ star formation. 
    \subsection{The Environment Around Sgr C}
    \label{sec:sgr_c_in_context}

        Concerning the origins of star formation in Sgr C, recently evidence has also been put forward to suggest that Sgr C is undergoing triggered star formation. This is asserted on the basis of the ``cometary" structure of the cloud, which points away from the Sgr C \Hii\:region, which contains many young, massive stars \citep{nogueras-lara24}, as well as the magnetic field orientations inferred from dust polarization measurements in this region, which appear to trace along the outline of the cloud, indicating an interface of interaction with the \Hii\:region \citep{pare24, lu24}. This proposition is also supported by the observed gradient of star formation occurring in the cloud, with the most massive star formation (e.g., G359.44a and G359.44b) occurring closest (in projection) to the interface between the \Hii\:region and the cloud where the compression on the cloud (and therefore its density) would be highest, and the lower-mass star formation tapering off going away from the \Hii\:region and deeper into the cloud.


        Ionized gas has been revealed in great detail in the NIRCam observations via the F405N filter (Br-$\alpha$; see Fig. \ref{fig:RGB}). This \Hii\: region hosts prominent filamentary striations that are also seen in longer-wavelength observations of other CMZ \Hii\: regions, i.e., those associated with Sgr B2, Sgr B1 \citep[e.g., 1.28 GHz MeerKAT observations presented in][]{heywood22}. These filaments are intriguing. Their linear morphologies indicate a potential role for strong magnetic fields in controlling the structure of the nebula. However, the wide variety of position angles of the features, including near orthogonal orientations at similar sky positions, indicates a potentially complicated 3D structure that we are viewing in projection. High resolution spectra of recombination lines associated with these structures will be helpful to better understand the dynamics of this ionized gas. A further analysis and discussion of these Br-$\alpha$ filaments and their potential formation mechanism will be presented in a forthcoming paper.
        

\section{Conclusions}\label{sec:conclusions}

    In this paper, we have presented NIRCam observations of the CMZ star-forming region Sgr C which, along with ancillary IR and millimeter data, provide a high level of detail into the star formation activity of the main cloud and its surroundings.
    We have characterized the two most massive protostars in the heart of the main Sgr C protocluster, G359.44a and G359.44b, and obtained their physical properties via SED-fitting; in particular, masses of $\sim20\,\mathrm{M_\odot}$ have been derived for each protostar.
    We have made a cross-match between JWST sources and ALMA cores in order to identify a sample of lower-mass protostars. From these sources, there are 5 matches that are redder than the overall population, which we take to be our strongest sample of low-mass YSOs in the cloud.
    We have carried out a census of the narrow-band NIRCam data, using the filters F212N, F405N, and F470N, which trace shocked molecular and atomic hydrogen emission from protostellar jets, to identify line-emitting features. We have identified
    88 features, which we believe to comprise protostellar outflows from over a dozen protostellar outflows in the NIRCam data. We attribute about a quarter of these outflow knots to the massive protostars G359.44a and G359.44b, forming an outflow axis for each protostar that agrees well with molecular line data from ALMA and archival IR data from Spitzer. There are $\sim40$ others likely originating from other, lower-mass star formation activity in the cloud. 
    
    The remaining outflow knots are attributed to a newly-discovered star-forming region, G359.42-0.104, located $\sim1\arcmin$ to the South of the main Sgr C protocluster.
    NIRCam data in G359.42-0.104 reveals a pair of prominent bow shocks in both F470N and F212N (tracing H$_2$ shocked emission),
    which we estimate to have originated from two protostars, G359.42a and G359.42b, the former of which we speculate may be a massive protostar due to its SED-fit inferred mass ($m_*>8 M_{\odot}$). However, the distance to this region, and whether it is inside the CMZ or foreground, is still uncertain.
    


\begin{acknowledgments}
    \textit{Acknowledgments:} This work is based on observations made with the NASA/ESA/CSA JWST and Hubble Space Telescopes. The data were obtained from the Mikulski Archive for Space Telescopes at the Space Telescope Science Institute, which is operated by the Association of Universities for Research in Astronomy, Inc., under NASA contract NAS 5-03127 for JWST. These observations are associated with program~4147. Support for program~4147 was provided by NASA through a grant from the Space Telescope Science Institute, which is operated by the Association of Universities for Research in Astronomy, Inc., under NASA contract NAS 5-03127. This paper makes use of the following ALMA data: ADS/JAO.ALMA~2016.1.00243.S, ADS/JAO.ALMA~2021.1.00172.L. ALMA is a partnership of ESO (representing its member states), NSF (USA) and NINS (Japan), together with NRC (Canada), NSTC and ASIAA (Taiwan), and KASI (Republic of Korea), in cooperation with the Republic of Chile. The Joint ALMA Observatory is operated by ESO, AUI/NRAO and NAOJ. The National Radio Astronomy Observatory is a facility of the National Science Foundation operated under cooperative agreement by Associated Universities, Inc.
    
    The authors would like to acknowledge X. Lu for providing access to the reduced ALMA Band 6 continuum and molecular data and for fruitful discussions. The authors would like to acknowledge M. Reid for insightful discussion and clarification. S.T.C. acknowledges support from the award JWST-GO-04147.003-A. R.F. acknowledges support from the grants Juan de la Cierva FJC2021-046802-I, PID2020-114461GB-I00, and PID2023-146295NB-I00. R.F., L.B.F., and R.S. acknowledge financial support from the Severo Ochoa grant CEX2021-001131-S funded by MCIN/AEI/ 10.13039/501100011033. Additionally, L.B.F. and R.S acknowledge support from grant EUR2022-134031 funded by MCIN/AEI/10.13039/501100011033 and by the European Union NextGenerationEU/PRTR. and grant PID2022-136640NB-C21 funded by MCIN/AEI 10.13039/501100011033 and by the European Union. Y.Z. acknowledges the support from the Yangyang Development Fund. Y.C. was partially supported by a Grant-in-Aid for Scientific Research (KAKENHI  number JP24K17103) of the JSPS. J.B. acknowledges support by National Science Foundation through grant No. AST-1910393. A.G. acknowledges support from the NSF under grants AAG 2008101, 2206511, and CAREER 2142300. Y.-L.Y. acknowledges support from Grant-in-Aid from the Ministry of Education, Culture, Sports, Science, and Technology of Japan (20H05845, 20H05844), and a pioneering project in RIKEN (Evolution of Matter in the Universe). Z.-Y.L. is supported in part by NSF AST- 2307199 and NASA 80NSSC20K0533.
\end{acknowledgments}

\bibliography{masterbib.bib}

\appendix
\restartappendixnumbering
    \section{Photometry Used for SED Fitting}\label{sec:source_photometry_table}
    Table \ref{tab:sed_flux_table} shows the integrated flux densities for all filters used for SED fitting in \S\ref{sec:SED_fitting}.
    \begin{sidewaystable*}       
        \centering
        \caption{\label{tab:sed_flux_table}Integrated Flux Densities. Filters in which a flux was not measured for a source, because it was either invisible or saturated, are marked with $\cdots$. $\mathrm{F_{1.15}}$, $\mathrm{F_{1.62}}$, $\mathrm{F_{1.82}}$, $\mathrm{F_{3.6}}$ (first appearance), and $\mathrm{F_{4.8}}$ refer to JWST-NIRCam data at 1.15, 1.62, 1.82, 3.60, and 4.80 $\mu$m, respectively. $\mathrm{F_{3.6}}$ (second appearance), $\mathrm{F_{4.5}}$, $\mathrm{F_{5.6}}$, and $\mathrm{F_{8.0}}$ refer to Spitzer-IRAC data at 3.6, 4.5, 5.6, and 8.0 $\mu$m, respectively. $\mathrm{F_{25.2}}$ and $\mathrm{F_{37.1}}$ refer to SOFIA-FORCAST data at 25.2 and 37.1 $\mu$m, respectively. $\mathrm{F_{70}}$, $\mathrm{F_{160}}$, $\mathrm{F_{250}}$, $\mathrm{F_{350}}$, and $\mathrm{F_{500}}$ refer to Herschel-PACS/SPIRE data at 70, 160, 250, 350, and 500 $\mu$m, respectively. The second row for each source refers to the error associated with the flux at each wavelength.}
        \begin{tabular}{c c c c c c c c c c c c c c c c c} 
        \hline\hline  
        Name & $\mathrm{F_{1.15}}$ & $\mathrm{F_{1.62}}$ & $\mathrm{F_{1.82}}$ & $\mathrm{F_{3.6}}$ & $\mathrm{F_{3.6}}$ & $\mathrm{F_{4.5}}$ & $\mathrm{F_{4.8}}$ & $\mathrm{F_{5.6}}$ & $\mathrm{F_{8.0}}$ & $\mathrm{F_{25.2}}$ & $\mathrm{F_{37.1}}$ & $\mathrm{F_{70}}$ & $\mathrm{F_{160}}$ & $\mathrm{F_{250}}$ & $\mathrm{F_{350}}$ & $\mathrm{F_{500}}$ \\
        & & & & (NIRCam) & (Spitzer) & & & & & & & & & & & \\
        &(Jy)&(Jy)&(Jy)&(Jy)&(Jy)&(Jy)&(Jy)&(Jy)&(Jy)&(Jy)&(Jy)&(Jy)&(Jy)&(Jy)&(Jy)&(Jy)\\
        \hline
        G359.44a & $\cdots$ & 0.0032 & 0.0040 & 0.022 & 0.018 & 0.12 & 0.15 & 0.23 & 0.21 & 14.3 & 43.1 & 340.1 & 380.9 & 90.1 & 17.7 & 2.1 \\
        & $\cdots$ & (0.0017) & (0.0023) & (0.011) & (0.022) & (0.05) & (0.05) & (0.06) & (0.02) & (1.5) & (4.5) & (34.1) & (217.1) & (109.9) & (47.0) & (13.4) \\
        G359.44b & $\cdots$ & 0.0024 & 0.0035 &0.015 &0.019 &0.035 &0.040 & 0.038 &0.0035 &2.21 &14.7 &113.5 &127.0 &30.0 &5.89 &0.70\\
        & $\cdots$ & (0.0006)& (0.0018)& (0.005)& (0.014)& (0.040)& (0.044)& (0.046)& (0.0128)& (1.11)& (3.9)& (11.4)& (72.4)& (36.6)& (15.67)& (4.48) \\
        G359.42a & 0.00010 & 0.00068 & 0.0014 &$\cdots$ & 0.11 & 0.28 & $\cdots$ & 0.54 & 0.72 & 3.9 & 4.2 & 2.4 & 0.51 & 0.11 & 0.014 & 0.0062\\
        & (0.00076) & (0.0018) & (0.0022) & $\cdots$ & (0.012) & (0.03) & $\cdots$ & (0.056) & (0.08) & (0.49) & (0.1) & (0.1) & (18.69) & (7.90) & (3.041) & (0.9873) \\
        \hline   
        \end{tabular}
    \end{sidewaystable*}
    \clearpage
    \section{Knot Feature Significance Level Maps}\label{sec:Knot_significance}

        \begin{figure*}[!htb]
            \centering
            \includegraphics[width=0.40\textwidth]{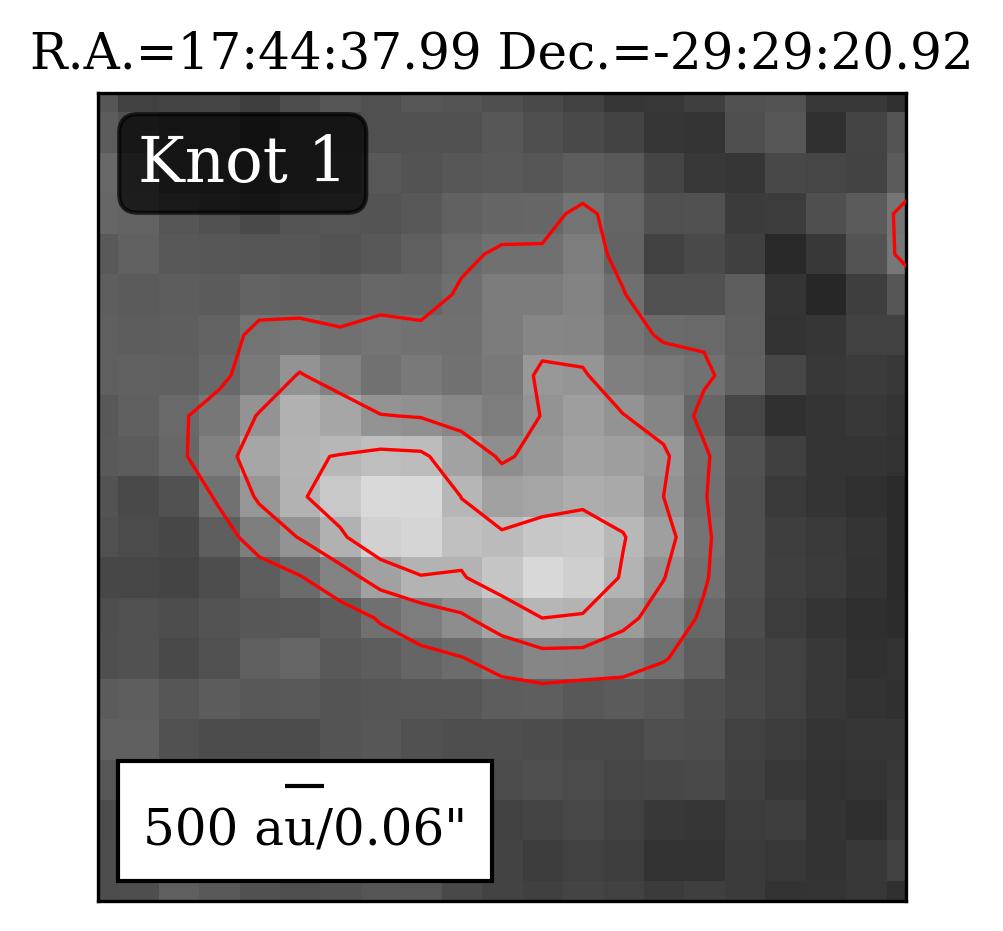}
            \includegraphics[width=0.40\textwidth]{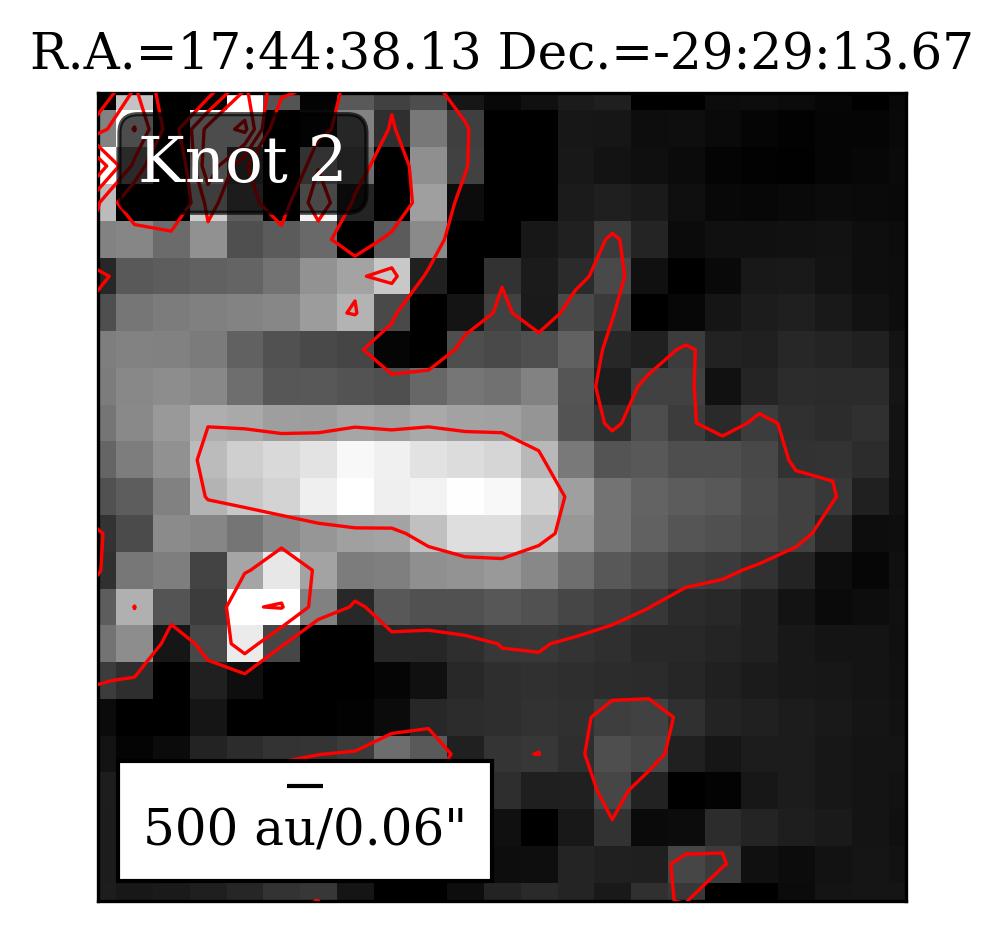}
            \includegraphics[width=0.40\textwidth]{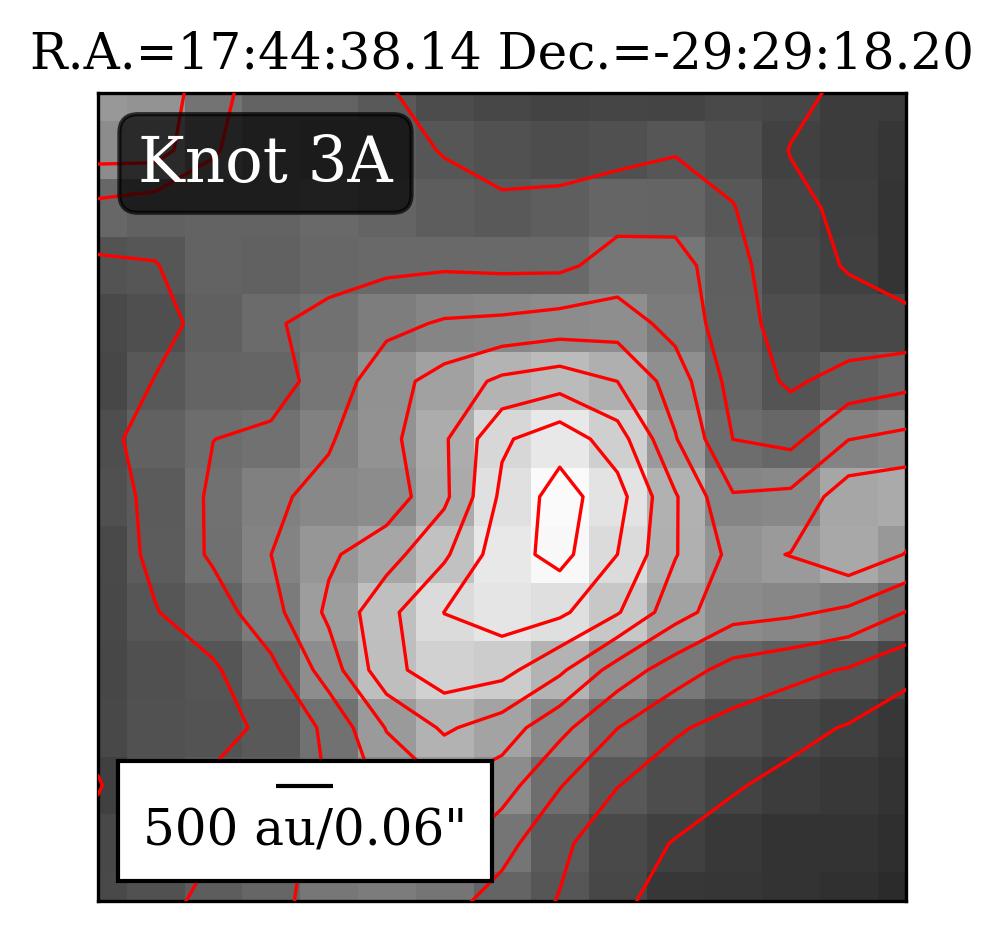}
            \includegraphics[width=0.40\textwidth]{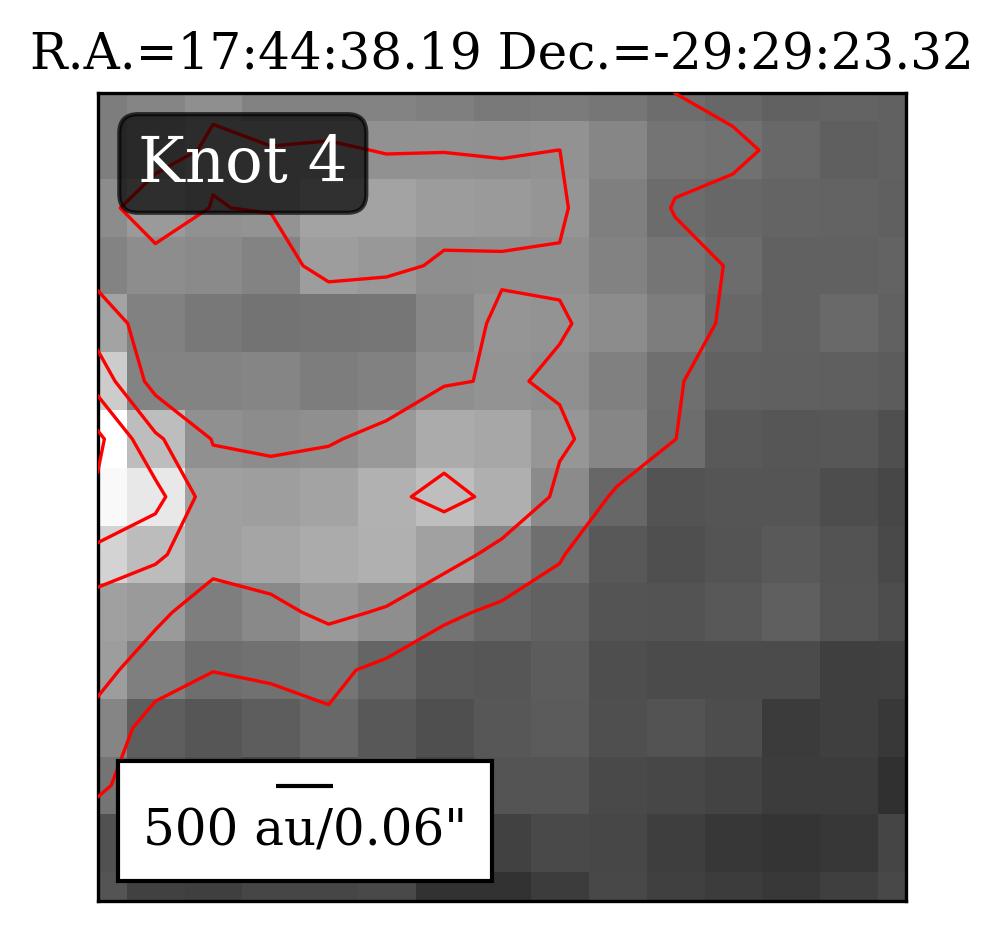}
            \includegraphics[width=0.40\textwidth]{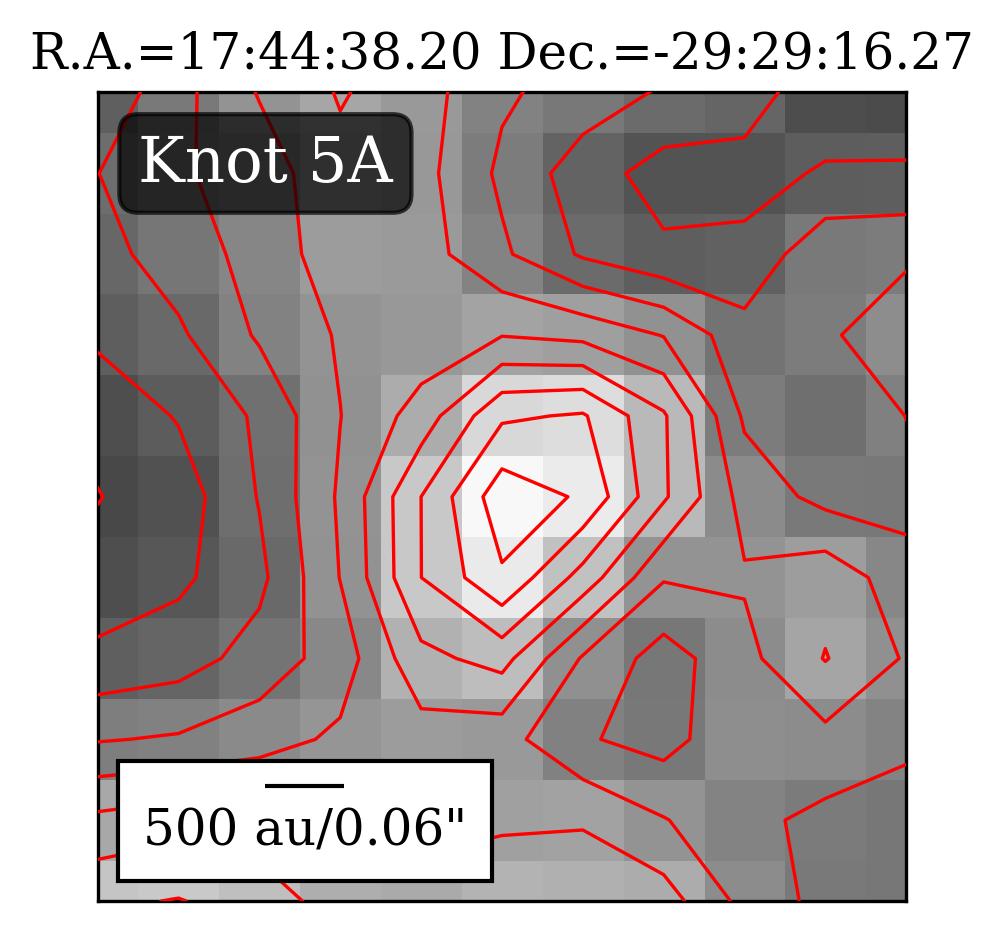}
            \includegraphics[width=0.40\textwidth]{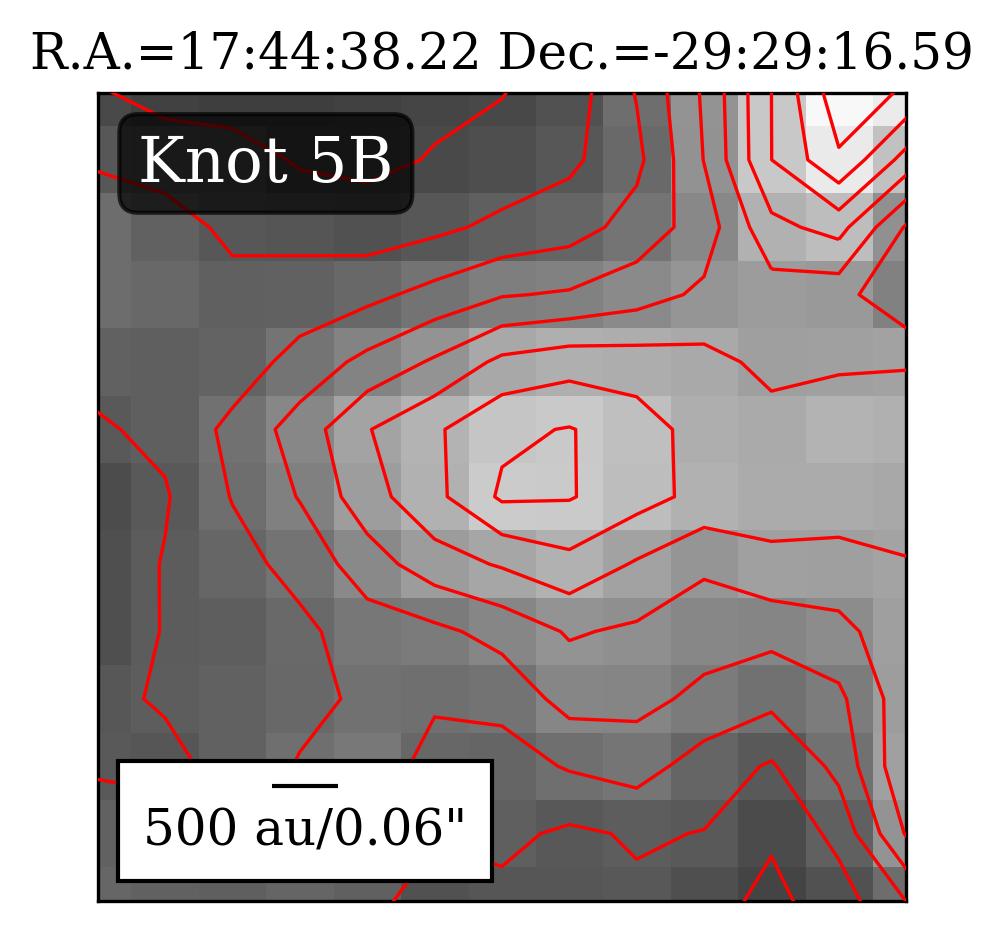}
            \caption{\label{fig:F470N_knots}Significance level contour maps of all knot features identified in the F470N  continuum-subtracted image and compiled in Table \ref{tab:knot_table}. The contour levels shown for knots 1, 3A, 4, 5A, and 5B represent 15 to 100$\mathrm{\sigma}$ in steps of 5$\sigma$ above the local background; those shown for knot 2 represent 25 to 300$\mathrm{\sigma}$ in steps of 55$\sigma$. The central coordinates of each knot determined from the peak pixel are given on the top of each panel. A physical scalebar of 500 au is given in the bottom-left corner of each panel, N is up and E is left in all panels.}
        \end{figure*}
        \renewcommand{\thefigure}{B\arabic{figure}}
        \addtocounter{figure}{-1}
        \begin{figure*}[!htb]
            \centering
            \includegraphics[width=0.40\textwidth]{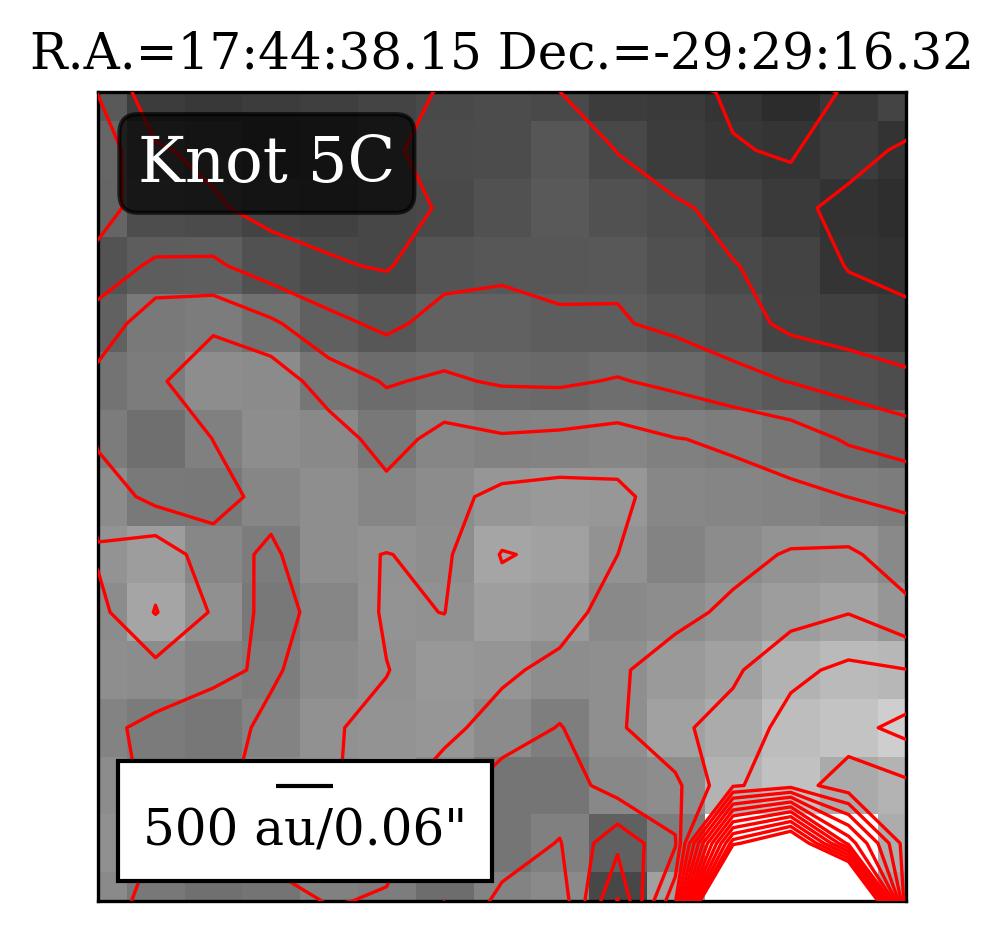}
            \includegraphics[width=0.40\textwidth]{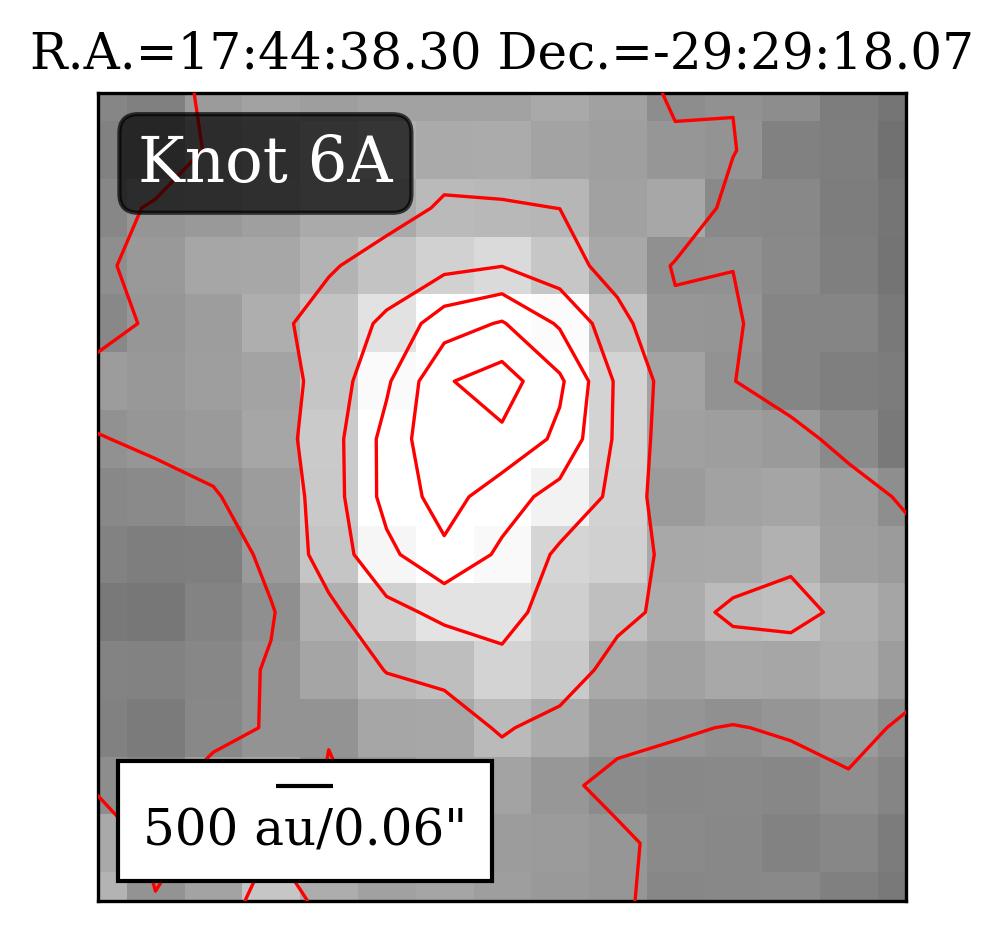}     
            \includegraphics[width=0.40\textwidth]{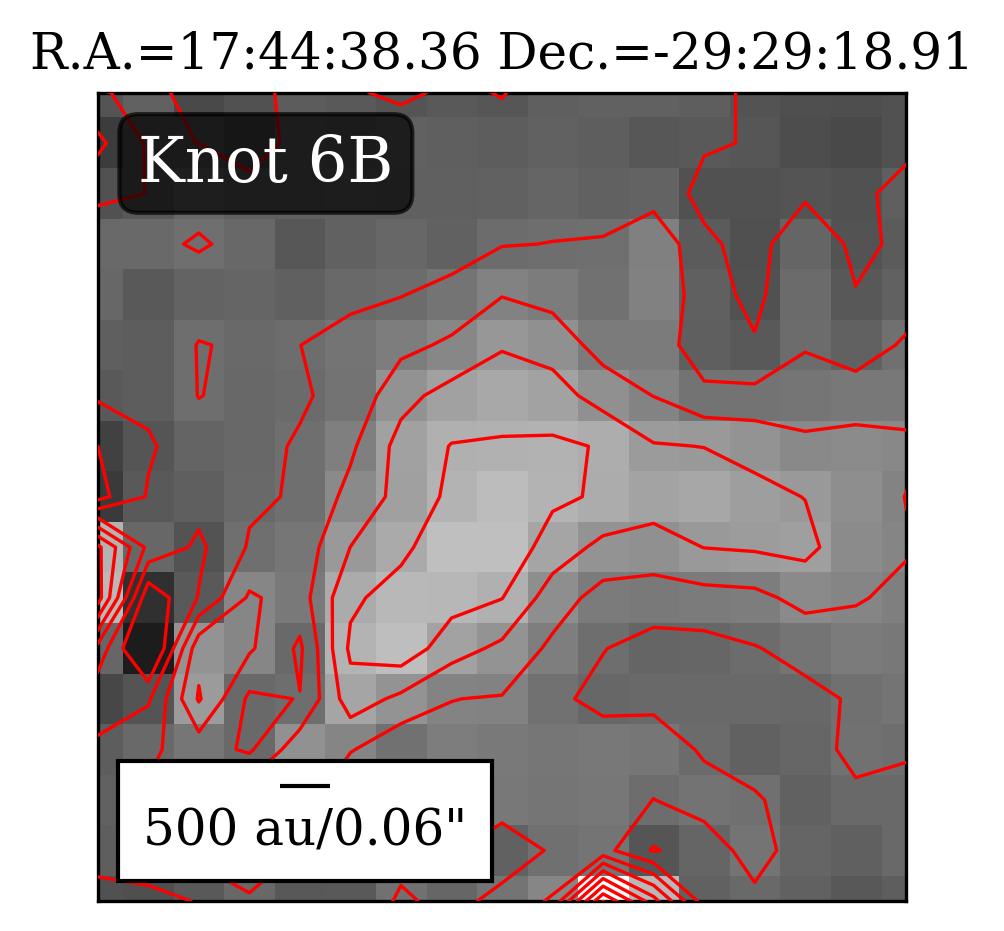}
            \includegraphics[width=0.40\textwidth]{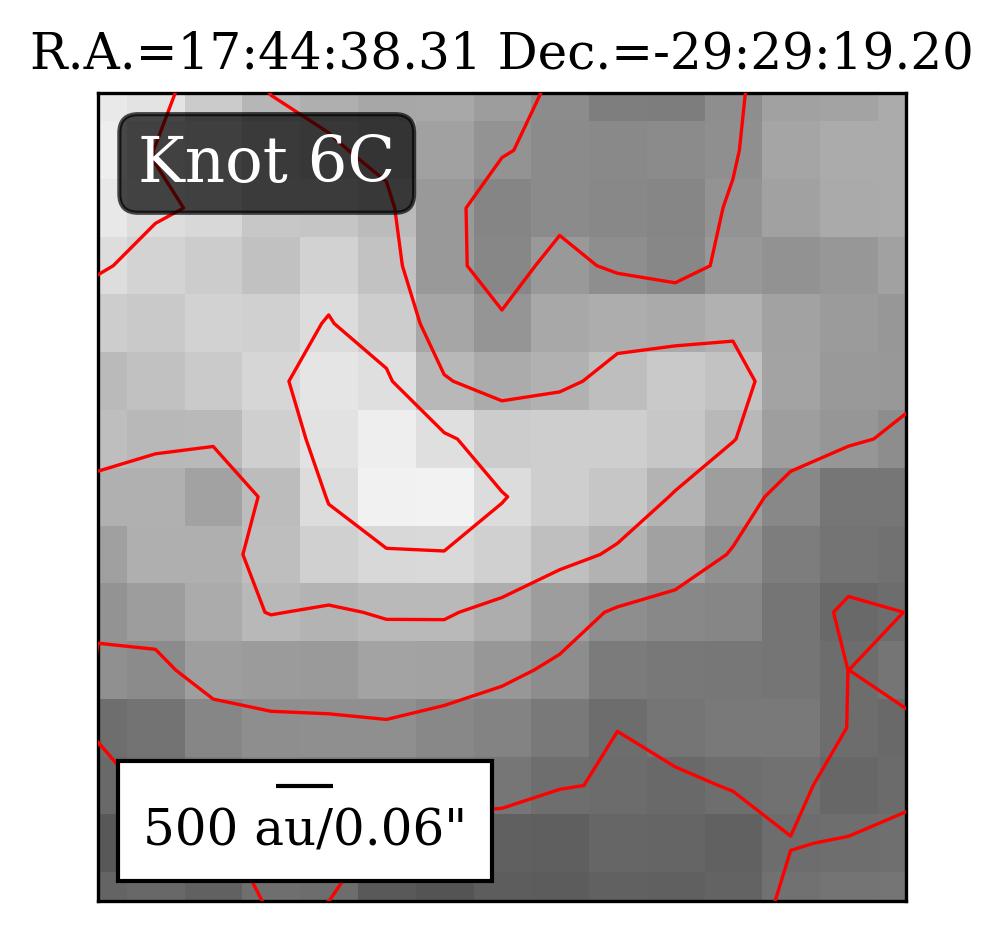}
            \includegraphics[width=0.40\textwidth]{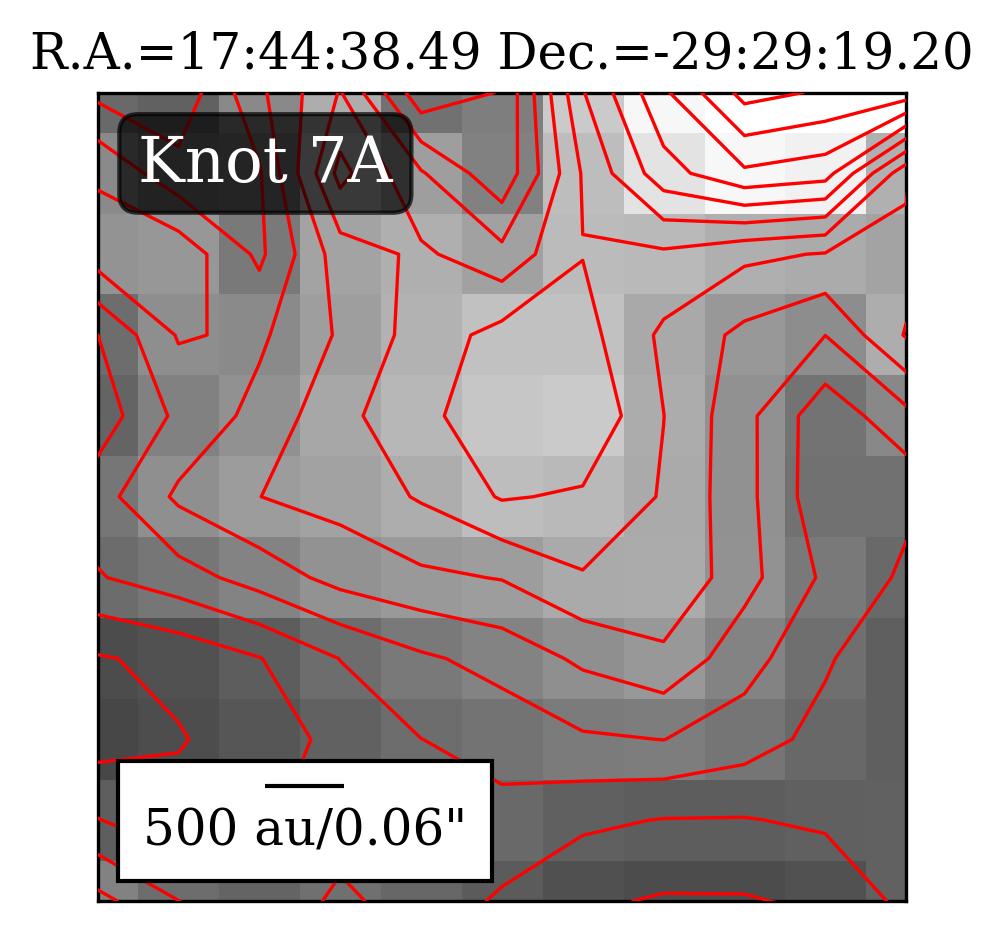}
            \includegraphics[width=0.40\textwidth]{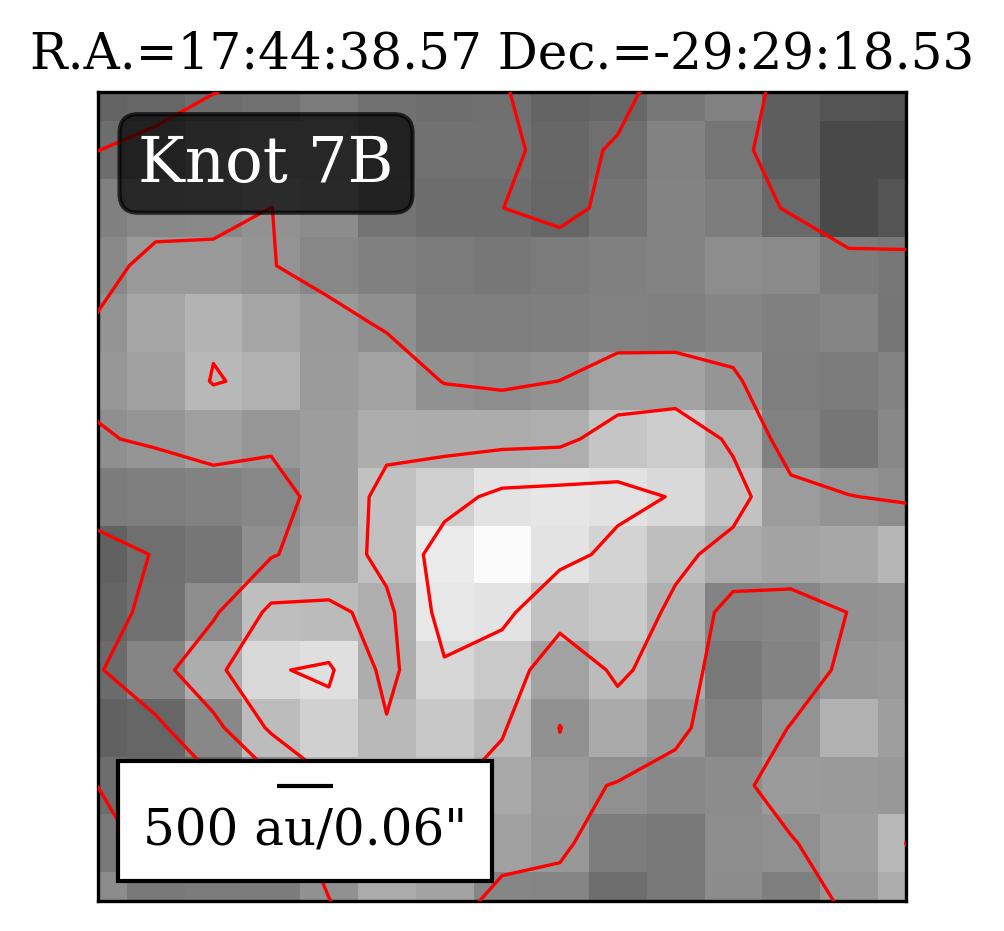}

            \caption{Continued. The contour levels shown represent 15 to 100$\mathrm{\sigma}$ in steps of 5$\sigma$ above the local background.}
        \end{figure*}
    
        \renewcommand{\thefigure}{B\arabic{figure}}
        \addtocounter{figure}{-1}
        \begin{figure*}[!htb]
            \centering
            \includegraphics[width=0.40\textwidth]{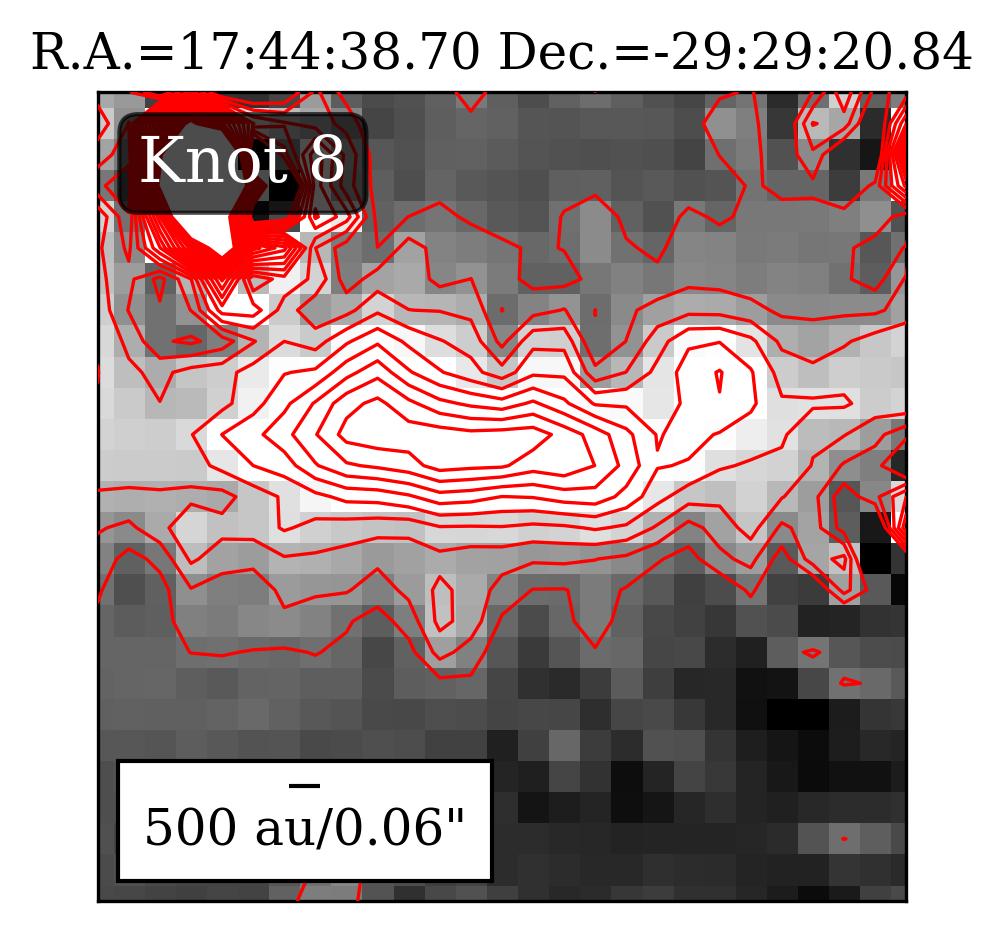}
            \includegraphics[width=0.40\textwidth]{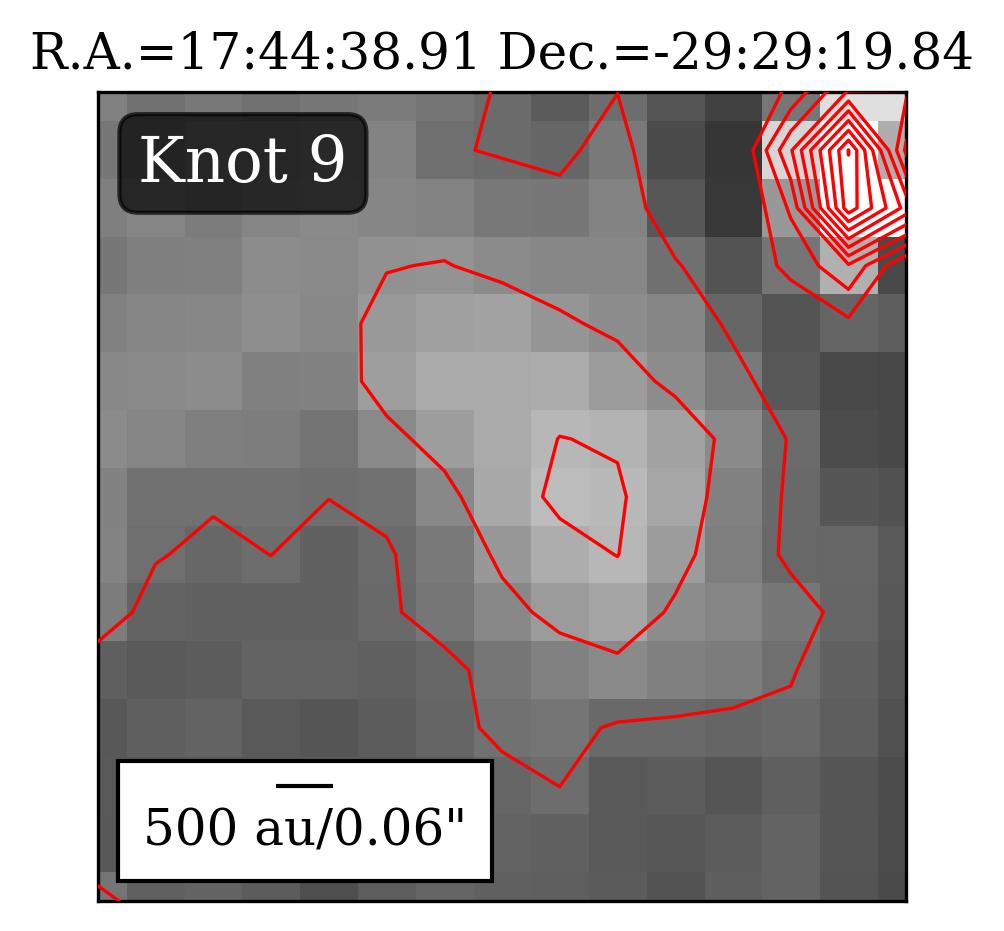}
            \includegraphics[width=0.40\textwidth]{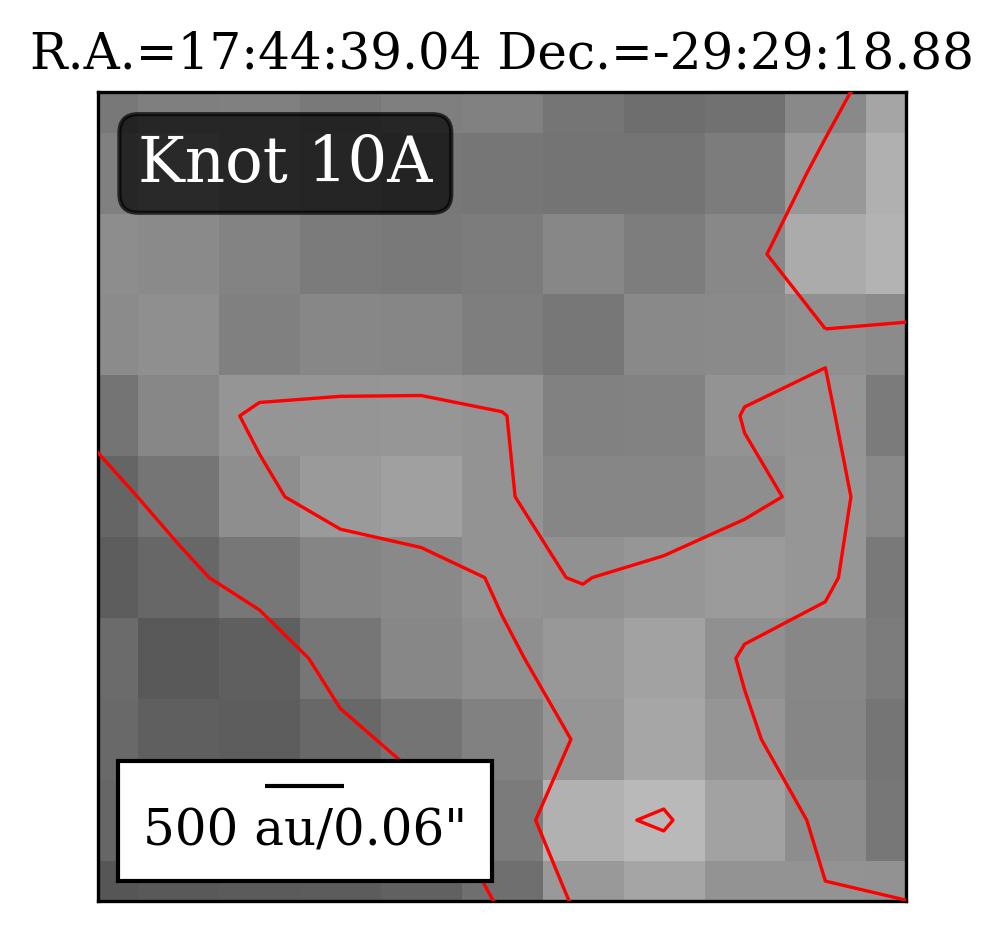}
            \includegraphics[width=0.40\textwidth]{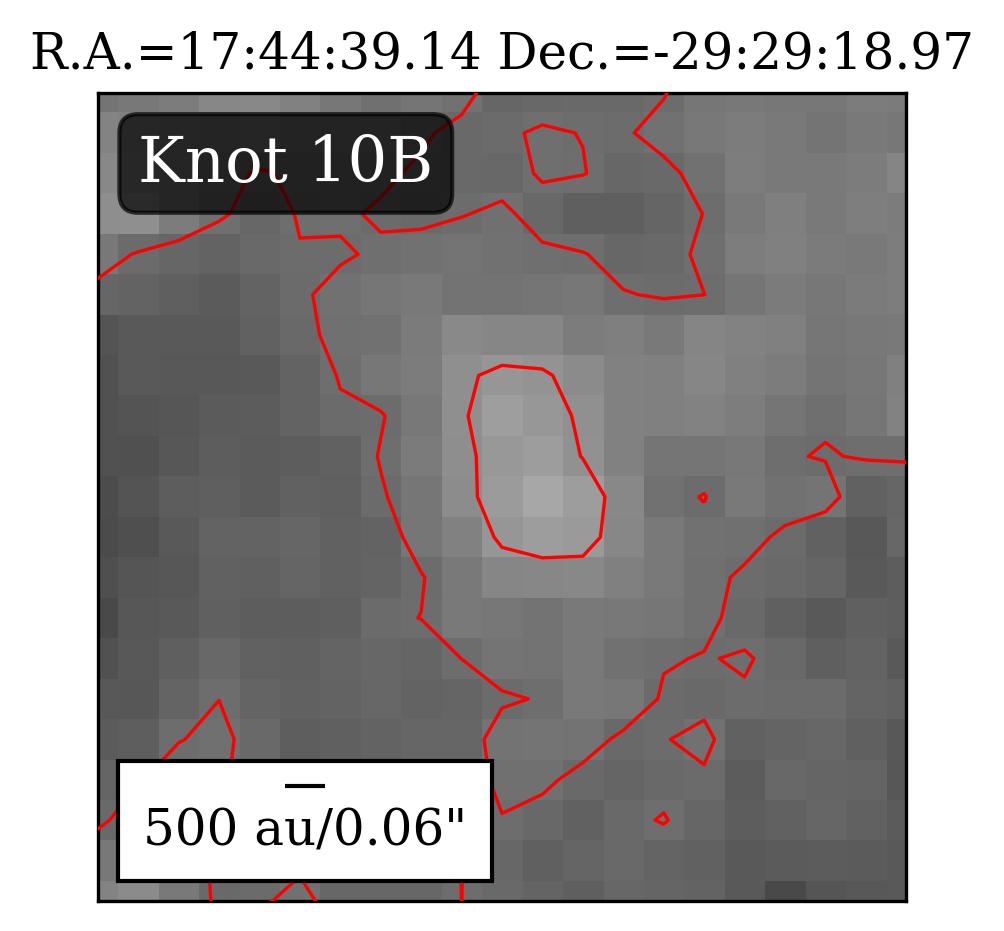}
            \includegraphics[width=0.40\textwidth]{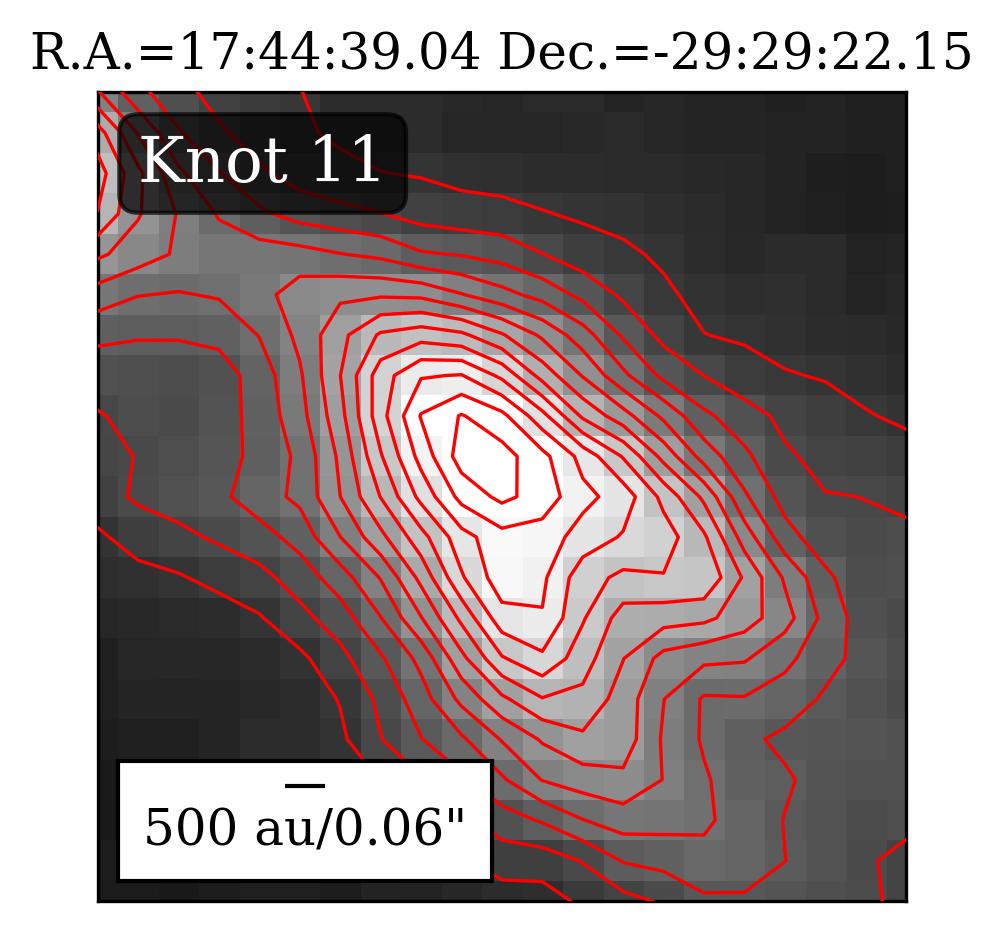}
            \includegraphics[width=0.40\textwidth]{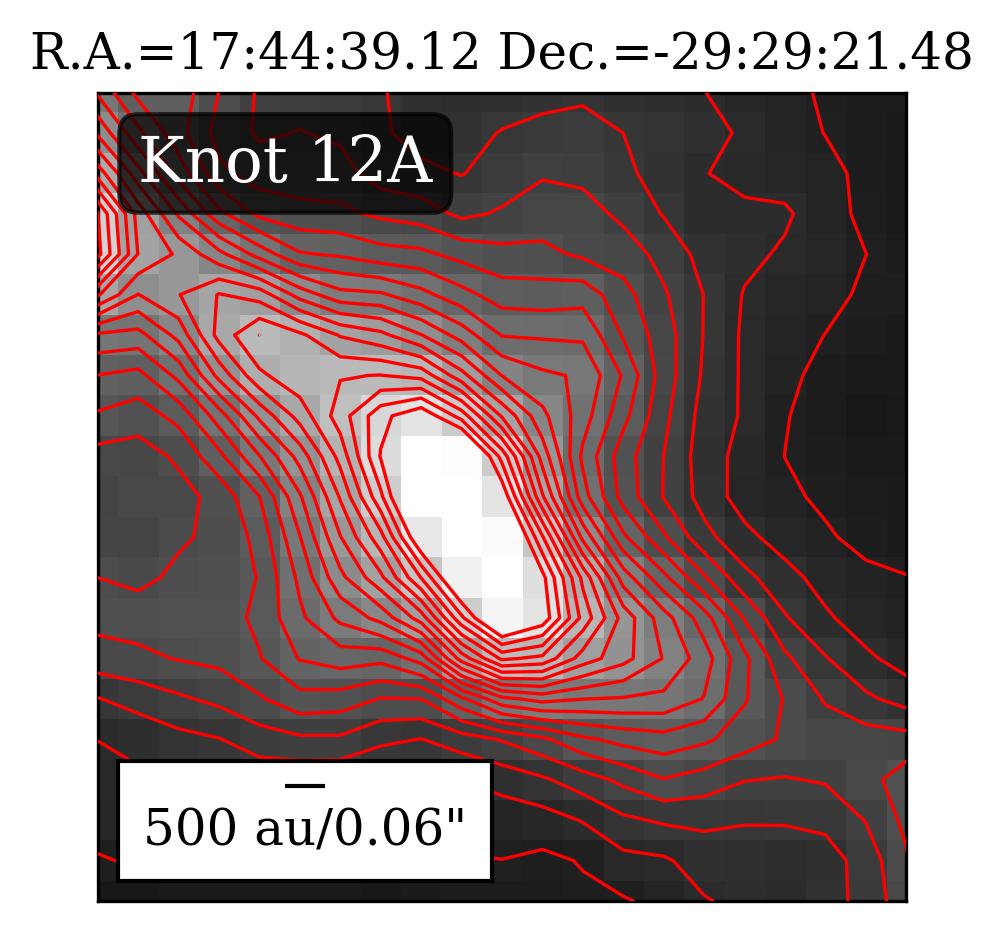}
            \caption{Continued. The contour levels shown represent 15 to 100$\mathrm{\sigma}$ in steps of 5$\sigma$ above the local background.}
        \end{figure*}
    
        \renewcommand{\thefigure}{B\arabic{figure}}
        \addtocounter{figure}{-1}
        \begin{figure*}[!htb]
            \centering
            \includegraphics[width=0.40\textwidth]{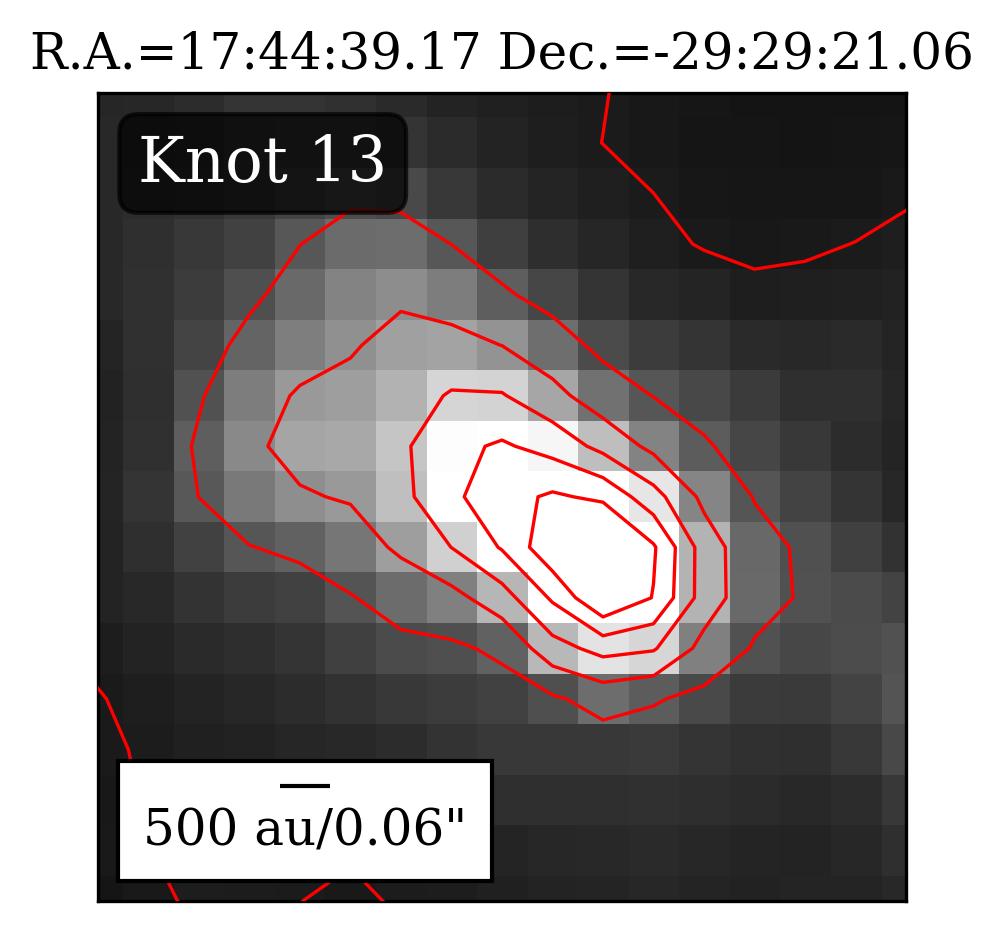}
            \includegraphics[width=0.40\textwidth]{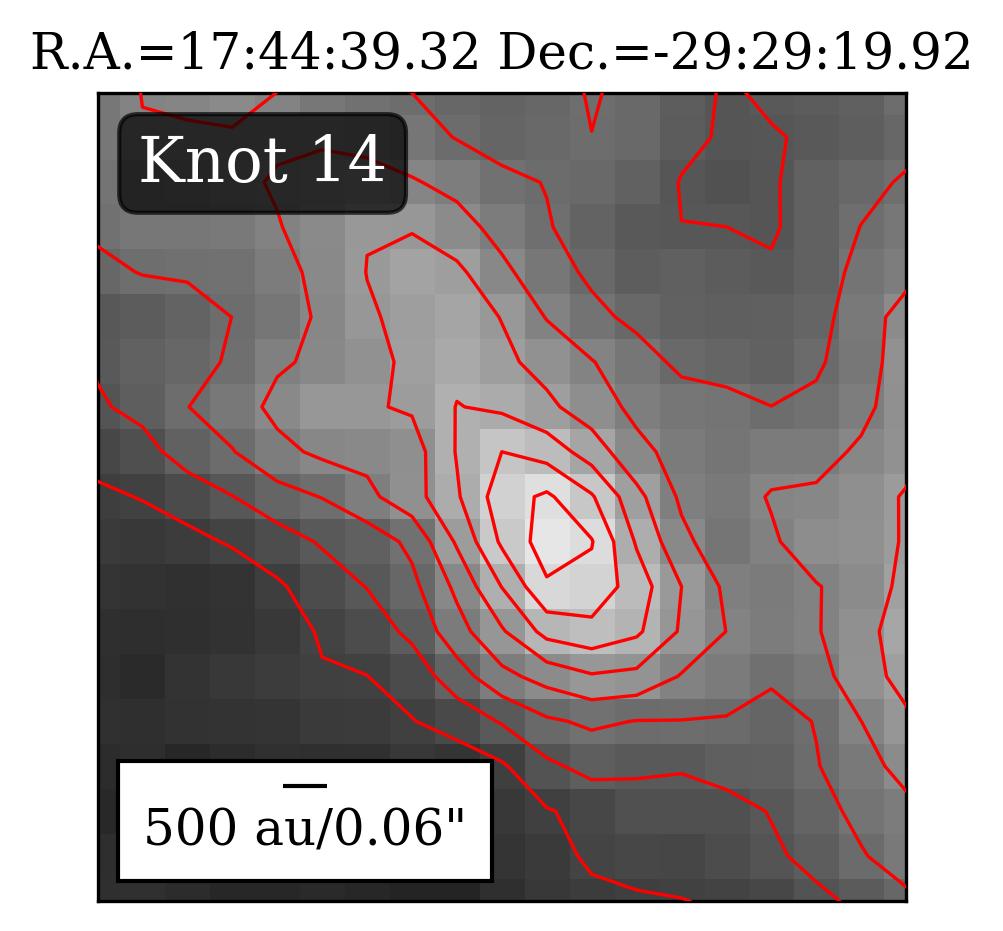}
            \includegraphics[width=0.40\textwidth]{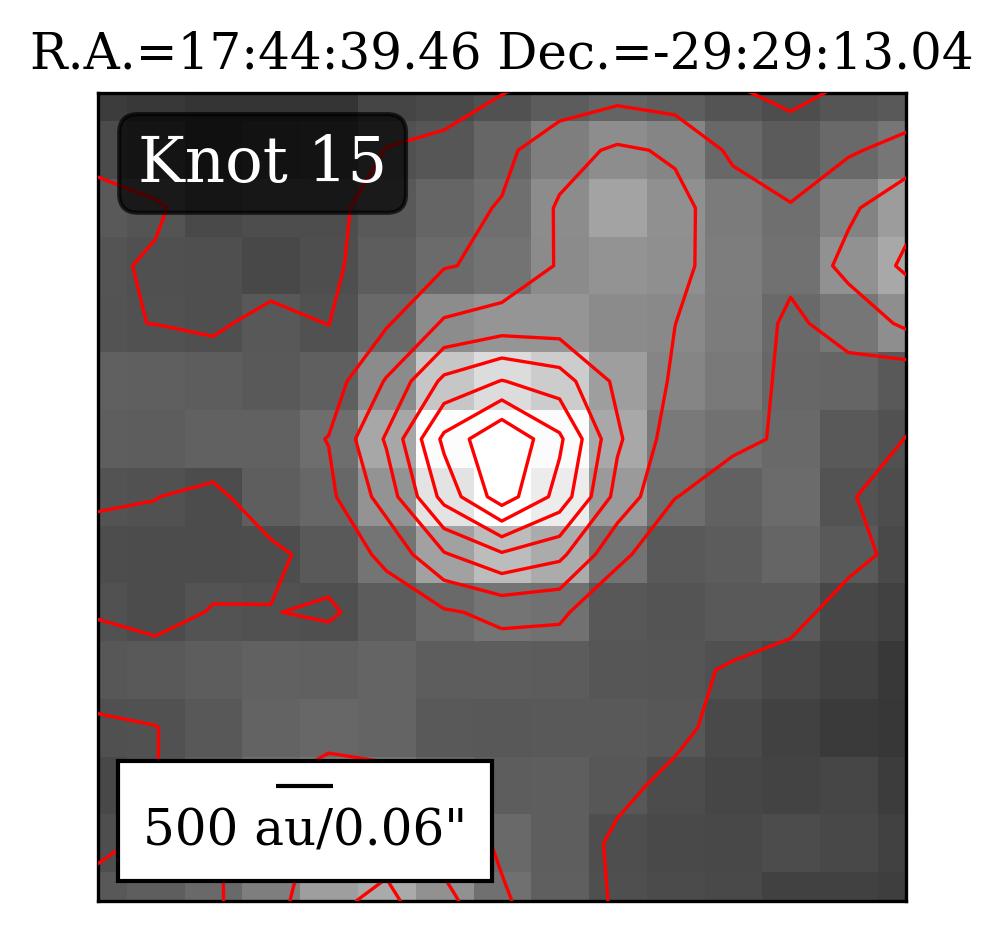}
            \includegraphics[width=0.40\textwidth]{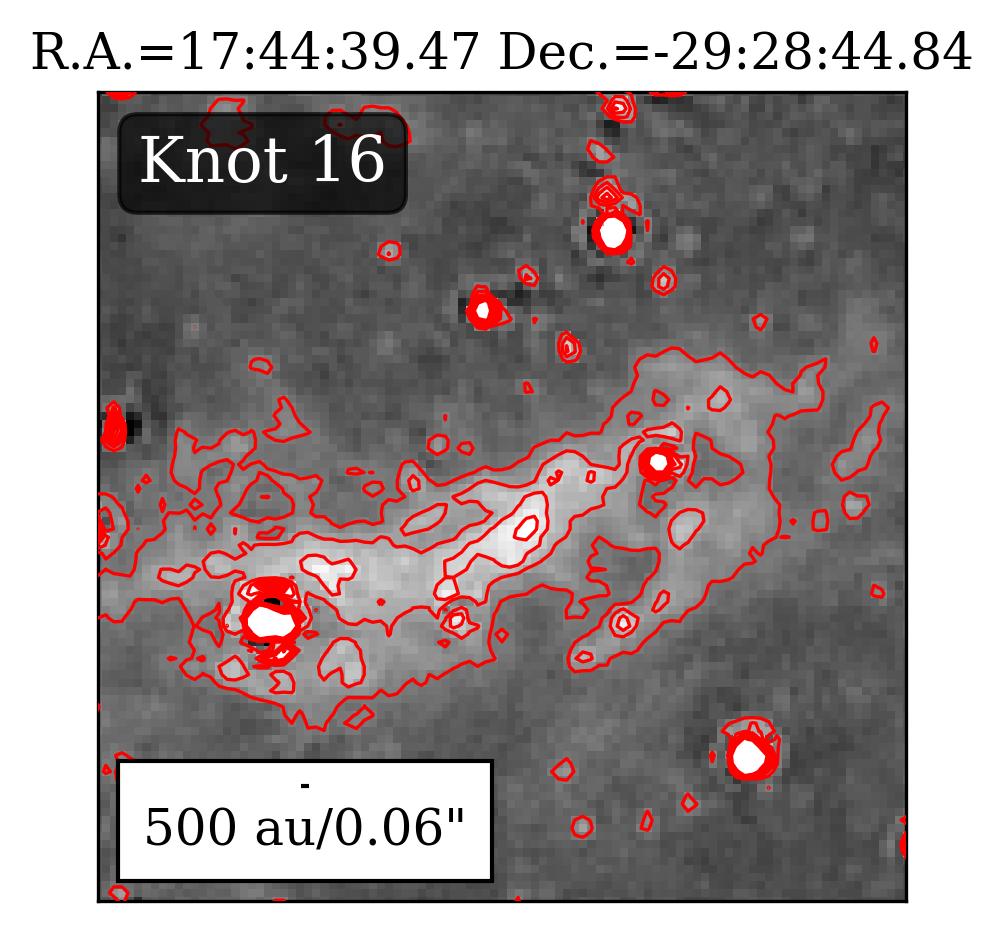}
            \includegraphics[width=0.40\textwidth]{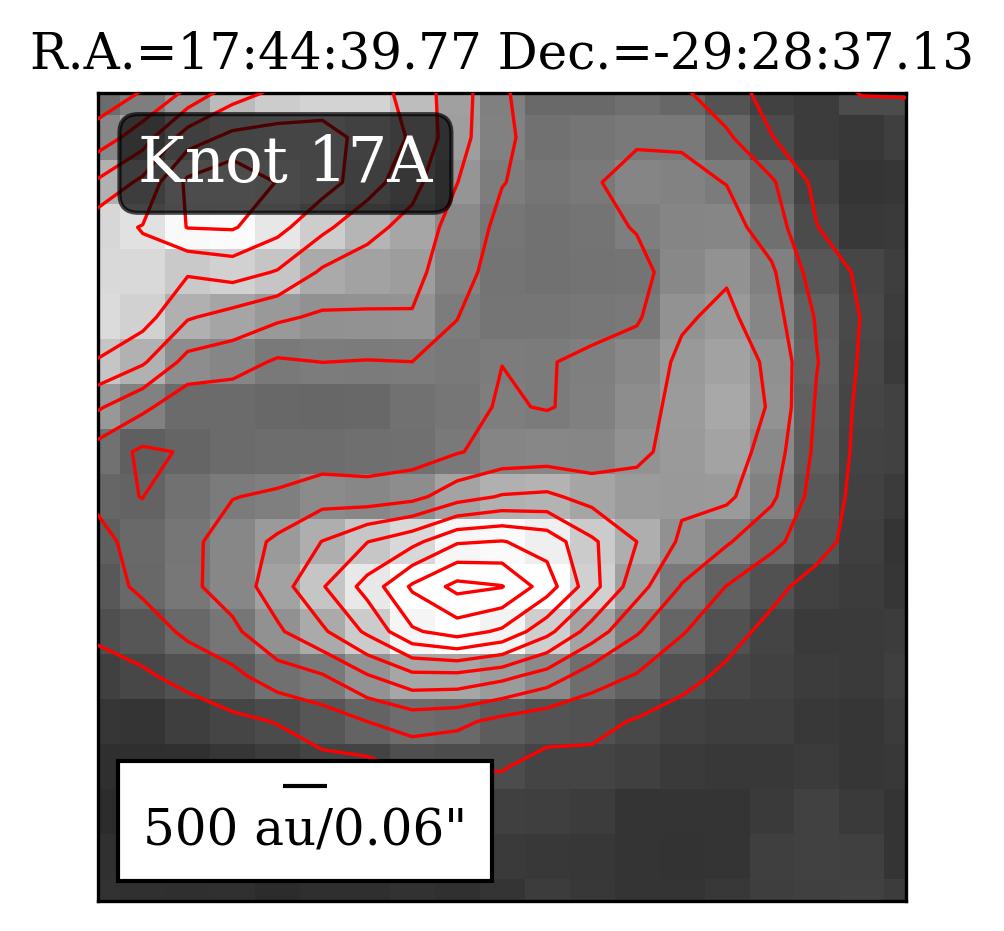}
            \includegraphics[width=0.40\textwidth]{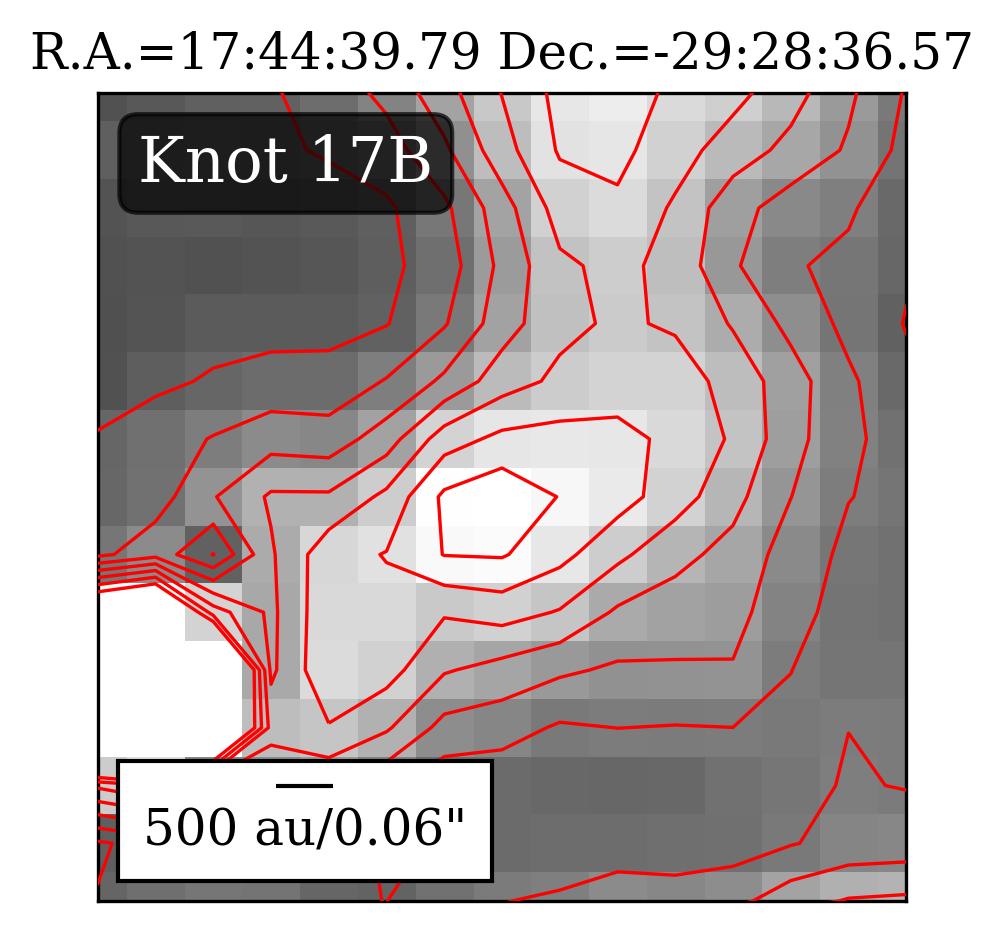}
            
            \caption{Continued. The contour levels shown for knots 14, 15, 16, 17A, and 17B represent represent 15 to 100$\mathrm{\sigma}$ in steps of 5$\sigma$ above the local background; those shown for knot 13 represent 25 to 300$\mathrm{\sigma}$ in steps of 55$\sigma$.} 
        \end{figure*}
        \renewcommand{\thefigure}{B\arabic{figure}}
        \addtocounter{figure}{-1}
        \begin{figure*}[!htb]
        \centering
            \includegraphics[width=0.40\textwidth]{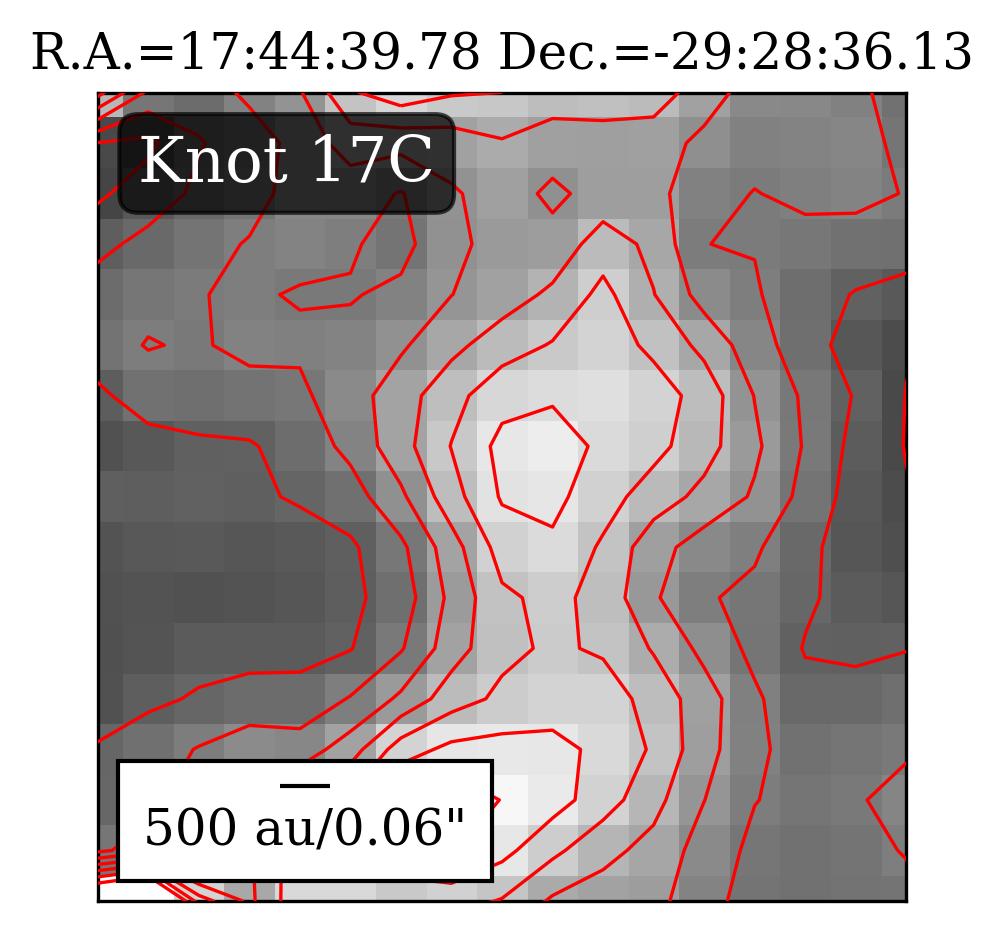}
            \includegraphics[width=0.40\textwidth]{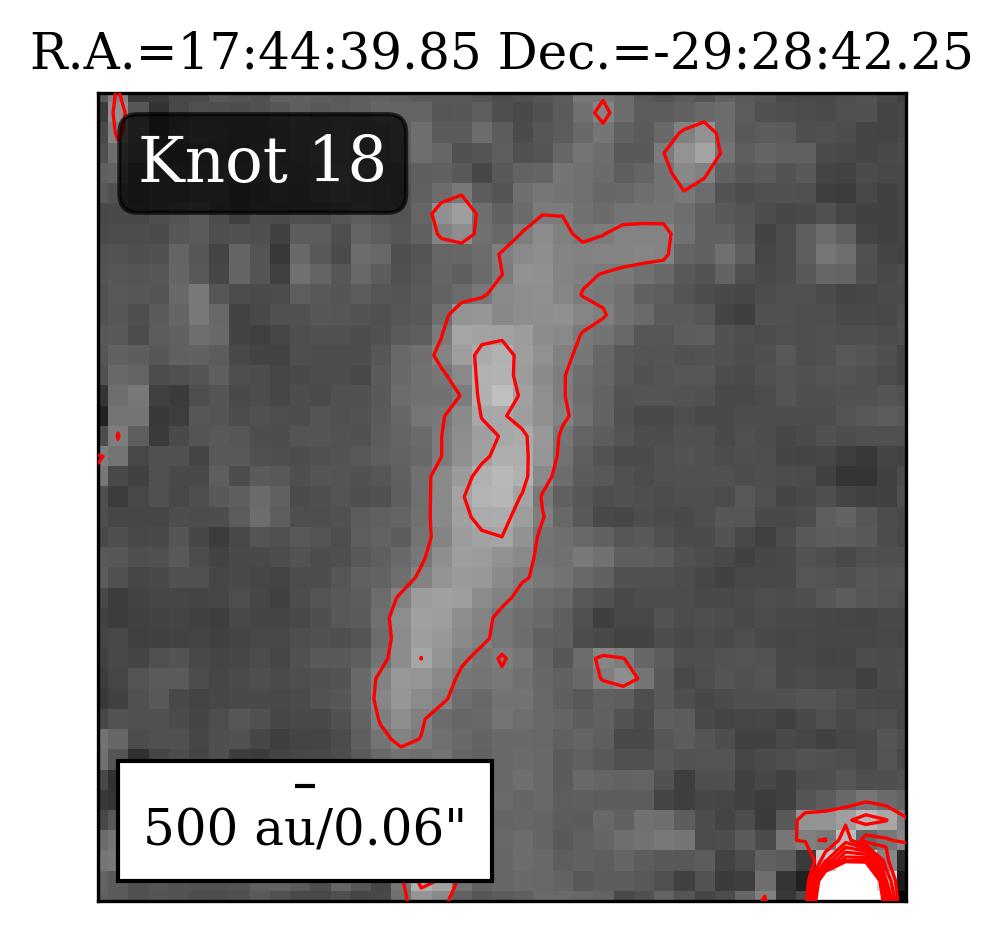}
            \includegraphics[width=0.40\textwidth]{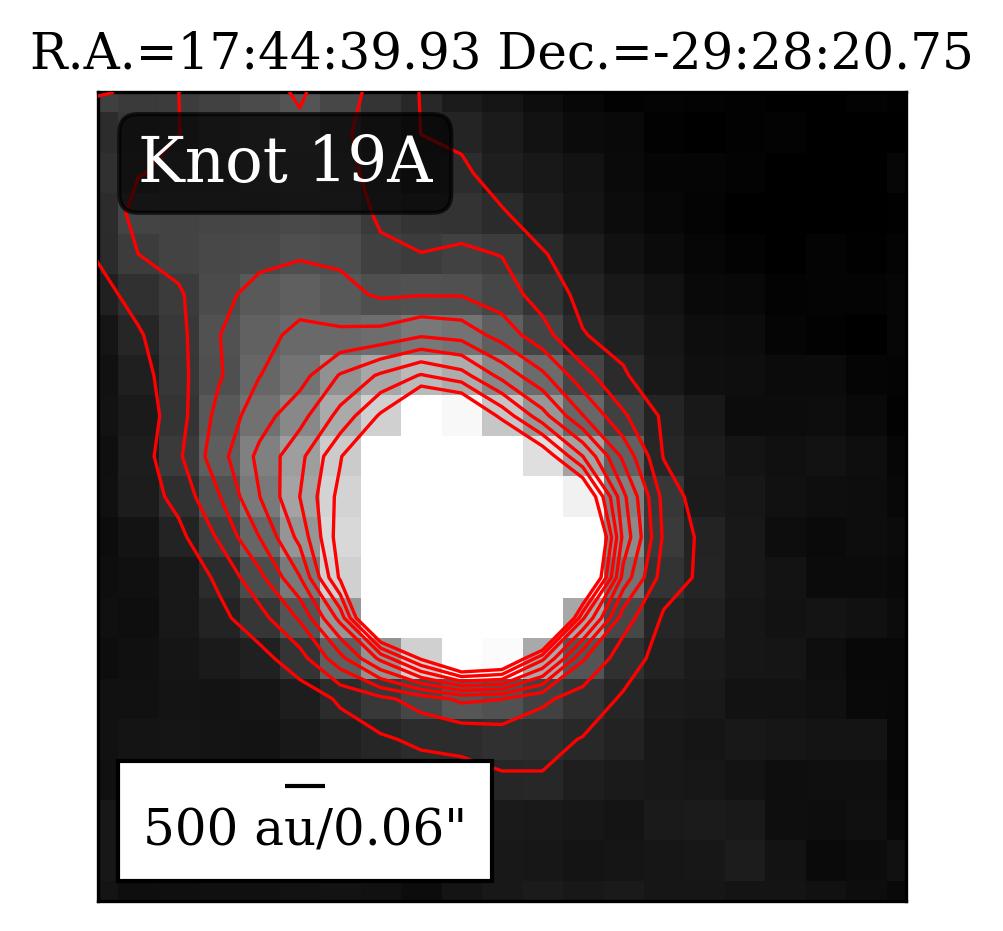}
            \includegraphics[width=0.40\textwidth]{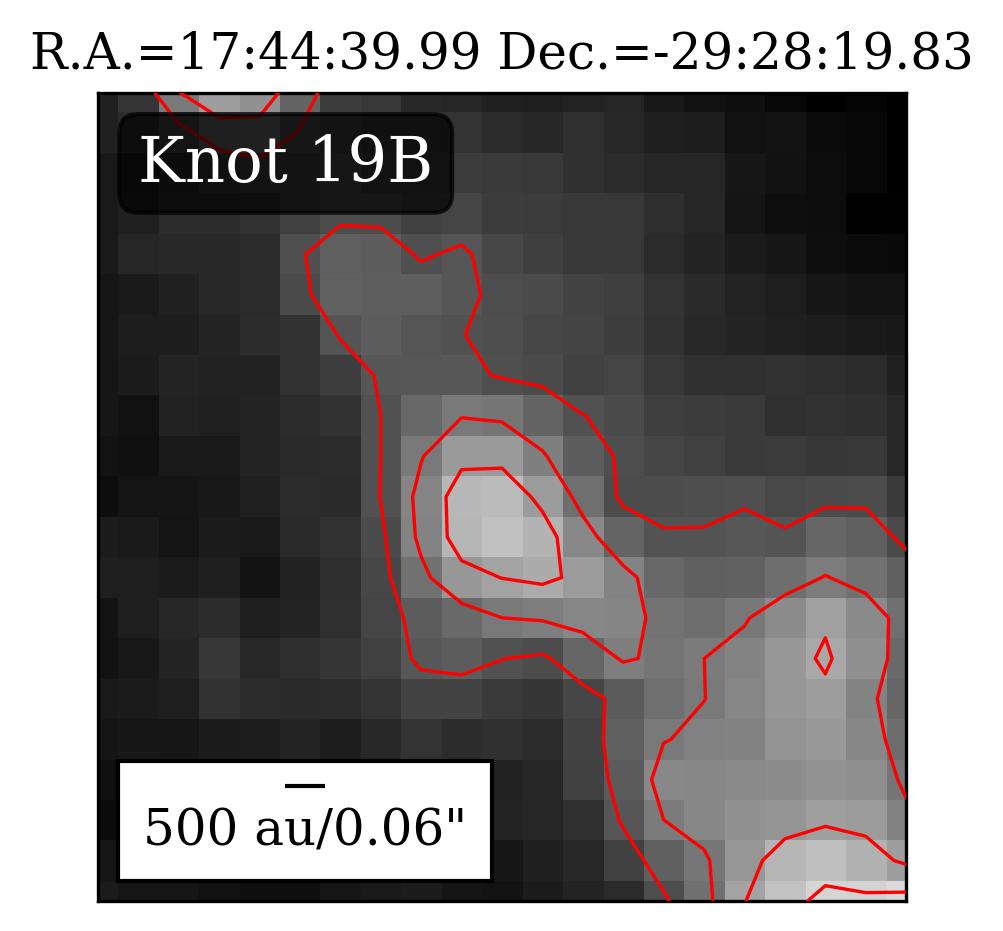}
            \includegraphics[width=0.40\textwidth]{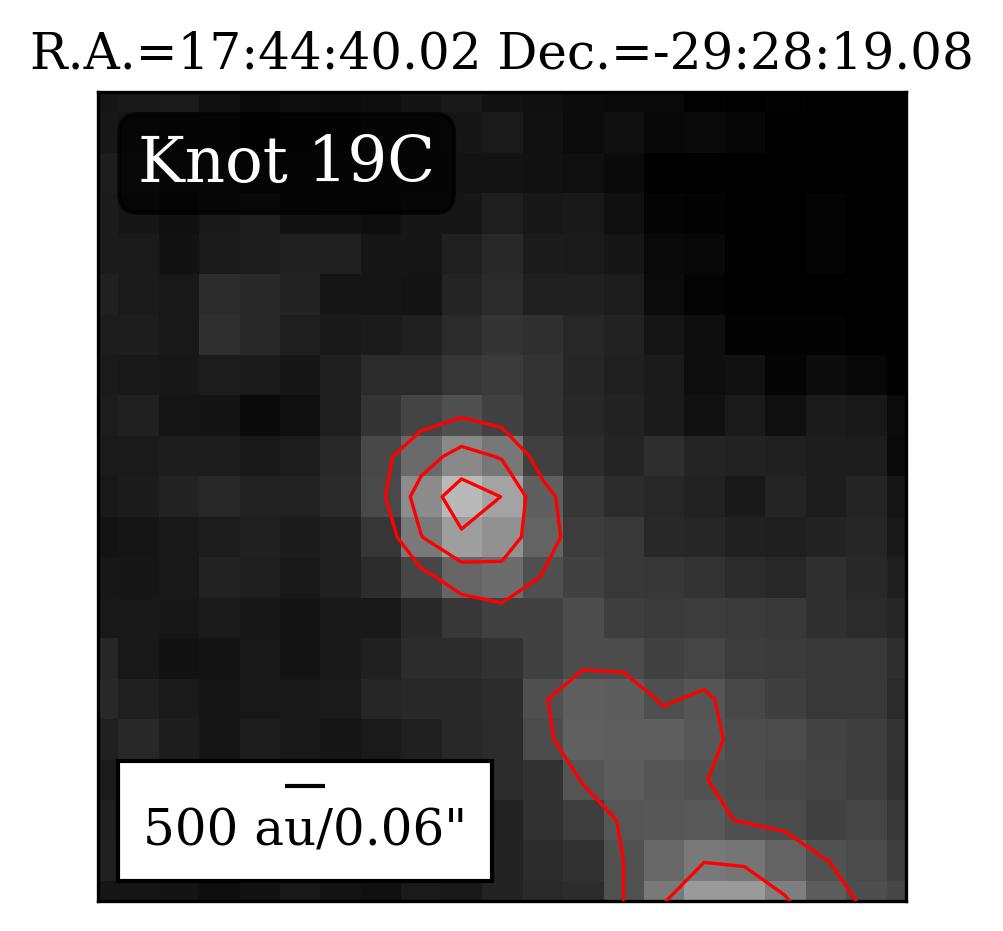}
            \includegraphics[width=0.40\textwidth]{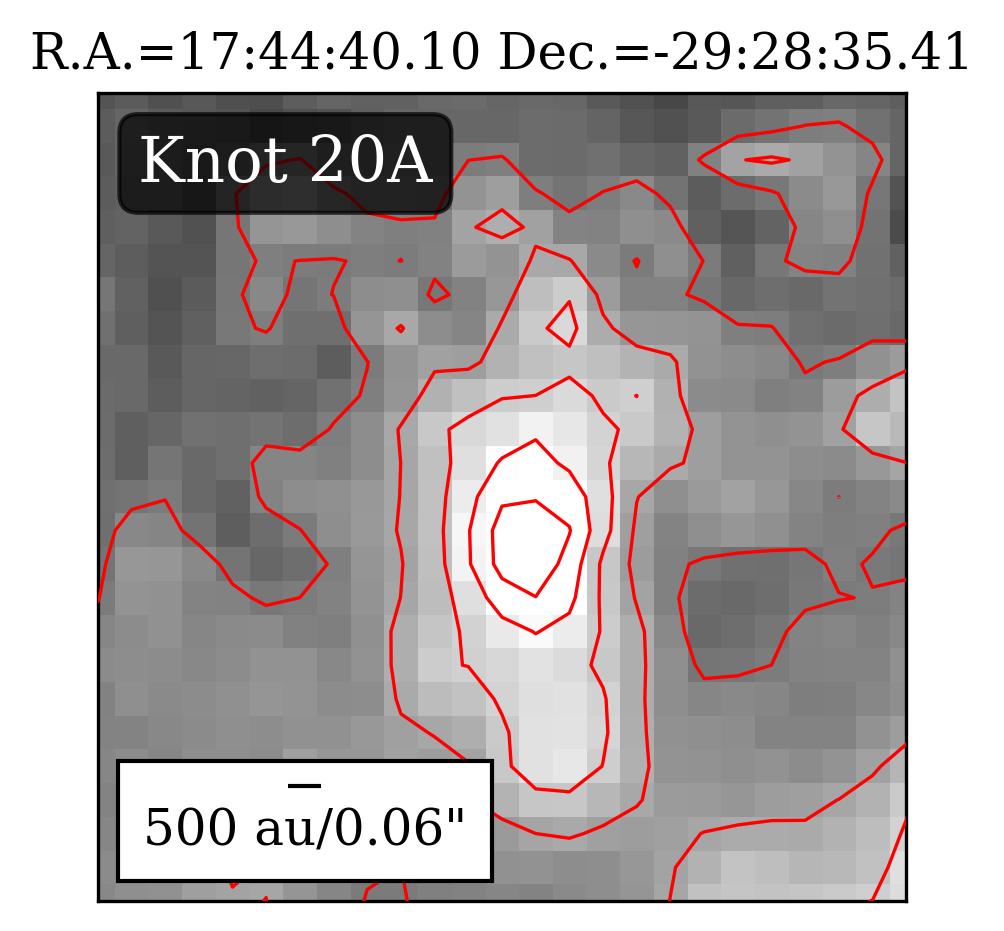}
            
            \caption{Continued. The contour levels shown for knots 17C, 18, and 20A represent 15 to 100$\mathrm{\sigma}$ in steps of 5$\sigma$ above the local background; those shown for knots 19A-C represent 10 to 50$\mathrm{\sigma}$ in steps of 5$\sigma$.}
        \end{figure*}
        \renewcommand{\thefigure}{B\arabic{figure}}
        \addtocounter{figure}{-1}
        \begin{figure*}[!htb]
        \centering
            \includegraphics[width=0.40\textwidth]{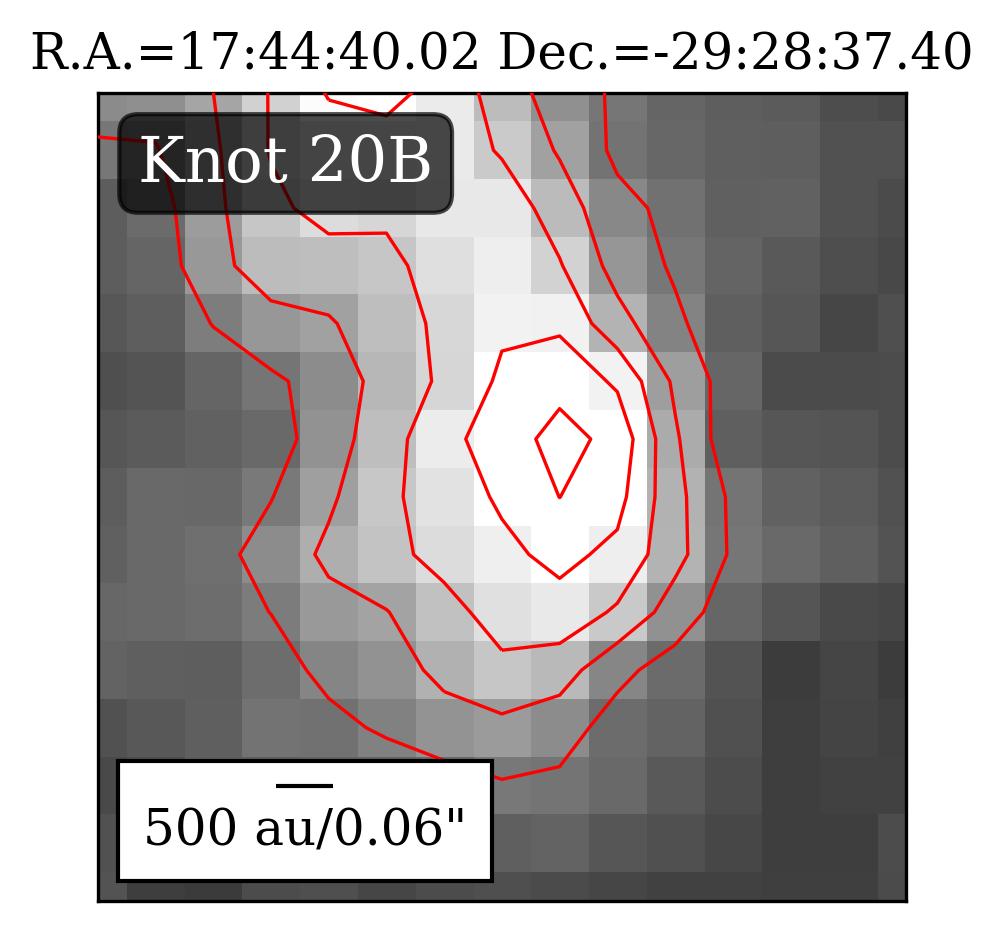}
            \includegraphics[width=0.40\textwidth]{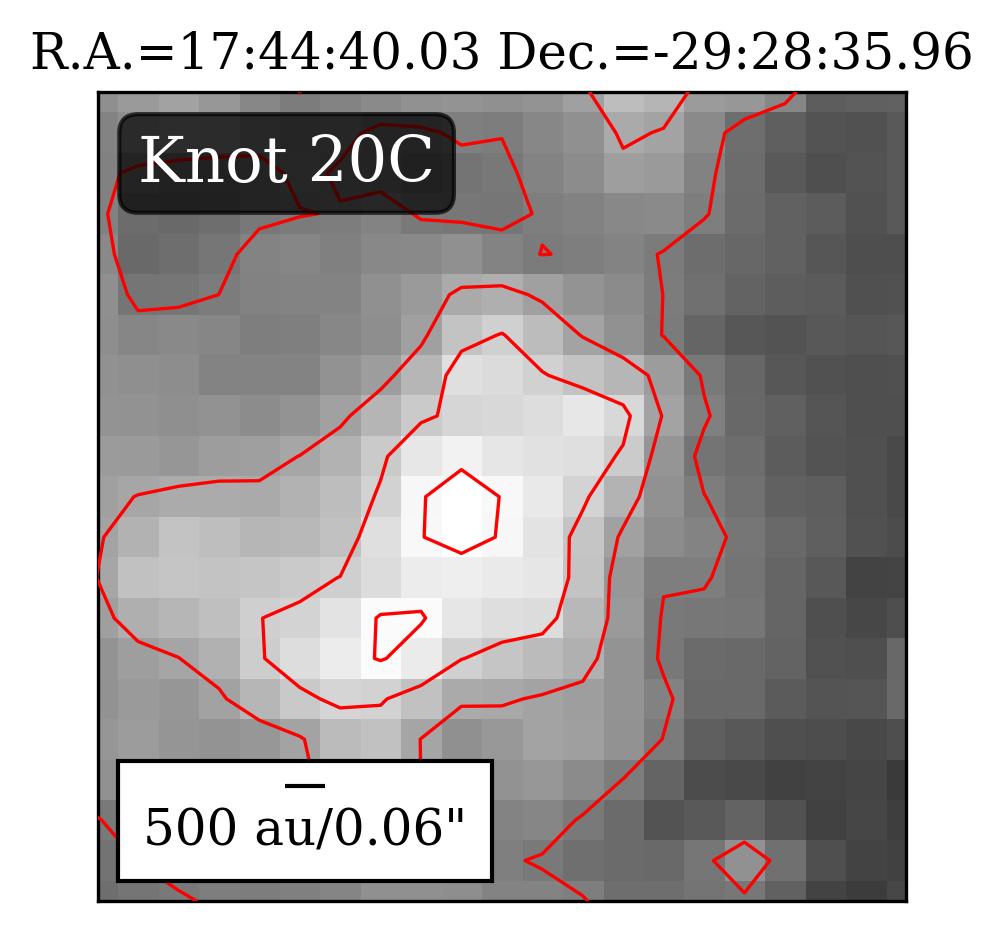}
            \includegraphics[width=0.40\textwidth]{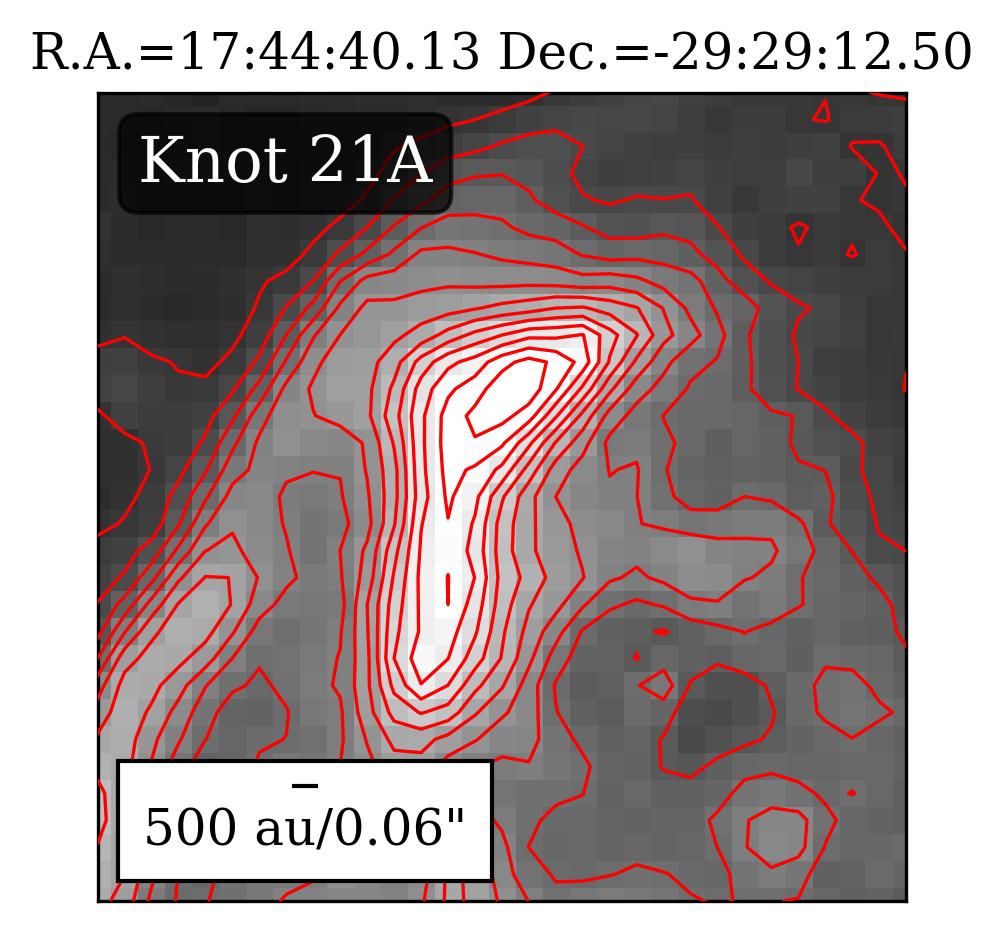}
            \includegraphics[width=0.40\textwidth]{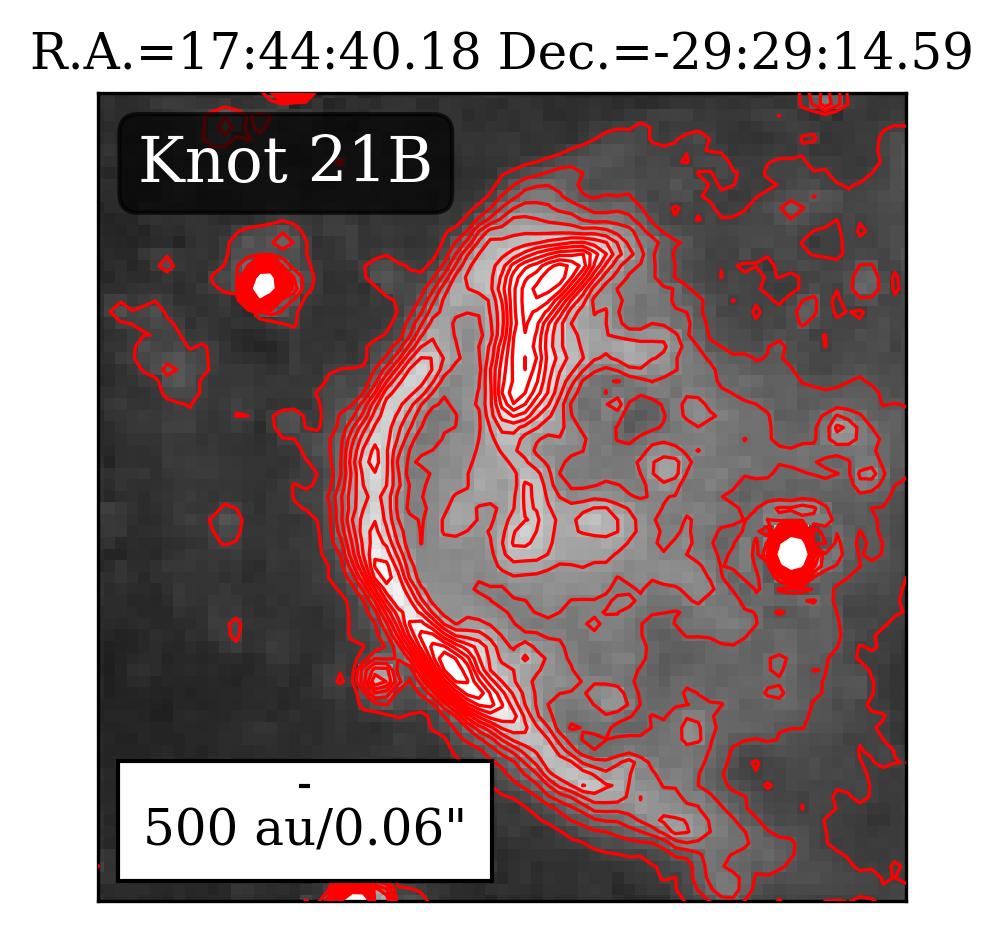}
            \includegraphics[width=0.40\textwidth]{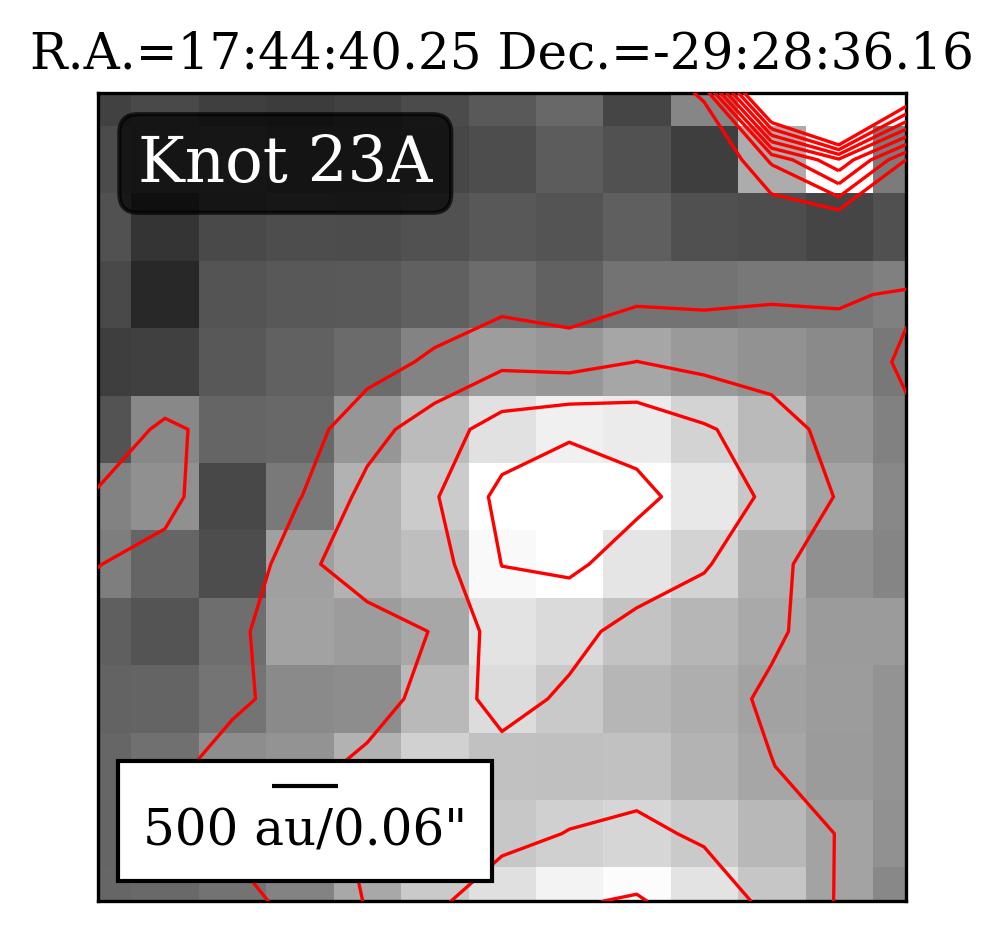}
            \includegraphics[width=0.40\textwidth]{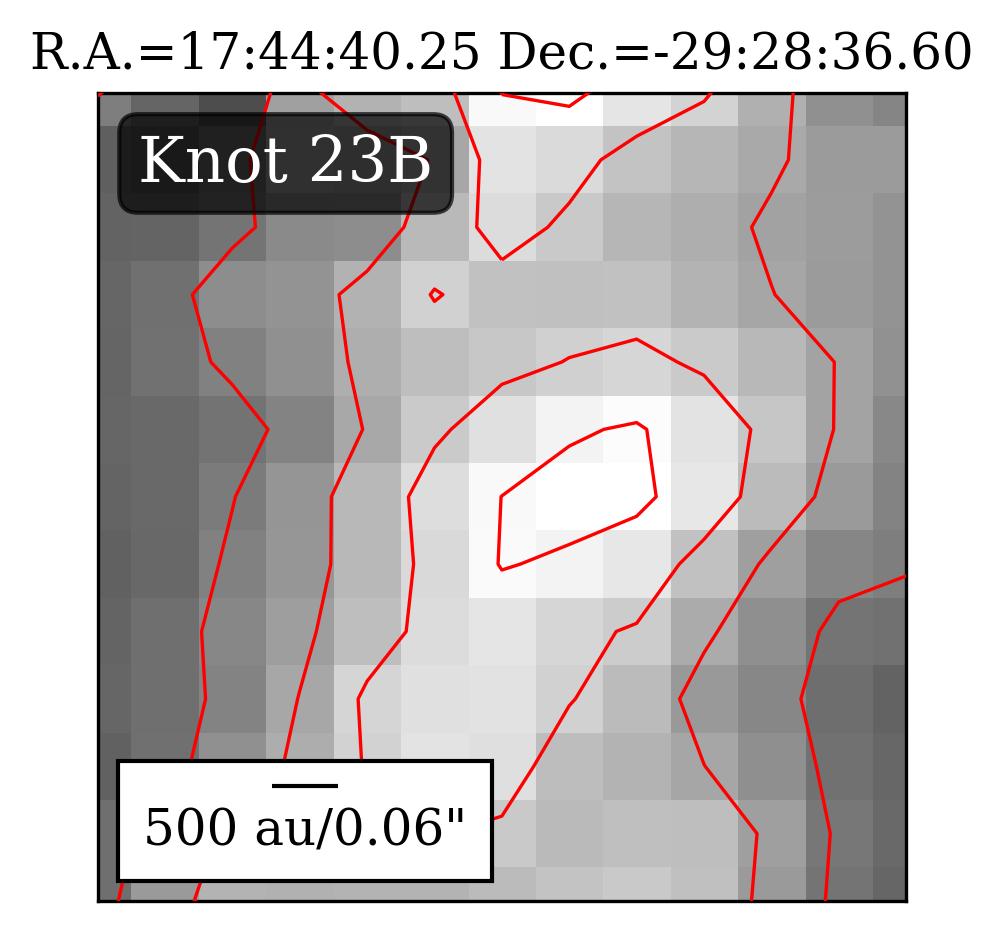}
            
            \caption{Continued. The contour levels shown represent 15 to 100$\mathrm{\sigma}$ in steps of 5$\sigma$ above the local background.}
        \end{figure*}
        \renewcommand{\thefigure}{B\arabic{figure}}
        \addtocounter{figure}{-1}
        \begin{figure*}[!htb]
        \centering
            \includegraphics[width=0.40\textwidth]{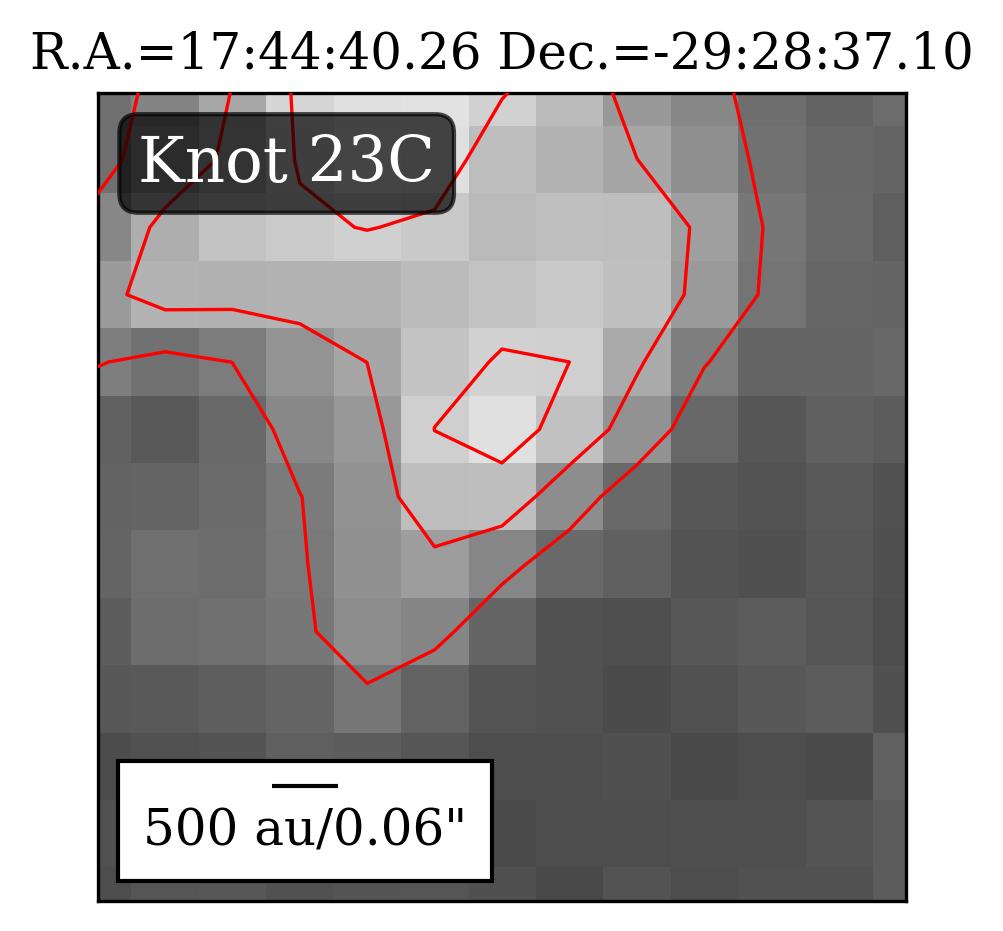}  
            \includegraphics[width=0.40\textwidth]{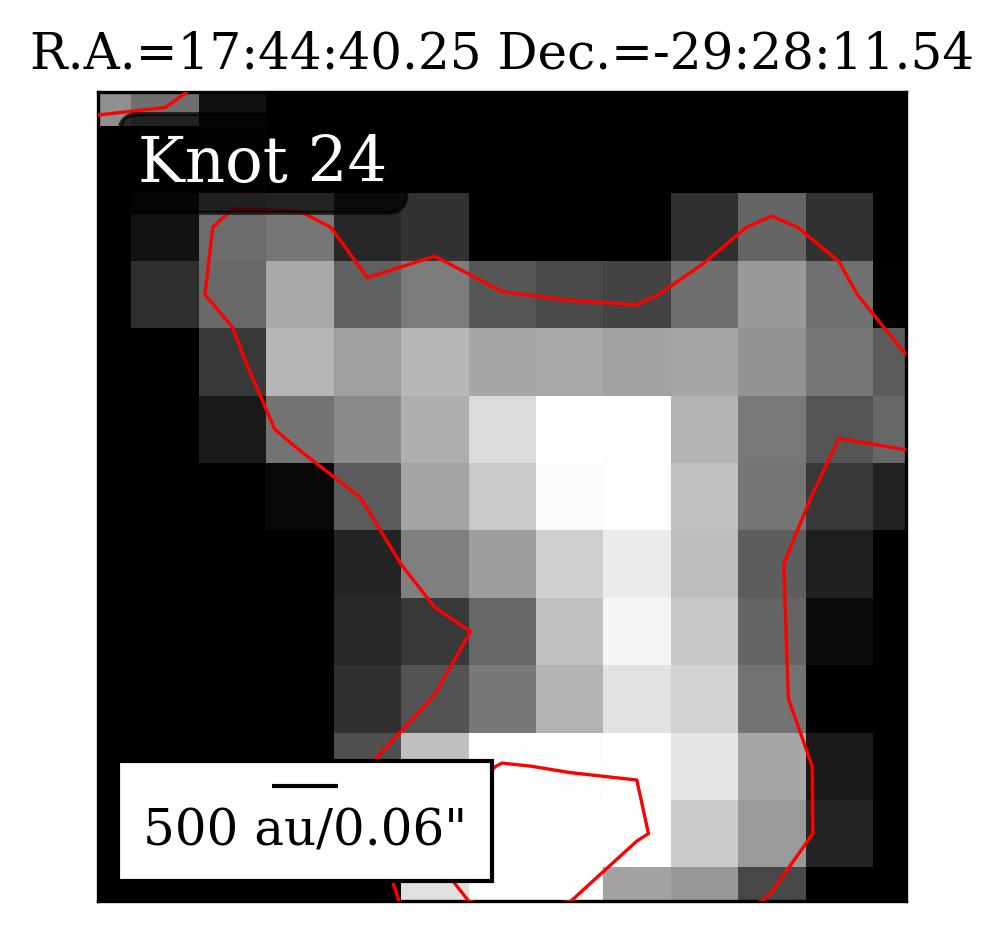}
            \includegraphics[width=0.40\textwidth]{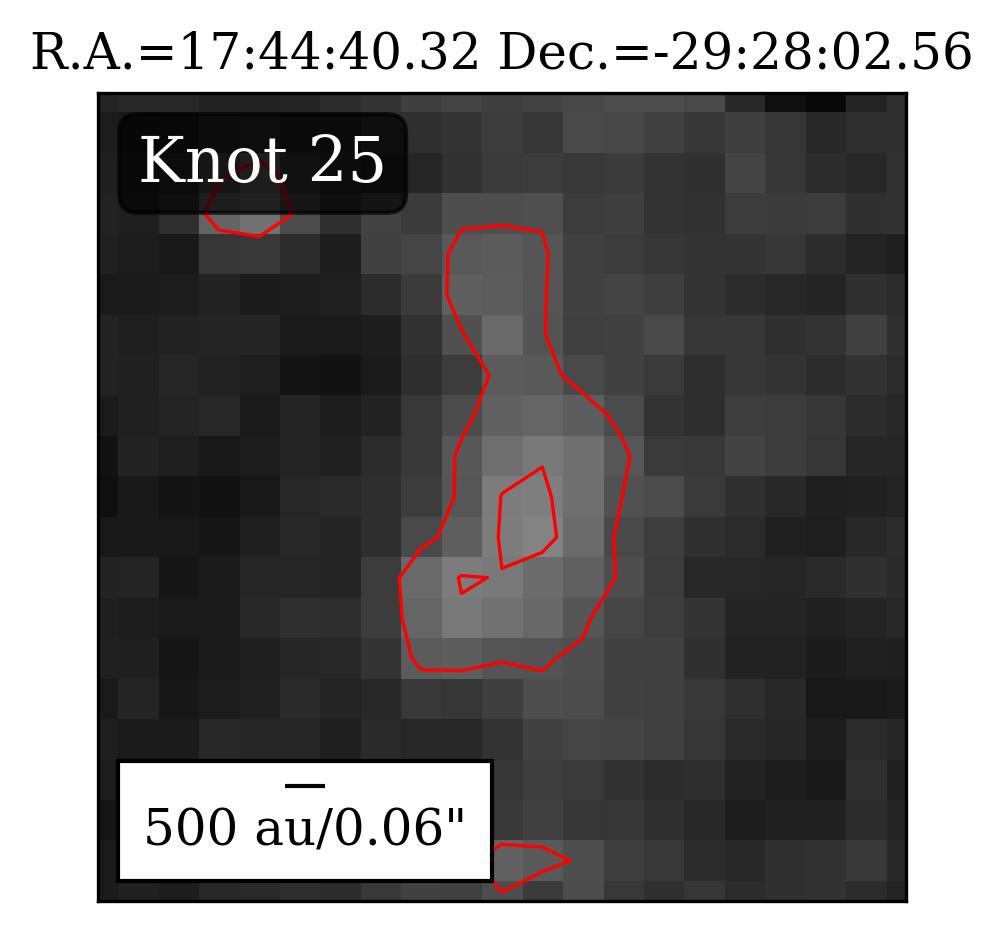}
            \includegraphics[width=0.40\textwidth]{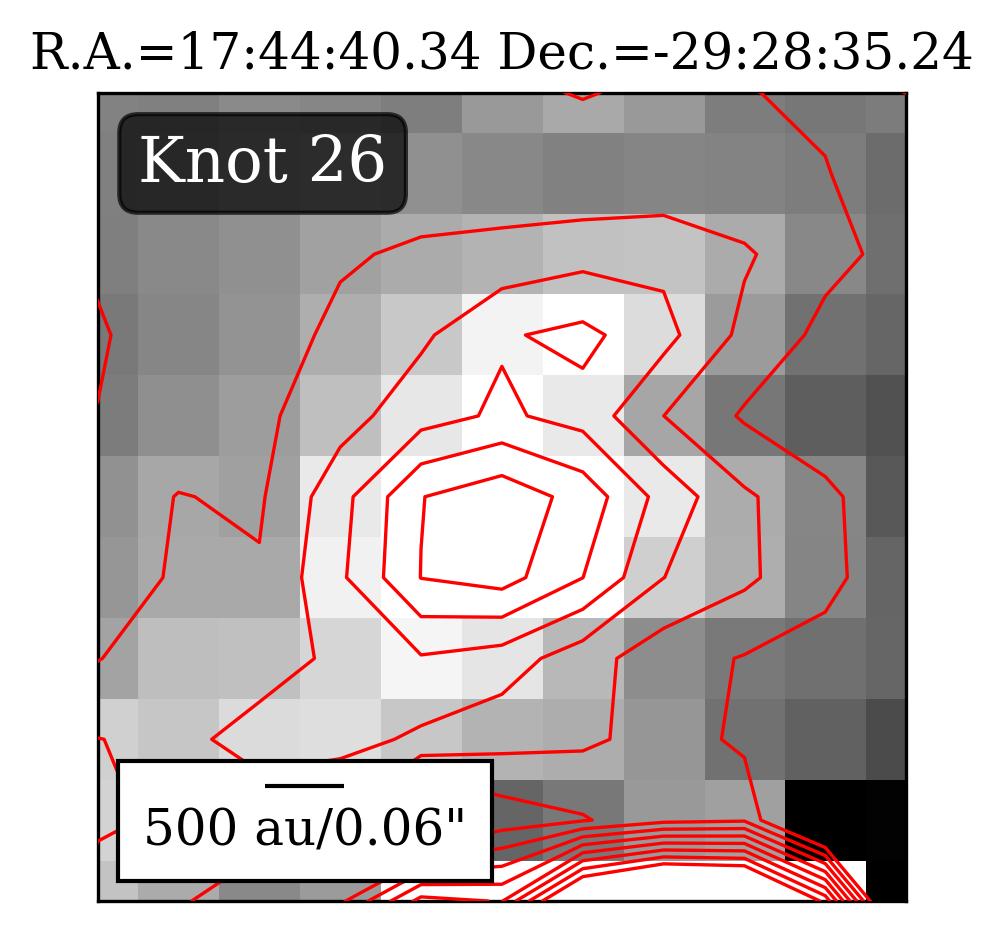}
            \includegraphics[width=0.40\textwidth]{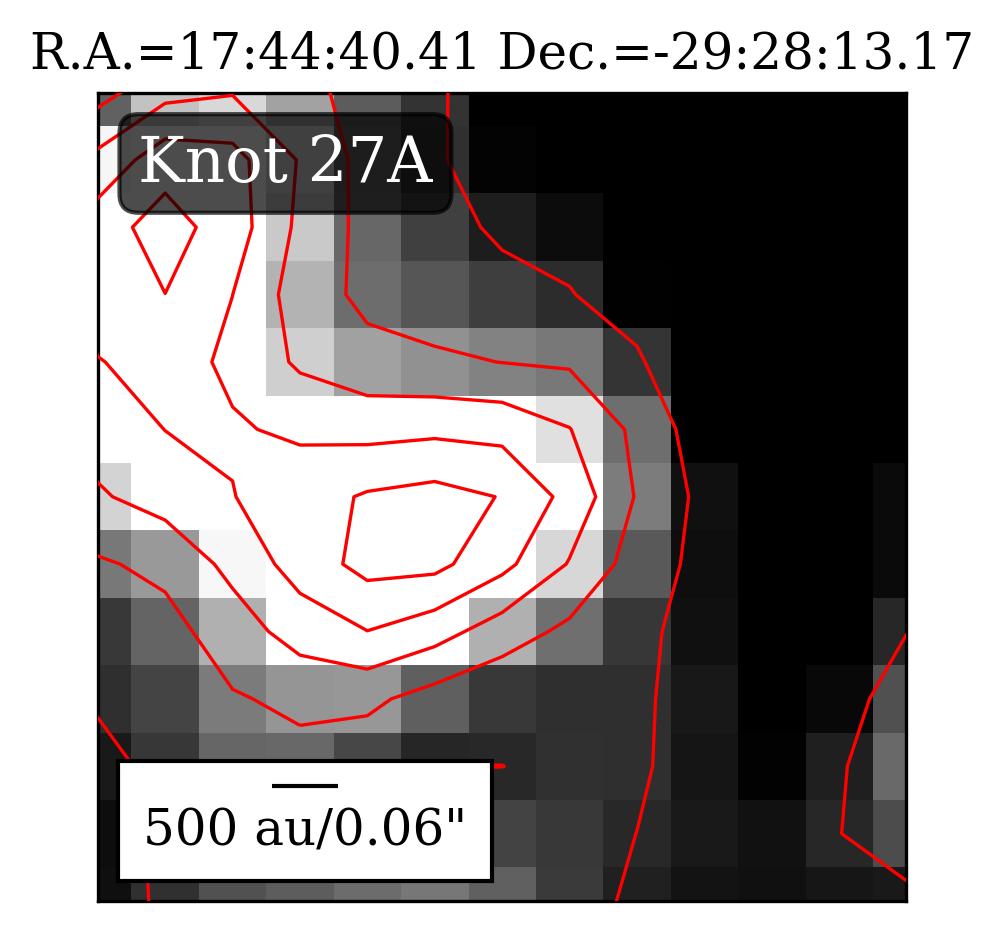}
            \includegraphics[width=0.40\textwidth]{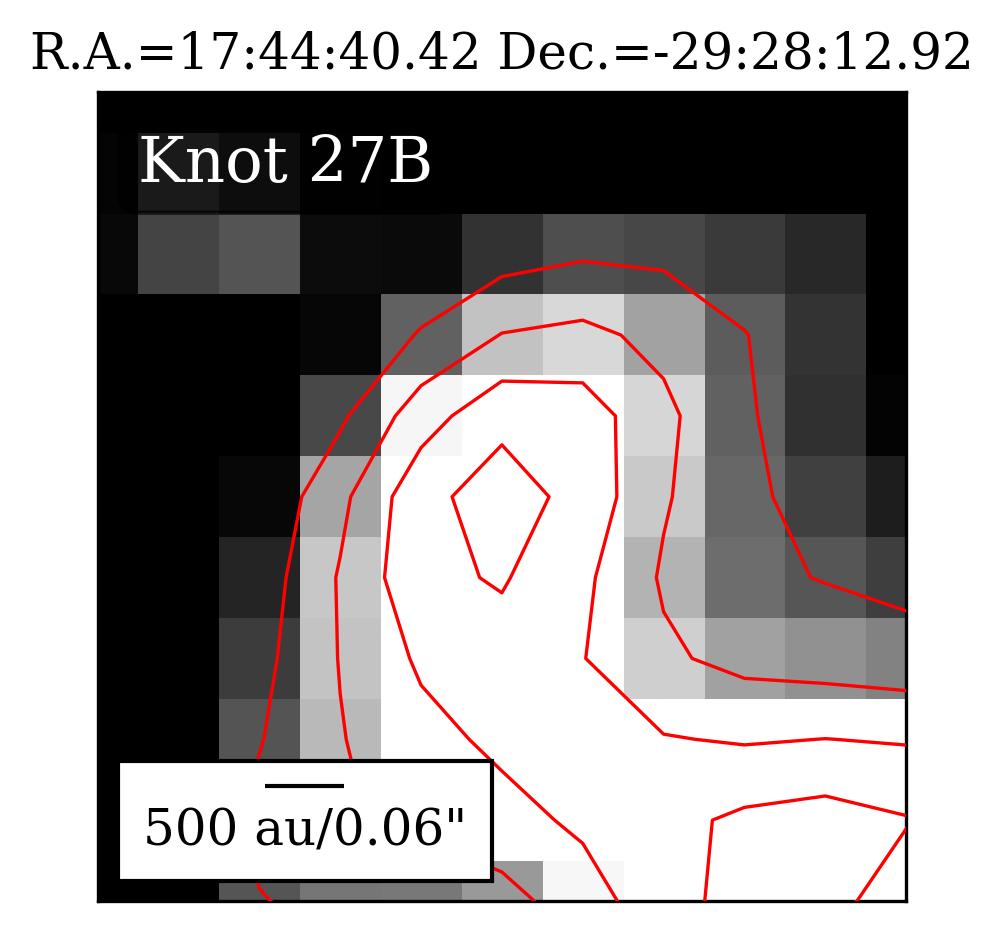}
            
            \caption{Continued. The contour levels shown for knots 23C and 26 represent 15 to 100$\mathrm{\sigma}$ in steps of 5$\sigma$ above the local background; those shown for knots 24, 27A, and 27B represent 25 to 300$\mathrm{\sigma}$ in steps of 55$\sigma$ above the local background; those shown for knot 25 represent represent 10 to 50$\mathrm{\sigma}$ in steps of 5$\sigma$.}
        \end{figure*}
        \renewcommand{\thefigure}{B\arabic{figure}}
        \addtocounter{figure}{-1}
        \begin{figure*}[!htb]
        \centering
            \includegraphics[width=0.40\textwidth]{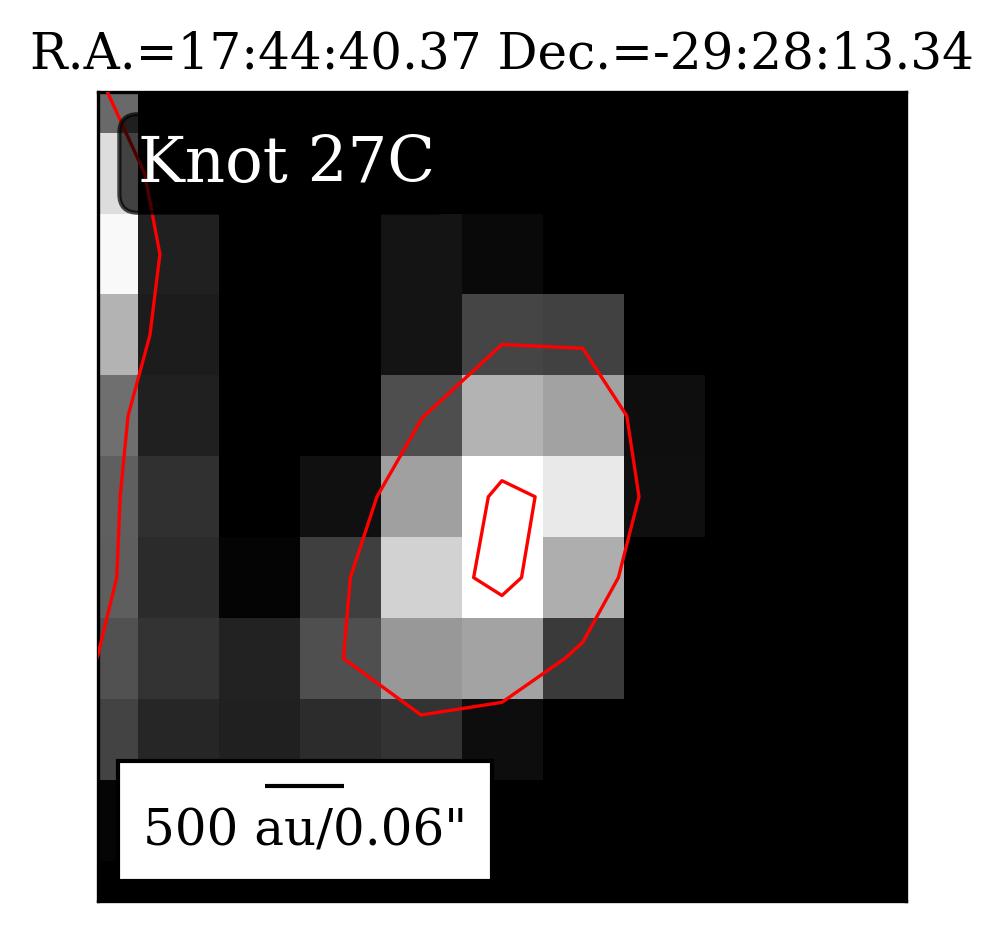} 
            \includegraphics[width=0.40\textwidth]{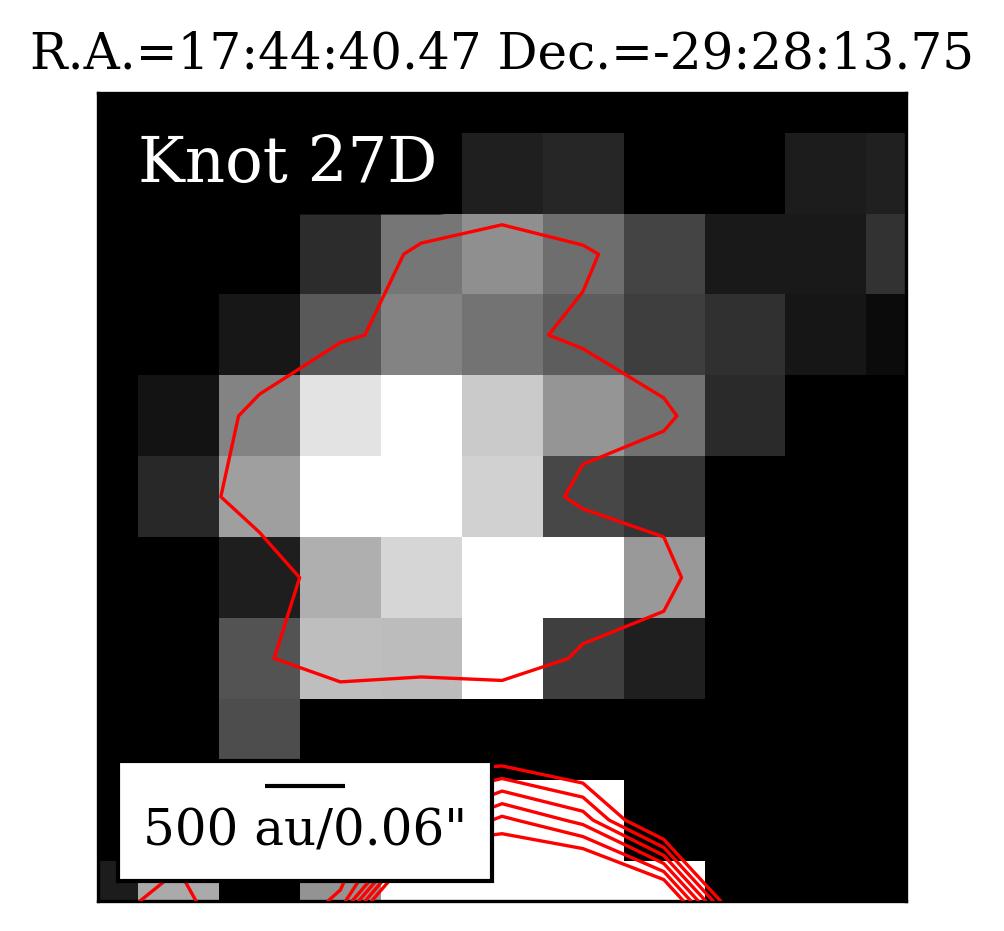}
            \includegraphics[width=0.40\textwidth]{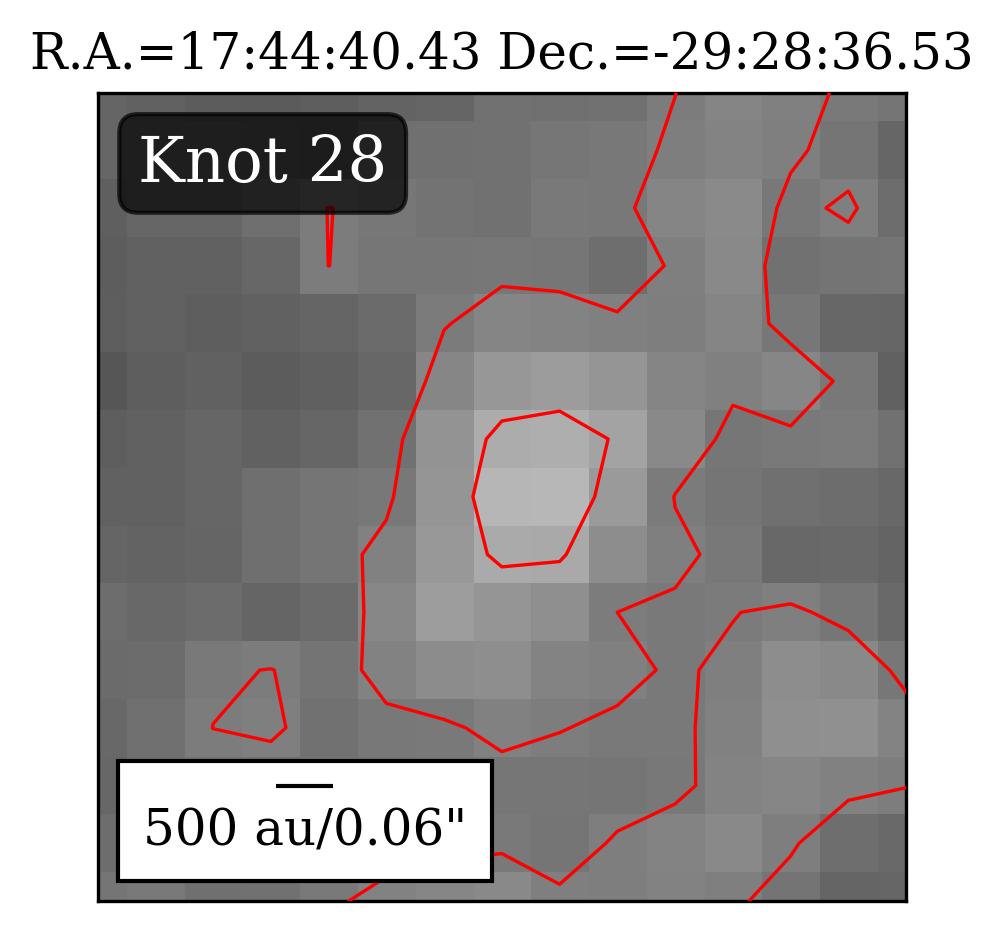}
            \includegraphics[width=0.40\textwidth]{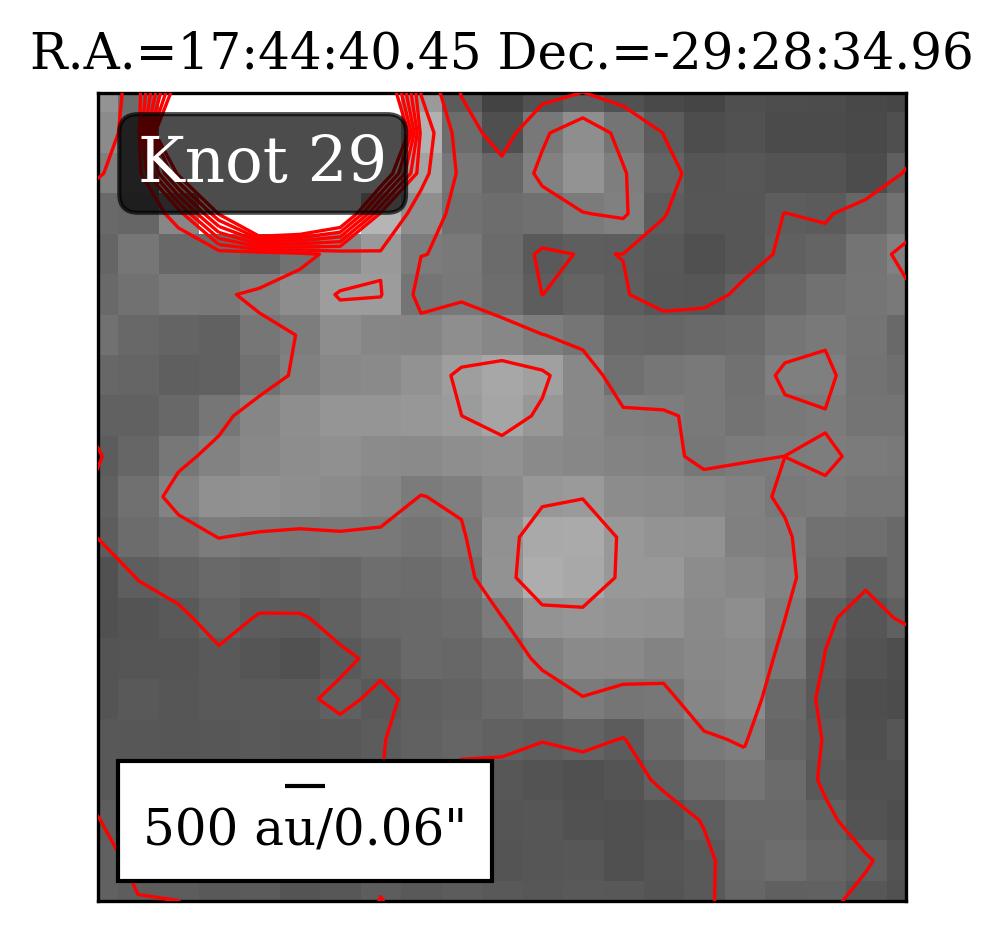}
            \includegraphics[width=0.40\textwidth]{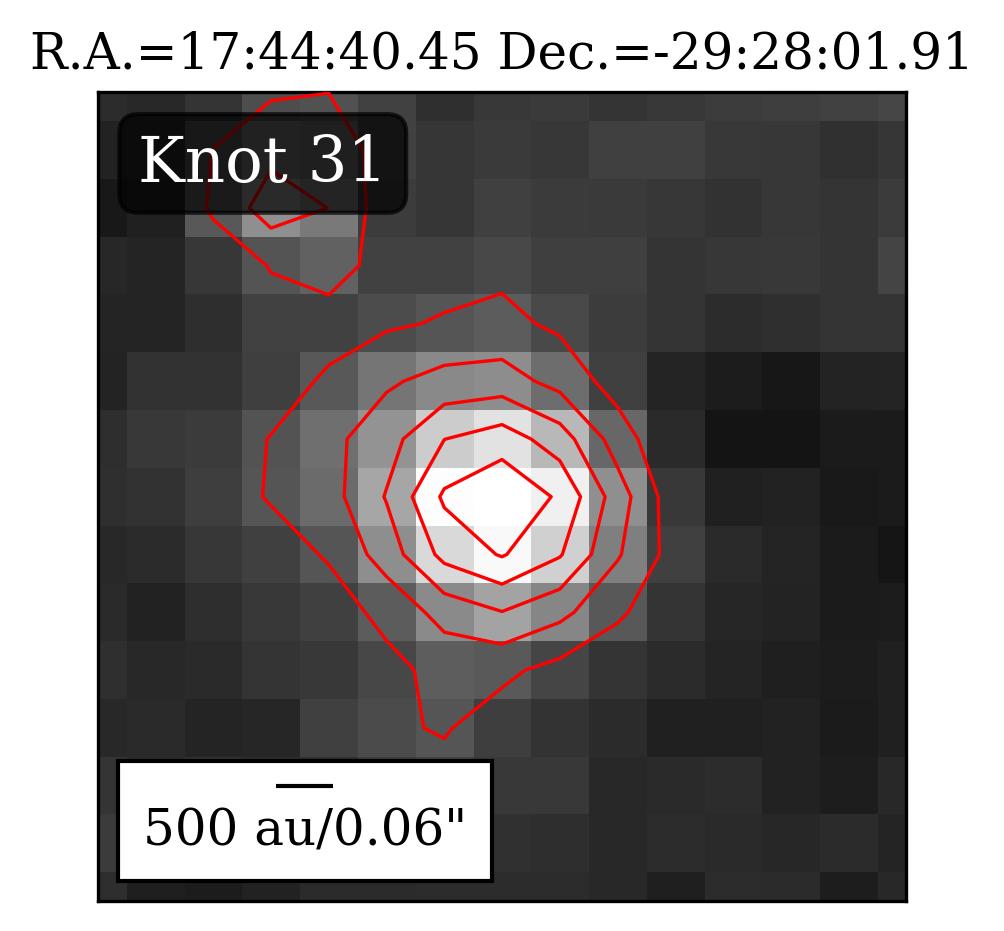}
            \includegraphics[width=0.40\textwidth]{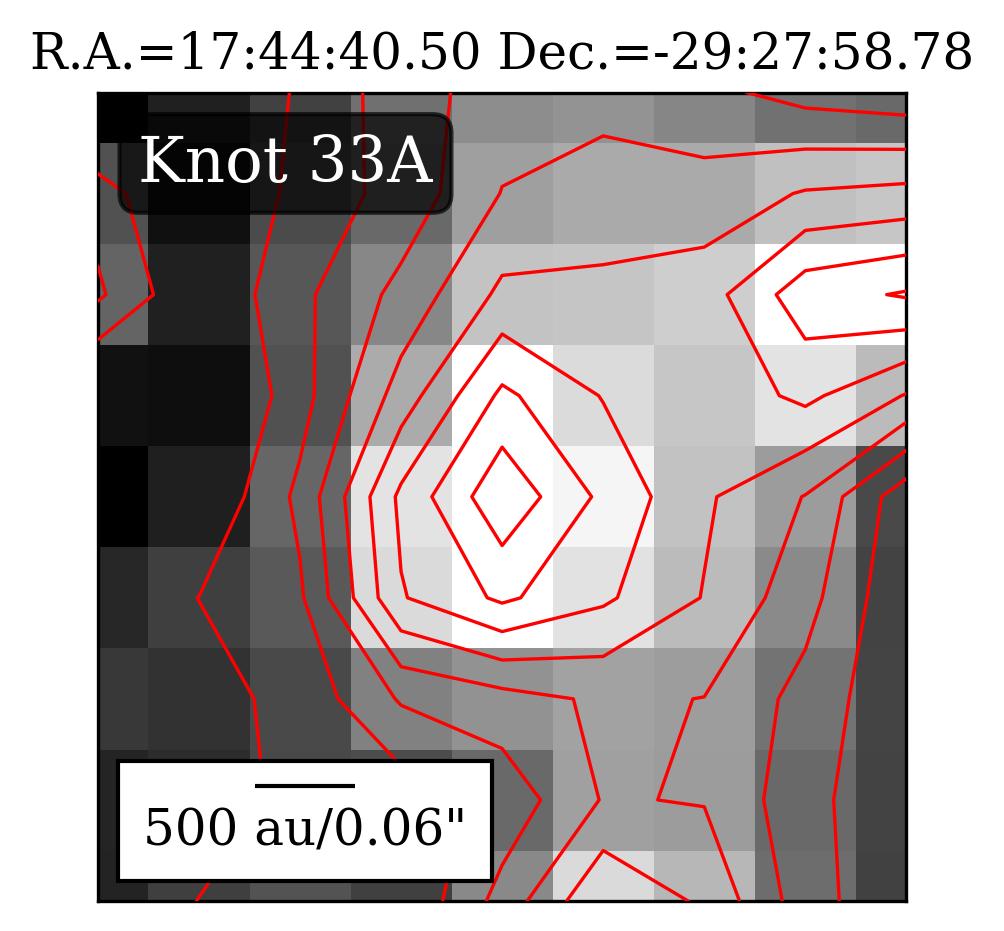}
            
            \caption{Continued. The contour levels shown for knots 27C and 27D represent 25 to 300$\mathrm{\sigma}$ in steps of 55$\sigma$ above the local background; those shown for knots 28 and 29 represent 15 to 100$\mathrm{\sigma}$ in steps of 5$\sigma$; those shown for knots 31 and 33A represent 10 to 50$\mathrm{\sigma}$ in steps of 5$\sigma$.}
        \end{figure*}
        \renewcommand{\thefigure}{B\arabic{figure}}
        \addtocounter{figure}{-1}
        \begin{figure*}[!htb]
        \centering
            \includegraphics[width=0.40\textwidth]{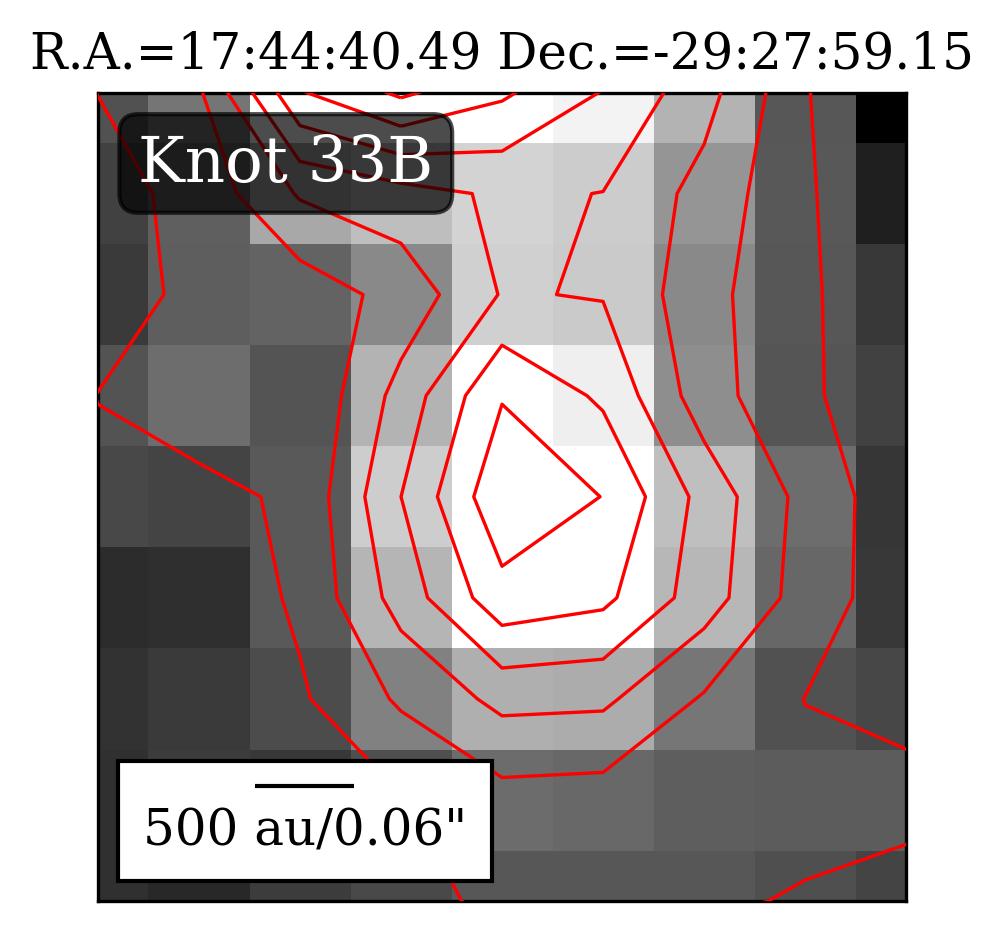} 
            \includegraphics[width=0.40\textwidth]{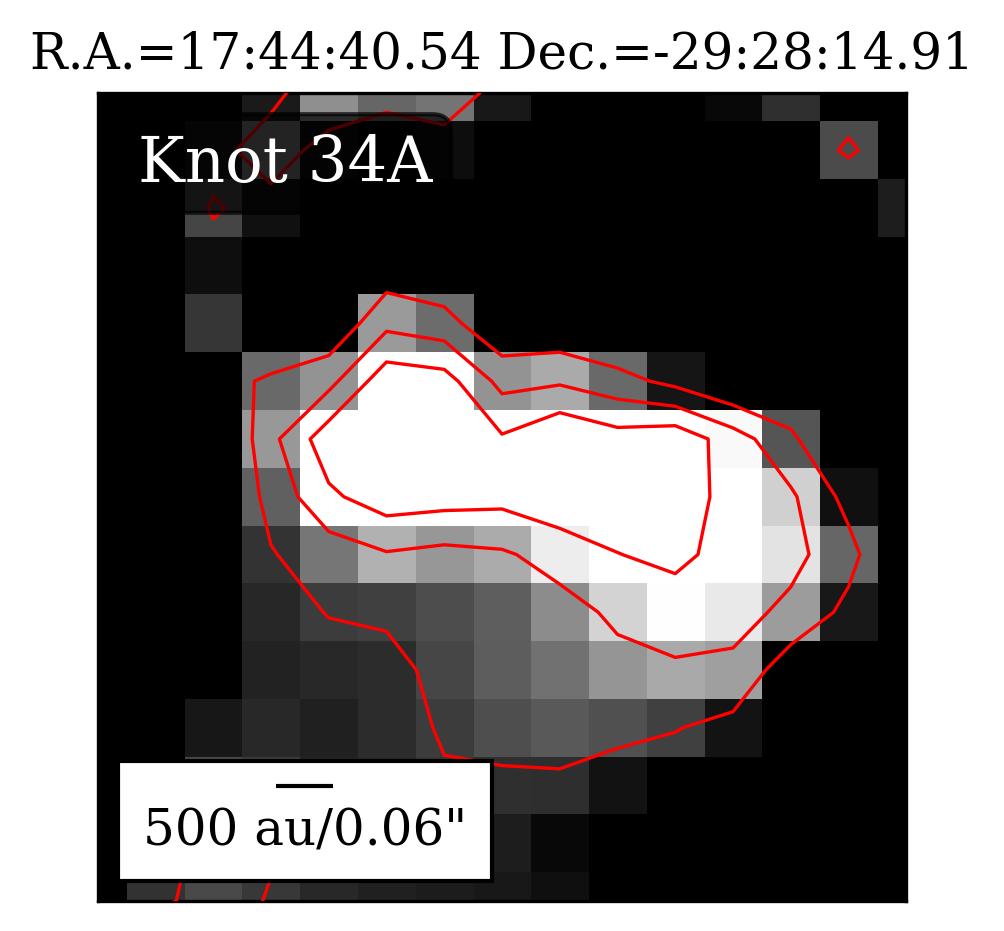}
            \includegraphics[width=0.40\textwidth]{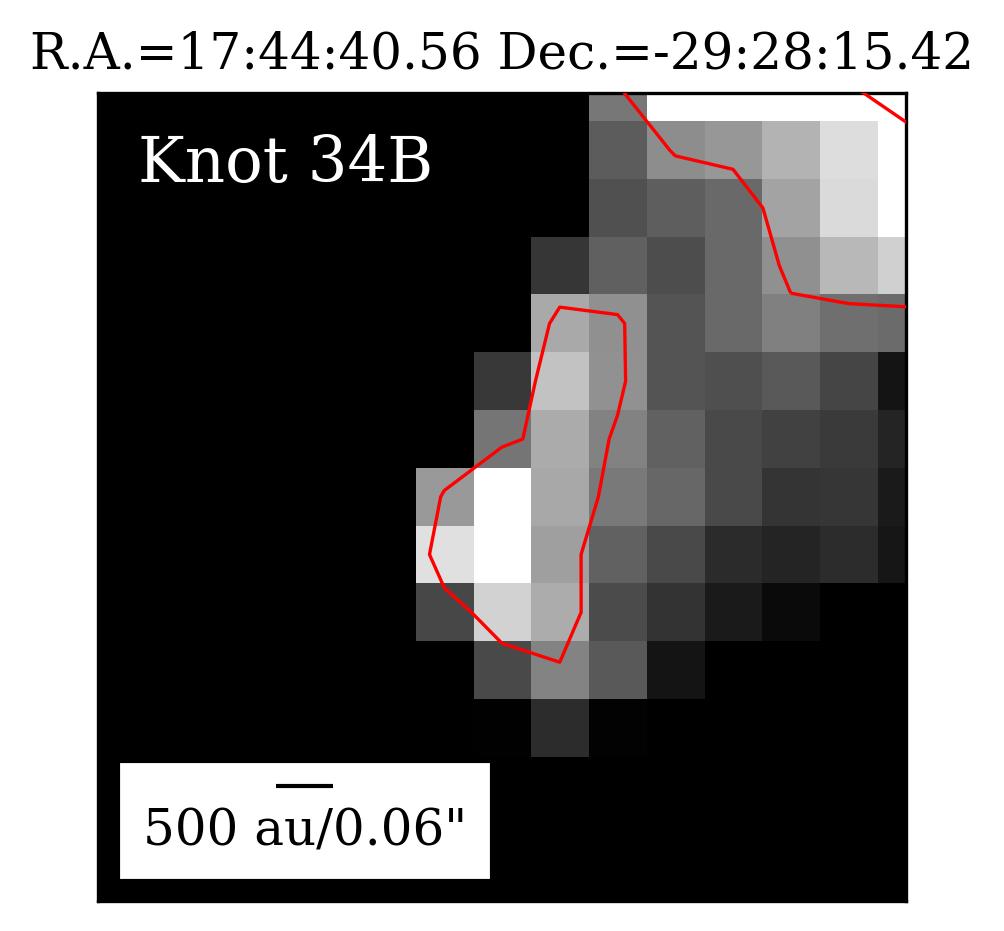}
            \includegraphics[width=0.40\textwidth]{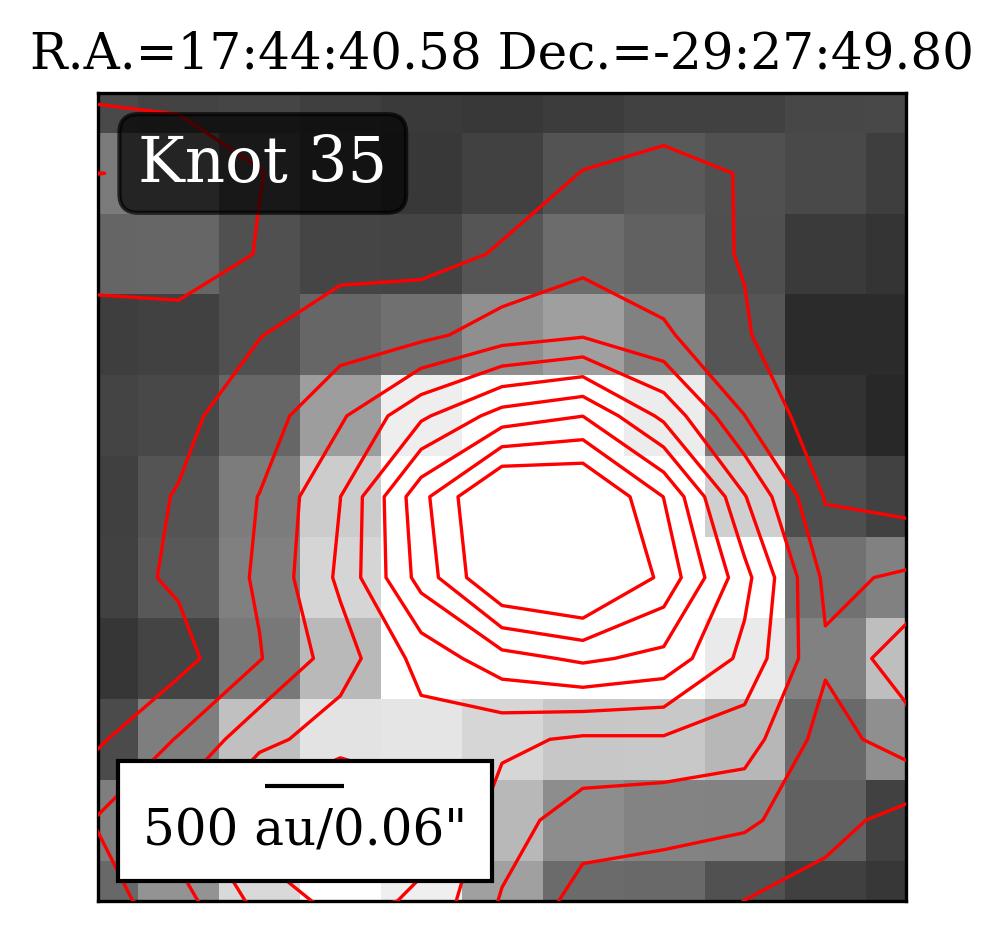}
            \includegraphics[width=0.40\textwidth]{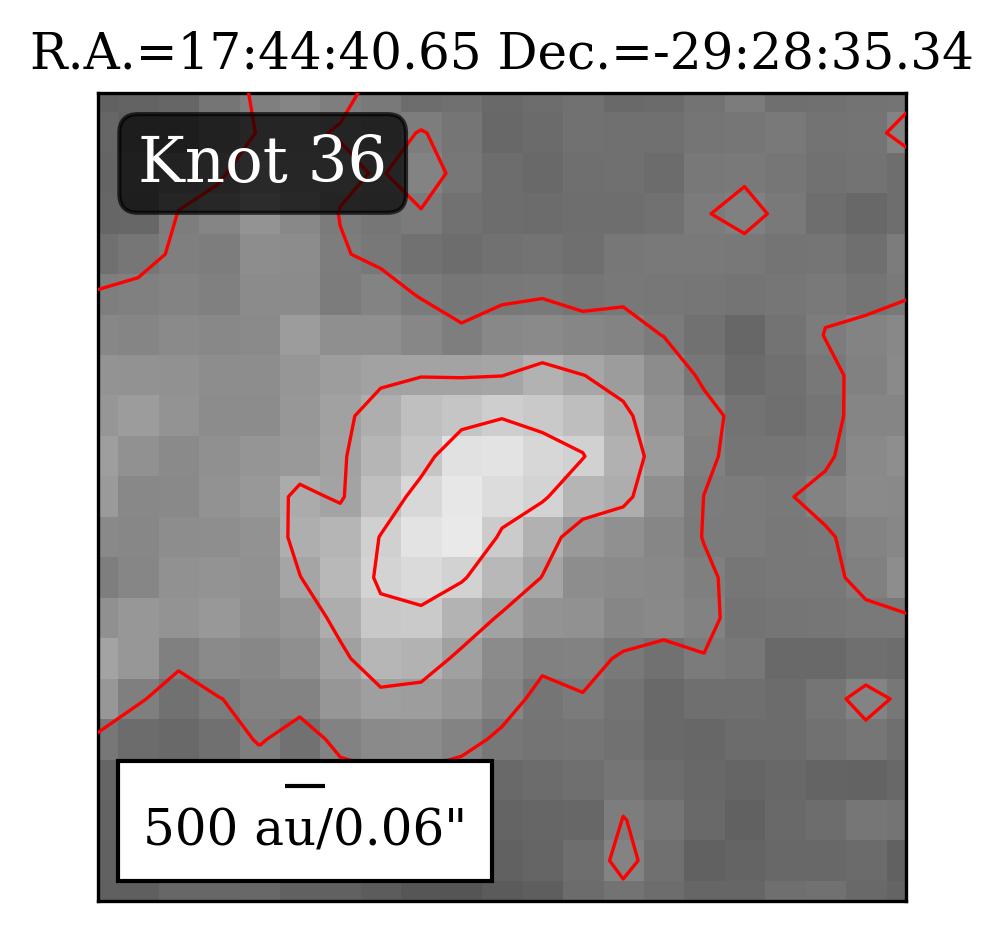}
            \includegraphics[width=0.40\textwidth]{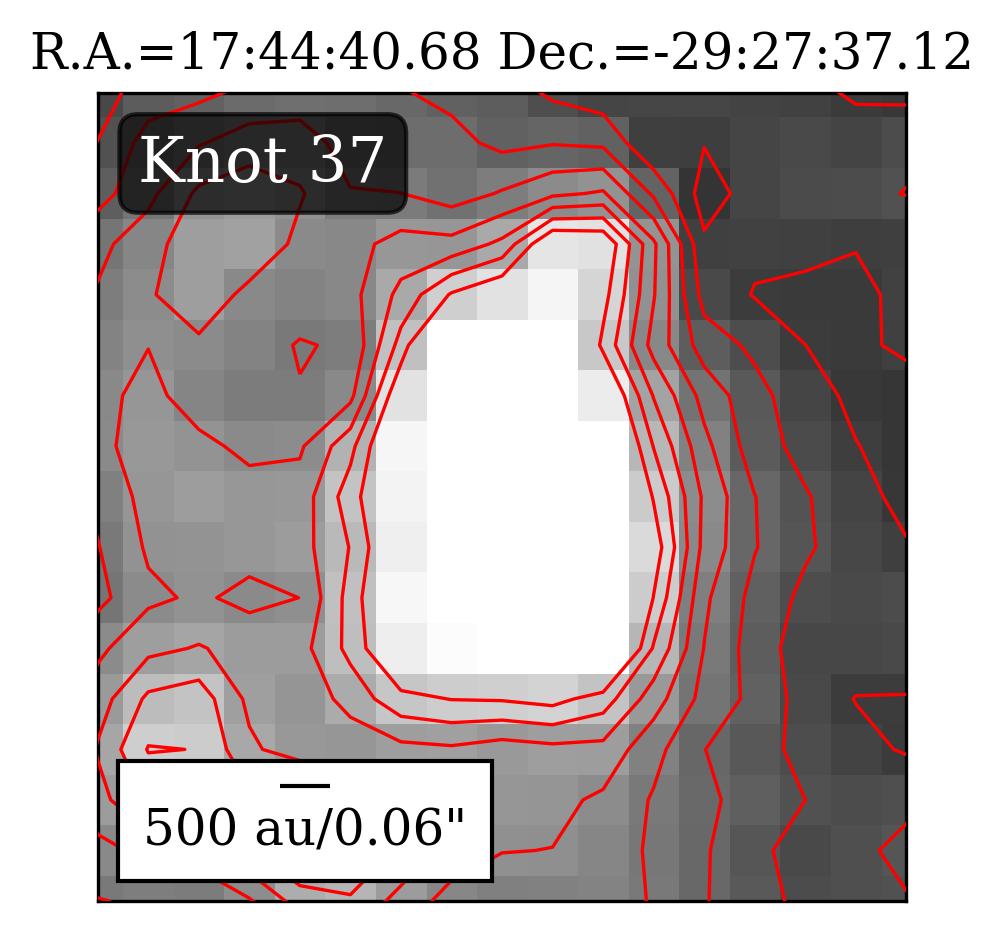}
             
            \caption{Continued. The contour levels shown for knots 33B, 35, and 37 represent 10 to 50$\mathrm{\sigma}$ in steps of 5$\sigma$ above the local background; those shown for knots 34A and 34B represent 25 to 300$\mathrm{\sigma}$ in steps of 55$\sigma$; those shown for knot 36 represent 15 to 100$\mathrm{\sigma}$ in steps of 5$\sigma$.}
        \end{figure*}
        \renewcommand{\thefigure}{B\arabic{figure}}
        \addtocounter{figure}{-1}
        \begin{figure*}[!htb]
        \centering
            \includegraphics[width=0.40\textwidth]{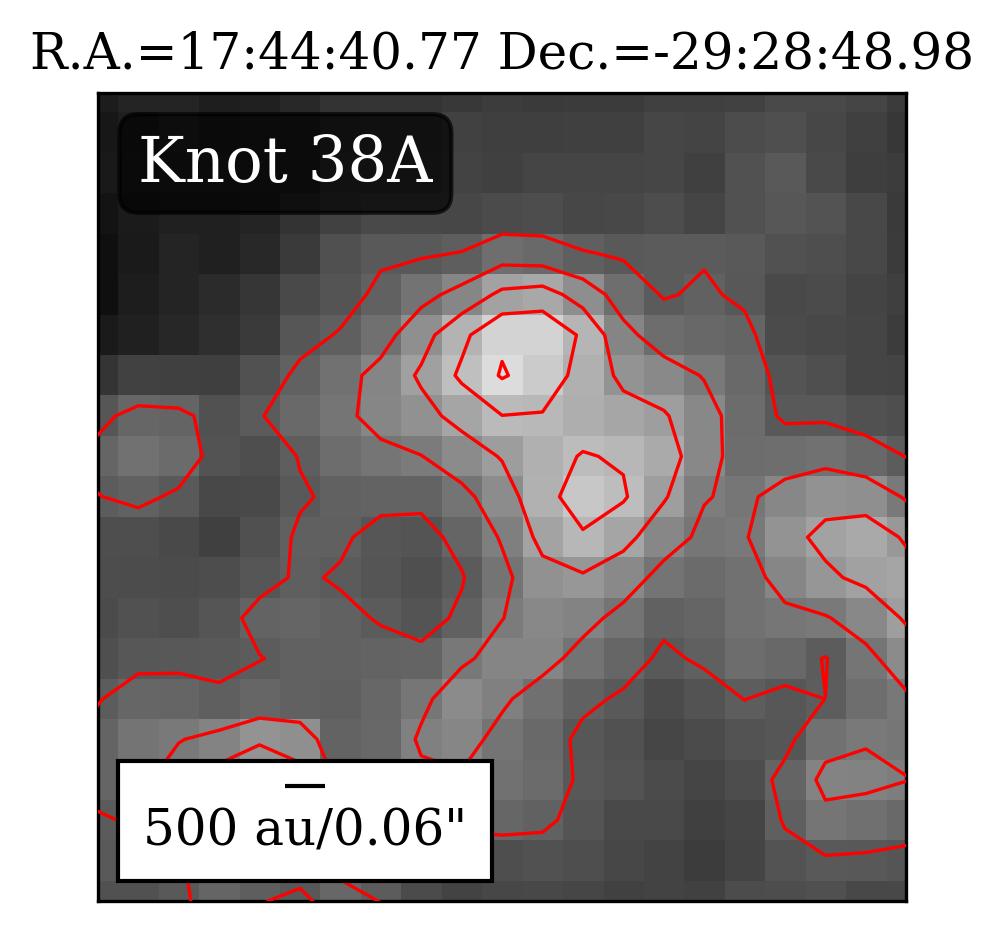}
            \includegraphics[width=0.40\textwidth]{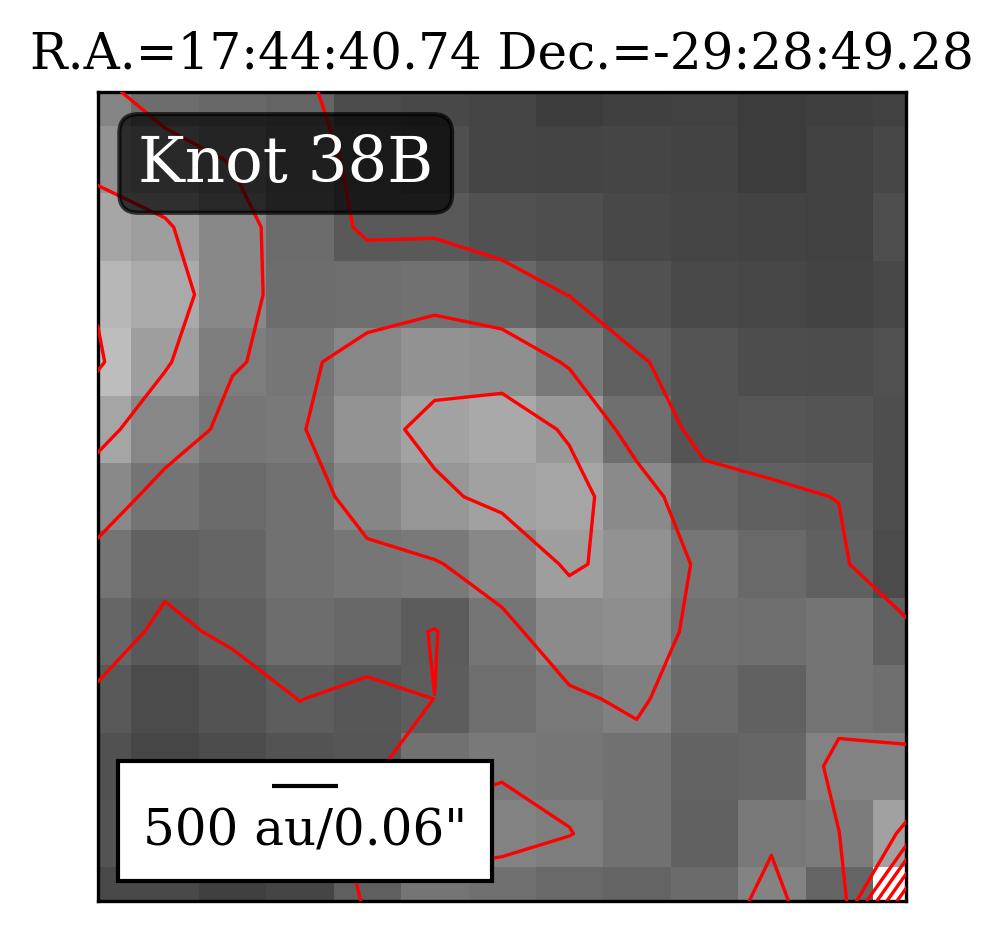}
            \includegraphics[width=0.40\textwidth]{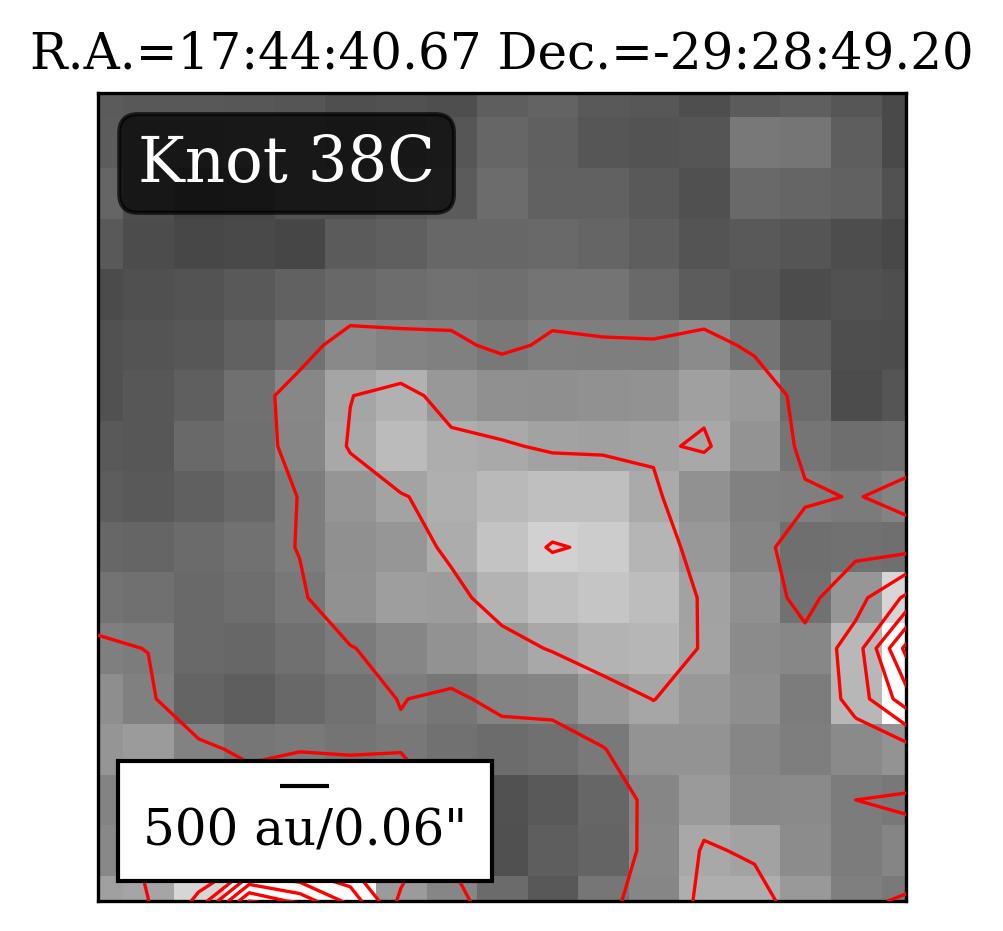}
            \includegraphics[width=0.40\textwidth]{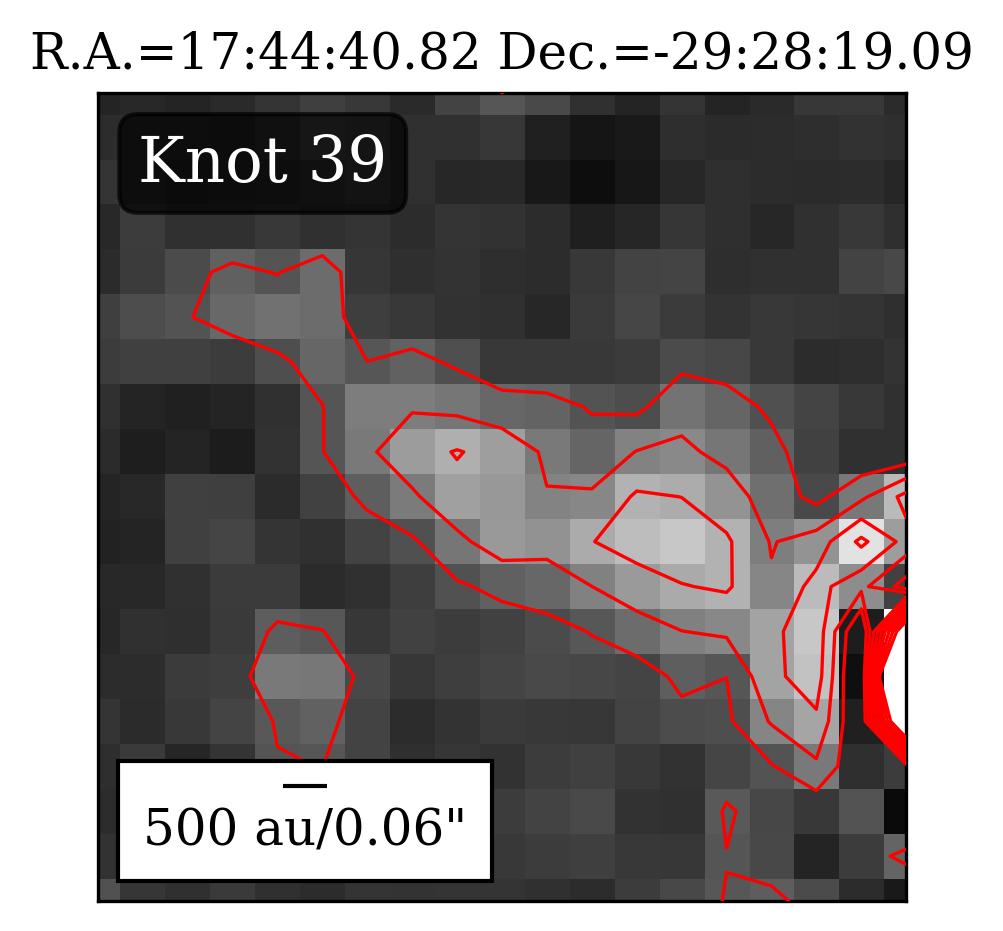}
            \includegraphics[width=0.40\textwidth]{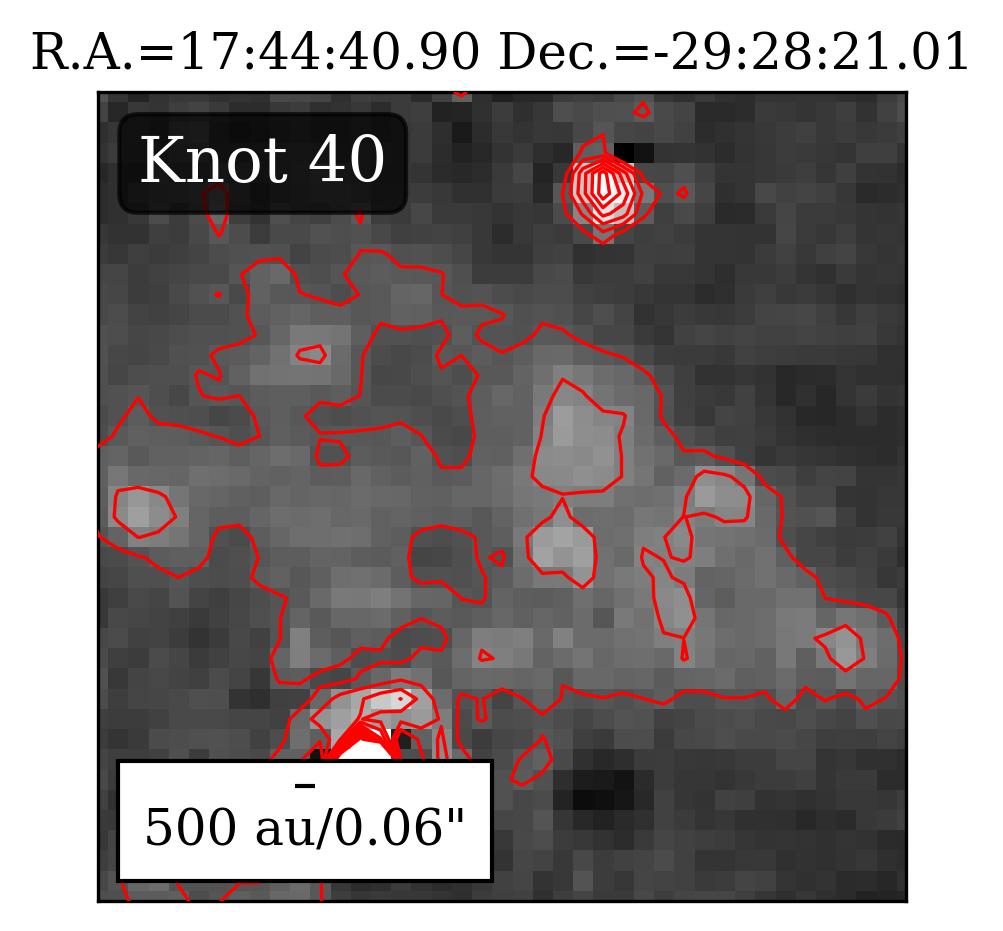}
            \includegraphics[width=0.40\textwidth]{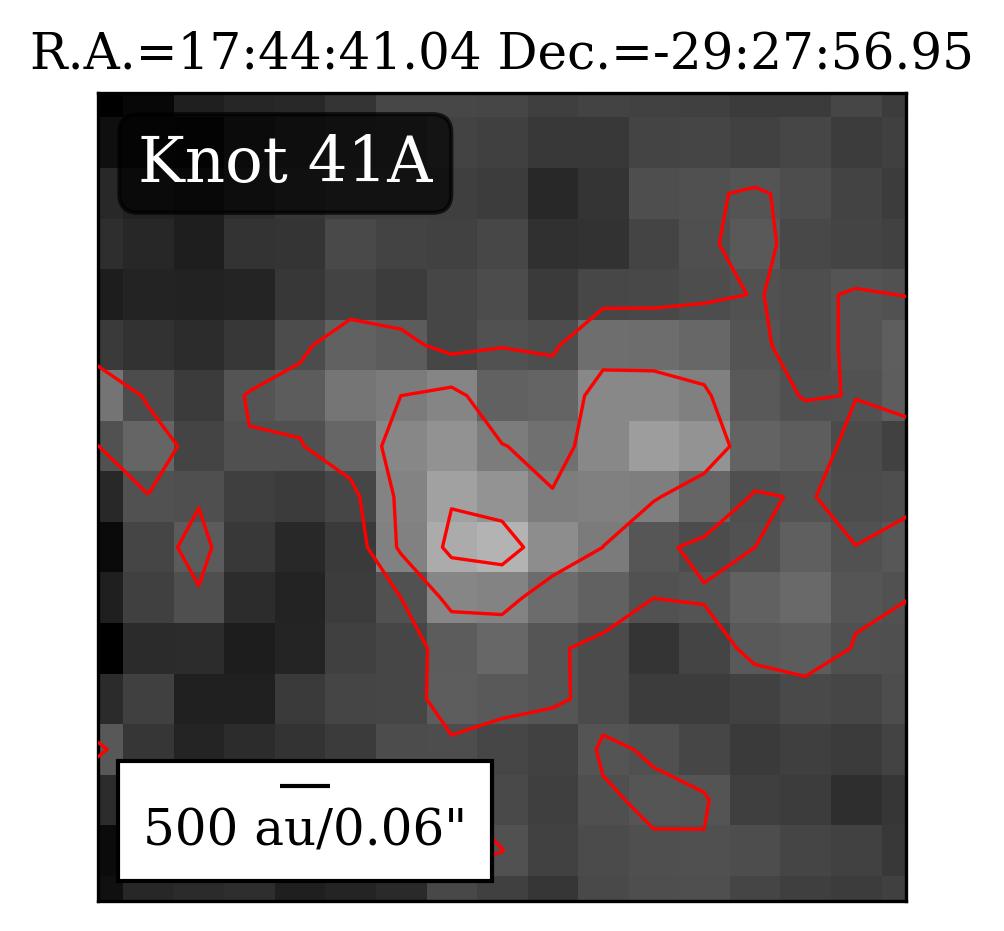}
            
            \caption{Continued. The contour levels shown for knots 38A, 38B, and 38C represent 15 to 100$\mathrm{\sigma}$ in steps of 5$\sigma$ above the local background; those shown for knots 39, 40, and 41A represent 10 to 50$\mathrm{\sigma}$ in steps of 5$\sigma$.}
        \end{figure*}
        \renewcommand{\thefigure}{B\arabic{figure}}
        \addtocounter{figure}{-1}
        \begin{figure*}[!htb]
        \centering
            \includegraphics[width=0.40\textwidth]{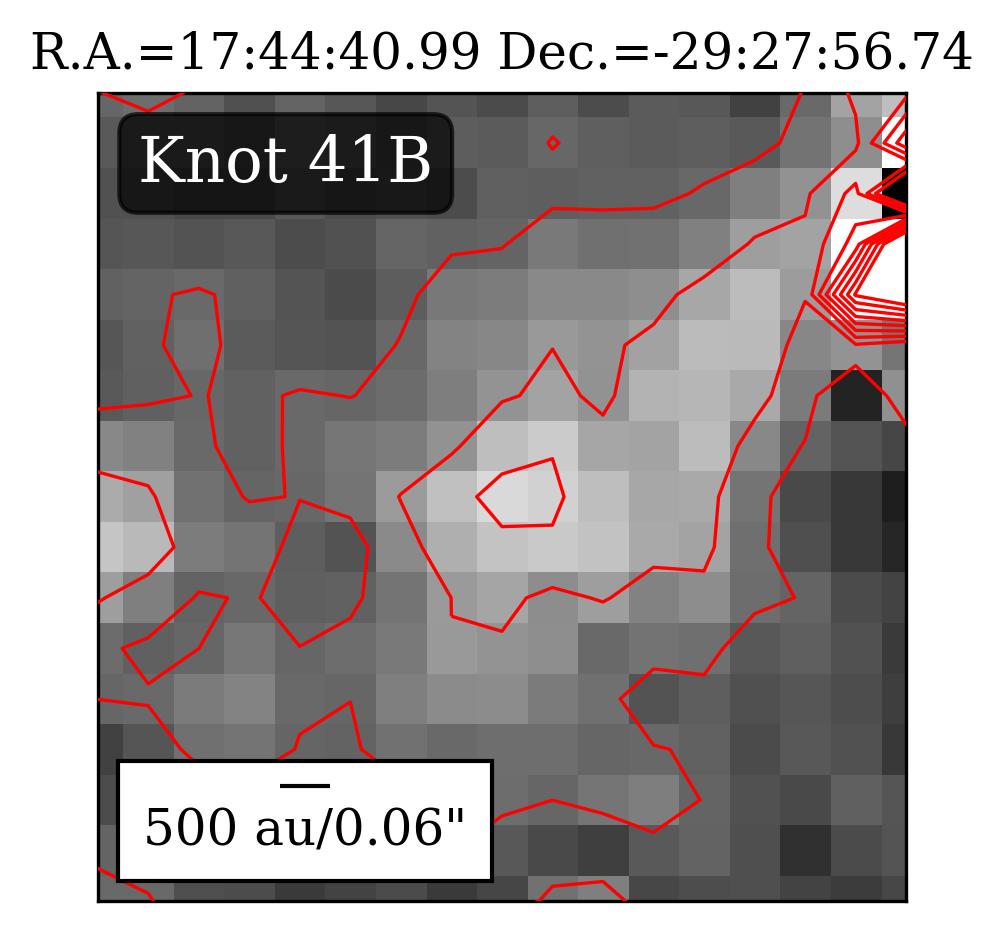} 
            \includegraphics[width=0.40\textwidth]{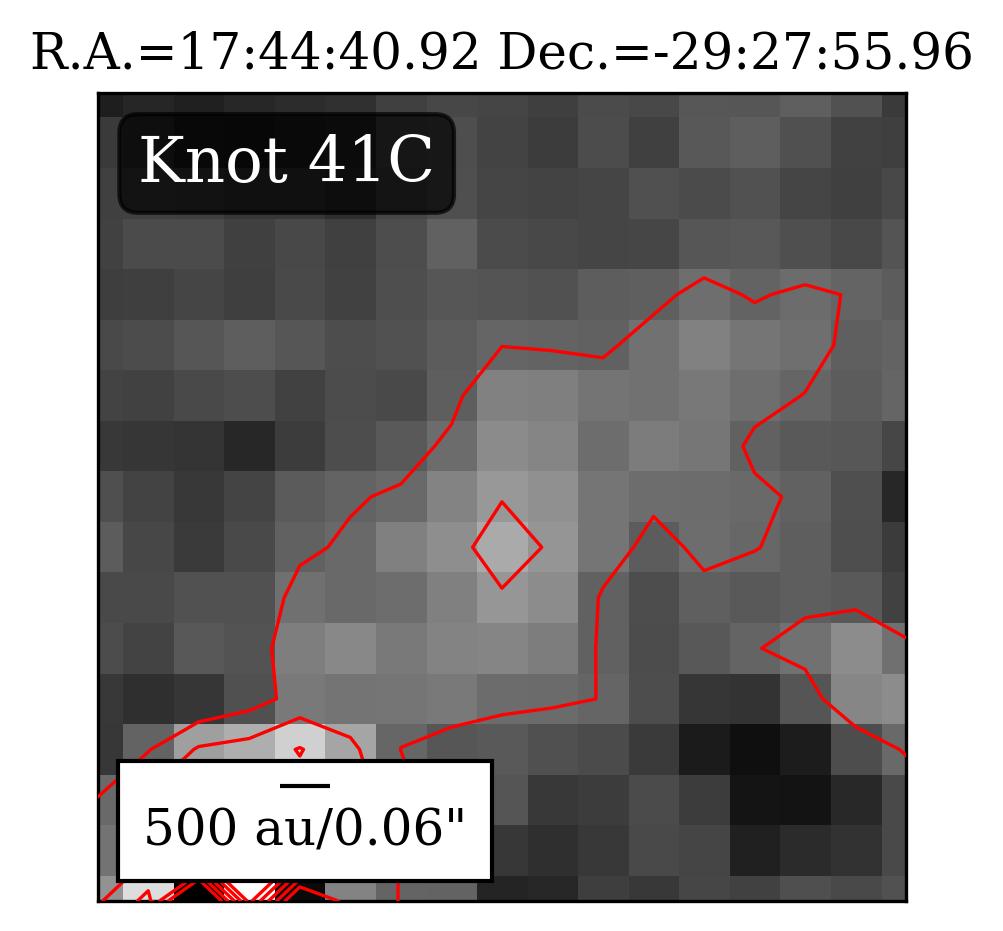}
            \includegraphics[width=0.40\textwidth]{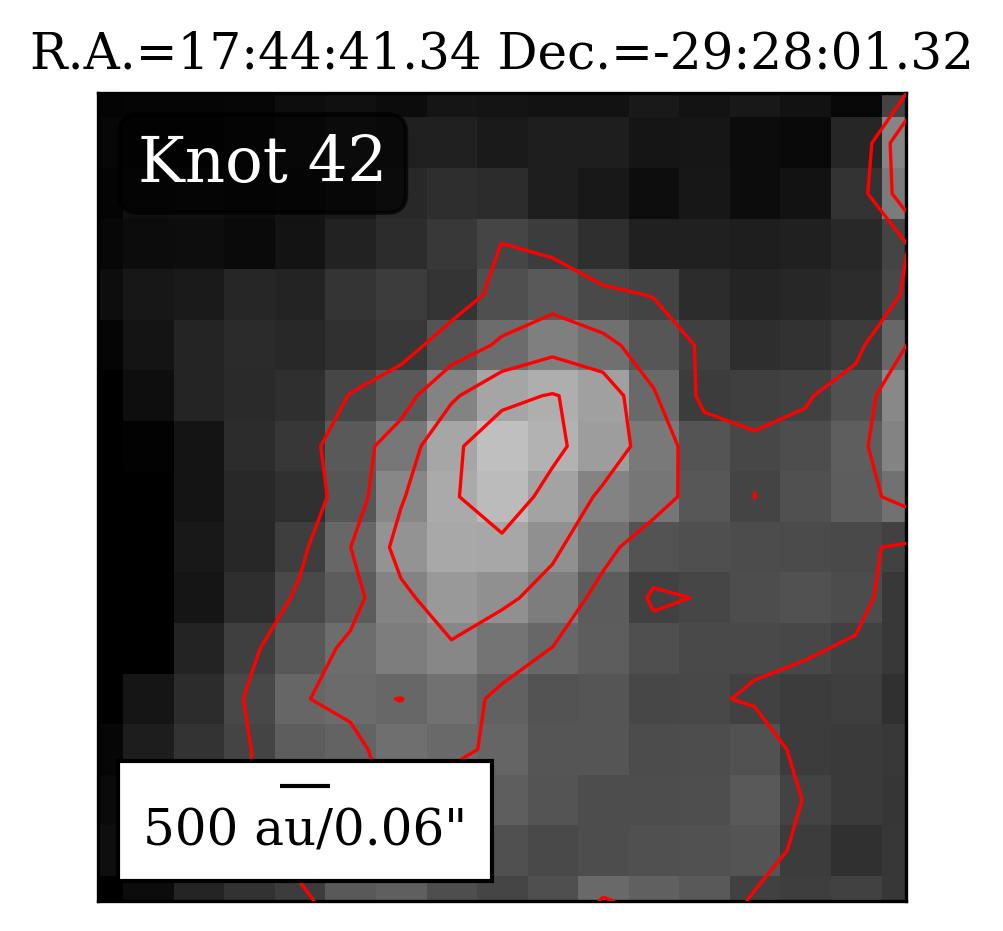}
            \includegraphics[width=0.40\textwidth]{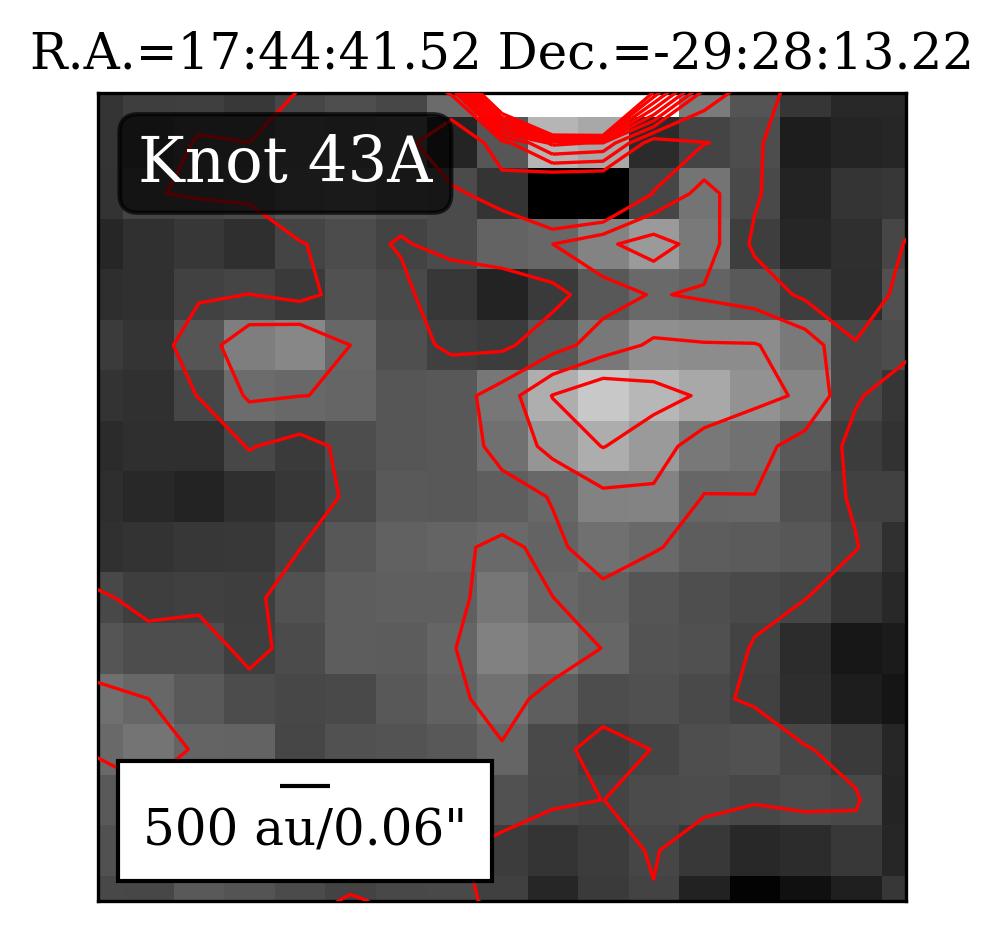}
            \includegraphics[width=0.40\textwidth]{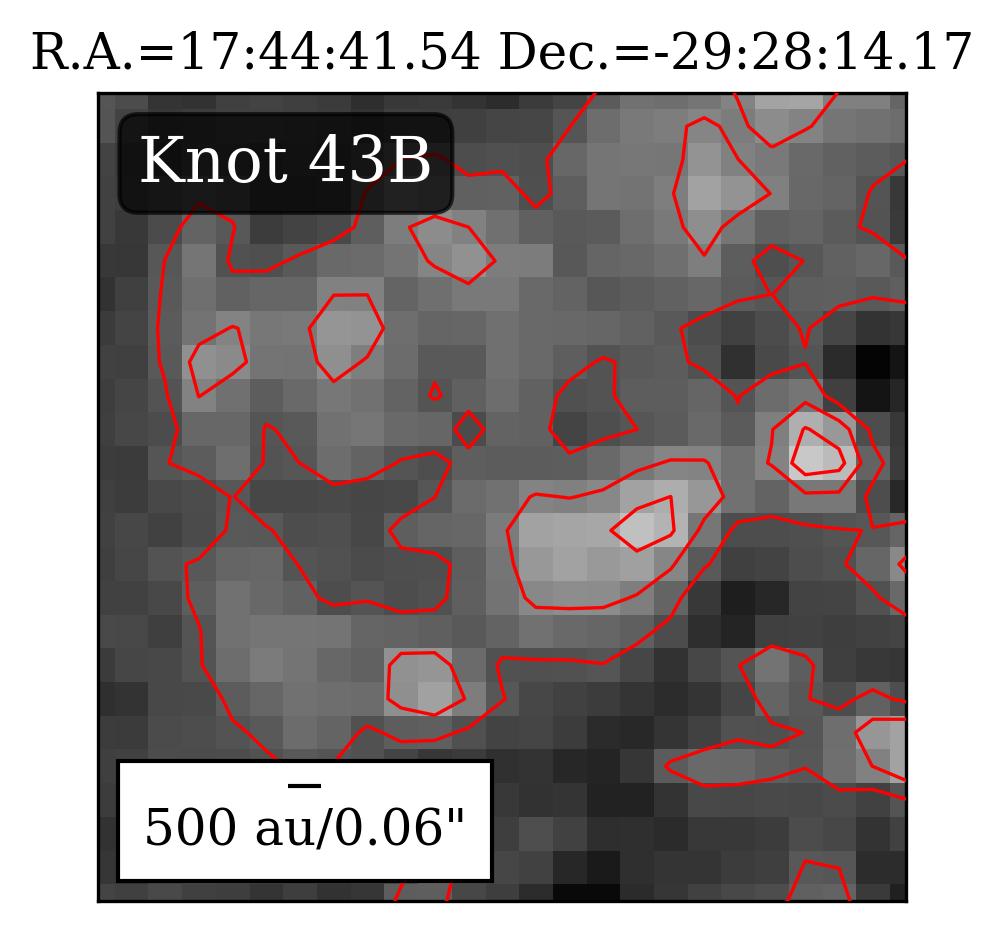}
            \includegraphics[width=0.40\textwidth]{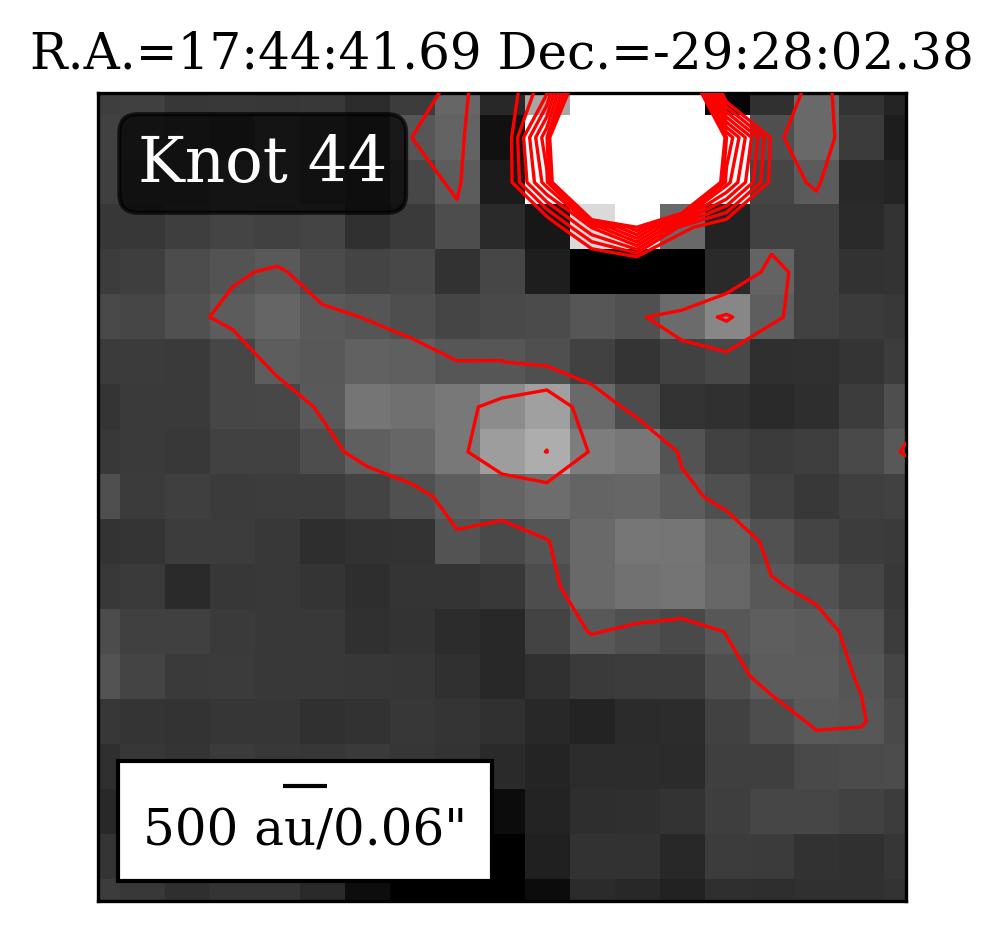}
            
            \caption{Continued. The contour levels shown represent 10 to 50$\mathrm{\sigma}$ in steps of 5$\sigma$ above the local background.}
        \end{figure*}
        \renewcommand{\thefigure}{B\arabic{figure}}
        \addtocounter{figure}{-1}
        \begin{figure*}[!htb]
        \centering
            \includegraphics[width=0.40\textwidth]{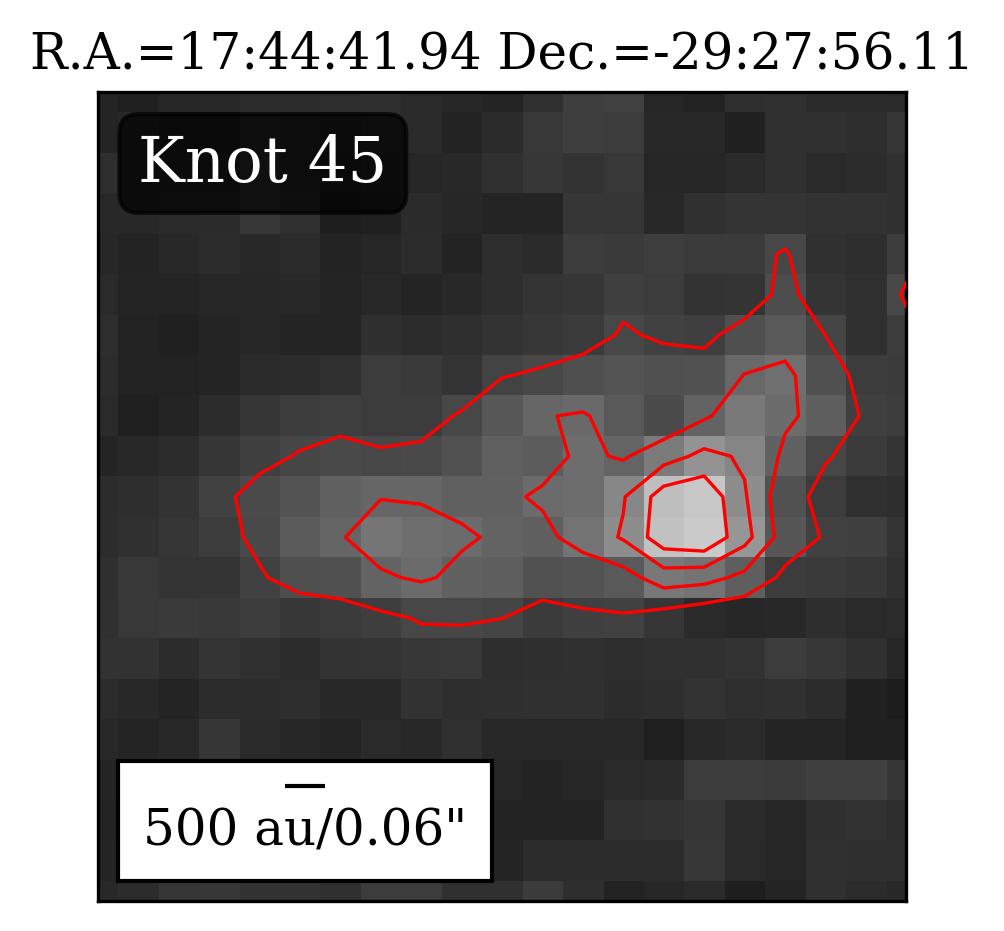}
            \includegraphics[width=0.40\textwidth]{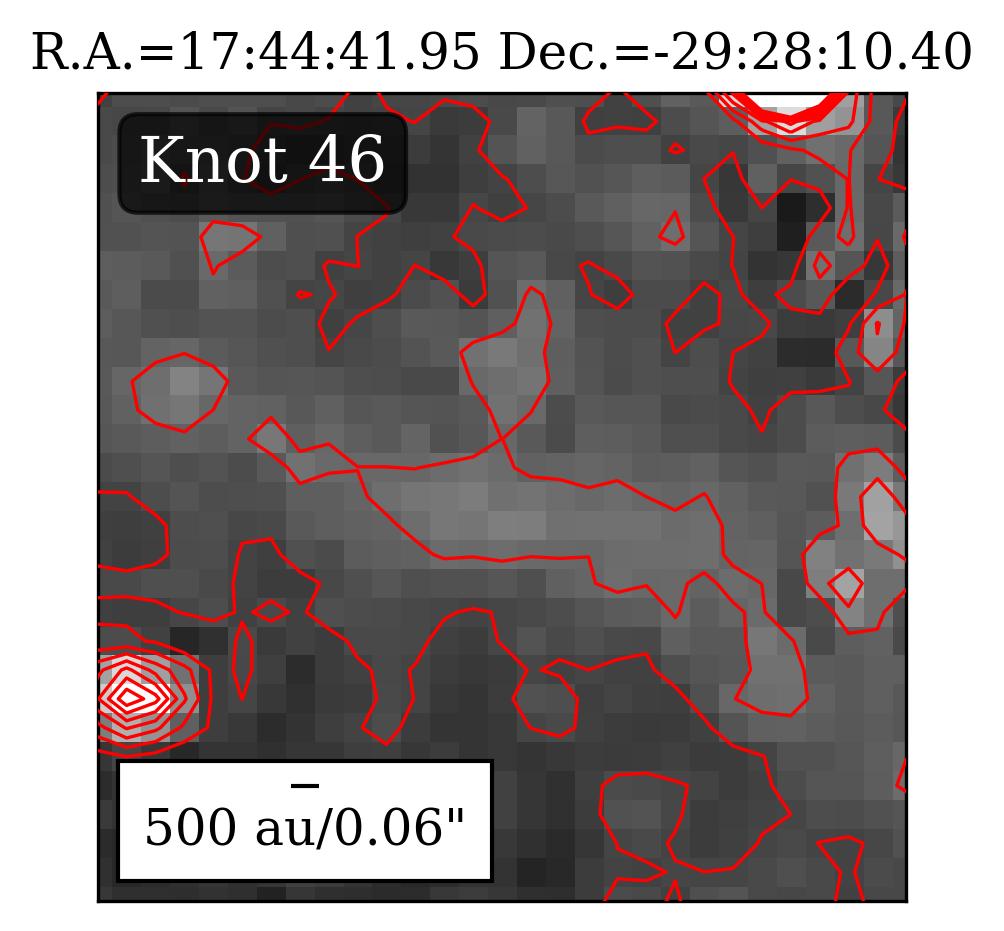}
            \includegraphics[width=0.40\textwidth]{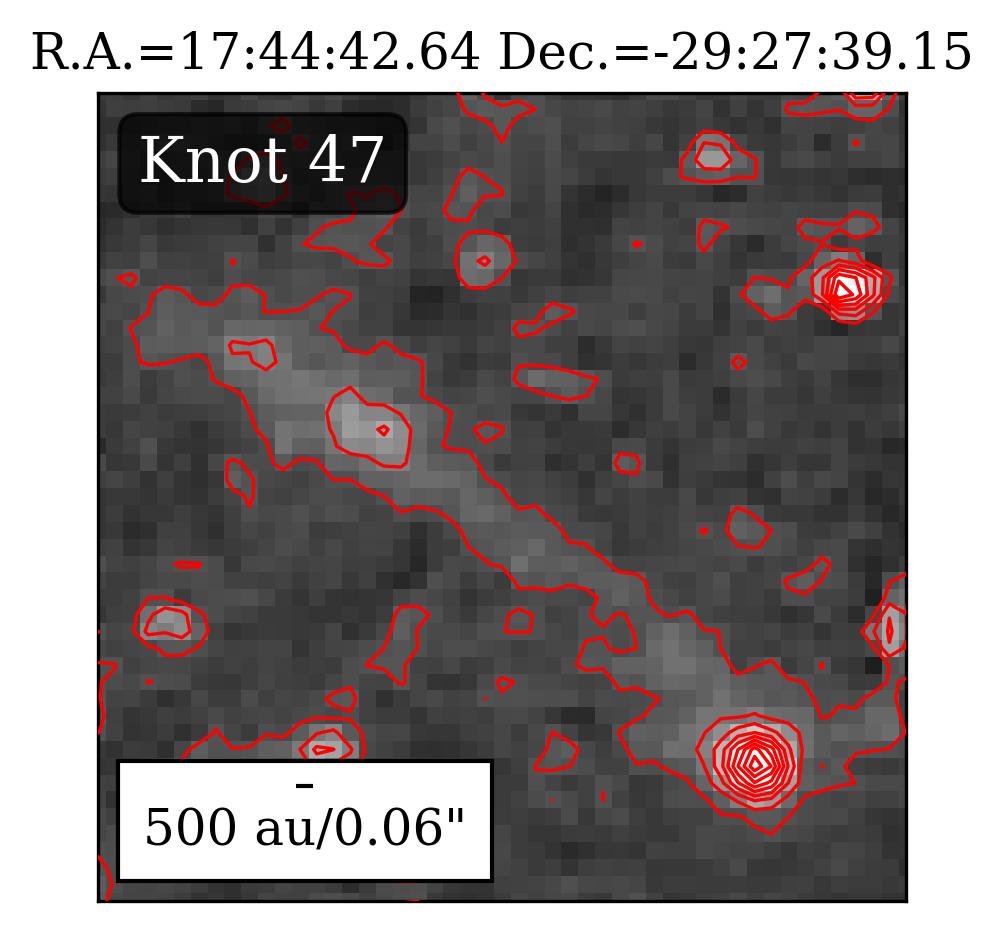}
            \includegraphics[width=0.40\textwidth]{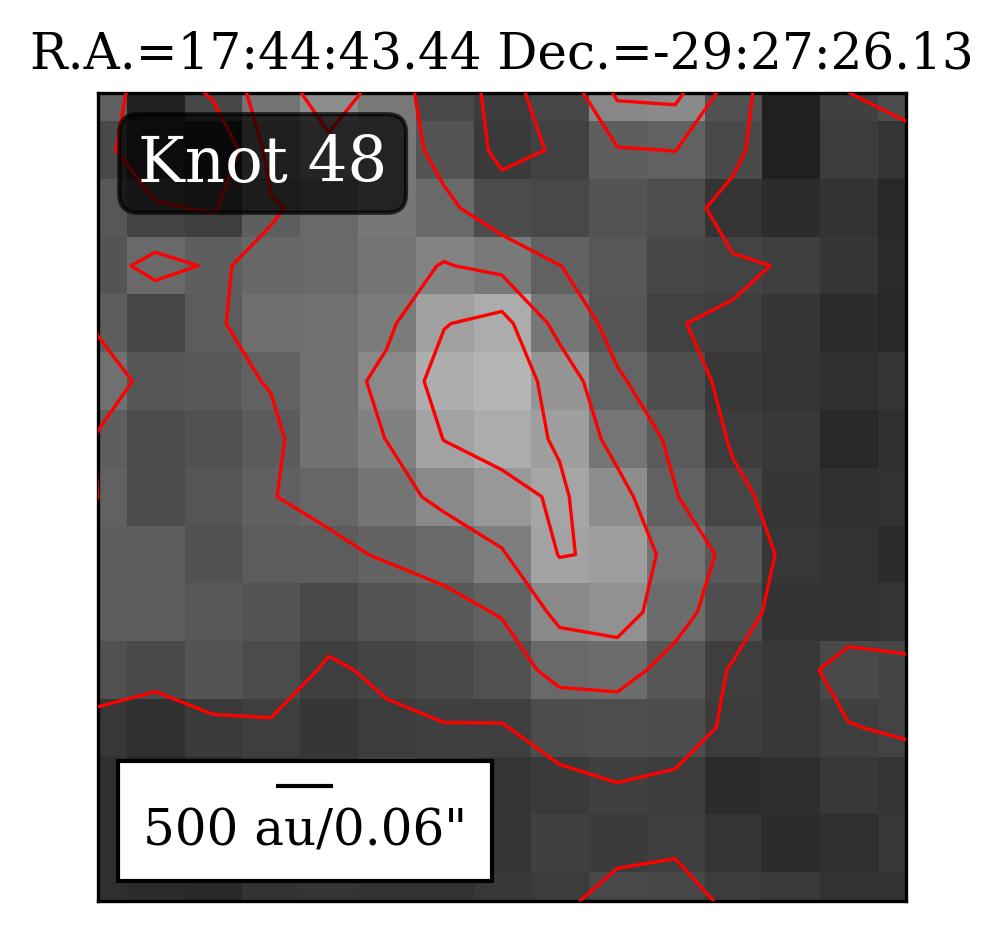}
            \includegraphics[width=0.40\textwidth]{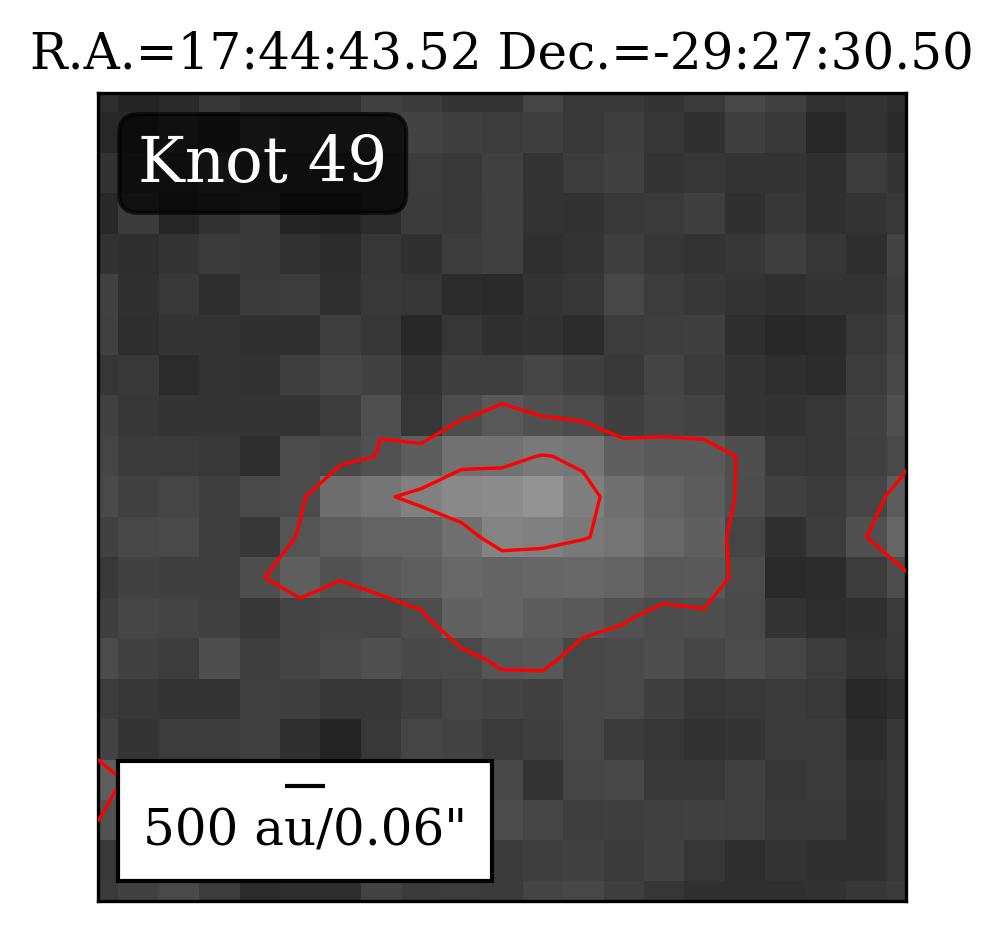}
            \includegraphics[width=0.40\textwidth]{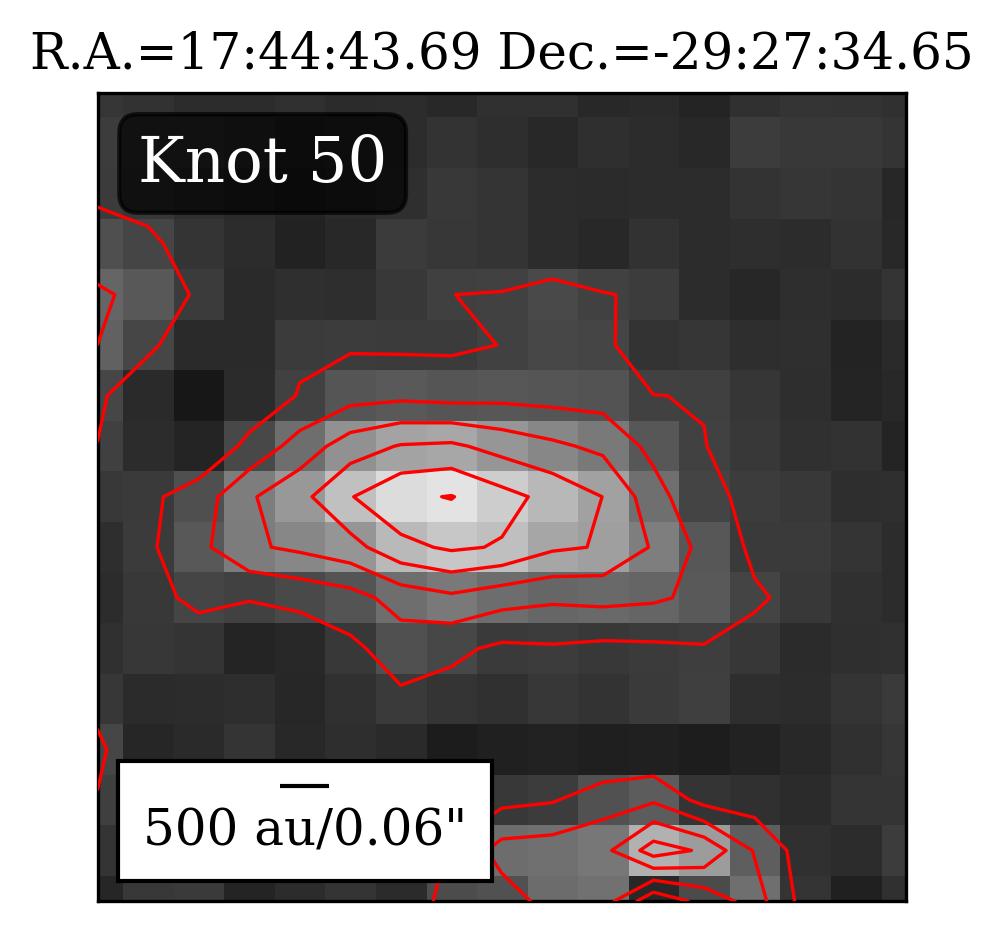}
            
            \caption{Continued. The contour levels shown represent 10 to 50$\mathrm{\sigma}$ in steps of 5$\sigma$ above the local background.}
        \end{figure*}
        \renewcommand{\thefigure}{B\arabic{figure}}
        \addtocounter{figure}{-1}
        \begin{figure*}[!htb]
        \centering
            \includegraphics[width=0.40\textwidth]{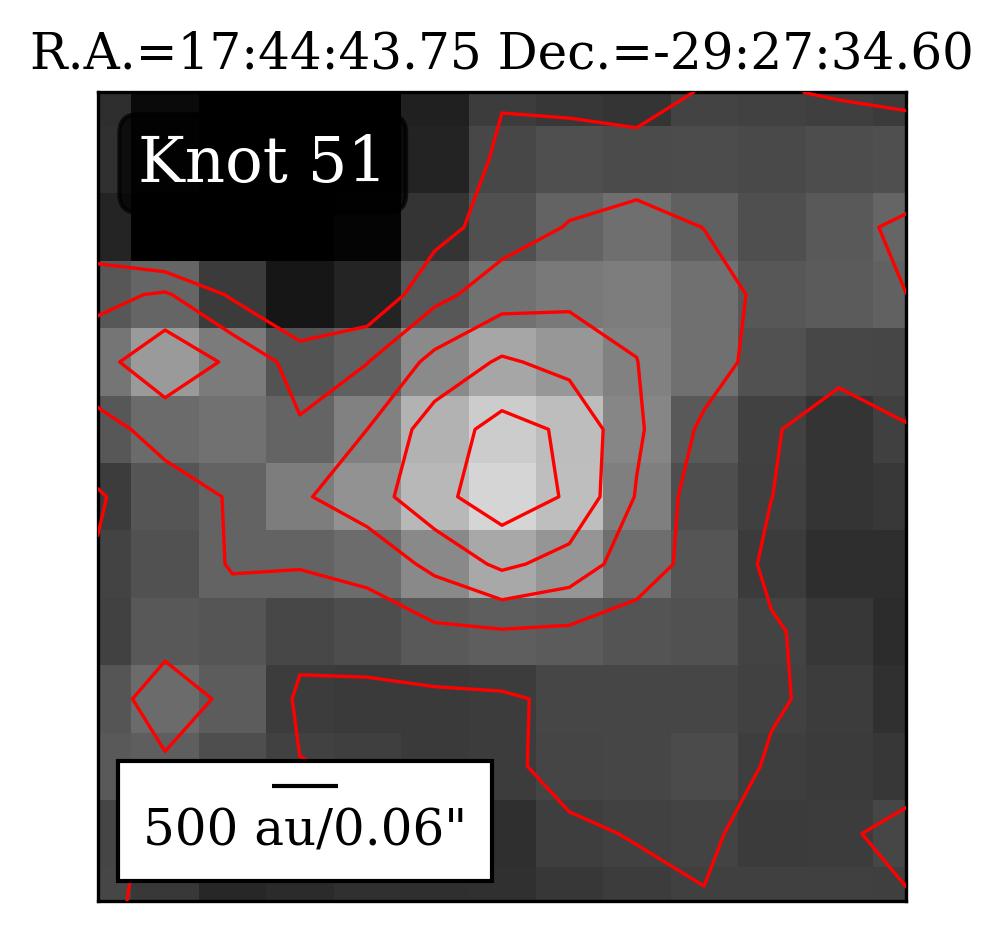}
            \includegraphics[width=0.40\textwidth]{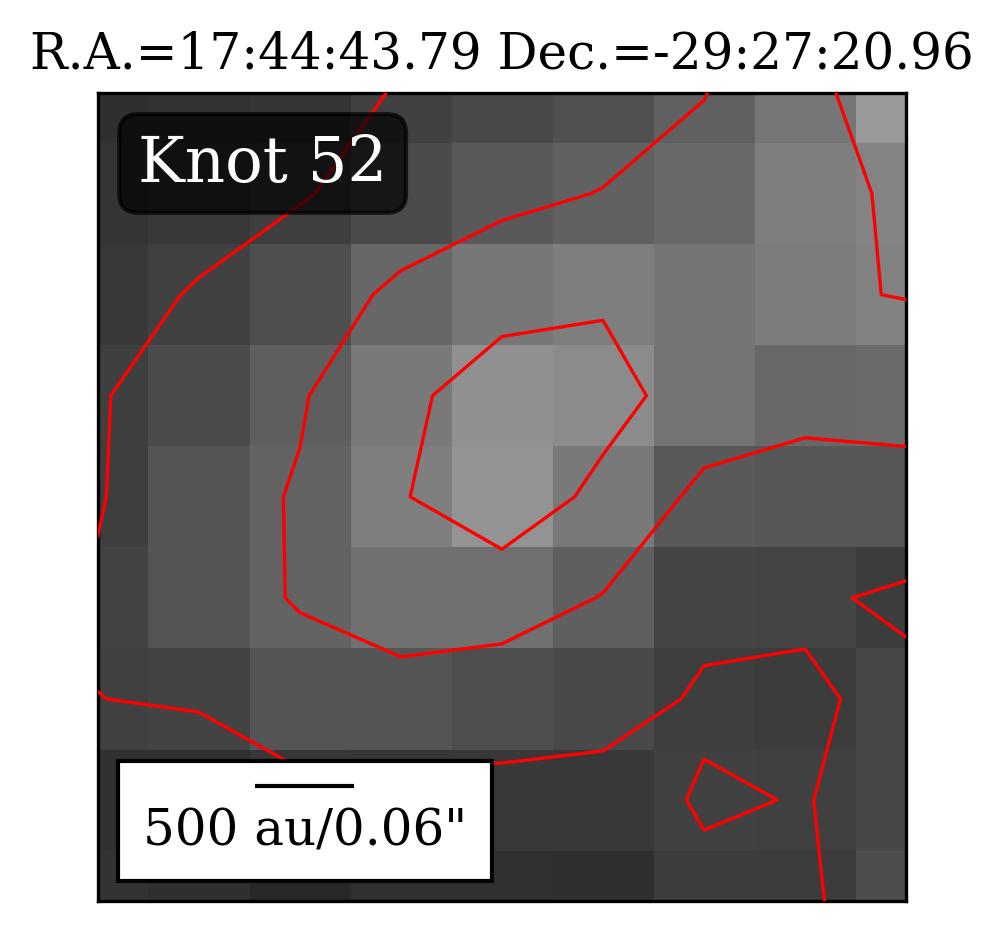}
            \includegraphics[width=0.40\textwidth]{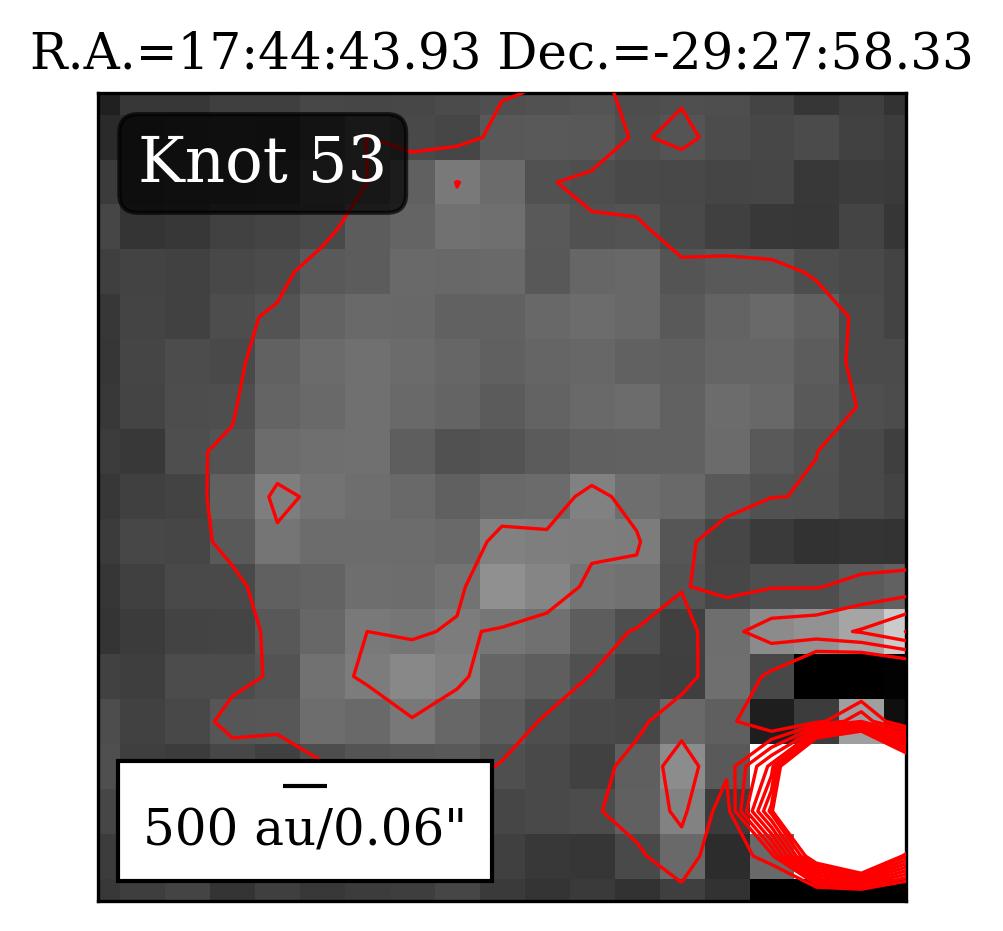}
            \includegraphics[width=0.40\textwidth]{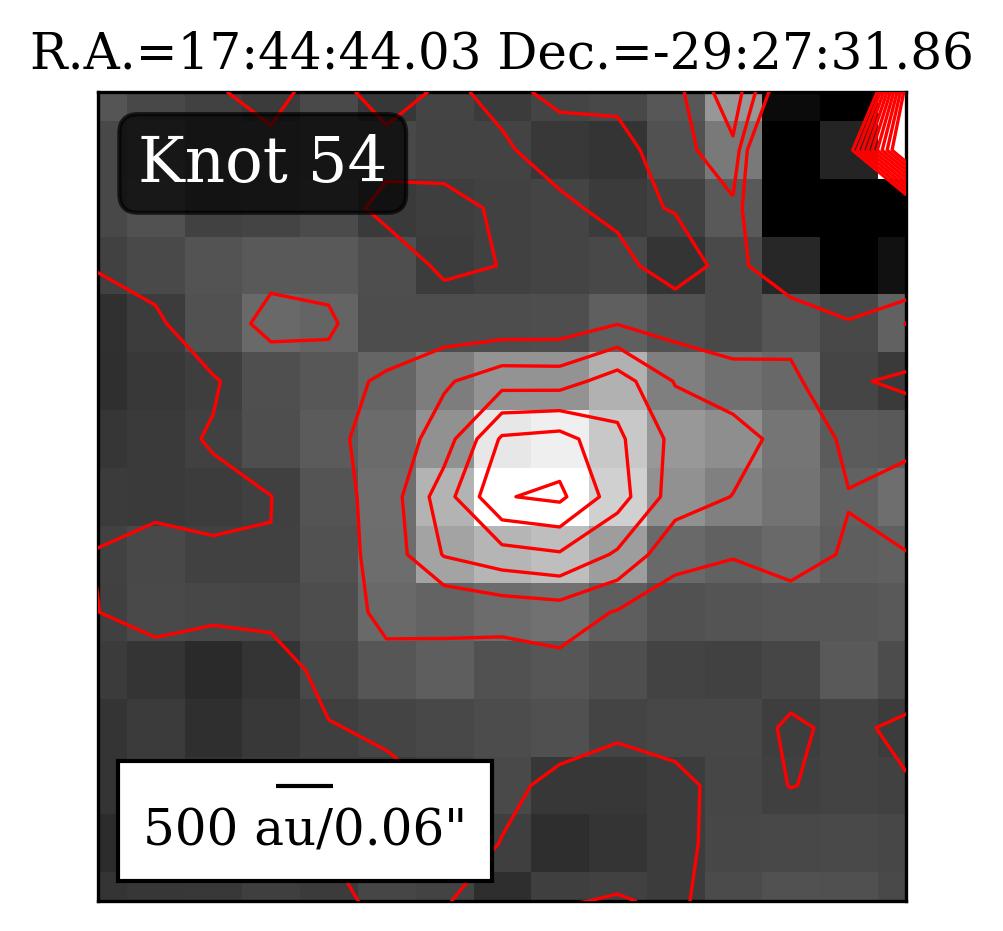}
            \includegraphics[width=0.40\textwidth]{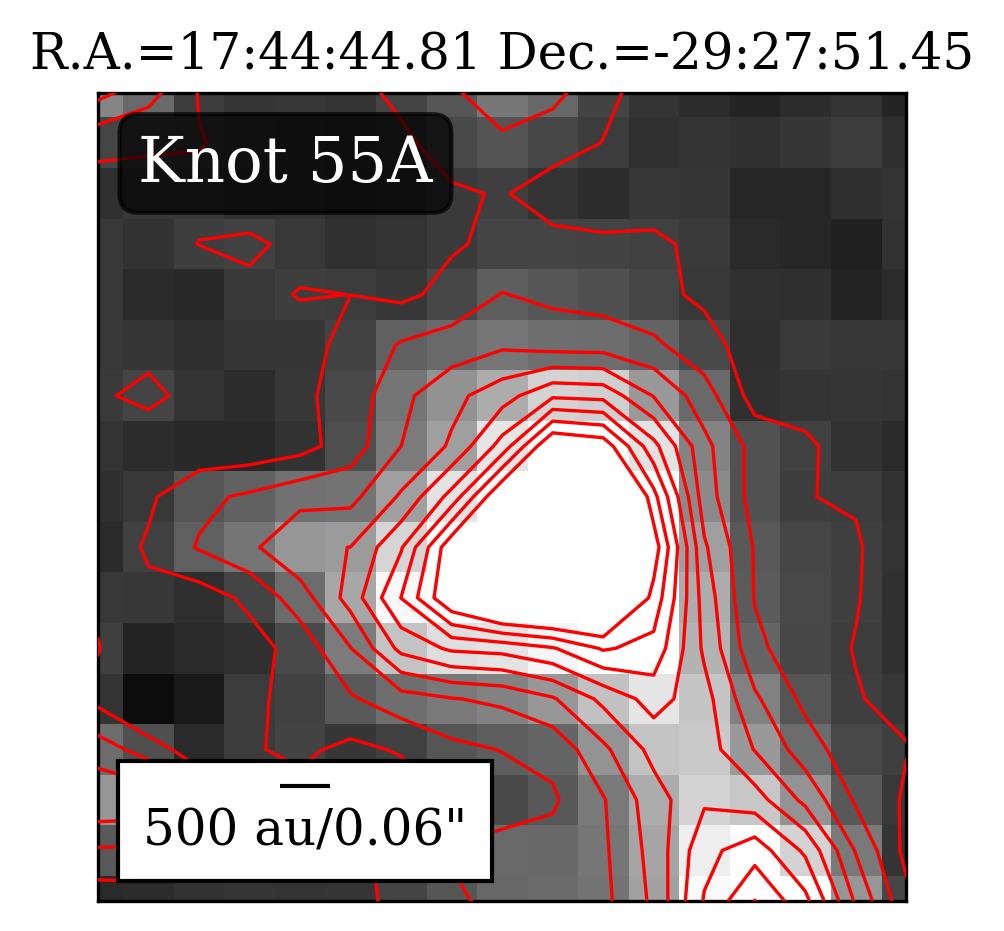}
            \includegraphics[width=0.40\textwidth]{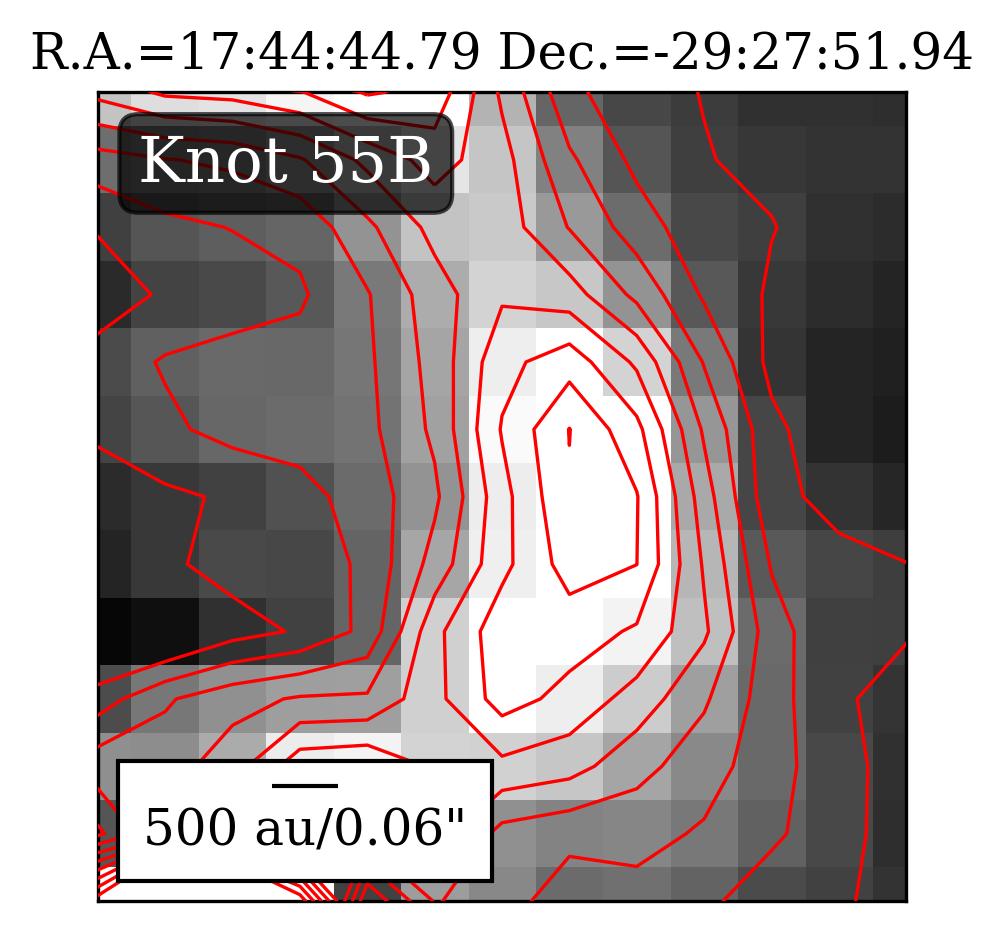}

            \caption{Continued. The contour levels shown represent 10 to 50$\mathrm{\sigma}$ in steps of 5$\sigma$ above the local background.}
        \end{figure*}
        \renewcommand{\thefigure}{B\arabic{figure}}
        \addtocounter{figure}{-1}
        \begin{figure*}[!htb]
        \centering
            \includegraphics[width=0.40\textwidth]{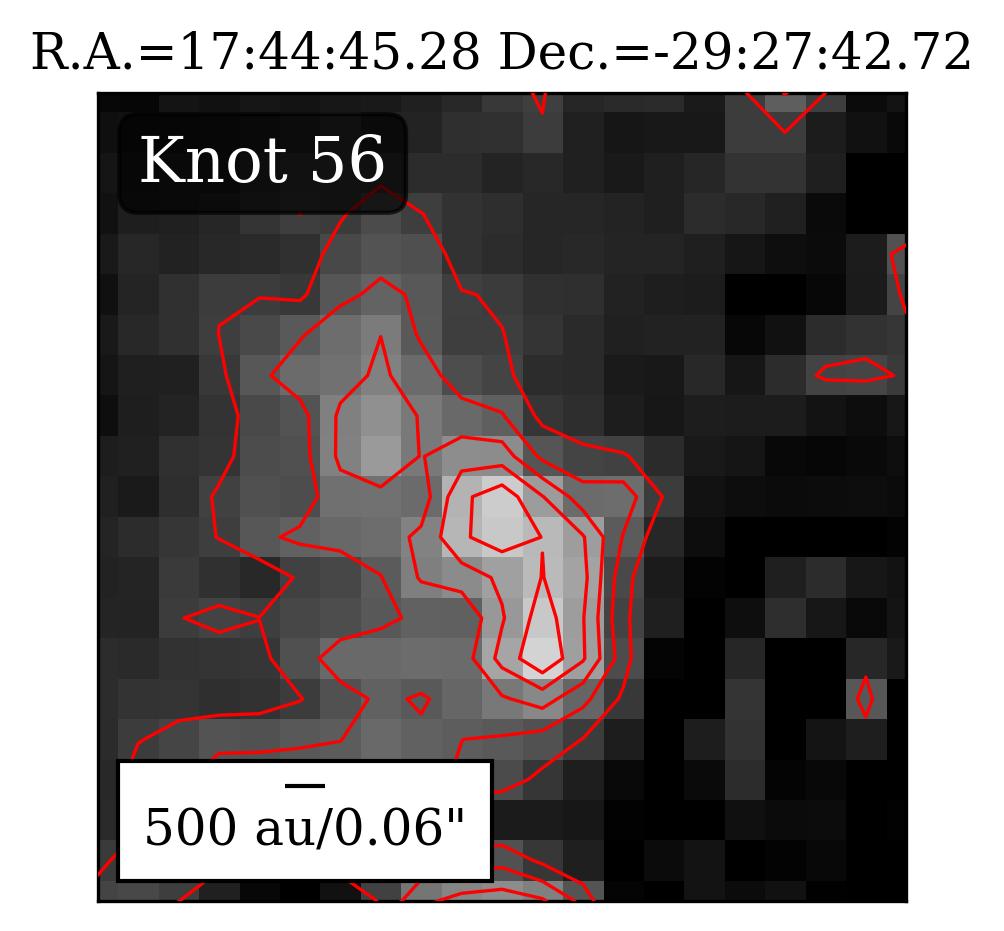}
            \includegraphics[width=0.40\textwidth]{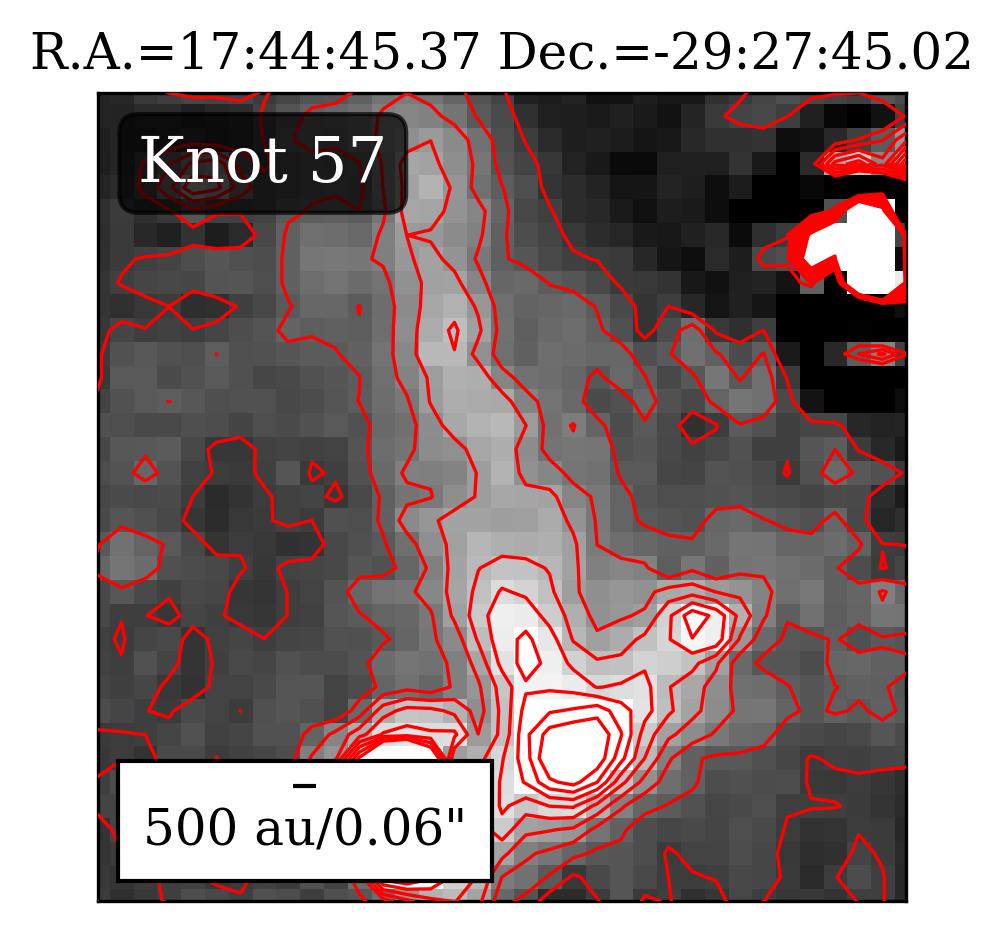}
            
            \caption{Continued. The contour levels shown represent 10 to 50$\mathrm{\sigma}$ in steps of 5$\sigma$ above the local background.}
        \end{figure*}
        \clearpage
        \begin{figure*}[!htb]
                \centering
                \includegraphics[width=0.40\textwidth]{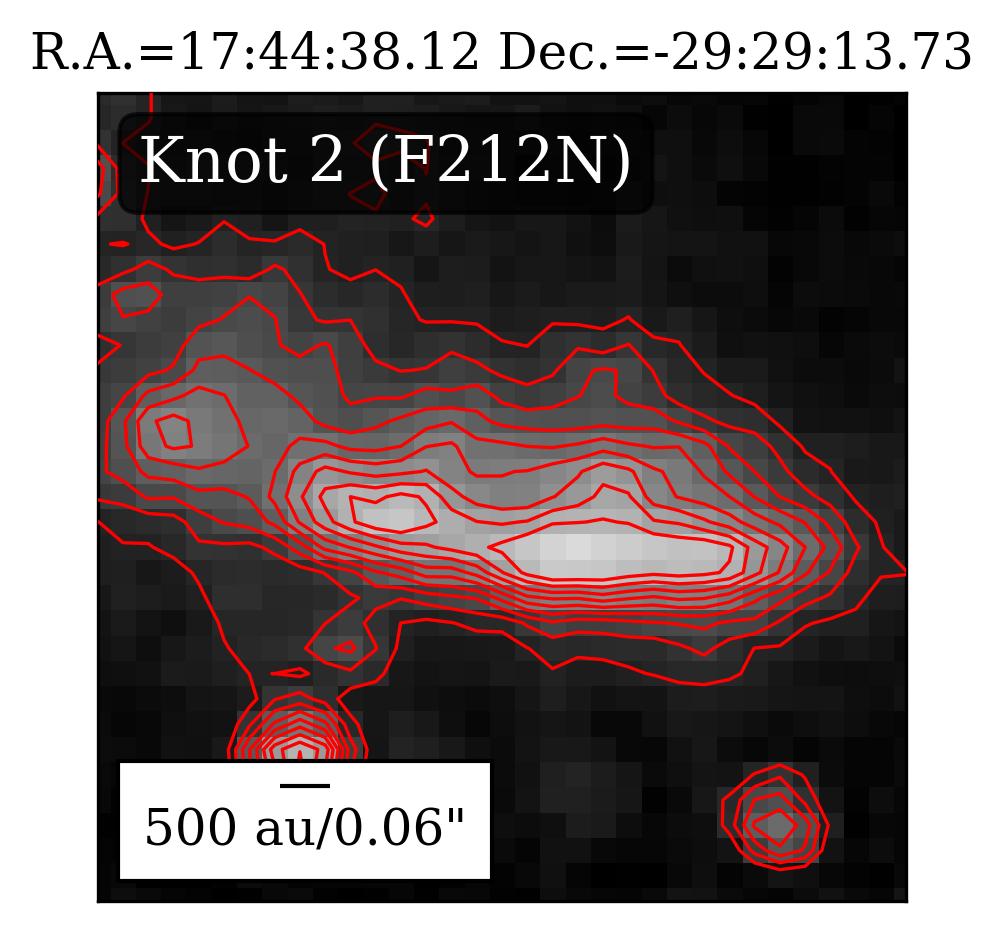}
                \includegraphics[width=0.40\textwidth]{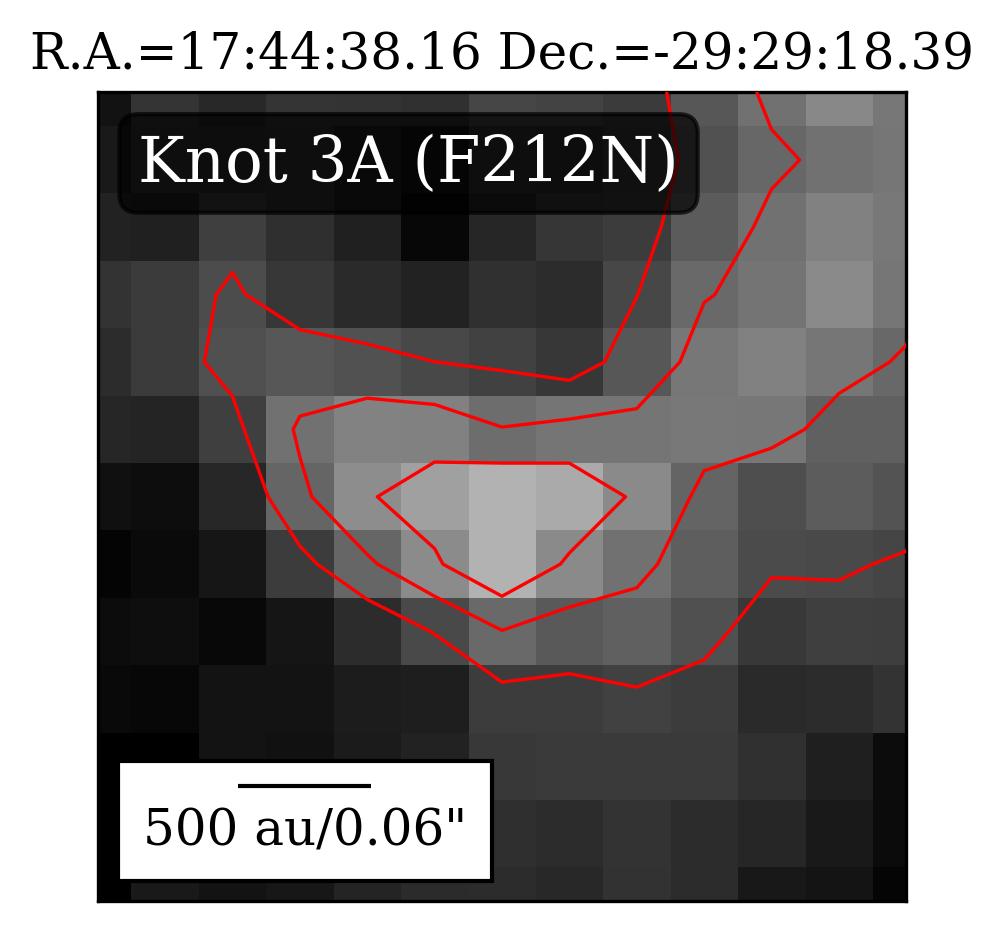}
                \includegraphics[width=0.40\textwidth]{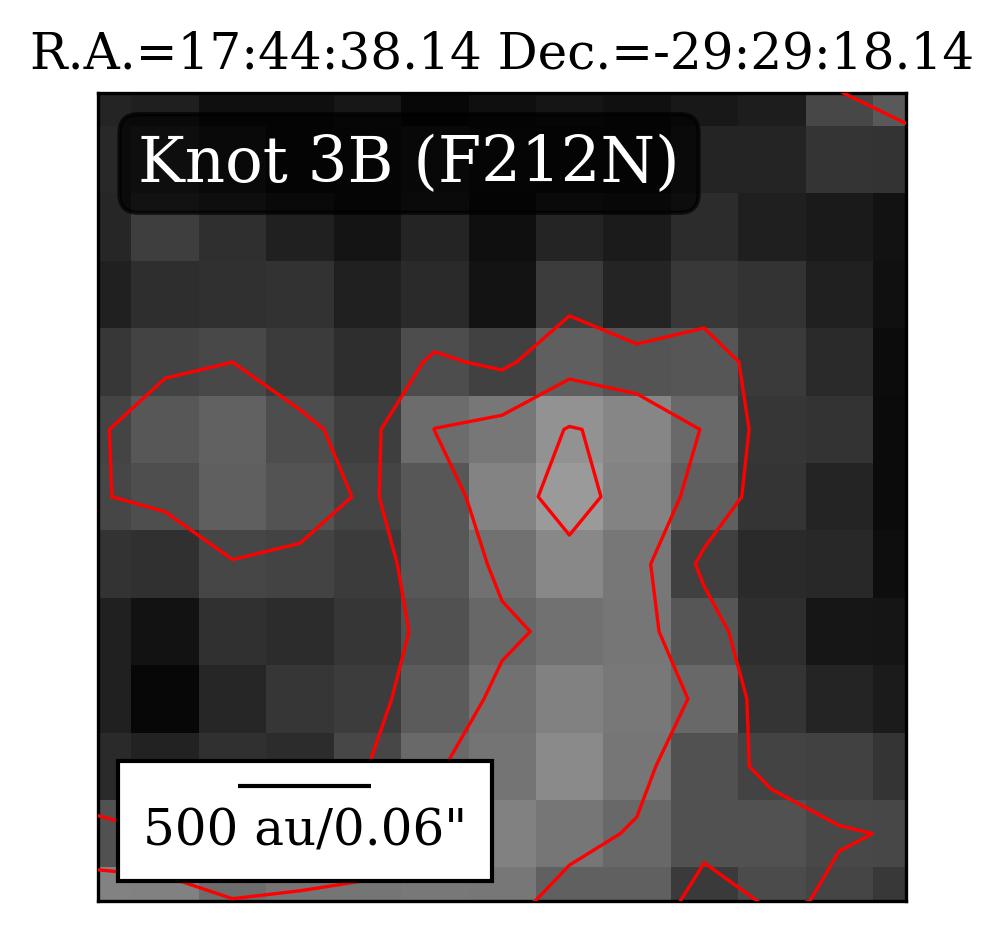}
                \includegraphics[width=0.40\textwidth]{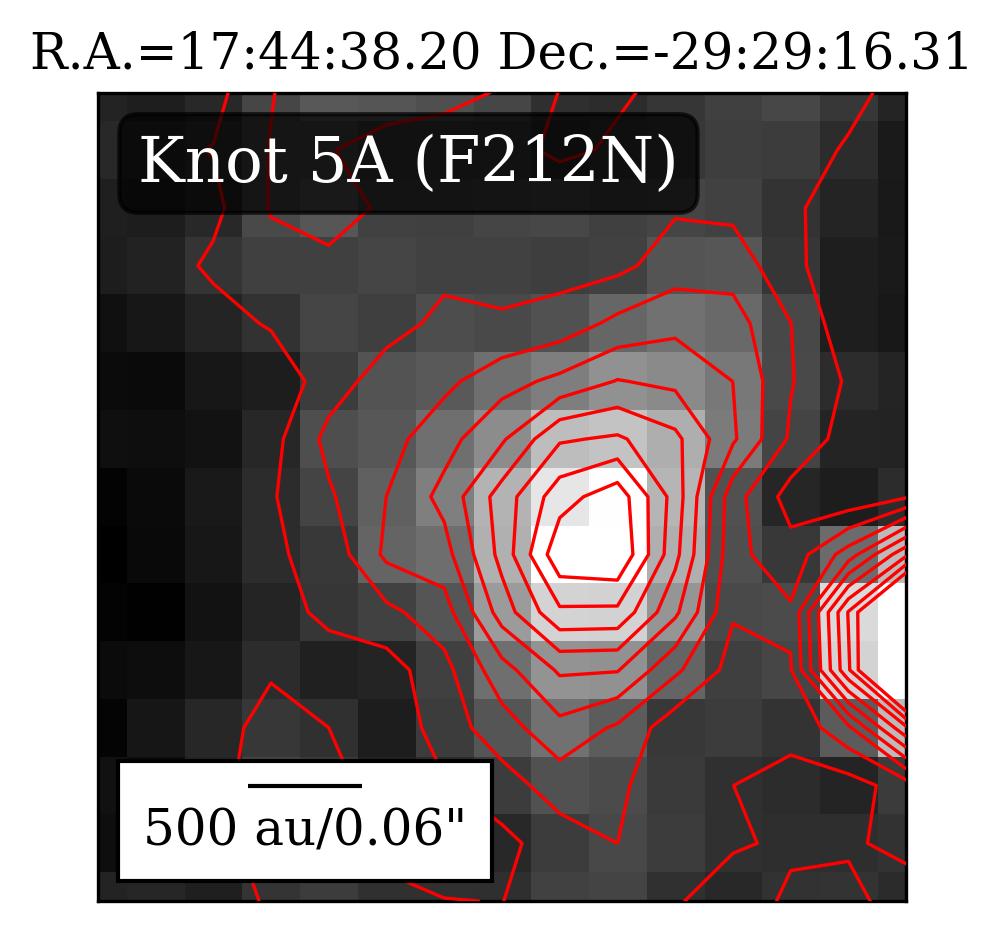}
                \includegraphics[width=0.40\textwidth]{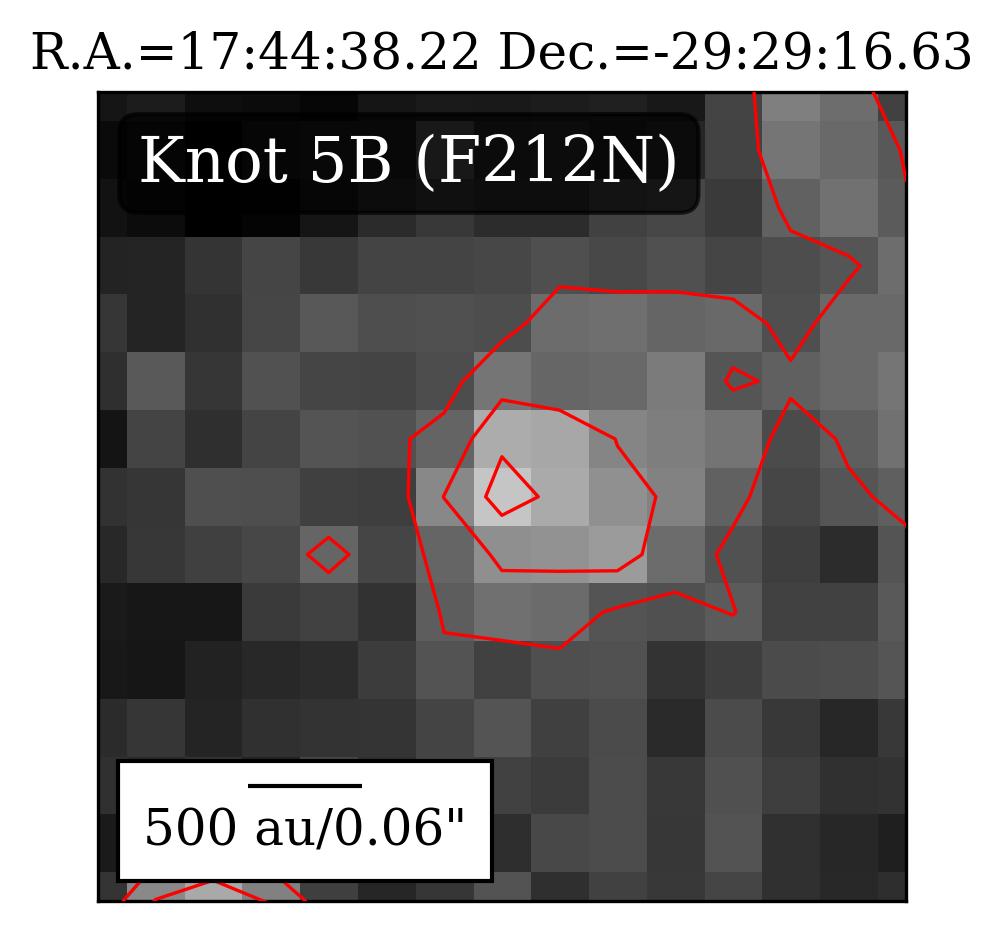}
                \includegraphics[width=0.40\textwidth]{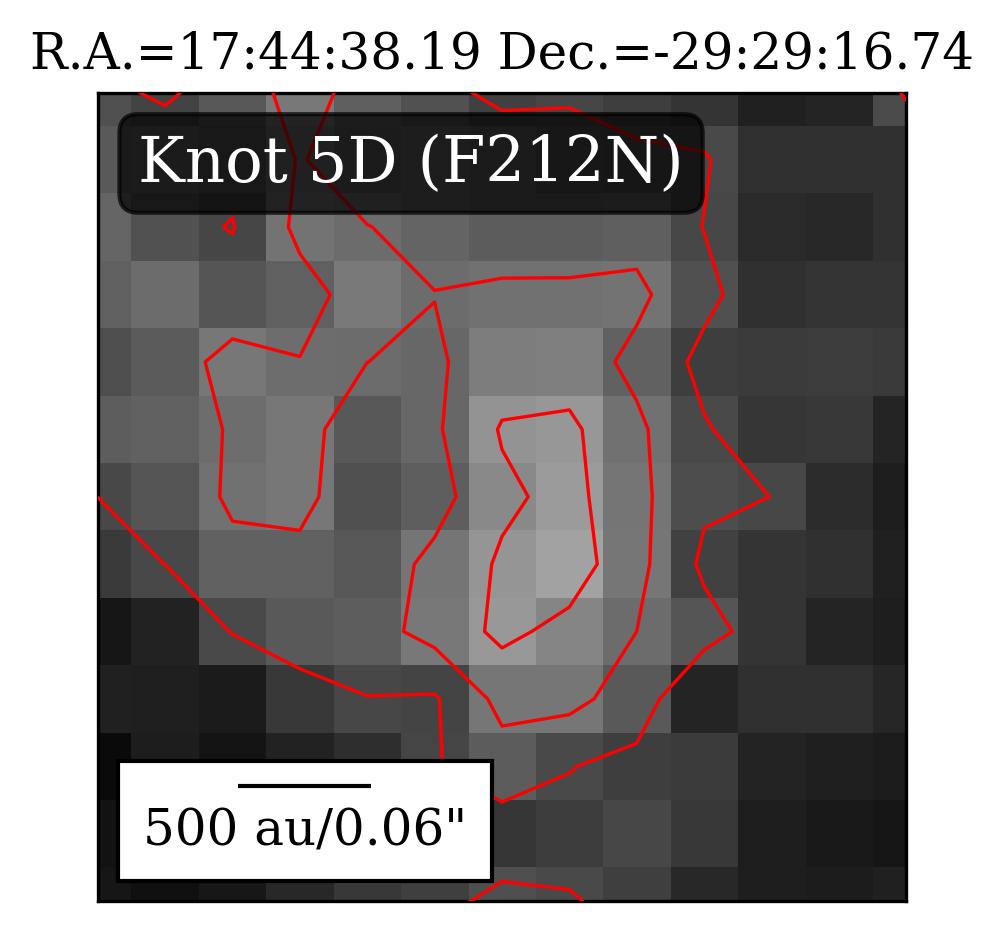}

          \caption{\label{fig:F212N_knots}Significance level contour maps of all knot features identified in the F212N  continuum-subtracted image and compiled in Table \ref{tab:knot_table}. The contour levels shown represent 10 to 50$\mathrm{\sigma}$ in steps of 5$\sigma$ above the local background. The central coordinates of each knot determined from the peak pixel are given on the top of each panel. A physical scalebar of 500 au is given in the bottom-left corner of each panel. N is up and E is left in all panels.}
        \end{figure*}
    
        \renewcommand{\thefigure}{B\arabic{figure}}
        \addtocounter{figure}{-1}
    
        \begin{figure*}[!htb]
                \centering
                \includegraphics[width=0.40\textwidth]{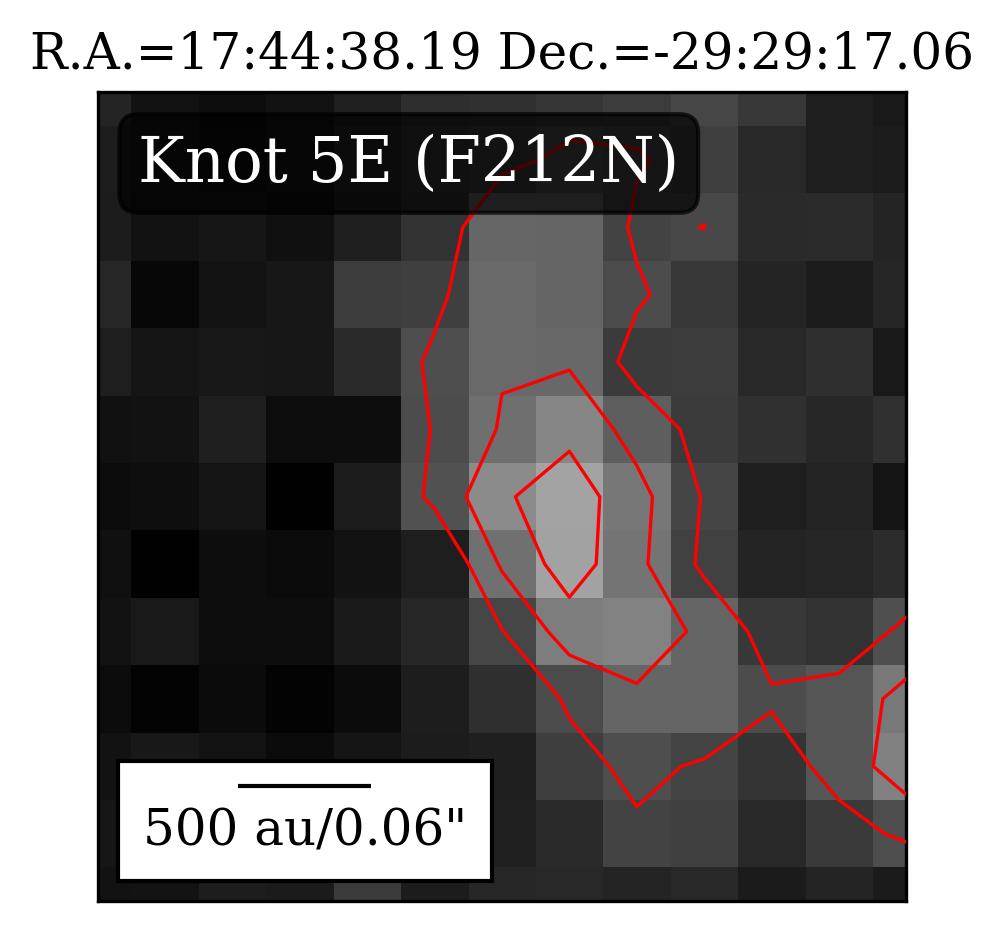}
                \includegraphics[width=0.40\textwidth]{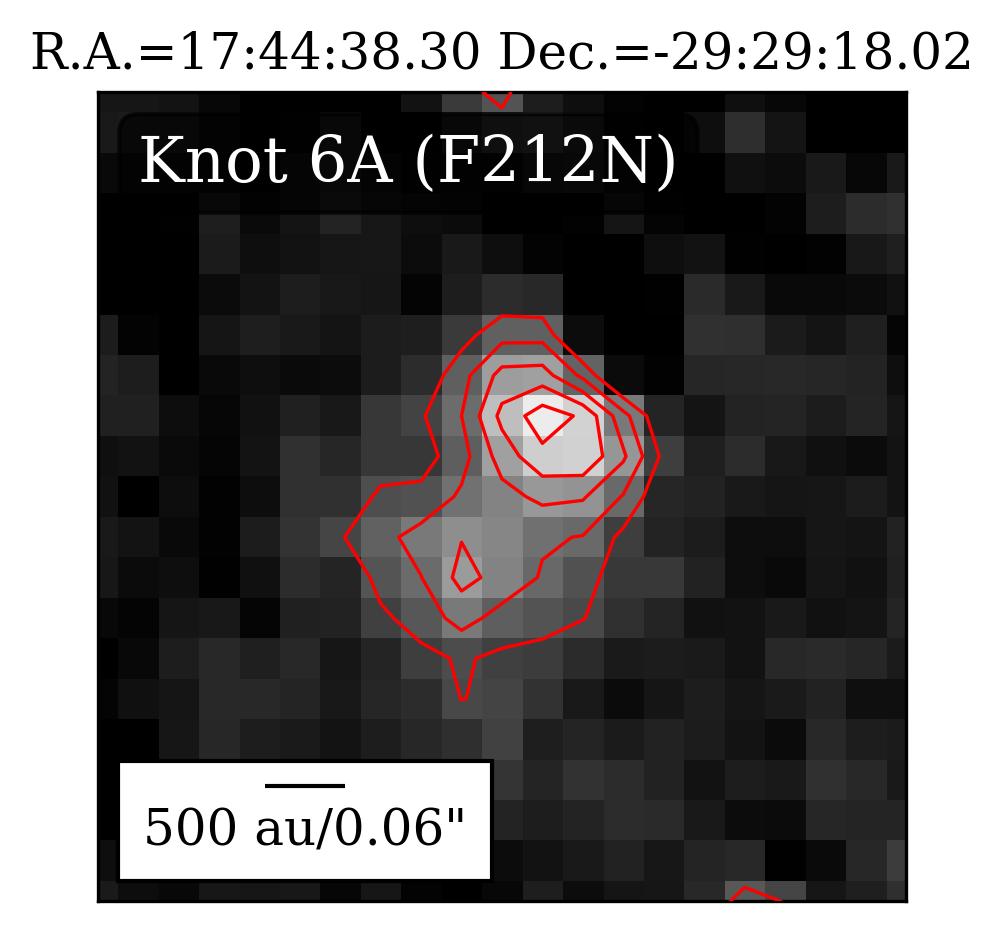}
                \includegraphics[width=0.40\textwidth]{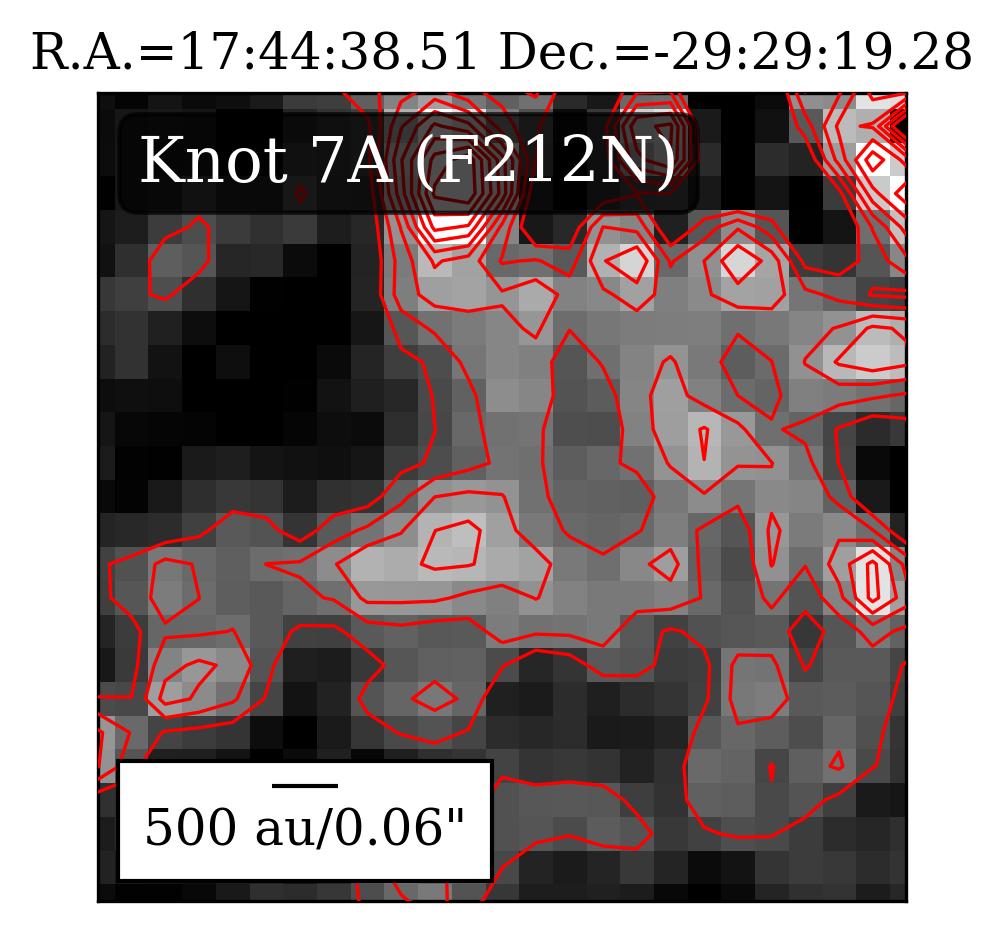}
                \includegraphics[width=0.40\textwidth]{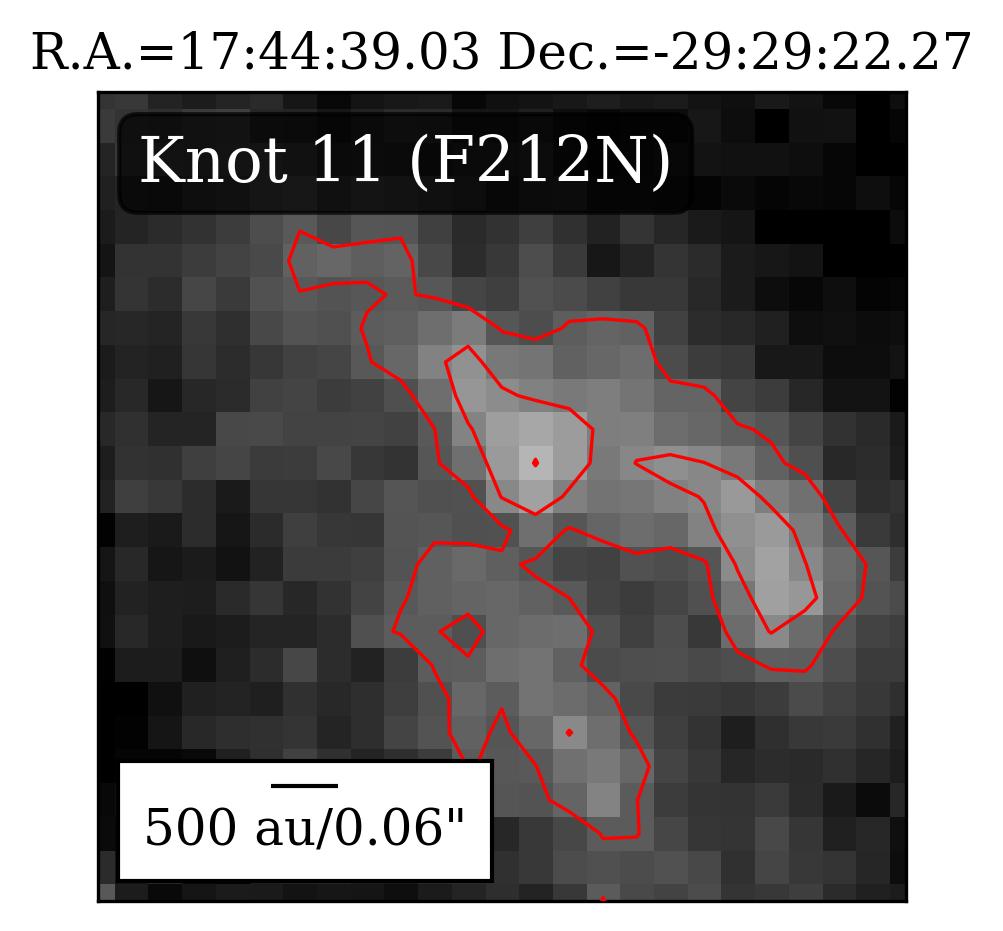}
                \includegraphics[width=0.40\textwidth]{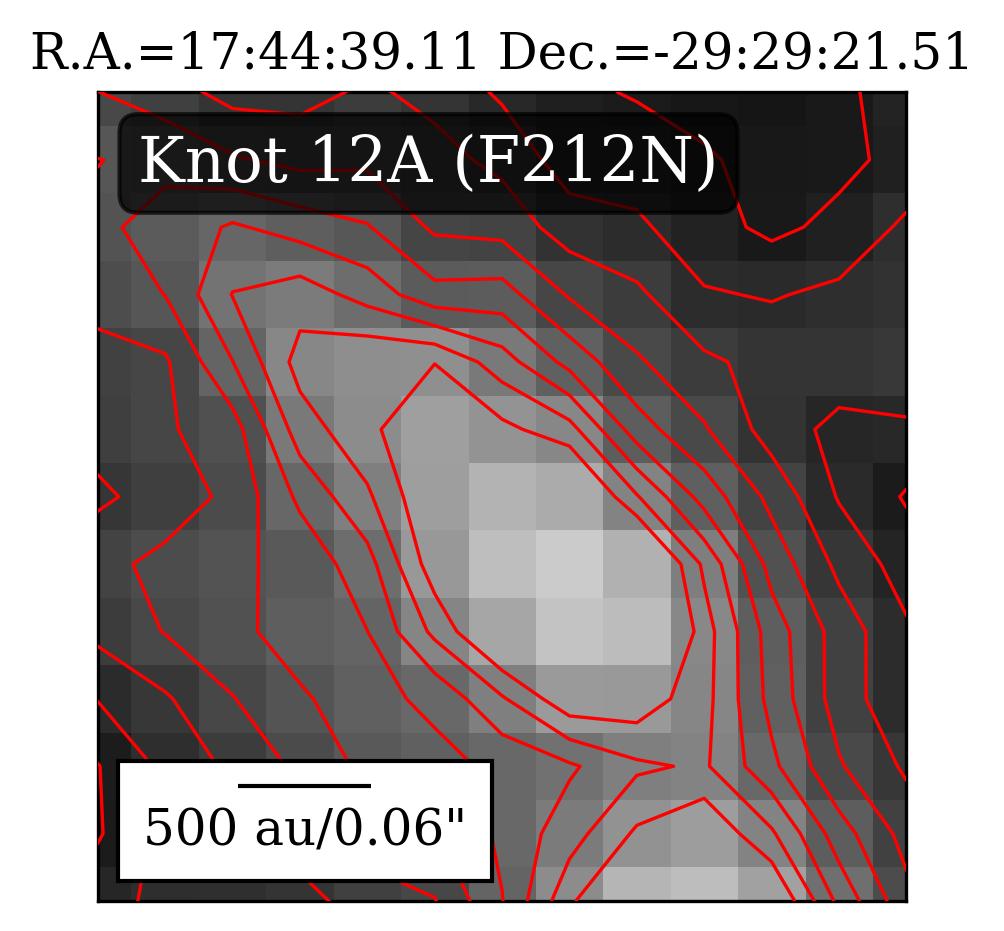}
                \includegraphics[width=0.40\textwidth]{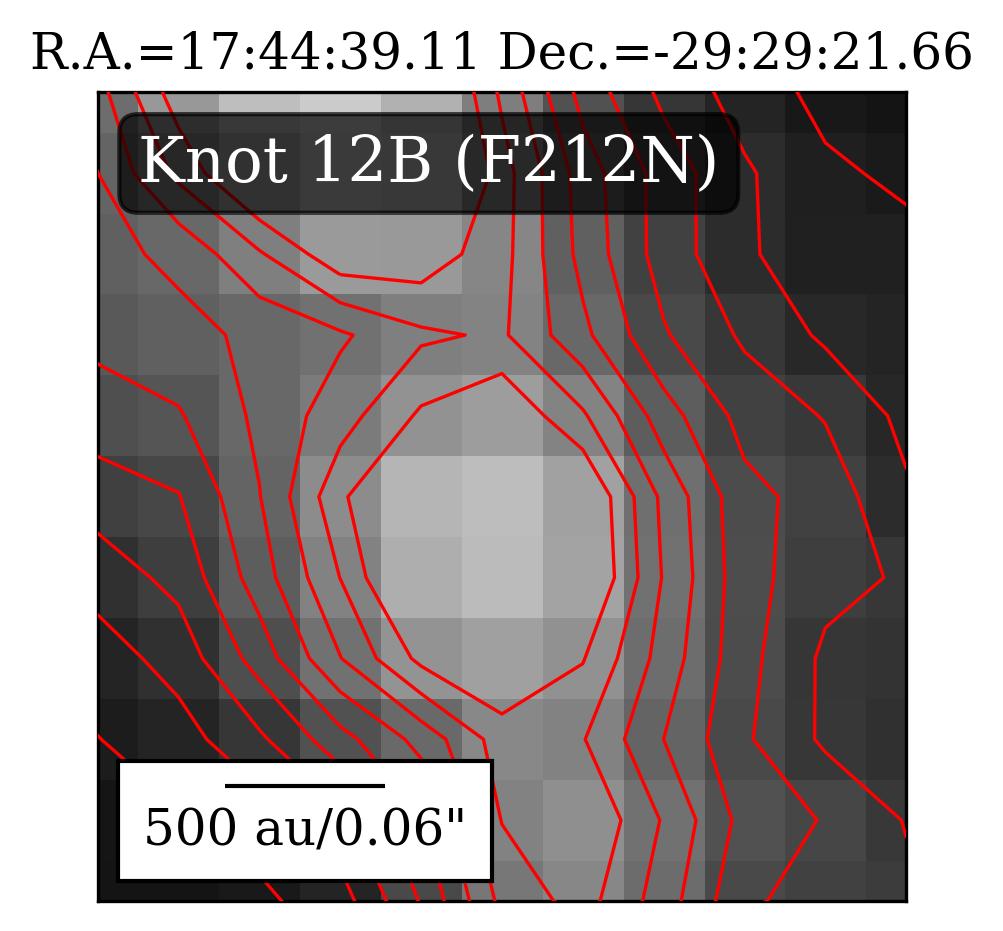}

          \caption{Continued. The contour levels shown represent 10 to 50$\mathrm{\sigma}$ in steps of 5$\sigma$ above the local background.}
        \end{figure*}
    
        \renewcommand{\thefigure}{B\arabic{figure}}
        \addtocounter{figure}{-1}
    
        \begin{figure*}[!htb]
                \centering
                \includegraphics[width=0.40\textwidth]{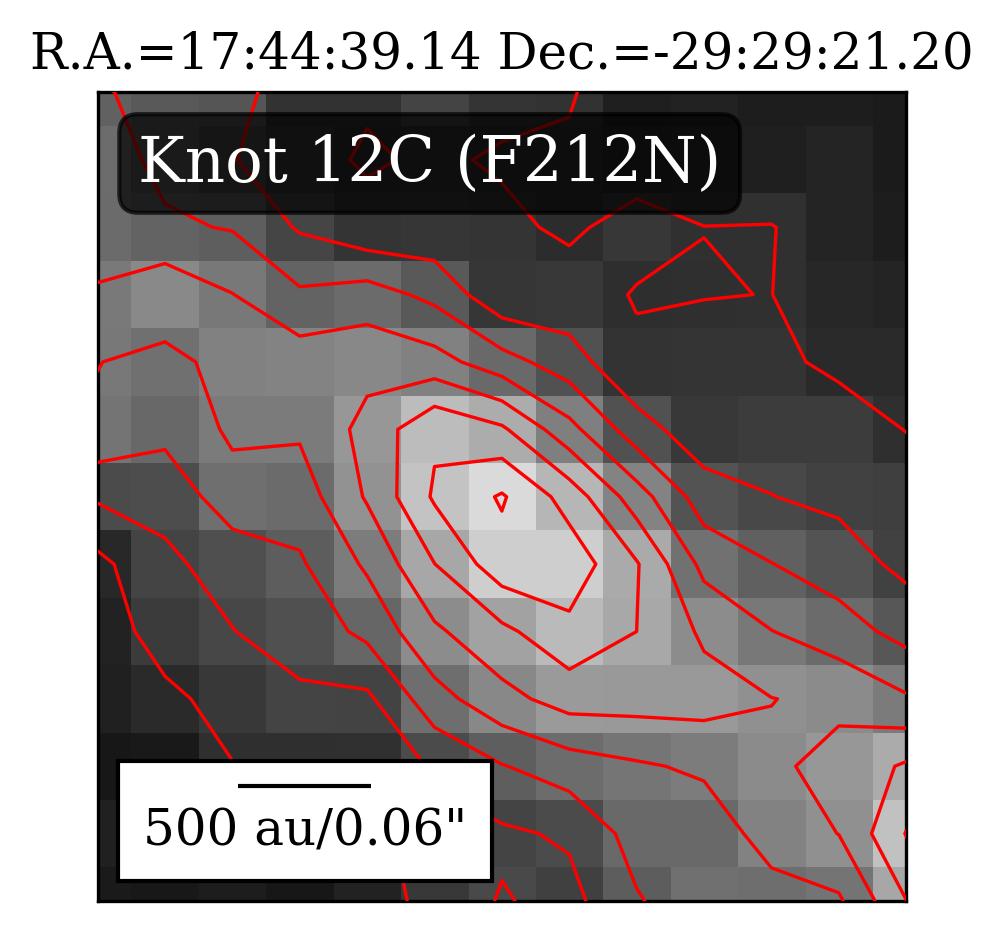}
                \includegraphics[width=0.40\textwidth]{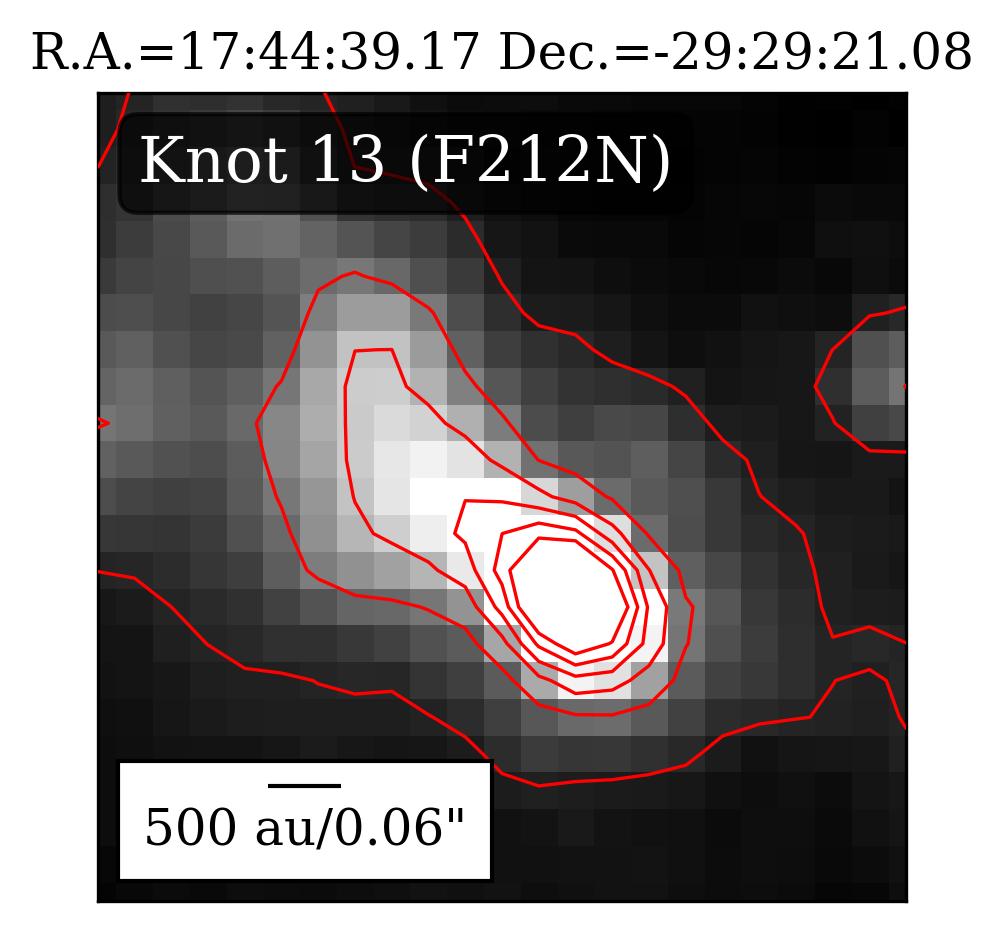}
                \includegraphics[width=0.40\textwidth]{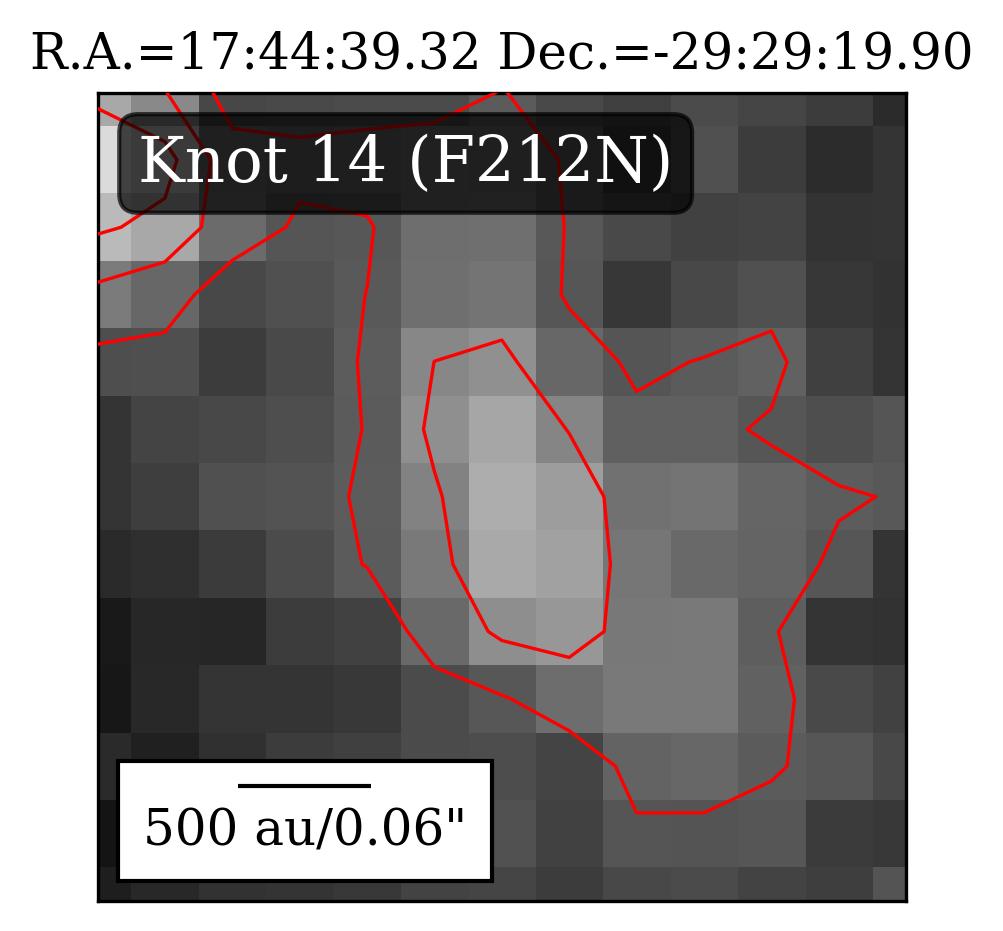}
                \includegraphics[width=0.40\textwidth]{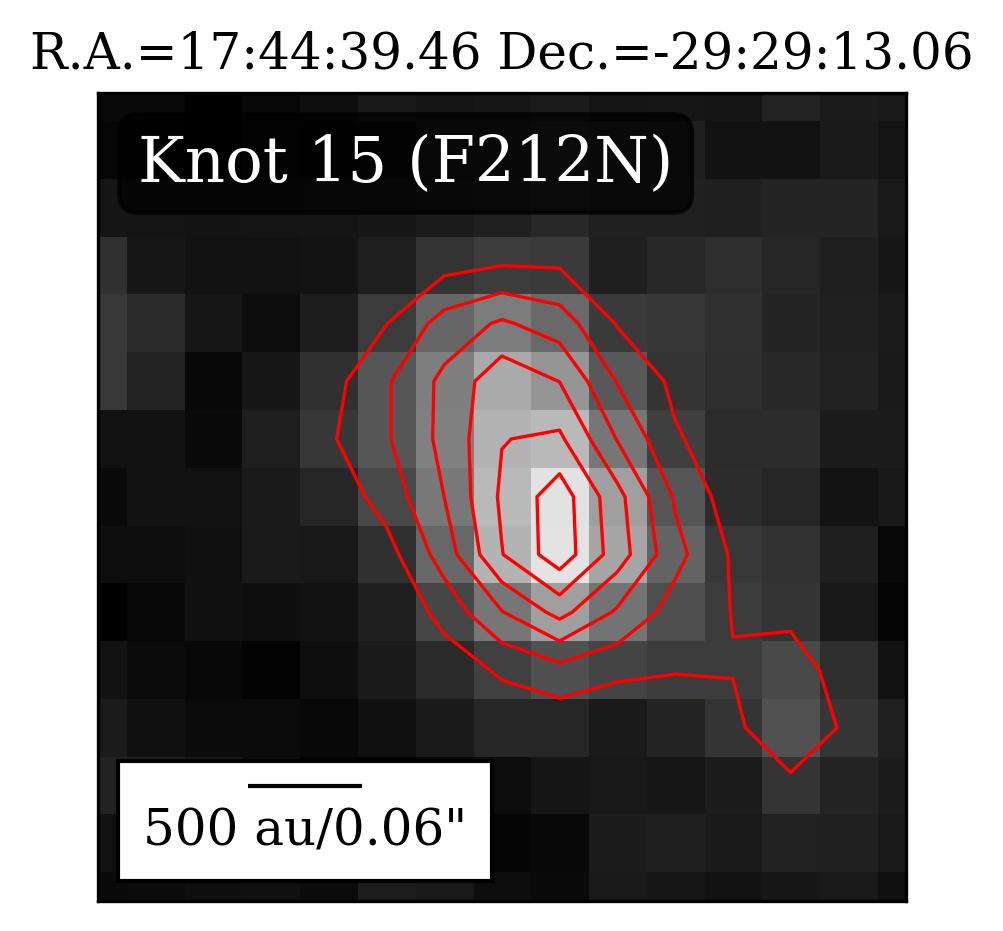}
                \includegraphics[width=0.40\textwidth]{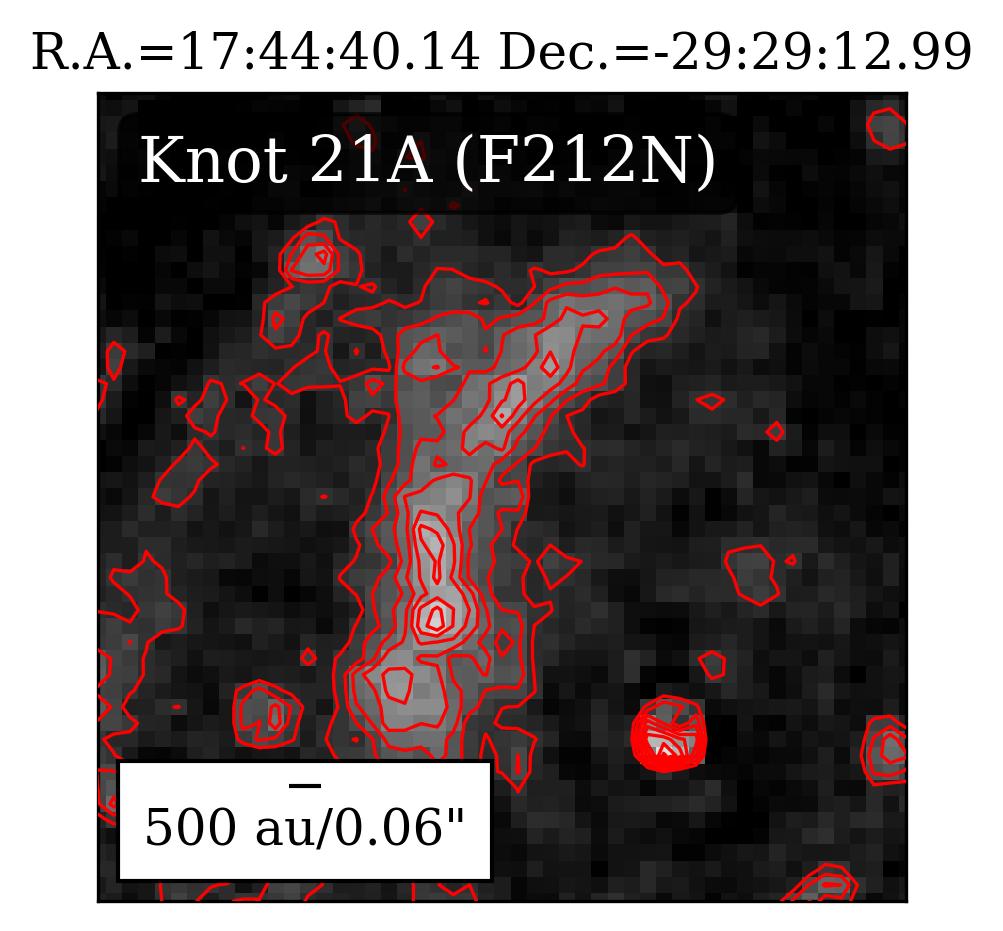}
                \includegraphics[width=0.40\textwidth]{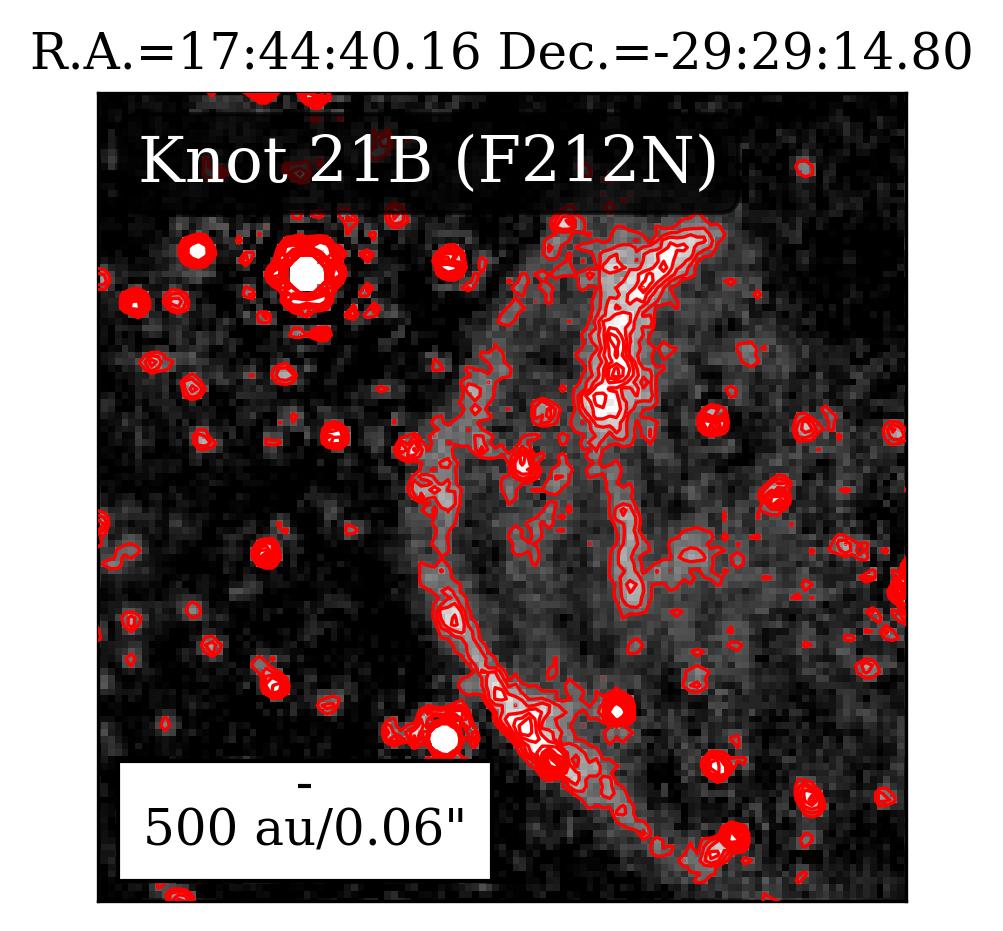}
        
          \caption{Continued. The contour levels shown for knots 12C, 14, 15, 21A, and 21B represent 10 to 50$\mathrm{\sigma}$ in steps of 5$\sigma$ above the local background; those shown for knot 13 represent 25 to 300$\mathrm{\sigma}$ in steps of 55$\sigma$.}
        \end{figure*}

        \renewcommand{\thefigure}{B\arabic{figure}}
        \addtocounter{figure}{-1}
    
        \begin{figure*}[!htb]
                \centering
                \includegraphics[width=0.40\textwidth]{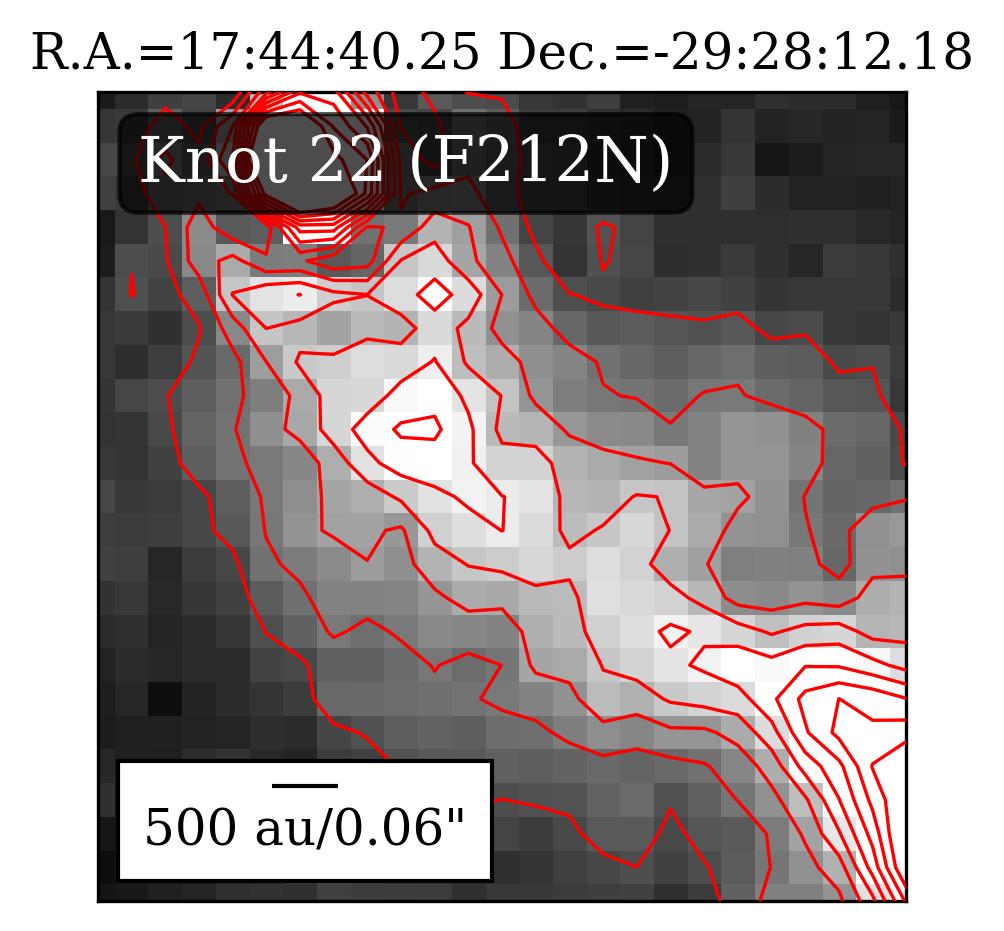}
                \includegraphics[width=0.40\textwidth]{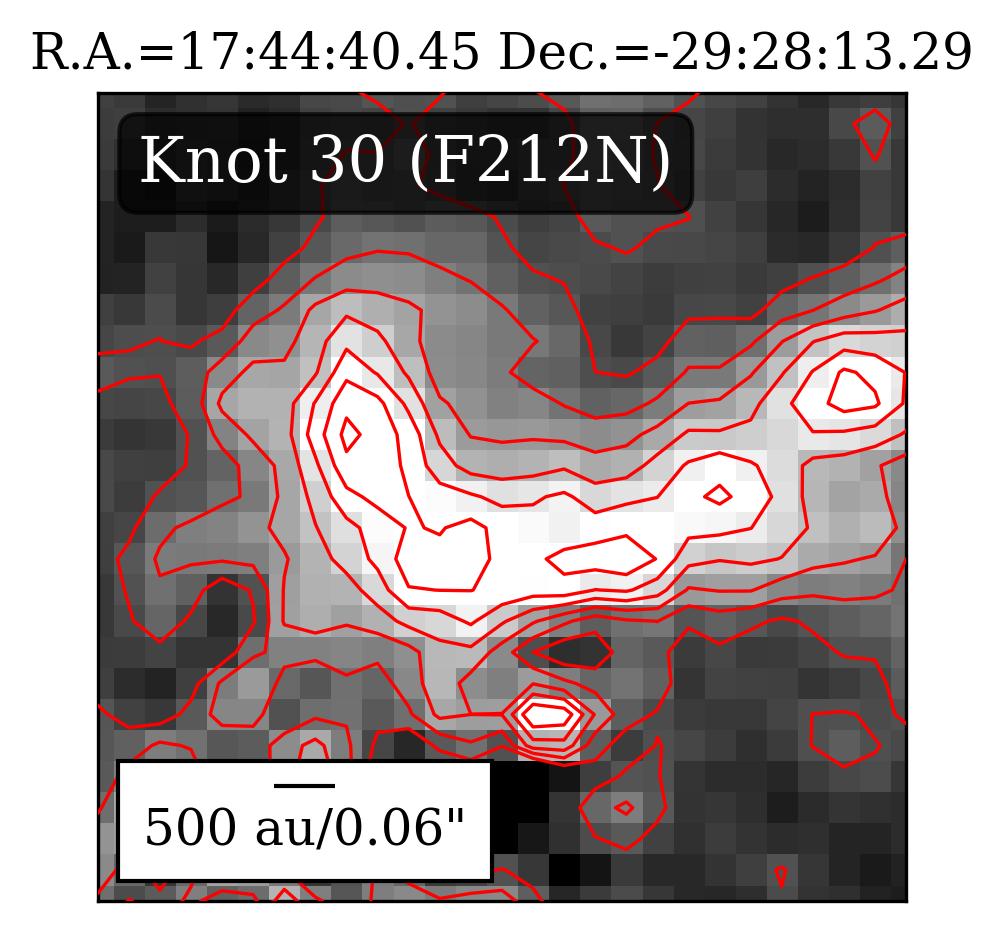}
                \includegraphics[width=0.40\textwidth]{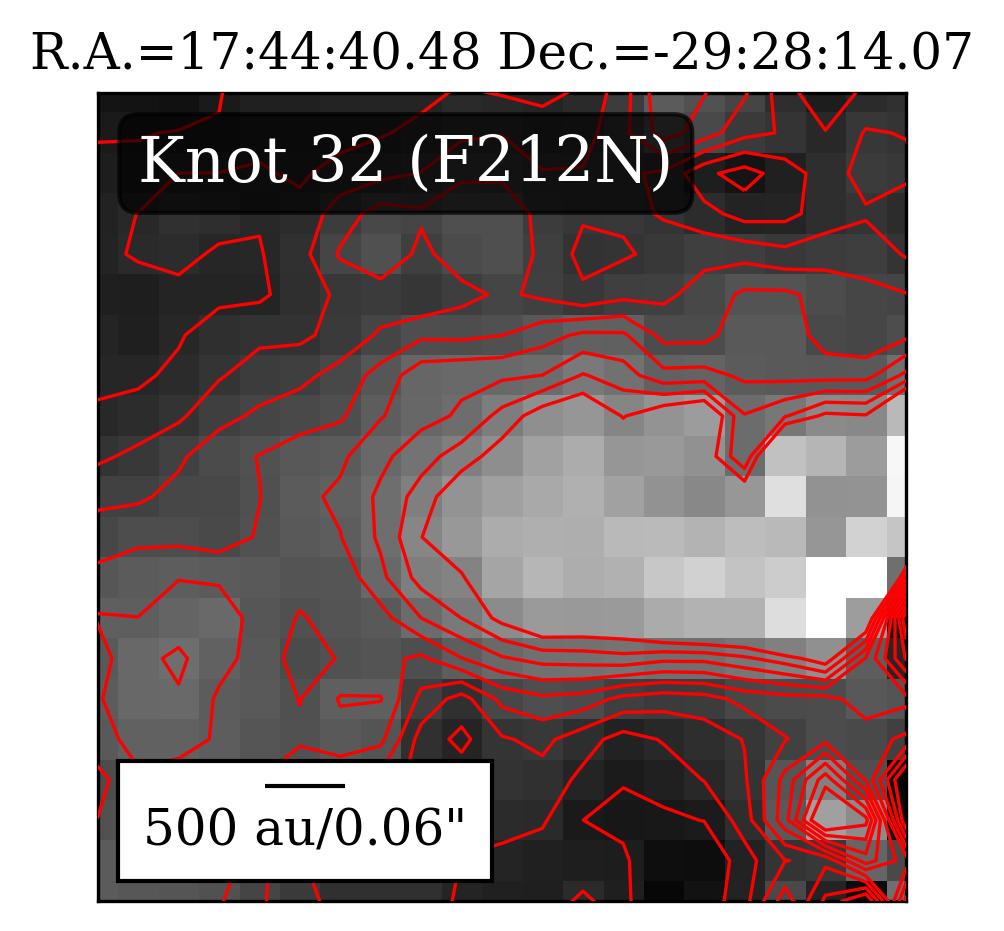}
                \includegraphics[width=0.40\textwidth]{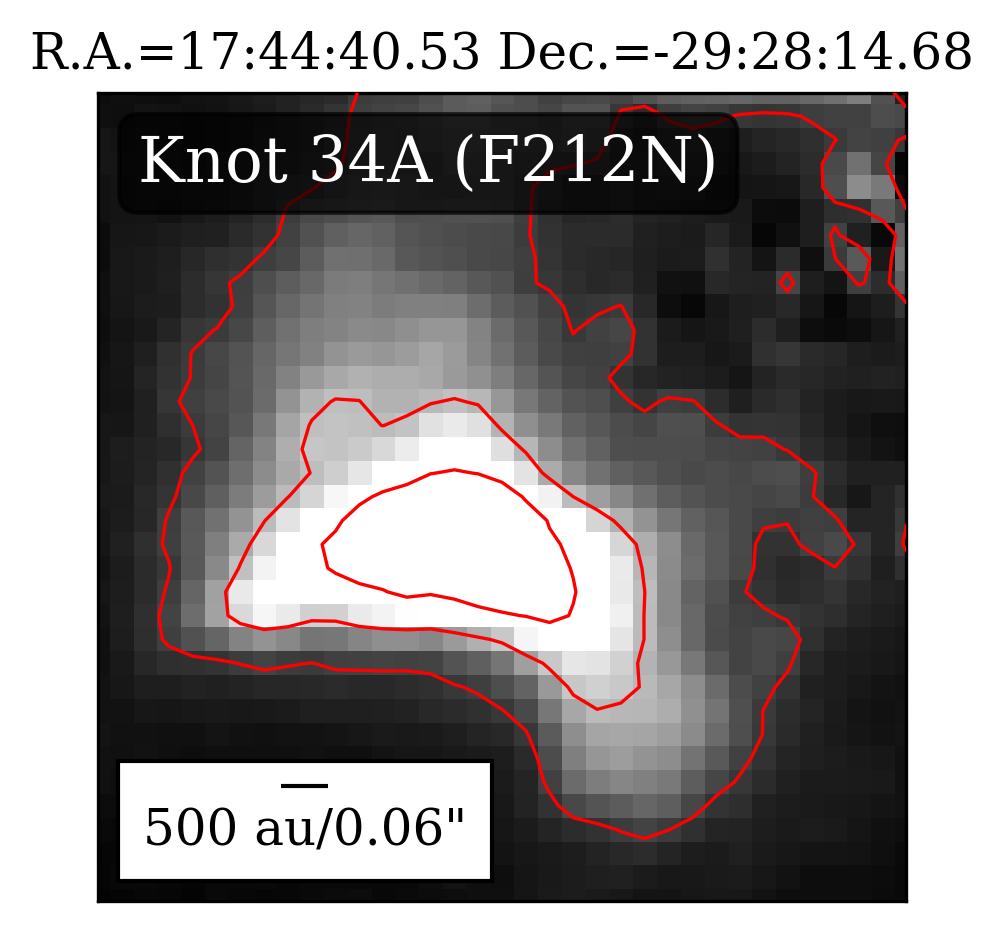}
        
          \caption{Continued. The contour levels shown for knots 22, 30, and 32 represent 10 to 50$\mathrm{\sigma}$ in steps of 5$\sigma$ above the local background; those shown for knot 34A represent 25 to 300$\mathrm{\sigma}$ in steps of 55$\sigma$.}
        \end{figure*}
        \clearpage
        \renewcommand{\thefigure}{B3}
        \begin{figure*}[!htb]
                \centering
                \includegraphics[width=0.40\textwidth]{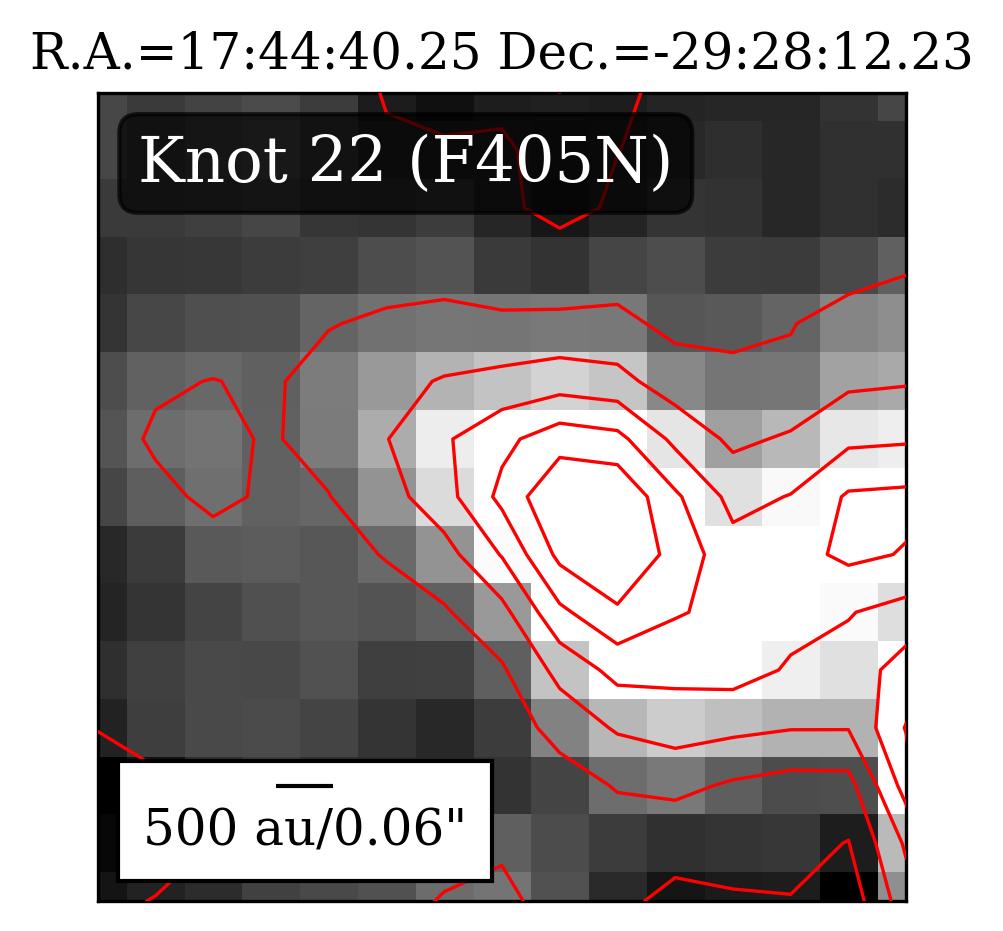}
                \includegraphics[width=0.40\textwidth]{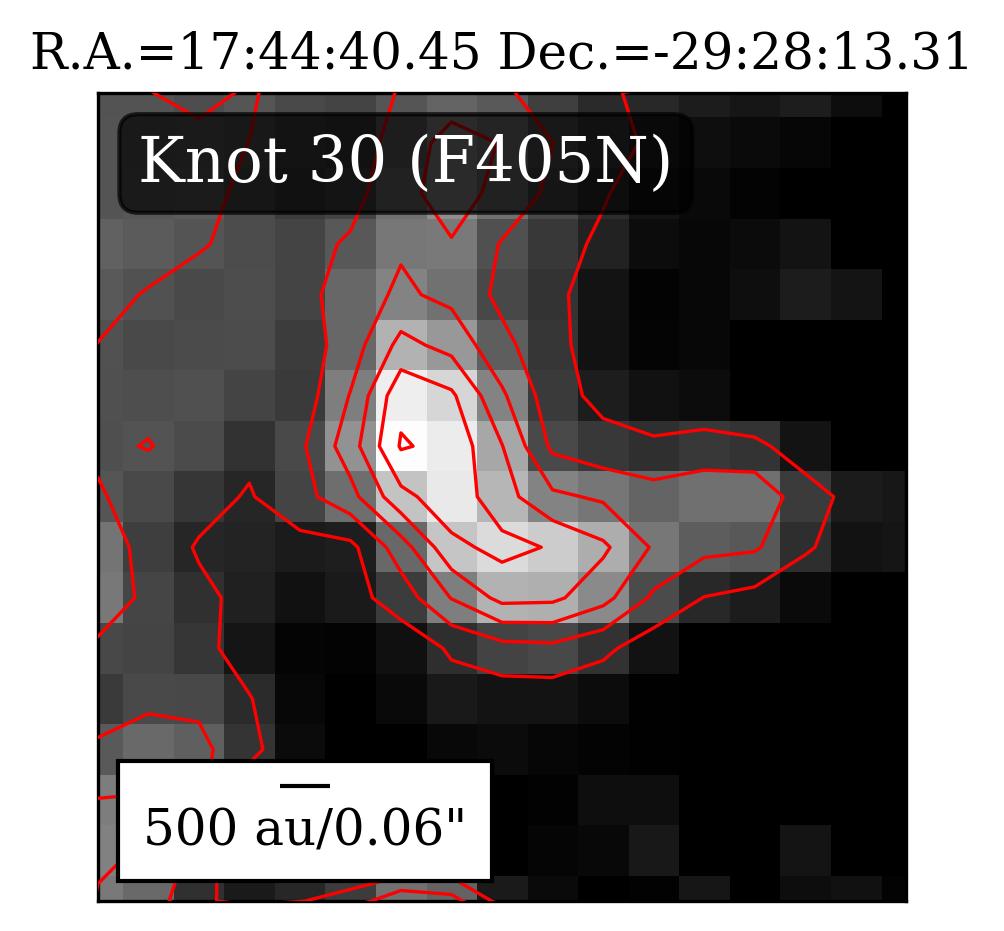}
                \includegraphics[width=0.40\textwidth]{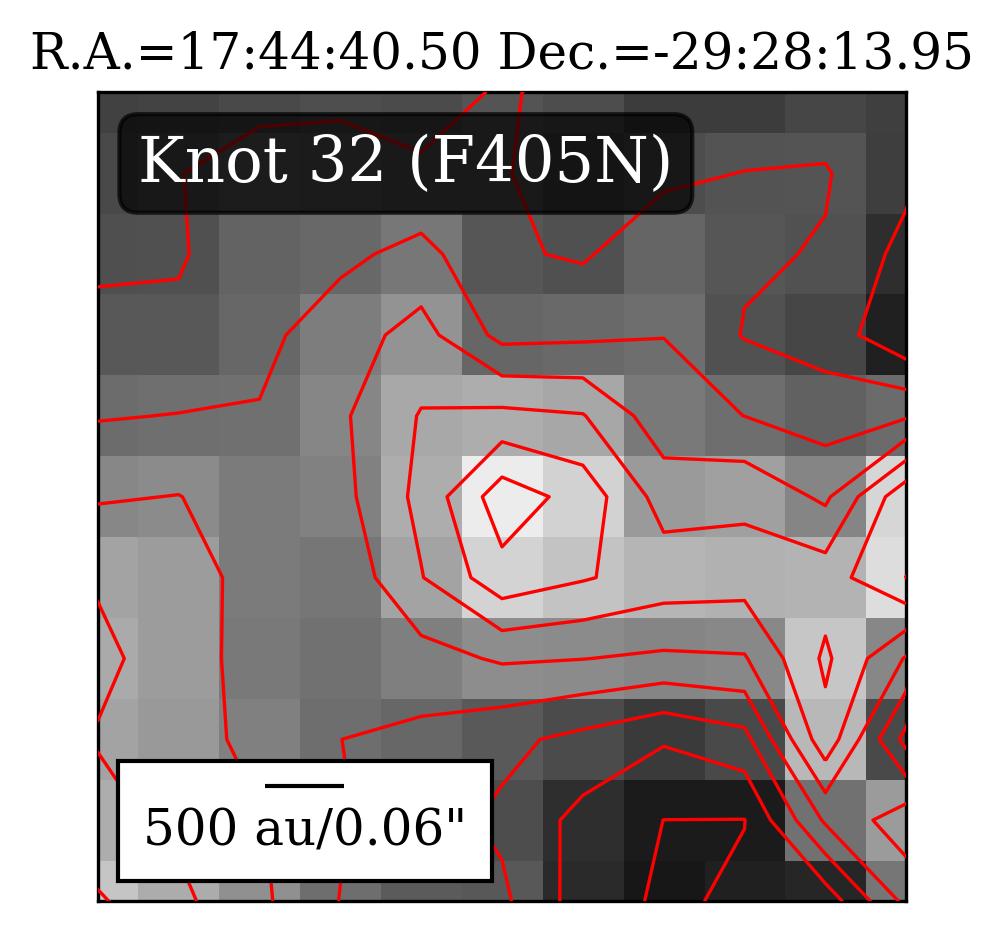}
                \includegraphics[width=0.40\textwidth]{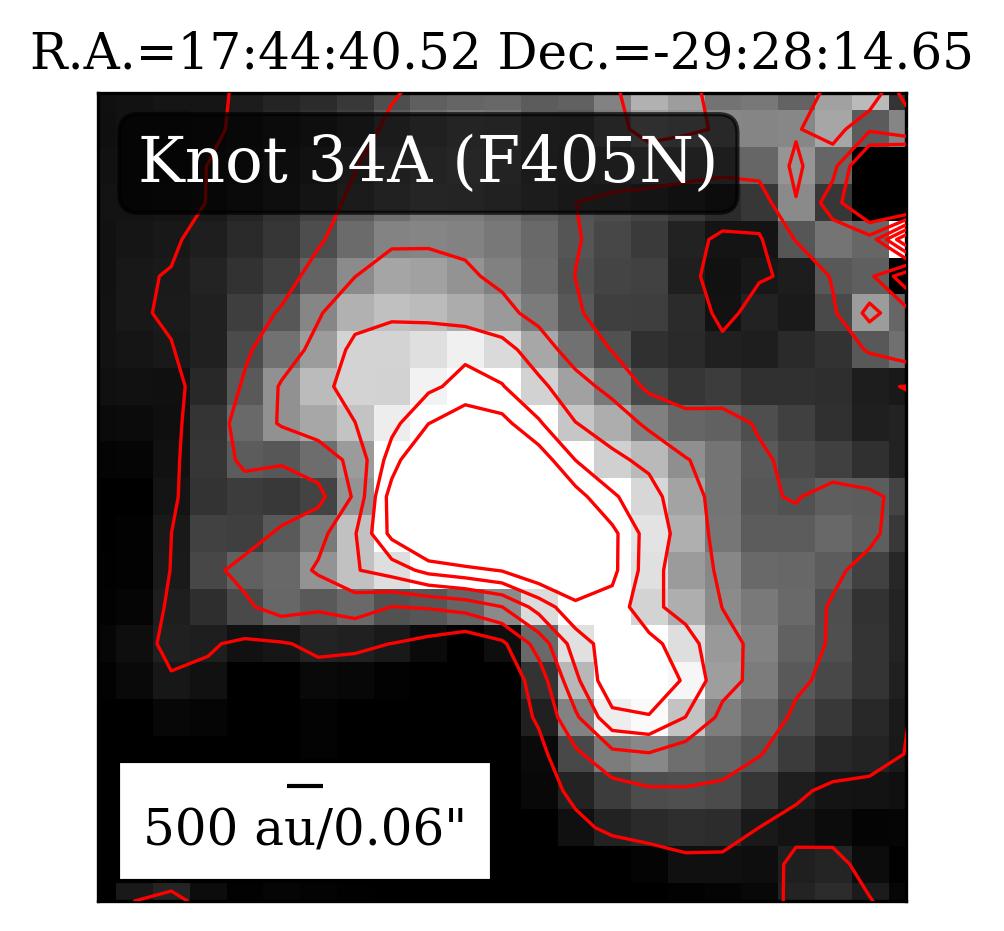}
          \caption{\label{fig:F405N_knots}Significance level contour maps of all knot features identified in the F405N  continuum-subtracted image and compiled in Table \ref{tab:knot_table}. The contour levels shown for knots 22 and 34A represent 25 to 300$\mathrm{\sigma}$ in steps of 55$\sigma$ above the local background; those shown for knots 30 and 32 represent 20 to 300$\mathrm{\sigma}$ in steps of 20$\sigma$. The central coordinates of each knot determined from the peak pixel are given on the top of each panel. A physical scalebar of 500 au is given in the bottom-left corner of each panel. N is up and E is left in all panels.}

        \end{figure*}
        \clearpage
    \section{Miscellaneous MHOs in the NIRCam Field}\label{sec:misc_mhos}

    In addition to the outflow knot candidates associated with star formation activity, a number of miscellaneous molecular hydrogen objects (MHOs) were also discovered in the NIRCam field. These are compiled in Table \ref{tab:misc_mhos}, and briefly shown and described here.

    \begin{table*}[h]
        
        \centering
        \caption{\label{tab:misc_mhos}Miscellaneous Molecular Hydrogen Objects (MHOs) found in the F470N continuum-subtracted NIRCam image. Approximately central coordinates for each feature are given.}           
        \begin{tabular}{c c c}     
        \hline\hline       
        Name &R.A.& Dec.\\
        &(J2000) & (J2000)\\
        \hline     
        Lobe 1 & 17:44:44.91 & -29:29:25.56\\
        Lobe 2 & 17:44:43.31 & -29:29:24.91\\
        Flame & 17:44:36.37 & -29:29:01.57\\
        Skein & 17:44:38.96 & -29:27:29.38\\
        Feathers & 17:44:38.75 & -29:27:06.61\\
        Thread 1 & 17:44:45.52 & -29:25:37.42\\
        Thread 2 & 17:44:44.54 & -29:25:15.12\\
        Thread 3 & 17:44:42.63 & -29:24:39.08\\
        Thread 4 & 17:44:40.95 & -29:25:52.61\\
        Thread 5 & 17:44:40.37 & -29:25:54.31\\
        Thread 6 & 17:44:39.86 & -29:25:50.67\\
        Streak & 17:44:41.33 & -29:24:33.82\\
        Lamp & 17:44:37.71 & -29:26:24.00\\
        \hline
        \end{tabular}
        \end{table*}

        \subsection{The Lobes}\label{sec:lobes}

            The Lobes are two relatively bright extended sources of emission to the far south-east of the NIRCam pointing. Their symmetric nature may indicate a bipolar outflow from a young star.

        \subsection{Filaments and the Flame}\label{sec:flame_filaments}

            The Flame is another prominent object located in the south-west corner of the NIRCam image, close to G359.42-0.104 and associated with a dark cloud. Notably, this structure is surrounded by filaments in the H$_2$ emission, the origins of which are unclear. Similar filaments can be seen in the eastern part of the Sgr C cloud (see Fig. \ref{fig:big_diagram} (b.2)). This region may host star formation, but it is notably lacking any associated emission in ALMA Band 3, VLA \citep{lu19b}, and MeerKAT \citep{heywood22} images. 

        \subsection{The Skein and the Feathers}\label{sec:skein_feathers}

            The Skein is a notable bright region of extended emission in the NIRCam data; this object is also visible in Spitzer and SOFIA mid-infrared images of the CMZ, indicating that it is not an artifact of the NIRCam data (e.g., a ``snowball"\footnote{https://jwst-docs.stsci.edu/depreciated-jdox-articles/data-artifacts-and-features/snowballs-and-shower-artifacts}). One may speculate that this object is a background galaxy or a YSO, but its origin is nonetheless unknown.

            The Feathers are another filamentary feature, similar to the filaments seen around the Flame.
    
        \subsection{The Streak}\label{sec:streak}

            The Streak is a very bright region of extended emission prominently visible in the JWST continuum images of Sgr C (see Fig. \ref{fig:RGB}), as well as in the F470N continuum-subtracted data.

            While the origin and nature of this object is uncertain, it is spatially coincident with a type II Cepheid star ($\mathrm{b333\_57\_97584}$) detected in \citep[][]{braga19} as well as a tentative young stellar object (YSO) candidate (SSTGC 368854) reported in \citep[][]{an11} that is negated by Spitzer-IRS spectroscopic follow-up. The NIRCam data does not provide any further clarity into the origin of this object.

        \subsection{The Threads and the Lamp}\label{sec:threads_lamp}

            The Threads (numbered 1-6) are filamentary structures seen in the NIRCam data, which appear to extend for as long as a parsec in length (if at the distance to the CMZ; angular size $\sim25\arcsec$). These filaments also appear similar in structure to those seen in the main Sgr C cloud (labeled in Fig. \ref{fig:big_diagram} (b.2)) and around the Flame (Fig. \ref{fig:misc_knots_mod_a}). Although such H$_2$ filaments are likely of different origin to the \Hi\:filaments seen in Br-$\alpha$ (F405N) in the Sgr C \Hii\:region, it is possible that the two share a similar formation mechanism.
            
            The Lamp is another bright emission feature, origin otherwise unknown, located in projection adjacent to a prominent dark cloud.
            
        \renewcommand{\thefigure}{C1}
        \begin{figure*}
            \centering
            \includegraphics[width=0.9\textwidth,keepaspectratio]{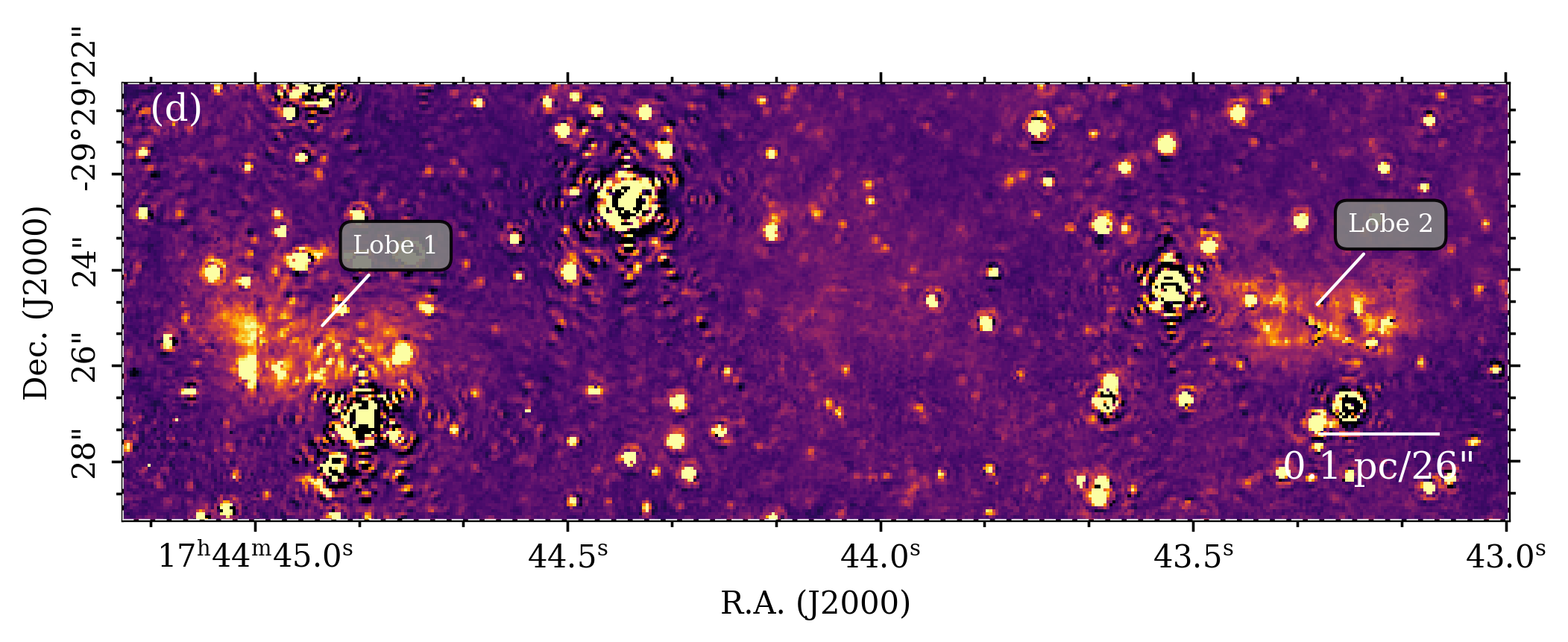}
            \includegraphics[width=0.9\textwidth]{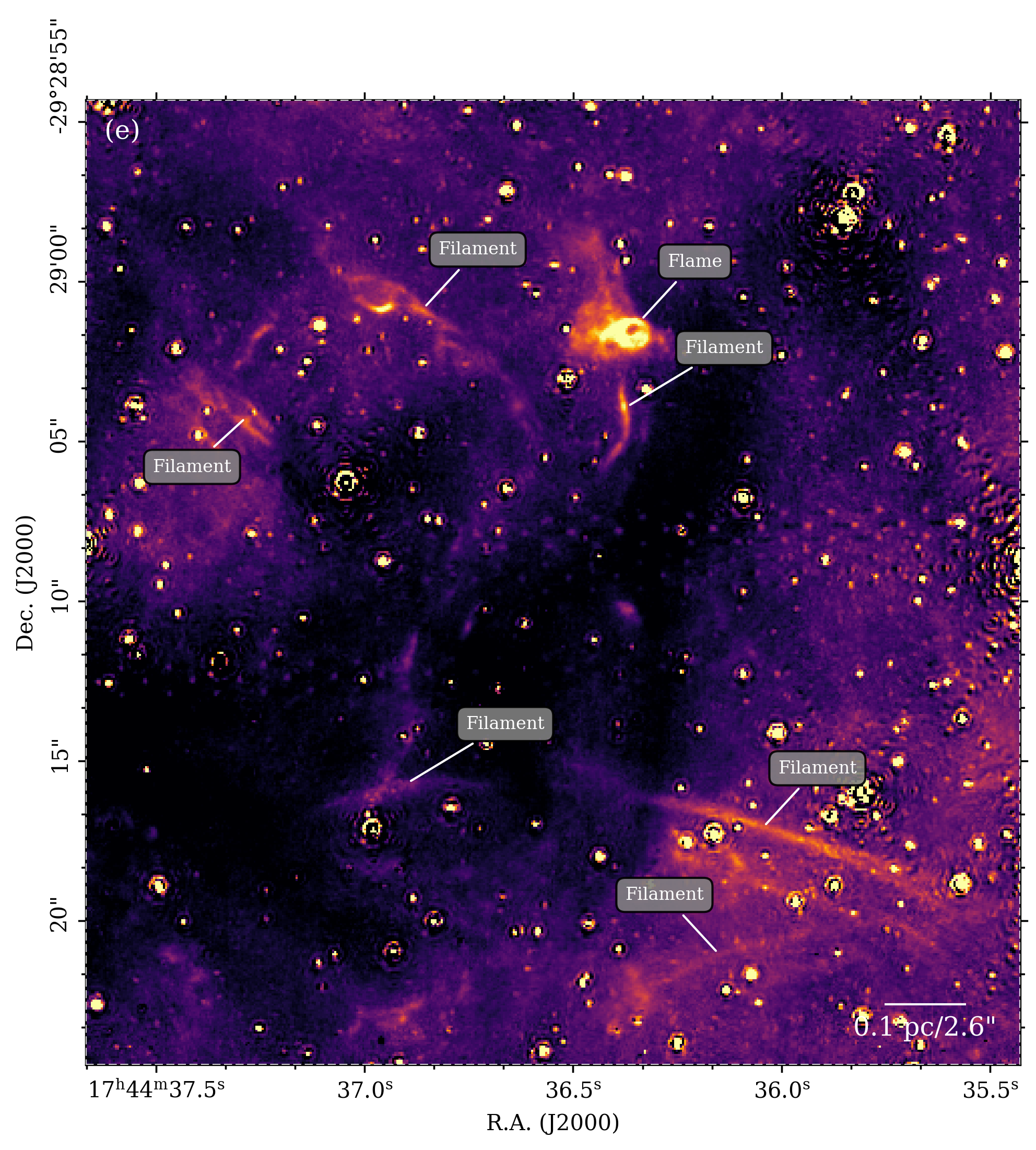}
            \caption{\label{fig:misc_knots_mod_a}  Compilation of miscellaneous MHOs found in the continuum-subtracted F470N NIRCam data.}
        \end{figure*}
        \renewcommand{\thefigure}{C1}
        \begin{figure*}
            \centering
            \includegraphics[width=0.36\textwidth,keepaspectratio]{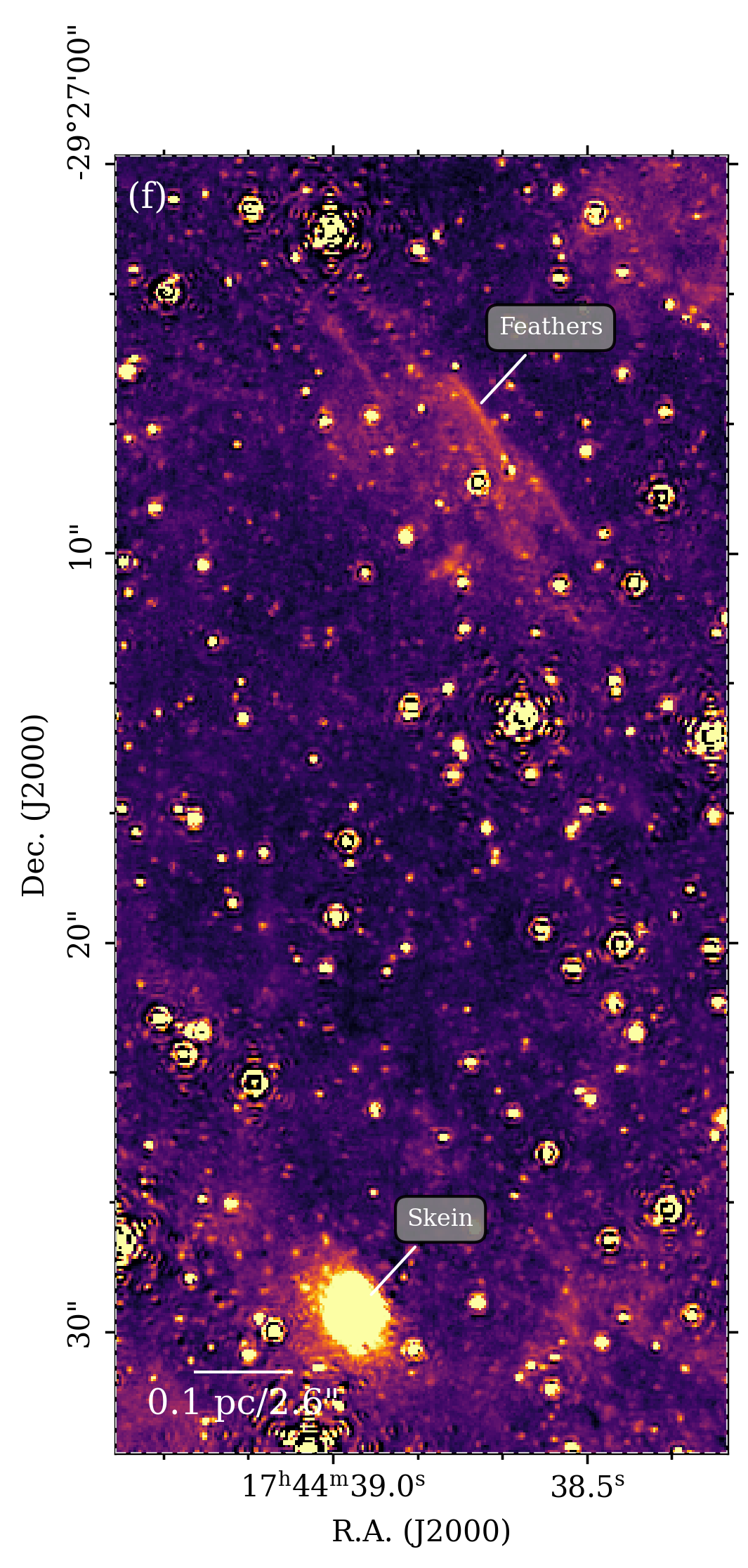}
            \includegraphics[width=0.63\textwidth,keepaspectratio]{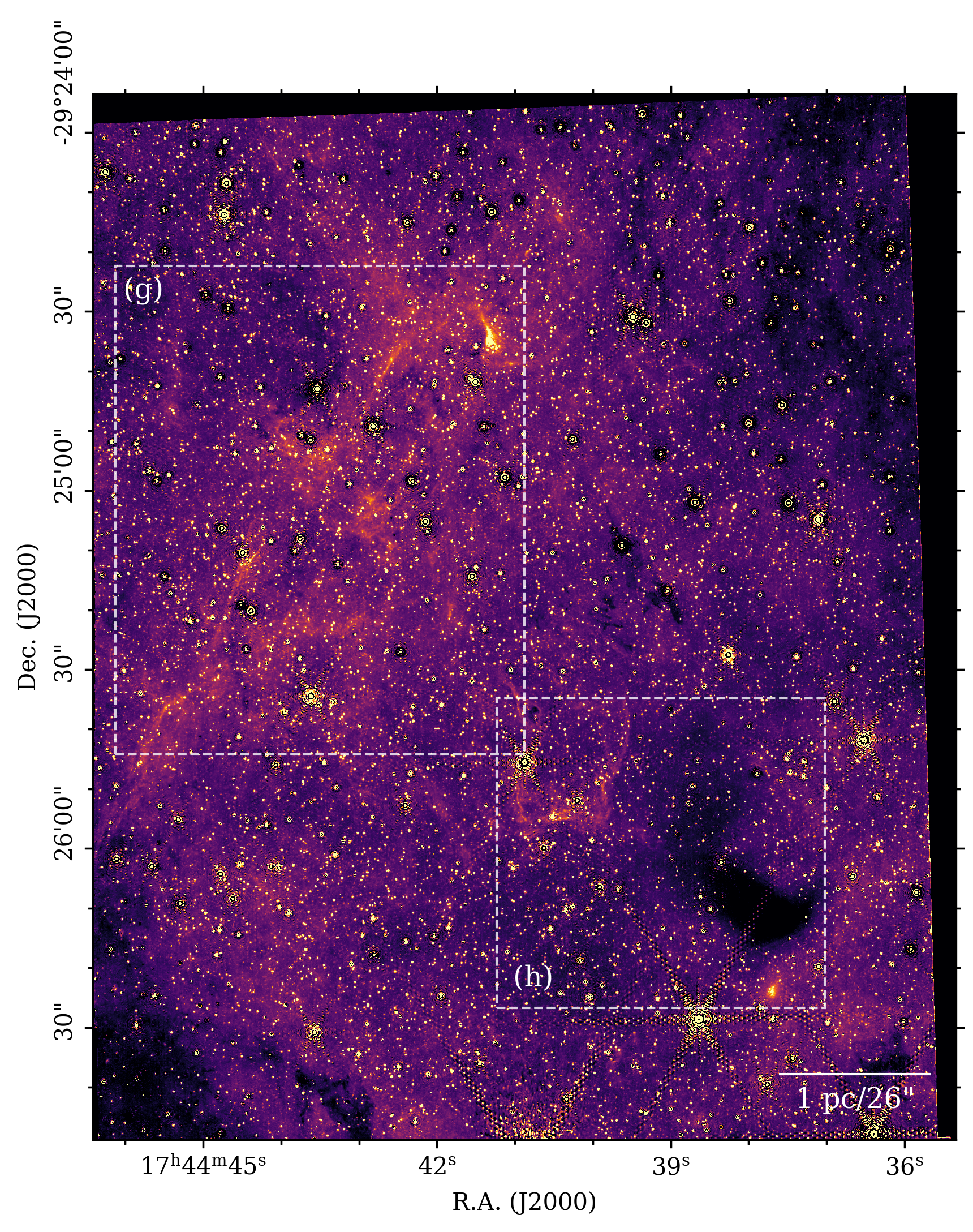}
            \caption{\label{fig:misc_knots_mod_b}Continued.}
        \end{figure*}
        \renewcommand{\thefigure}{C1}
        \begin{figure*}
            \centering
            \includegraphics[width=0.6\textwidth,keepaspectratio]{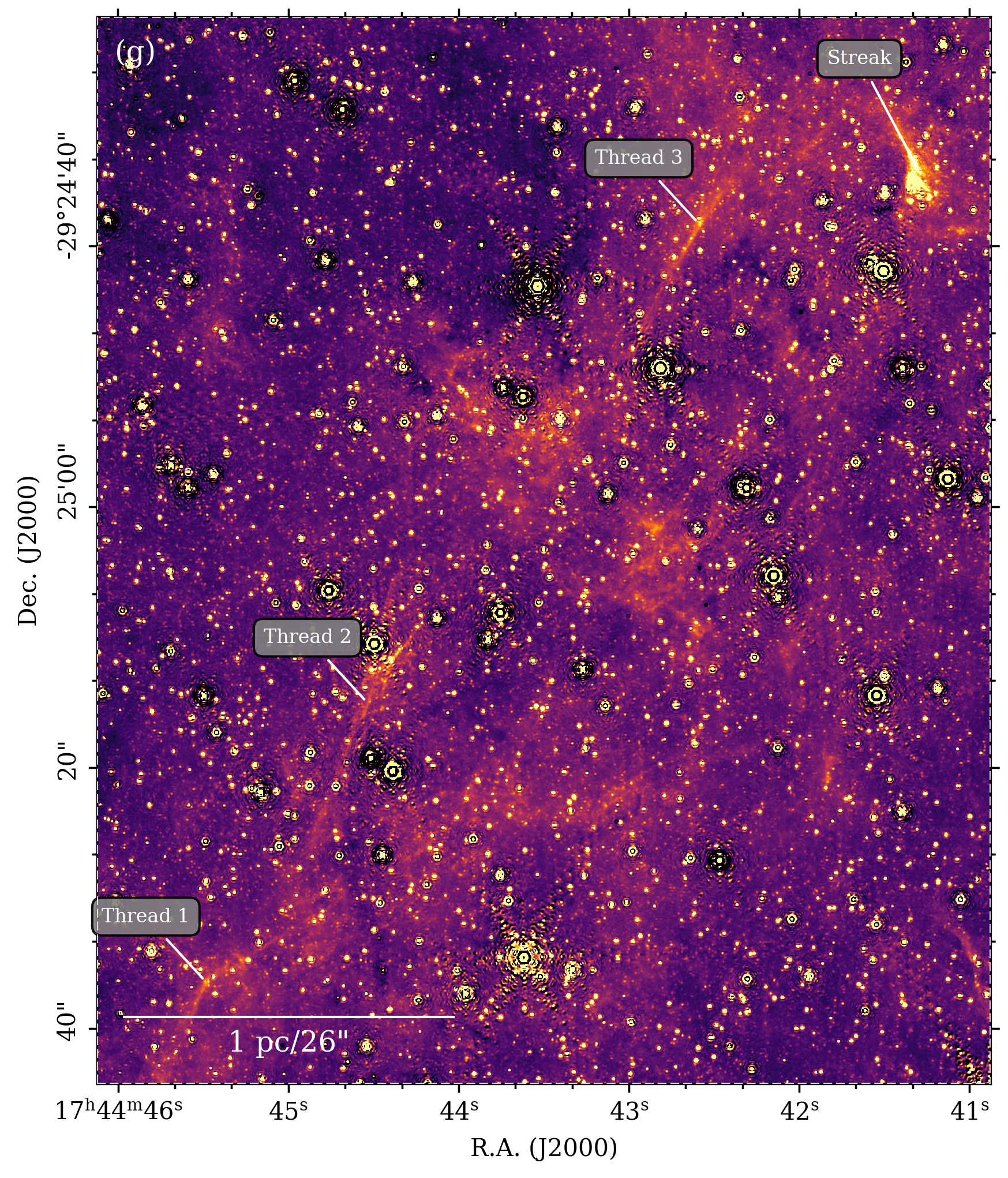}
            \includegraphics[width=0.6\textwidth,keepaspectratio]{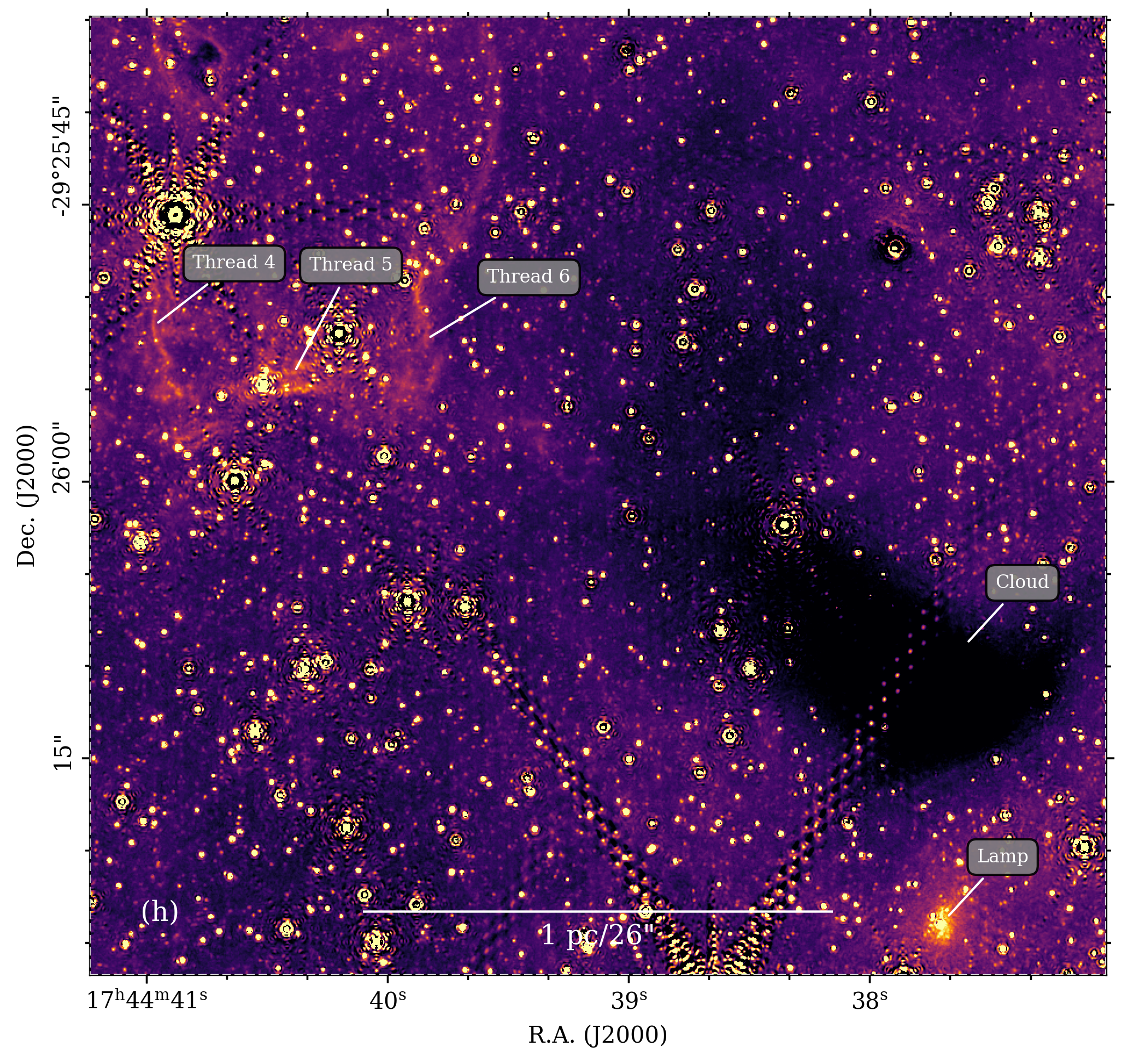}
            \caption{\label{fig:misc_knots_mod_b_cont}Continued.}
        \end{figure*}
\end{document}